%% file: main.tex
\documentclass[11pt]{book}

\usepackage{amsmath}
\usepackage{makeidx}
\usepackage{latexsym}
\usepackage{theorem}
\usepackage{amssymb}

\newcommand{\ly}{}
\newcommand{\I}{}
\newcommand{\at}{}

\input{prooftree}
\input{macros}

\makeindex

\begin{document}

\pagestyle{empty}
\input{title}

\newpage~\newpage
\input{thanks}

\newpage~\newpage
\tableofcontents

\pagestyle{headings}
\input{intro}

\input{prelim}

\input{tsm}

\input{tsm-prop}

\input{rts}

\input{ats}

\input{conditions}

\input{examples}

\input{begin-sn}

\input{candidates}

\input{interpret}

\input{int-const}

\input{ordering}

\input{int-def}

\input{end-sn}

\input{future}

\addtocontents{toc}{\protect\vspace*{1cm}}
\addcontentsline{toc}{chapter}
  {\numberline{}\hspace*{-6mm}{\Large Bibliography}}

\input{biblio}
\newpage~\newpage
\addtocontents{toc}{\protect\vspace*{1cm}}
\addcontentsline{toc}{chapter}
  {\numberline{}\hspace*{-6mm}{\Large Index}}
\printindex

\end{document}

%% file: prooftree.tex
\newdimen\proofrulebreadth \proofrulebreadth=.05em
\newdimen\proofdotseparation \proofdotseparation=1.25ex
\newdimen\proofrulebaseline \proofrulebaseline=2ex
\newcount\proofdotnumber \proofdotnumber=3
\let\then\relax
\def\hfi{\hskip0pt plus.0001fil}
\mathchardef\squigto="3A3B
\newif\ifinsideprooftree\insideprooftreefalse
\newif\ifonleftofproofrule\onleftofproofrulefalse
\newif\ifproofdots\proofdotsfalse
\newif\ifdoubleproof\doubleprooffalse
\let\wereinproofbit\relax
\newdimen\shortenproofleft
\newdimen\shortenproofright
\newdimen\proofbelowshift
\newbox\proofabove
\newbox\proofbelow
\newbox\proofrulename
\def\shiftproofbelow{\let\next\relax\afterassignment\setshiftproofbelow\dimen0 }
\def\shiftproofbelowneg{\def\next{\multiply\dimen0 by-1 }%
\afterassignment\setshiftproofbelow\dimen0 }
\def\setshiftproofbelow{\next\proofbelowshift=\dimen0 }
\def\setproofrulebreadth{\proofrulebreadth}

\def\prooftree{
\ifnum  \lastpenalty=1
\then   \unpenalty
\else   \onleftofproofrulefalse
\fi
\ifonleftofproofrule
\else   \ifinsideprooftree
        \then   \hskip.5em plus1fil
        \fi
\fi
\bgroup
\setbox\proofbelow=\hbox{}\setbox\proofrulename=\hbox{}%
\let\justifies\proofover\let\leadsto\proofoverdots\let\Justifies\proofoverdbl
\let\using\proofusing\let\[\prooftree
\ifinsideprooftree\let\]\endprooftree\fi
\proofdotsfalse\doubleprooffalse
\let\thickness\setproofrulebreadth
\let\shiftright\shiftproofbelow \let\shift\shiftproofbelow
\let\shiftleft\shiftproofbelowneg
\let\ifwasinsideprooftree\ifinsideprooftree
\insideprooftreetrue
\setbox\proofabove=\hbox\bgroup$\displaystyle 
\let\wereinproofbit\prooftree
\shortenproofleft=0pt \shortenproofright=0pt \proofbelowshift=0pt
\onleftofproofruletrue\penalty1
}

\def\eproofbit{
\ifx    \wereinproofbit\prooftree
\then   \ifcase \lastpenalty
        \then   \shortenproofright=0pt  
        \or     \unpenalty\hfil         
        \or     \unpenalty\unskip       
        \else   \shortenproofright=0pt  
        \fi
\fi
\global\dimen0=\shortenproofleft
\global\dimen1=\shortenproofright
\global\dimen2=\proofrulebreadth
\global\dimen3=\proofbelowshift
\global\dimen4=\proofdotseparation
\global\count255=\proofdotnumber
$\egroup  
\shortenproofleft=\dimen0
\shortenproofright=\dimen1
\proofrulebreadth=\dimen2
\proofbelowshift=\dimen3
\proofdotseparation=\dimen4
\proofdotnumber=\count255
}

\def\proofover{
\eproofbit 
\setbox\proofbelow=\hbox\bgroup 
\let\wereinproofbit\proofover
$\displaystyle
}%
\def\proofoverdbl{
\eproofbit 
\doubleprooftrue
\setbox\proofbelow=\hbox\bgroup 
\let\wereinproofbit\proofoverdbl
$\displaystyle
}%
\def\proofoverdots{
\eproofbit 
\proofdotstrue
\setbox\proofbelow=\hbox\bgroup 
\let\wereinproofbit\proofoverdots
$\displaystyle
}%
\def\proofusing{
\eproofbit 
\setbox\proofrulename=\hbox\bgroup 
\let\wereinproofbit\proofusing
\kern0.3em$
}

\def\endprooftree{
\eproofbit 
  \dimen5 =0pt
\dimen0=\wd\proofabove \advance\dimen0-\shortenproofleft
\advance\dimen0-\shortenproofright
\dimen1=.5\dimen0 \advance\dimen1-.5\wd\proofbelow
\dimen4=\dimen1
\advance\dimen1\proofbelowshift \advance\dimen4-\proofbelowshift
\ifdim  \dimen1<0pt
\then   \advance\shortenproofleft\dimen1
        \advance\dimen0-\dimen1
        \dimen1=0pt
        \ifdim  \shortenproofleft<0pt
        \then   \setbox\proofabove=\hbox{%
                        \kern-\shortenproofleft\unhbox\proofabove}%
                \shortenproofleft=0pt
        \fi
\fi
\ifdim  \dimen4<0pt
\then   \advance\shortenproofright\dimen4
        \advance\dimen0-\dimen4
        \dimen4=0pt
\fi
\ifdim  \shortenproofright<\wd\proofrulename
\then   \shortenproofright=\wd\proofrulename
\fi
\dimen2=\shortenproofleft \advance\dimen2 by\dimen1
\dimen3=\shortenproofright\advance\dimen3 by\dimen4
\ifproofdots
\then
        \dimen6=\shortenproofleft \advance\dimen6 .5\dimen0
        \setbox1=\vbox to\proofdotseparation{\vss\hbox{$\cdot$}\vss}%
        \setbox0=\hbox{%
                \advance\dimen6-.5\wd1
                \kern\dimen6
                $\vcenter to\proofdotnumber\proofdotseparation
                        {\leaders\box1\vfill}$%
                \unhbox\proofrulename}%
\else   \dimen6=\fontdimen22\the\textfont2 
        \dimen7=\dimen6
        \advance\dimen6by.5\proofrulebreadth
        \advance\dimen7by-.5\proofrulebreadth
        \setbox0=\hbox{%
                \kern\shortenproofleft
                \ifdoubleproof
                \then   \hbox to\dimen0{%
                        $\mathsurround0pt\mathord=\mkern-6mu%
                        \cleaders\hbox{$\mkern-2mu=\mkern-2mu$}\hfill
                        \mkern-6mu\mathord=$}%
                \else   \vrule height\dimen6 depth-\dimen7 width\dimen0
                \fi
                \unhbox\proofrulename}%
        \ht0=\dimen6 \dp0=-\dimen7
\fi
\let\doll\relax
\ifwasinsideprooftree
\then   \let\VBOX\vbox
\else   \ifmmode\else$\let\doll=$\fi
        \let\VBOX\vcenter
\fi
\VBOX   {\baselineskip\proofrulebaseline \lineskip.2ex
        \expandafter\lineskiplimit\ifproofdots0ex\else-0.6ex\fi
        \hbox   spread\dimen5   {\hfi\unhbox\proofabove\hfi}%
        \hbox{\box0}%
        \hbox   {\kern\dimen2 \box\proofbelow}}\doll%
\global\dimen2=\dimen2
\global\dimen3=\dimen3
\egroup 
\ifonleftofproofrule
\then   \shortenproofleft=\dimen2
\fi
\shortenproofright=\dimen3
\onleftofproofrulefalse
\ifinsideprooftree
\then   \hskip.5em plus 1fil \penalty2
\fi
}


%% file: macros.tex
\newenvironment{ded}
   {\begin{center}\begin{prooftree}}
   {\end{prooftree}\end{center}}
\newcommand{\pra}[3][]{\[#2\justifies #3\using\mbox{(#1)}\]}

\newcommand{\qed}[2][]{\justifies #2\using\mbox{(#1)}}

\newcommand{\comment}[1]{}
\newcommand{\bu}{$\bullet$}
\newcommand{\hs}[1][3ex]{\hspace*{#1}}
\newcommand{\vs}[1][1mm]{\vspace*{#1}}
\newcommand{\moins}{\setminus}
\newcommand{\vide}{\emptyset}
\newcommand{\alaligne}{\\}

\renewcommand{\l}[2]{[#1\!:\!#2]}
\newcommand{\p}[2]{(#1\!:\!#2)}
\newcommand{\lx}{\l{x}}
\newcommand{\px}{\p{x}}
\renewcommand{\ly}{\l{y}}
\newcommand{\py}{\p{y}}
\newcommand{\lz}{\l{z}}
\newcommand{\pz}{\p{z}}
\newcommand{\lX}{\l{X}}
\newcommand{\pX}{\p{X}}

\newcommand{\lfp}{\mr{lfp}}

\newcommand{\dom}{\mr{dom}}

\newcommand{\FV}{\mr{FV}}
\newcommand{\pos}{\mr{Pos}}

\renewcommand{\a}{\rightarrow}
\newcommand{\A}{\Rightarrow}
\renewcommand{\aa}{\leftrightarrow}
\renewcommand{\AA}{\Leftrightarrow}

\newcommand{\ad}{\downarrow}

\renewcommand{\to}{\mapsto}

\newcommand{\ab}{\a_\b}
\newcommand{\ai}{\a_\io}
\renewcommand{\ae}{\a_\eta}
\newcommand{\abe}{\a_{\b\eta}}
\newcommand{\abi}{\a_{\b\io}}

\newcommand{\ar}{\a_\cR}
\newcommand{\abr}{\a_{\b\cR}}

\newcommand{\al}[1][]{{}_{#1}\!\!\leftarrow}
\newcommand{\als}[1][]{{}_{#1}^*\!\!\leftarrow}
\newcommand{\alb}{\al[\b]}

\newcommand{\ps}[1]{{\langle #1\rangle}}

\renewcommand{\I}[1]{[\![#1]\!]}

\newcommand{\ex}{\exists}
\newcommand{\all}{\forall}
\newcommand{\ou}{\vee}

\newcommand{\biget}{\bigwedge}
\newcommand{\et}{\wedge}
\newcommand{\non}{\neg}
\newcommand{\st}{\star}
\newcommand{\B}{\Box} 
\renewcommand{\th}{\vdash}
\renewcommand{\TH}{\!\th\!}
\newcommand{\IN}{\!\in\!}

\newcommand{\sle}{\subseteq}
\newcommand{\sge}{\supseteq}

\newcommand{\tlt}{\lhd}
\newcommand{\tgt}{\rhd}

\newcommand{\cge}{\succeq}

\newcommand{\cgt}{\succ}

\newcommand{\qge}{\sqsupseteq}

\newcommand{\qgt}{\sqsupset}

\newcommand{\lex}{_\mr{lex}}
\newcommand{\mul}{_\mr{mul}}

\renewcommand{\o}[1]{{\overline{#1}}}
\renewcommand{\u}[1]{{\underline{#1}}}

\renewcommand{\b}{\beta}
\newcommand{\g}{\gamma}
\newcommand{\G}{\Gamma}
\renewcommand{\d}{\delta}
\newcommand{\D}{\Delta}

\newcommand{\vep}{\varepsilon}
\newcommand{\z}{\zeta}
\renewcommand{\t}{\theta}
\newcommand{\T}{\Theta}
\newcommand{\io}{\iota}
\renewcommand{\k}{\kappa}
\newcommand{\la}{\lambda}
\renewcommand{\L}{\Lambda}
\renewcommand{\r}{\rho}
\newcommand{\s}{\sigma}
\renewcommand{\S}{\Sigma}

\newcommand{\Up}{\Upsilon}
\newcommand{\vp}{\varphi}
\newcommand{\w}{\omega}

\newcommand{\vpi}{\varpi}

\newcommand{\mc}{\mathcal}

\newcommand{\mr}{\mathrm}
\newcommand{\mb}{\mathbb}
\newcommand{\mg}{\mathbf}

\newcommand{\cA}{\mc{A}}
\newcommand{\cB}{\mc{B}}
\newcommand{\cC}{\mc{C}}
\newcommand{\cD}{\mc{D}}
\newcommand{\cE}{\mc{E}}
\newcommand{\cF}{\mc{F}}
\newcommand{\cG}{\mc{G}}

\newcommand{\cI}{\mc{I}}
\newcommand{\cJ}{\mc{J}}
\newcommand{\cK}{\mc{K}}

\newcommand{\cM}{\mc{M}}
\newcommand{\cN}{\mc{N}}
\newcommand{\cO}{\mc{O}}
\newcommand{\cP}{\mc{P}}

\newcommand{\cR}{\mc{R}}
\newcommand{\cS}{\mc{S}}
\newcommand{\cT}{\mc{T}}

\newcommand{\cW}{\mc{W}}
\newcommand{\cX}{\mc{X}}

\newcommand{\fa}{\mathfrak{a}} 

\newenvironment{rew}[1][\a]%
  {$\begin{array}{r@{~~#1~~}l}}%
  {\end{array}$}
\newenvironment{rewc}[1][\a]%
  {\begin{center}\begin{rew}[#1]}%
  {\end{rew}\end{center}}

\newcounter{counter}

{\theorembodyfont{\rmfamily} 
  \newtheorem{dfn}[counter]{Definition}
  \newtheorem{lem}[counter]{Lemma}
  \newtheorem{thm}[counter]{Theorem}
  \newtheorem{cor}[counter]{Corollary}
  \newtheorem{rem}[counter]{Remark}
  
  \newtheorem{conj}[counter]{Conjecture}
  
}

\newcommand{\cqfd}{\hfill$\blacksquare$} 
\newcommand{\cqfdd}{\vs[-4mm]{\flushright\hfill$\blacksquare$}} 
\newenvironment{prf}{{\bf Proof.}}{}

\newenvironment{lstgeneric}[2]
  {\begin{list}{#1}{\topsep=.5mm\itemsep=.5mm\parsep=0mm%
    \itemindent=-3ex\labelsep=1ex\labelwidth=0ex #2}}
  {\end{list}}

\newenvironment{lst}[1]
  {\begin{lstgeneric}{#1}{\itemindent=-1ex}}
  {\end{lstgeneric}}

\newenvironment{enumi}[1]
  {\begin{lstgeneric}{}{\usecounter{enumi}\leftmargin=7mm%
    }}
  {\end{lstgeneric}}

\newenvironment{bfenumi}[1]
  {\begin{lstgeneric}{}{\usecounter{enumi}\leftmargin=7mm%
    }}
  {\end{lstgeneric}}

\newenvironment{enumii}[1]
  {\begin{lstgeneric}{}{\usecounter{enumii}%
    }}
  {\end{lstgeneric}}

\newenvironment{bfenumii}[1]
  {\begin{lstgeneric}{}{\usecounter{enumii}\leftmargin=7mm%
    }}
  {\end{lstgeneric}}

\newenvironment{enumalphai}
  {\begin{lstgeneric}{}{\usecounter{enumi}\leftmargin=7mm%
    }}
  {\end{lstgeneric}}

\newenvironment{bfenumalphai}
  {\begin{lstgeneric}{}{\usecounter{enumi}\leftmargin=7mm%
    }}
  {\end{lstgeneric}}

\newenvironment{enumalphaii}
  {\begin{lstgeneric}{}{\usecounter{enumii}\leftmargin=10mm%
    }}
  {\end{lstgeneric}}



%% file: title.tex



\begin{titlepage}

\centering
\large

Universit\'e Paris XI\\
Orsay, France

\vs[2cm]

Ph. D. Thesis

\vs[2cm]

{\Huge\bf Type Theory and Rewriting}

\vs[2cm]

{\Large\bf Fr\'ed\'eric BLANQUI}

\vs[2cm]

28 september 2001

\vs[2cm]

Committee:

\vs[2mm]

Mr Thierry COQUAND, referee\\
Mr Gilles DOWEK, president of the jury\\
Mr Herman GEUVERS, referee\\
Mr Jean-Pierre JOUANNAUD, supervisor\\
Mme Christine PAULIN\\
Mr Miklos SANTHA

\end{titlepage}


%% file: thanks.tex



\vs[2cm]

The 19th of March 1998 was an important date for at least two
reasons. The first one was personal. The second one was that
Jean-Pierre Jouannaud agreed to supervise my master thesis on
``extending the Calculus of Constructions with a new version of the
General Schema'' which he had roughed out with Mitsuhiro Okada. This
did not mean much to me then. However, I was very happy with the
idea of studying both $\la$-calculus and rewriting, and
their interaction. This work results of this enthusiasm.

~

This is why I will begin by thanking Jean-Pierre Jouannaud, for the
honor he made to me, the trust, the help, the advice and the support
that he gave me during these three years. He taught me a lot and I
will be always grateful to him.

I also thank Mitsuhiro Okada for the discussions we had together and
the support he gave me. It was a great honor to have the opportunity
to work with him. I hope we will have other numerous fruitful
collaborations.

I also thank Maribel Fern\'andez who helped me at the beginning of my
thesis by supervising my work with Jean-Pierre Jouannaud.

I also thank Gilles Dowek who supported me in my work and helped me on
several important occasions. His work was (and still is !) an
important source of reflexion and inspiration.

I also thank Daria Walukiewicz with whom I had many fruitful
discussions. I thank her very much for having read in detail an
important part of this thesis and for having helped me to correct
errors and lack of precision.

I also thank every person in the D\'EMONS team from the LRI and the
Coq team (newly baptized LogiCal) from INRIA Rocquencourt, in
particular Christine Paulin and Claude March\'e who helped me several
times. These two teams are a privileged research place and have a pleasant
atmosphere.

~

I also thank the referees of this thesis, Thierry Coquand and Herman
Geuvers, for their interest in my work and the remarks they made for
improving it.

Finally, I thank the members of the jury and the president of the jury
for the honor they made to me by accepting to consider my work.


%% file: intro.tex



\chapter{Introduction}

What is good programming? Apart from writing programs which are
understandable and reusable by other people, above all it is being
able to write programs without errors. But how do you know whether a
program has no error? By proving it. In other words, to program well
requires doing mathematics.

But how do you know whether a proof that a program has no error itself
has no error? By writing a proof which can be checked by a
computer. In other words, to program well requires doing formal
mathematics.

This is the subject of our thesis~: defining a formal system in which
one can program and prove that a program is correct.

~

However, it is not the case that work is duplicated~: programming and
proving. In fact, from a proof that a program specification is
correct, one can extract an error-free program! This is due to the
termination of ``cut elimination'' in intuitionist logic discovered by
G. Gentzen in 1933 \cite{gentzen33thesis}.

~

More precisely, we will consider a particular class of formal systems,
the type systems. We will study their properties when they are
extended with definitions by rewriting. Rewriting is a simple and
general computation paradigm based on rules like $x+0 \a x$, that is,
if one has an expression of the form $x+0$, then one can simplify it
to $x$.

But for such a system to serve in proving the correctness of programs,
one must make sure that the system itself is correct, that is, that
one cannot prove something which is false. This is why we will give
conditions on the rewriting rules and prove that these conditions
indeed ensure the correctness of the system.

~

First, let us see how type systems appeared, what results are already
known and what our contributions are (they are summarized in
Section~\ref{sec-contrib}).




\section{Some history}

This section is not intended to provide an absolutely rigorous
historical summary. We only want to recall the basic concepts on which
our work is based (type theory, $\la$-calculus, etc.) and show how our
work takes place in the continuation of previous works aiming at
introducing more programming into logic, or dually, more logic into
programming. We will therefore take some freedom with the formalisms
used.

The reader familiar with these notions (in particular the Calculus of
Constructions and the Calculus of Inductive Constructions) can
directly go to Section~\ref{sec-motiv} where we present our
motivations for adding rewriting in the Calculus of Constructions both
at the object-level and at the predicate-level.


\subsection*{Set theory}
\index{set theory|indpage}

One of the first formal system enabling one to describe all
mathematics was the {\em set theory\,} of E. Zermelo (1908) later
extended by A. Fraenkel (1922). It was followed by the {\em type
theory\,} of A. Whitehead and B. Russell (1911)
\cite{whitehead11book}, also called {\em higher-order logic\,}. These
two formal systems were introduced to avoid the inconsistency of the
{\em set theory\,} of G. Cantor (1878).

~

In first-order logic, in which the set theory of E. Zermelo and
A. Fraenkel is generally expressed, the objects of the discourse are
defined from constants and function symbols $(0,+,\ldots)$. Then, some
predicate symbols ($\in$, \ldots), the logical connectors
$(\ou,\et,\A$, $\ldots)$ and the universal and existential quantifiers
$(\all,\ex)$ enable one to express propositions in these objects.

~

One of the axioms of G. Cantor's set theory is the {\em Comprehension
Axiom\,} which says that every proposition defines a set~:

\begin{center}
$(\ex x) (\all y) ~y\in x \AA P(y)$
\end{center}

From this axiom, one can express Russell's paradox (1902). By taking
$P(x)= x\notin x$, one can define the set $R$ of the $x$'s which do
not belong to themselves. Then, $R\in R \AA R\notin R$ and one can
deduce that any proposition is true. To avoid this problem, E. Zermelo
proposed to restrict the Comprehension Axiom as follows~:

\begin{center}
$(\all z) (\ex x) (\all y) ~y\in x \AA y\in z \et P(y)$
\end{center}

\noindent
that is, one can define by comprehension only subsets of previously
well-defined sets.


\subsection*{Type theory}
\index{type!theory|indpage}

In type theory, instead of restricting the Comprehension Axiom, the
idea is to forbid expressions like $x\notin x$ or $x\in x$ by
restricting the application of a predicate to an object. To this end,
one associates to each function symbol and predicate symbol (except
$\in$) a {\em type\,} as follows~:

\begin{lst}{--}
\item to a constant, one associate the type $\io$,
\index{iota@$\io$|indpage}
\item to a function symbol taking one argument, one associates the
type $\io\a\io$,
\item to a function symbol taking two arguments, one associates the
type $\io\a\io\a\io$,
\item \ldots
\item to a proposition, one associates the type $o$,
\index{omicron@$o$|indpage}
\item to a predicate symbol taking one argument, one associates the
type $\io\a o$,
\item to a predicate symbol taking two arguments, one associates the
type $\io\a\io\a o$,
\item \ldots
\end{lst}

Then, one can apply a function $f$ taking $n$ arguments to $n$ objects
$t_1,\ldots,t_n$ if the type of $f$ is $\io\a\ldots\a\io\a\io$ and
every $t_i$ is of type $\io$. And one can say that $n$ objects
$t_1,\ldots,t_n$ satisfy a predicate symbol $P$ taking $n$ arguments
if the type of $P$ is $\io\a\ldots\a\io\a o$ and every $t_i$ is of
type $\io$.

Finally, one considers that a set is not an object anymore (that is,
an expression of type $\io$) but a predicate (an expression of type
$\io\a o$). And, for representing $x\in E$, which means that $x$
satisfies $E$, one writes $Ex$ (application of $E$ to $x$). Hence, one
can easily verify that it is not possible to express Russell's
paradox~: one cannot write $xx$ for representing $x\in x$ since then
$x$ is both of the type $\io\a o$ and of the type $\io$ which is not
allowed. In the following, we write $t:\tau$ for saying that $t$ is of
type $\tau$.

~

Now, to represent natural numbers, there are several
possibilities. However it is always necessary to state an axiom of
infinity for $\io$ and to be able to express the set of natural
numbers as the smallest set containing zero and stable by
incrementation. To this end, one must be able to quantify on sets,
that is, on expressions of type $\io\a o$.

Now, one is not restricted to objects and predicate expressions as
described before, but can consider all the expressions that can be
formed by applications which respect types~:

\begin{lst}{--}
\item The set of the {\em simple types\,} is the smallest set $T$
containing $\io$, $o$ and $\s\a \tau$ whenever $\s$ and $\tau$ belong
to $T$.
\item The set of {\em terms of type $\tau$\,} is the smallest set
containing the constants of type $\tau$ and the applications $tu$
whenever $t$ is a term of type $\s\a\tau$ and $u$ a term of type $\s$.
\end{lst}

Finally, one introduces an explicit notation for functions and sets,
the {\em $\la$-abstrac\-tion\,}, and considers logical connectors and
quantifiers as predicate symbols by giving them the following
respective types~: $\ou:o\a o\a o$, $\et:o\a o\a o$, $\all_\tau:
(\tau\a o)\a o$, \ldots For example, if $\io$ denotes the set of
natural numbers then one can represent the predicate ``is even'' (of
type $\io\a o$) by the expression $pair=\la x\!:\!\io.  \ex_\io(\la
y\!:\!\io. x=2\times y)$ that we will abbreviate by $\la
x\!:\!\io. \ex y\!:\!\io. x=2\times y$. The language we obtain is
called the {\em simply-typed $\la$-calculus\,} $\la^\a$.

~

\newcommand{\xu}{\{x\to u\}}

But what can we say about $(pair\,2)$ and $\ex y\!:\!\io. 2=2\times y$
?  The second expression can be obtained from the first one by
substituting $x$ by $2$ in the body of $pair$. This operation of
substitution is called {\em
$\b$-reduction\,}. \index{reduction!beta@$\b$|indpage}
\index{beta-reduction@$\b$-reduction|indpage} More generally, $\la
x\!:\!\tau.\,t$ applied to $u$ $\b$-reduces to $t$ where $x$ is
substituted by $u$~: $\la x\!:\!\tau.\,t ~u \ab t\xu$.

It is quite natural to consider theses two expressions as denoting the
same proposition. This is why one adds the following {\em Conversion
Axiom\,}~:

\begin{center}
$P \AA Q$ ~if~ $P\ab Q$
\end{center}

One then gets the {\em type theory\,} of A. Church (1940)
\cite{church40jsl}.

~

In this theory, it is possible to quantify over all propositions~:
$\all P\!:\!o. P\A P$. In other words, a proposition can be defined by
quantifying over all propositions, including itself. If one allows
such quantifications, the theory is said to be {\em impredicative\,},
otherwise it said {\em predicative\,}.


\subsection*{Mathematics as a programming language}

The $\b$-reduction corresponds to the evaluation process of a
function. When one has a function $f$ defined by an expression $f(x)$
and wants its value on $5$ for example, one substitutes $x$ by $5$ in
$f(x)$ and simplify the expression until one gets the value of $f(5)$.

One can wonder which functions are definable in Church's type
theory. In fact, very few. With Peano's natural numbers ({\em i.e.\,}
by taking $0:\io$ for zero and $s:\io\a\io$ for the successor
function), one can express only constant functions or functions adding
a constant to one of its arguments. With Church's numerals, where $n$
is represented by $\la x\!:\!\io. \la f\!:\!\io\a\io. f\ldots fx$ with
$n$ occurrences of $f$, H. Schwichtenberg \cite{schwichtenberg76aml}
proved that one can express only extended polynomials (smallest set of
functions closed under composition and containing the null function,
the successor function, the projections, the addition, the
multiplication and the test for zero).

Of course, it is possible to prove the existence of numerous
functions, that is, to prove a proposition of the form $(\all x) (\ex
y)\, Pxy$ where $P$ represents the graph of the function. In the
intuitionist type theory for example ({\em i.e.\,} without using
the Excluded-middle Axiom $P\ou\neg P$), it is possible to prove the
existence of any primitive recursive function. But there is no term
$f:\io\a\io$ enabling us to {\em compute\,} the powers of $2$ for
example, that is, such that $fn \ab \ldots \ab 2^n$.


\subsection*{Representation of proofs}

G. Frege and D. Hilbert proposed to represent a proof of a proposition
$Q$ as a sequence $P_1, \ldots, P_n$ of propositions such that $P_n=Q$
and, for every $i$, either $P_i$ is an axiom, either $P_i$ is a
consequence of the previous propositions by {\em modus ponens\,} (from
$P$ and $P \A Q$ one can deduce $Q$) or by {\em generalization\,}
(from $P(x)$ with $x$ arbitrary one can deduce $(\all
x)P(x)$). However, to do such proofs, it is necessary to consider many
axioms, independent of any theory, which express the sense of logical
connectors.

Later, in 1933, G. Gentzen \cite{gentzen33thesis} proposed a new
deduction system, called Natural Deduction, \index{Natural
Deduction|indpage} where logical axioms are replaced by {\em
introduction rules\,} \index{introduction rule|indpage}
\index{rule!introduction|indpage} and {\em elimination rules\,}
\index{elimination rule|indpage} \index{rule!elimination|indpage} for
the connectors and the quantifiers~:

\begin{center}
(axiom) $\cfrac{}{\G,P,\G'\th P}$

~

($\et$-intro) $\cfrac{\G\th P \quad \G\th Q}{\G\th P\et Q}$
\hs[1cm] ($\et$-\'elim1) $\cfrac{\G\th P\et Q}{\G\th P}$
\hs[1cm] ($\et$-\'elim2) $\cfrac{\G\th P\et Q}{\G\th Q}$

~

($\A$-intro) $\cfrac{\G,P\th Q}{\G\th P\A Q}$
\hs[1cm] ($\A$-\'elim) $\cfrac{\G\th P\A Q \quad \G\th P}{\G\th Q}$

~

($\ex$-intro) $\cfrac{\G\th P(t)}{\G\th (\ex x)P(x)}$ \hs[1cm]
($\ex$-\'elim)\footnote{If $x$ does not occur neither in $\G$ nor in
$Q$.} $\cfrac{\G\th (\ex x)P \quad \G,P\th Q}{\G\th Q}$ \hs[1cm]
\ldots
\end{center}

\vs[2mm]
\noindent
where $\G$ is a set of propositions (the hypothesis). A pair $\G\th Q$
made of a set of hypothesis $\G$ and a proposition $Q$ is called a
{\em sequent\,}. \index{sequent|indpage} Then, a proof of a sequent
$\G\th Q$ is a tree whose root is $\G\th Q$, whose nodes are instances
of the deduction rules and whose leaves are applications of the rule
(axiom).


\subsection*{Cut elimination}
\index{cut elimination|indpage}

G. Gentzen remarked that some proofs can be simplified. For example,
this proof of $Q$~:

\begin{ded}
  \pra[$\A$-intro]
  {\G,P\th Q}
  {\G\th P\A Q}   {\G\th P}
\qed[$\A$-\'elim]{\G\th Q}
\end{ded}

\noindent
does a detour which can be eliminated. It suffices to replace in the
proof of $\G,P\th Q$ all the leaves (axiom) giving $\G,P,\G'\th P$
($\G'$ are additional hypothesis that may be introduced for proving
$Q$) by the proof of $\G\th P$ where $\G$ is also replaced by
$\G,\G'$. In fact, at every place where there is a {\em cut\,}, that
is, an introduction rule followed by an elimination rule for the same
connector, it is possible to simplify the proof. G. Gentzen proved the
following remarkable fact~: the cut-elimination process terminates.

Hence, any provable proposition has a cut-free proof. But, in
intuitionist logic, any cut-free proof of a proposition $(\ex
x)P(x)$ must terminate by an introduction rule whose premise is of the
form $P(t)$. Therefore, the cut-elimination process gives us a
witness $t$ of an existential proposition. In other words, any
function whose existence is provable is computable.

If one can express the proofs themselves as objects of the theory,
then it becomes possible to express many more functions than those
allowed in the simply-typed $\la$-calculus.


\subsection*{The isomorphism of Curry-de Bruijn-Howard}
\index{isomorphim of Curry-Howard|indpage}

In 1958, Curry \cite{curry58book} remarked that there is a
correspondence between the types of the simply-typed $\la$-calculus
and the propositions formed from the implication $\A$ (one can
identify $\a$ and $\A$), and also between the terms of type $\tau$ and
the proofs of the proposition corresponding to $\tau$. In other words,
the simply-typed $\la$-calculus enables one to represent the proofs of
the minimal propositional logic. To this end, one associates to each
proposition $P$ a variable $x_P$ of type $P$. Then, one defines the
$\la$-term associated to a proof by induction on the size of the
proof~:

\begin{lst}{--}
\item the proof of $\G\th P$ obtained by (axiom) is associated to $x_P$;

\item the proof of $\G\th P\A Q$ obtained by ($\A$-intro) from a proof
$\pi$ of $\G,P\th Q$ is associated to the term $\la x_P\!:\!P.\,t$
where $t$ is the term associated to $\pi$;
  
\item the proof of $\G\th Q$ obtained by ($\A$-elim) from a proof
$\pi$ of $\G\th P\A Q$ and a proof $\pi'$ of $\G\th P$ is associated
to the term $tu$ where $t$ is the term associated to $\pi$ and $u$ the
term associated to $\pi'$.
\end{lst}

~

The set of $\la$-terms that we obtain can be directly defined as
follows. We call an {\em environment\,} any set $\G$ of pairs $x:P$
made of a variable $x$ and a type $P$ (representing a
proposition). Then, a term $t$ is of type $P$ (a proof of $P$) in the
environment $\G$ (under the hypothesis $\G$) if $\G\th t:T$ can be
deduced by the following inference rules~:

\begin{center}
\begin{tabular}{c}
(axiom) $\cfrac{}{\G,x\!:\!P,\G'\th x:P}$\\
\\
($\A$-intro) $\cfrac{\G,x\!:\!P\th t:Q}{\G\th \la x\!:\!P.\,t:P\A Q}$\\
\\
($\A$-elim) $\cfrac{\G\th t:P\A Q \quad \G\th u:P}{\G\th tu:Q}$\\
\end{tabular}
\end{center}

~

In 1965, W. W. Tait \cite{tait65book} remarked that $\b$-reduction
corresponds to cut-elimina\-tion. Indeed, if one annotates the example
of cut previously given then one gets~:

\begin{ded}
  \pra[$\A$-intro]
  {\G,x\!:\!P\th t:Q}
  {\G\th \la x\!:\!P.\,t:P\A Q}   {\G\th u:P}
\qed[$\A$-elim]{\G\th \la x\!:\!P.\,t~u:Q}
\end{ded}

If one $\b$-reduces $\la x\!:\!P.\,t~u$ to $t\xu$, then one exactly
obtains the term corresponding to the cut-free proof of $\G\th
Q$. Hence, the existence of a cut-free proof corresponds to the {\em
weak normalization\,} of $\b$-reduction, that is, the existence for
any typable $\la$-term $t$ of a sequence of $\b$-reductions resulting
in a $\b$-irreducible term (we also say in {\em normal form\,}). This
is why normalization has such an important place in the study of type
systems.


~

In 1968, N. de Bruijn \cite{debruijn68} proposed a system of {\em
dependent types\,} \index{type!dependent|indpage}
\index{dependent!type|indpage} extended the simply-typed
$\la$-calculus and in which it was possible to express the
propositions and the proofs of intuitionist first-order logic. This
system was the basis of one of the first programs for doing formal
proofs~: AUTOMATH. A dependent type is simply a function which
associates a type expression to each object. It enables one to
represent predicates and quantifiers. In 1969, W. A.  Howard
\cite{howard69} considered a similar system but without considering it
as a logical system in its own right.

~

In a dependent type system, the well-formedness of types depends on
the well-formedness of terms. It is then necessary to consider
environments with type variables and to add typing rules for types and
environments (the order of variables now matters). Finally, it is
necessary to add a conversion rule for identifying $\b$-equivalent
propositions. One then gets a set of typing rules similar to the ones
of Figure~\ref{fig-th-lap} (this is a modern presentation which
emerged at the end of the 80's only).

\newcommand{\GxT}{\G,x\!:\!T}
\newcommand{\GxU}{\G,x\!:\!U}
\newcommand{\GxV}{\G,x\!:\!V}
\newcommand{\GXK}{\G,X\!:\!K}

\newcommand{\GpxU}{\G',x\!:\!U}
\newcommand{\GpxT}{\G',x\!:\!T}

\begin{figure}[ht]
centering
\caption{Typing rules of $\la$P\label{fig-th-lap}}
\index{typing!lambdaP@$\la$P|indfig[fig-th-lap]}
\index{lambdaP@$\la$P|indfig[fig-th-lap]}

\begin{tabular}{rcl}
\\ (ax) & $\cfrac{}{\th \st:\B}$\\

\\ (var) & $\cfrac{\G\th T:s\in\{\st,\B\}}{\GxT \th x:T}$\\

\\ (weak) & $\cfrac{\G\th t:T \quad \G\th U:s\in\{\st,\B\}}
{\GxU \th t:T}$\\

\\ (prod-$\la^\a$) & $\cfrac{\G\th T:\st \quad \GxT \th U:\st}
{\G\th \px{T}U:\st}$\\

\\ (prod-$\la P$) & $\cfrac{\G\th T:\st \quad \GxT \th U:\B}
{\G\th \px{T}U:\B}$\\

\\ (abs) & $\cfrac{\GxT \th u:U \quad \G\th \px{T}U:s\in\{\st,\B\}}
{\G\th \la x\!:\!T.\,u:\px{T}U}$\\

\\ (app) & $\cfrac{\G\th t:\px{U}V \quad \G\th u:U}
{\G\th tu:V\xu}$\\

\\ (conv) & $\cfrac{\G\th t:T \quad T \aa^*_\b T' \quad \G\th T':\st}
{\G\th t:T'}$\\

\\
\end{tabular}
\end{figure}

In this system, $\st$ \index{$\st,\B$|indpage} is the type of
propositions and of the sets of the discourse (natural numbers, etc.),
and $\B$ is the type of predicate types (of which $\st$ is). For
example, the set of natural numbers $nat$ has the type $\st$, the
predicate $even$ has the type $\p{n}{nat}\st$ that we abbreviate by
$nat\a\st$ since $n$ does not occur in $\st$ (non-dependent product)
and $nat\a\st$ has the type $\B$. Starting from the rule (ax), the
rules (var) and (weak) enables one to build environments. The rule
(prod-$\la^\a$) enables one to build propositions and the rule
(prod-$\la$P) enables one to build predicate types. In the case of a
proposition, if the product is not dependent ($x$ does not occur in
$U$) then it is an implication, otherwise it is a universal
quantification. In other words, without the rule (prod-$\la$P), we get
the simply-typed $\la$-calculus. The rule (abs) enables one to build a
function (if $s=\st$) or a predicate (if $s=\B$). Finally, the rule
(app) enables the application of a function or a predicate to an
argument. In other words, the rules (abs) and (app) generalize the
rules ($\A$-intro) and ($\A$-elim) of the simply-typed $\la$-calculus.

~

From the point of view of programming, dependent types enables one to
have more information about data and hence to reduce the risk of
error. For example, one can define the type $(list~n)$ of lists of
natural numbers of length $n$ by declaring $list:nat\a\st$.  Then, the
empty list $nil$ has the $(list~0)$ and the function $cons$ which adds
a natural number $x$ at the head of a list $\ell$ of length $n$ has
the type $nat\a \p{n}{nat}(list~n) \a (list~(s~n))$. One can then
verify if a list does not exceed some given length.


\subsection*{Inductive definitions}
\index{inductive!type|indpage}

In higher-order logic, the induction principle for natural numbers can
be proved only if natural numbers are impredicatively defined. In
other words, if one prefers to stay in a predicative framework, it is
necessary to state the induction principle for natural numbers as an
axiom.

This is why, in 1971, P. Martin-L\"of \cite{martinlof71} extended the
calculus of N. de Bruijn by including expressions for representing
inductive types and their induction principles. For example, the type
of natural numbers is represented by the symbol $nat:\st$, zero by
$0:nat$, the successor function by $s:nat\a nat$ and a proof by
induction for a predicate $P:nat\a o$ by $rec^P: P0 \a (\all
n\!:\!nat. Pn \a Ps(n)) \a \all n\!:\!nat.  Pn$. In the conversion
rule (conv), to the $\b$-reduction, P. Martin-L\"of adds the {\em
$\io$-reduction\,} which corresponds to the elimination of induction
cuts~:

\begin{center}
  (conv) \quad $\cfrac{\G\th t:T \quad T \aa^*_{\b\io} T' \quad \G\th
    T':\st} {\G\th t:T'}$
\end{center}

In the case of $nat$, the rules defining the $\io$-reduction
\index{$\ai$|indpage} \index{iota-reduction@$\io$-reduction|indpage}
\index{reduction!iota@$\io$} are those of K. G\"odel's System T
\index{System T|indpage} \cite{godel58}~:

\begin{rewc}[\ai]
rec^P(p_0,p_s,0) & p_0\\
rec^P(p_0,p_s,s(n)) & p_s~ n~ rec^P(p_0,p_s,n)\\
\end{rewc}

\noindent
where $p_0$ is a proof of $P0$ and $p_s$ a proof of $\all
n\!:\!nat. Pn \a Ps(n)$. From these rules, by taking $P= \la
x\!:\!nat. nat$, it is possible to define functions on natural numbers
like the addition and the multiplication~:

\begin{rewc}[=]
x+y & rec^P(y,\la u.\la v.s(v),x)\\
x\times y & rec^P(0,\la u.\la v.v+y,x)\\
\end{rewc}

To convince oneself, let $f=\la u.\la v.s(v)$ and let us show that
$2+2$ rewrites to $4$~: $2+2= rec(2,f,2) \ai f\, 2~ rec(2,f,1) \ab
s(rec(2,f,1)) \ai s(f\, 2~ rec(2,f,0)) \ab s(s(rec(2,f,0))) \ai
s(s(2))=4$.

In fact, in this theory, it is possible to express by a term any
function whose existence is provable in predicative intuitionist
higher-order arithmetic (and these functions are also those that are
expressible in K. G\"odel's System T).


\subsection*{Polymorphism}
\index{polymorphism|indpage}

The termination problem of cut-elimination in intuitionist
impredicative higher-order arithmetic was solved by J.-Y. Girard
\cite{girard71} in 1971. To this end, he introduced a {\em
polymorphic\,} type system F$\w$ (J. Reynolds \cite{reynolds74}
introduced independently a similar system for second-order
quantifications only). A polymorphic type is a function which, to a
type expression, associates a type expression. And for representing
the proofs of impredicative propositions, one also needs the terms
themselves to be polymorphic, that is, it must be possible for a term
to be applied to a type expression. Formally, for second-order
quantifications ({\em i.e.\,} on propositions), this requires the
replacement of the rule (prod-$\la$P) in Figure~\ref{fig-th-lap} by
the rule~:

\begin{center}
(prod-F) \quad $\cfrac{\G\th T:\B \quad \GxT \th U:\st}
{\G\th \px{T}U:\st}$
\end{center}

\noindent
which for example allows one to build the type $\p{P}\st P\a P$
corresponding to the proposition $\all P\!:\!o. P\A P$ in higher-order
logic. For higher-order quantifications, one must add the following
rule~:

\begin{center}
(prod-F$\w$) \quad $\cfrac{\G\th T:\B \quad \GxT \th U:\B}
{\G\th \px{T}U:\B}$
\end{center}

\noindent
which allows the formation of predicate types like for example
$\st\a\st$ which corresponds to $o\A o$ in higher-order logic.

~

In this system, it is then possible to express any function whose
existence is provable in impredicative intuitionist higher-order
arithmetic.

From the point of view of programming, polymorphism enables one to
formalize generic algorithms with respect to data types. For example,
one can speak of the type $(list~A)$ of lists of elements of type $A$,
for any type $A$, by declaring $list:\st\a\st$.

~

In 1984, T. Coquand and G. Huet \cite{coquand84} defined a system, the
Calculus of Constructions (CC), \index{CC|indpage}
\index{Calculus of Constructions|indpage}
which makes the synthesis of the systems of
N. de Bruijn and J.-Y. Girard (it contains all the product-formation
rules we have seen) and in which it is then possible to express all
the higher-order logic (but it does not allow to express more
functions than F$\w$). This system served as a basis for the proof
assistant Coq \cite{coq01}.


\subsection*{Pure Types Systems (PTS)}
\index{PTS|indpage}
\index{Type System!Pure|indpage}

\begin{figure}[ht]
\setlength{\unitlength}{.8mm}
\begin{center}
\caption{Barendregt's Cube\label{fig-cube}}
\index{Barendregt's cube|indfig[fig-cube]}

\begin{picture}(100,90)
\put(10,10){\framebox(50,50)}
\put(30,30){\framebox(50,50)}

\put(10,60){\line(1,1){20}}
\put(60,60){\line(1,1){20}}
\put(60,10){\line(1,1){20}}

\put(10,10){\vector(1,1){20}}
\put(10,10){\vector(0,1){50}}
\put(10,10){\vector(1,0){50}}

\put(5,5){$\la^\a$}
\put(5,0){$(\st,\st)$ simple types}

\put(60,5){LF, $\la$P}
\put(60,0){$(\st,\B)$ dependent types}

\put(32,35){$(\B,\B)$ type constructors}

\put(5,65){$(\B,\st)$ polymorphic types}
\put(5,60){F}

\put(30,83){F$\w$}

\put(80,83){CC}
\end{picture}
\end{center}
\vs[3mm]
\end{figure}
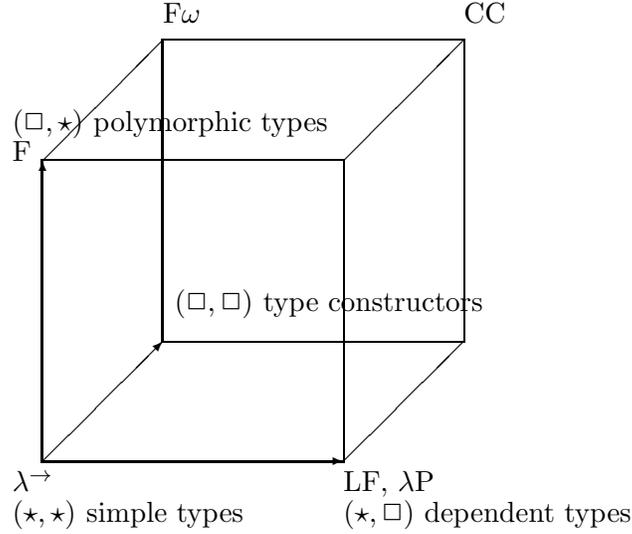

At the end of the 80's, H. Barendregt \cite{barendregt91jfp} remarked
that many type systems ($\la^\a$, $\la$P, F, F$\w$, CC, etc.) can be
characterized by their product-formation rules. This led to the
presentation we adopted here. By considering the following general
typing rule parametrized by two {\em sorts\,} $s_1,s_2\in\{\st,\B\}$~:

\begin{center}
$(s_1,s_2)$ \quad $\cfrac{\G\th T:s_1 \quad \GxT \th U:s_2}
{\G\th \px{T}U:s_2}$
\end{center}

\noindent
it is possible to have 4 different rules ($(\st,\st)$ corresponds to
(prod-$\la^\a$), $(\st,\B)$ to (prod-$\la$P), $(\B,\st)$ to (prod-F),
and $(\B,\B)$ to (prod-F$\w$)) and hence to have, from $(\st,\st)$, 8
different systems that one can organize in a cube whose directions
correspond to the presence or the absence of dependent types (rule
$(\st,\B)$), polymorphic types (rule $(\B,\st)$) or type constructors
(rule $(\B,\B)$) (see Figure~\ref{fig-cube}).


\begin{lst}{--}
\item $\la^\a$ denotes the simply-typed $\la$-calculus of A. Church
\cite{church40jsl},
\item LF (Logical Framework) denotes the system of R. Harper,
F. Honsell and G. Plotkin \cite{harper87lics},
\item $\la$P denote the AUTOMATH system of N. de Bruijn
\cite{debruijn68},
\item F and F$\w$ respectively denote the second-order and
higher-order polymorphic $\la$-calculus of J.-Y. Girard
\cite{girard88book},
\item CC denotes the Calculus of Constructions of T. Coquand and
G. Huet \cite{coquand88ic}.
\end{lst}

~

This led S. Berardi \cite{berardi88tr} and J. Terlouw
\cite{terlouw89tr} to a more systematic study of type systems with
respect to expressible types until the definition by H. Geuvers and
M.-J. Nederhof of the {\em Pure Type Systems\,} (PTS)
\cite{geuvers91jfp} which are type systems parametrized by~:

\begin{lst}{--}
\item a set of {\em sorts\,} $\cS$ representing the different
universes of discourse ($\{\st,\B\}$ in the Cube),

\item a set of {\em axioms\,} $\cA\sle\cS^2$ representing how these
universes are included in one another ($\{(\st,\B)\}$ in the Cube) and
the rule~:

\begin{center}
(ax) \quad $\cfrac{}{\th s_1:s_2}$ \quad $((s_1,s_2)\in\cA)$\\[3mm]
\end{center}

\item a set of product-formation {\em rules\,} $\cB\sle\cS^3$
representing the possible quantifications ($\{(\st,\st,\st),
(\st,\B,\B), (\B,\st,\st), (\B,\B,\B)\}$ in the Cube) and the rule~:\\

\begin{center}
(prod) \quad $\cfrac{\G\th T:s_1 \quad \GxT \th U:s_2}
{\G\th \px{T}U:s_3}$ \quad $((s_1,s_2,s_3)\in\cB)$\\
\end{center}
\end{lst}


\subsection*{Calculus of Inductive Constructions (CIC)}
\index{CIC|indpage}
\index{Calculus of Constructions!Inductive|indpage}

We have seen that the Calculus of Constructions is a very powerful
system in which it is possible to express many functions. However,
these functions cannot be defined as one would like. For example, it
does not seem possible to program the predecessor function on natural
numbers such that its evaluation takes a constant time
\cite{girard88book}. This is not the case in P. Martin-L\"of's system
where natural numbers and their induction principle are first-class
objects while, in the Calculus of Constructions, natural numbers are
impredicatively defined.

This is why, in 1988, T. Coquand and C. Paulin proposed the Calculus
of Inductive Constructions (CIC) \cite{coquand88colog} which makes the
synthesis between the Calculus of Constructions and P. Martin-L\"of's
type theory, and hence enable one to write more efficient programs. In
1994, B. Werner \cite{werner94thesis} proved the termination of
cut-elimination in this system. (In 1993, T.  Altenkirch
\cite{altenkirch93thesis} proved also this property but for a
presentation of the calculus with equality judgments.)

But, even in this system, some algorithms may be
inexpressible. L. Colson \cite{colson98tcs} proved for example that,
if one uses a call-by-value evaluation strategy (reduction of the
arguments first) then the minimum function of two natural numbers
cannot be implemented by a program whose evaluation time is relative
to the minimum of the two arguments.




\section{Motivations}
\label{sec-motiv}

We said at the beginning that rewriting is a simple and general
computation para\-digm based on {\em rewrite rules\,}. This notion is
of course very old but it was seriously studied from the 70's with the
works of D. Knuth and D. Bendix \cite{bendix70book}. They studied
rewriting for knowing whether, in a given equational theory, an
equation is valid or not. Then, rewriting was quickly used as a
programming paradigm
\cite{odonnell77,goguen79,futatsugi85popl,jouannaud92jlp,maude99}
since any computable function can be defined by rewrite rules
(Turing-completeness).

Let us see the case of addition and multiplication on natural numbers
defined from $0$ for zero and $s$ for the successor function~:

\begin{rewc}
0+x & x\\
s(x)+y & s(x+y)\\
0\times x & 0\\
s(x)\times y & (x\times y)+y\\
\end{rewc}

These rules completely define these two arithmetic operations~:
starting from two arbitrary natural numbers $p$ and $q$ (expressed
with $0$ and $s$), $p+q$ and $p\times q$ rewrite in a finite number of
steps to a term which cannot be further rewritten, that is, to a
number representing the value of $p+q$ and $p\times q$ respectively.

~

\noindent{\bf Higher-order rewriting}
\index{higher-order!rewriting|indpage}

We can also imagine definitions using functional parameters or
abstractions~: this is {\em higher-order\,} rewriting as opposed to
{\em first-order\,} rewriting which does not allow functional
parameters or abstractions. For example, the function $map$ which to a
function $f$ and a list of natural numbers $(a_1,\ldots,a_n)$
associates the list $(f(a_1),\ldots$, $f(a_n))$, can be defined by the
following rules~:

\begin{rewc}
map(f,nil) & nil\\
map(f,cons(x,\ell)) & cons(fx,map(f,\ell))\\
\end{rewc}

\noindent
where $nil$ stand for the empty list and $cons$ for the function which
adds an element at the head of a list.

Hence, the rules defining the recursor of an inductive type (the
$\io$-reduction) are a particular case of higher-order rewriting.

~

\noindent{\bf Easier definitions}

One can see that definitions by rewriting are more natural and easier
to write than the ones based on recursors like in P. Martin-L\"of type
theory or in the Calculus of Inductive Constructions. For example, the
definition by recursion of the function $\le$ on natural numbers
requires two levels of recursion~:

\begin{center}
$\la x.rec(x,\la y.true,\la nzy.rec(y,false,\la n'z'.zn',y))$
\end{center}

\noindent
while the definition by rewriting is~:

\begin{rewc}
0 \le y & true\\
s(x) \le 0 & false\\
s(x) \le s(y) & x \le y\\
\end{rewc}

~

\noindent{\bf More efficient definitions}

From a computing point of view, definitions by rewriting can be made
more efficient by adding rules. For example, with a definition by
recursion on its first argument, $n+0$ requires $n+1$ reduction
steps. By simply adding the rule $x+0\a x$, this takes only one step.

However, it can become more difficult to ensure that, for any sequence
of arguments, the definition always leads, in a finite number of steps
(property called{\em strong normalization\,}), to a unique result
(property called {\em confluence\,}) that we call the {\em normal
form\,} of the starting expression.

~

\noindent{\bf Quotient types}
\index{type!quotient|indpage}
\index{quotient type|indpage}

Until now we have always spoken of natural numbers but never of
integers. Yet they have an important place in mathematics. One way to
represent integers is to add a predecessor function $p$ beside $0$ and
$s$. Hence, $p(p(0))$ represents $-2$.  Unfortunately, in this case,
an integer can have several representations~: $p(s(0))$ or $s(p(0))$
represent both $0$. In fact, integers are equivalent modulo the
equations $p(s(x))=x$ and $s(p(x))=x$.

However, it is possible to orient these equations so as to have a
confluent and strongly normalizing rewrite system~: $p(s(x))\a x$ and
$s(p(x))\a x$. Then, each integer has a unique normal form. Therefore,
we see that rewriting enables us to models quotient types without
using further extensions \cite{barthe95csl}.

~

\noindent{\bf More typable terms}

The introduction of rewriting in a dependent type system allows one to
type more terms and therefore to formalize more propositions. Let us
consider, in the Calculus of Inductive Constructions, the type $listn:
nat\a\st$ of lists of natural numbers of length $n$ with the
constructors $niln: listn(0)$ for the empty list and $consn: nat\a
\p{n}{nat} listn(n)\a listn(s(n))$ for adding an element at the head
of a list. Let $appn: \p{n}{nat} listn(n)\a \p{n'}{nat} listn(n')\a
listn(n+n')$ be the concatenation function. Like $+$, $appn$ can be
defined by using the recursor associated to the type $listn$. Assume
furthermore that $+$ and $appn$ are defined by induction on their
first argument. Then, the following propositions are not typable~:

\begin{rewc}[=]
appn(n,\ell,0,\ell') & \ell\\
appn(n+n',appn(n,\ell,n',\ell'),n'',\ell'') &
appn(n,\ell,n'+n'',appn(n',\ell',n'',\ell''))\\
\end{rewc}

In the first rule, the left hand-side is of type $listn(n+0)$ and the
right hand-side is of type $listn(n)$. We can prove that $\p{n}{nat}
n+0= n$ by induction on $n$ but $n+0$ is not $\b\io$-convertible to
$n$ since $+$ is defined by induction on its first
argument. Therefore, we cannot apply the (conv) rule for typing the
equality.

In the second rule, the left hand-side is of type $listn((n+n')+n'')$
and the right hand-side is of type $listn(n+(n'+n''))$. Again,
although we can prove that $(n+n')+n''= n+(n'+n'')$ (associativity of
$+$), the two terms are not $\b\io$-convertible. Therefore, we cannot
apply the (conv) rule for typing the equality.

This shows some limitation of the definitions by recursion. The use of
rewriting, that is, the replacement in the (conv) rule of the
$\io$-reduction by a reduction relation $\ar$ generated from a
user-defined set $\cR$ of arbitrary rewrite rules~:

\begin{center}
  (conv) \quad $\cfrac{\G\th t:T \quad T \aa^*_{\b\cR} T' \quad \G\th
    T':\st} {\G\th t:T'}$,
\end{center}

\noindent
allows us to type the previous propositions which are not typable in
the Calculus of Inductive Constructions.

~

\noindent{\bf Automatic equational proofs}

Another motivation for introducing rewriting in type systems is that
it makes equational proofs much easier, which is the reason why
rewriting was studied initially. Indeed, in the case of a confluent
and strongly normalizing rewrite system, to check whether two terms
are equal, it suffices to check whether they have the same normal
form.

Moreover, it is not necessary to keep a trace of the rewriting steps
since this computation can be done again (if the equality is
decidable). This reduces the size of proof-terms and enable us to deal
with bigger proofs, which is a critical problem now in proof
assistants.

~

\noindent{\bf Integration of decision procedures}
\index{decision procedure|indpage}

\newcommand{\xor}{\,\mbox{xor}\,}

One can also imagine defining predicates by rewriting or having
simplification rules on propositions, hence generalizing the
definitions by {\em strong elimination\,} of the Calculus of Inductive
Constructions \cite{paulin93tlca}. For example, one can consider the
set of rules of Figure~\ref{fig-hsiang} \cite{hsiang82thesis} where
$\xor$ (exclusive ``or'') and $\et$ are commutative and associative
symbols, $\bot$ represents the proposition always false and $\top$ the
proposition always true (by taking a constant $I$ of type $\top$).

\begin{figure}[ht]
\begin{center}
\caption{Decision procedure for classical propositional
tautologies\label{fig-hsiang}}

~

\begin{tabular}{cc}
\begin{rew}
P \xor \bot & P\\
P \xor P & \bot\\[2mm]

P \et \top & P\\
P \et \bot & \bot\\
P \et P & P\\
P \et (Q \xor R) & (P \et Q) \xor (P \et R)\\
\end{rew}
&
\begin{rew}
\neg P & P \xor \top\\
P \ou Q & (P \et Q) \xor P \xor Q\\
P \A Q & (P \et Q) \xor P \xor \top\\
P \AA Q & (P \xor Q) \xor \top\\
\end{rew}
\end{tabular}
\end{center}
\end{figure}

J. Hsiang \cite{hsiang82thesis} showed that this system is confluent
and strongly normalizing and that a proposition $P$ is a tautology
({\em i.e.\,} is always true) if $P$ reduces to $\top$. This system is
therefore a decision procedure for classical propositional
tautologies.

Hence, type-level rewriting allows the integration of decision
procedures. Indeed, thanks to the conversion rule (conv), if $P$ is a
tautology then $I$, the canonical proof of $\top$, is a proof of
$P$. In other words, if the typing relation is decidable, to know
whether a proposition $P$ is a tautology, it is sufficient to propose
$I$ to the verification program.

~

We can also imagine simplification rules for equality like the ones of
Figure~\ref{fig-int-ring} where $+$ and $\times$ are associative and
commutative, and $=$ is commutative.

\begin{figure}[ht]
\begin{center}
\caption{Simplification rules for equality\label{fig-int-ring}}

~

\begin{tabular}{cc}
\begin{rew}
x + 0 & x\\
x + s(y) & s(x+y)\\
x \times 0 & 0\\
x \times s(y) & (x\times y) + x\\
x \times (y+z) & (x\times y) + (x\times z)\\
\end{rew}
&
\begin{rew}
x = x & \top\\
s(x) = s(y) & x=y\\
s(x) = 0 & \bot\\
x+y = 0 & x=0 \et y=0\\
x\times y=0 & x=0 \ou y=0\\
\end{rew}
\end{tabular}
\end{center}
\end{figure}




\section{Previous works}

The first works on the combination of typed $\la$-calculus and
(first-order) rewriting were due to V. Breazu-Tannen in 1988
\cite{breazu88lics}. They showed that the combination of simply-typed
$\la$-calculus and first-order rewriting is confluent if the rewriting
is confluent. In 1989, V. Breazu-Tannen and J. Gallier
\cite{breazu89icalp}, and M. Okada \cite{okada89issac} independently,
showed that the strong normalization is also preserved. These results
were extended by D. Dougherty \cite{dougherty91rta} to any ``stable''
set of pure $\la$-terms.

In 1991, J.-P. Jouannaud and M. Okada \cite{jouannaud91lics} extended
the result of V. Breazu-Tannen and J. Gallier to higher-order rewrite
systems satisfying the {\em General Sche\-ma\,}, a generalization of
the primitive recursion schema. With higher-order rewriting, strong
normalization becomes more difficult to prove since there is a strong
interaction between rewriting and $\b$-reduction, which is not the
case with first-order rewriting.

In 1993, M. Fern\'andez \cite{fernandez93thesis} extended this
method to the Calculus of Constructions with object level rewriting
and simply typed symbols. The methods used for first-order rewriting
and non dependent systems \cite{breazu89icalp,dougherty91rta} cannot
be applied to this case since rewriting is not just a syntactic
addition~: since rewriting is included in the type conversion rule
(conv), it is a component of typing (in particular, it allows more
terms to be typed).

Other methods for proving strong normalization appeared. In 1996,
J. van de Pol \cite{vandepol96thesis} extended to the simply-typed
$\la$-calculus the use of monotone interpretations. In 1999,
J.-P. Jouannaud and A. Rubio \cite{jouannaud99lics} extended to the
simply-typed $\la$-calculus the method RPO (Recursive Path Ordering)
\cite{plaisted78tr,dershowitz79focs}. This method (HORPO) is more
powerful than the General Schema since it is a recursively defined
ordering.

In all these works, even the ones on the Calculus of Constructions,
function symbols are always simply typed. It was T. Coquand
\cite{coquand92types} in 1992 who initiated the study of rewriting on
dependent and polymorphic symbols. He studied the completeness of
definitions on dependent types. For the strong normalization, he
proposed a schema more general than the schema of J.-P. Jouannaud and
M. Okada since it allows recursive definitions on strictly positive
types \cite{coquand88colog} but it does not necessary imply strong
normalization. In 1996, E. Gim\'enez \cite{gimenez96thesis} defined a
restriction of this schema for which he proved strong
normalization. In 1999, J.-P. Jouannaud, M. Okada and I
\cite{blanqui02tcs,blanqui99rta} extended the General Schema, keeping
simply typed symbols, in order to deal with strictly positive
types. Finally, in 2000, D. Walukiewicz \cite{walukiewicz00lfm}
extended J.-P. Jouannaud and A. Rubio's HORPO to the Calculus of
Constructions with dependent and polymorphic symbols.

~

But there still is a common point between all these works~: rewriting
is always confined to the object level.

~

In 1998, G. Dowek, T. Hardin and C. Kirchner \cite{dowek98trtpm}
proposed a new approach to deduction for first-order logic~: the {\em
Natural Deduction Modulo\,} (NDM) \index{NDM|indpage}
\index{Natural Deduction!Modulo|indpage} a congruence
$\equiv$ on propositions. This deduction system consists of replacing
the rules of usual Natural Deduction by rules equivalent modulo
$\equiv$. For example, the elimination rule for $\A$ ({\em modus
ponens\,} )is replaced by~:

\begin{center}
  ($\A$-elim-modulo) \quad $\cfrac{\G\th R \quad \G\th P}{\G\th Q}$
  \quad if $R \equiv P\et Q$
\end{center}

They proved that the simple theory of types and the skolemized set
theory can be seen as first-order theory modulo some congruences using
{\em explicit substitutions\,}. In \cite{dowek98types}, G. Dowek and
B. Werner gave several conditions ensuring the strong normalization of
cut elimination.




\section{Contributions}
\label{sec-contrib}

Our main contribution was establishing very general conditions for
ensuring the strong normalization of the Calculus of Constructions
extended with type level rewriting \cite{blanqui01lics}. We showed
that our conditions are satisfied by a large subsystem of the Calculus
of Inductive Constructions (CIC) and by Natural Deduction Modulo (NDM)
a large class of equational theories.

Our work can be seen as an extension of both the Natural Deduction
Modulo and the Calculus of Constructions, where the congruence not
only includes first-order rewriting but also higher-order rewriting
since, in the Calculus of Constructions, functions and predicates can
be applied to functions and predicates.

It can therefore serve as the basis of a powerful extension of proof
assistants like Coq \cite{coq01} and LEGO \cite{lego92} which allow
definitions by recursion only. Indeed, strong normalization not only
ensures the logical consistency (if the symbols are consistent) but
also the decidability of type checking, that is, the verification that
a term is the proof of a proposition.

For deciding particular classes of problems, it may be more efficient
to use specialized rewriting-based applications like CiME \cite{cime},
ELAN \cite{elan00} or Maude \cite{maude99}. Furthermore, for program
extraction \cite{paulin89popl}, we can use rewriting-based languages
and hence get more efficient extracted programs.

~

To consider type-level rewriting is not completely new~: a particular
case is the ``strong elimination'' of the Calculus of Inductive
Constructions, that is, the ability to define predicates by induction
on some inductive data type. The main novelty here is to consider any
set of user-defined rewrite rules.

The strong normalization proofs with strong elimination of B. Werner
\cite{werner94thesis} and T. Altenkirch \cite{altenkirch93thesis} use
in an essential way the fact that the definitions are inductive.

Moreover, the methods used in case of first-order rewriting
\cite{breazu89icalp,barbanera90ctrs,dougherty91rta} cannot be applied
here. Firstly, we consider higher-order rewriting which has a strong
interaction with $\b$-reduction. Secondly, rewriting is part of the
type conversion rule, which implies that more terms are typable.

~

For establishing our conditions and proving their correctness, we have
adapted the method of reductibility candidates of Tait and Girard
\cite{girard88book} also used by F.  Barbanera, M. Fern\'andez and
H. Geuvers \cite{barbanera94lics,barbanera93icalp,barbanera93tlca} for
object-level rewriting and by B. Werner and T. Altenkirch for strong
elimination. As candidates, they all use sets of pure (untyped)
$\la$-terms and, except T. Altenkirch, they all use intermediate
languages of type systems. By using a work by T. Coquand and J.
Gallier \cite{coquand90lf}, we use candidates made of well-typed terms
and do not use intermediate languages. We therefore get a simpler and
shorter proof for a more general result.

~

We also mention other contributions.

For allowing quotient types (rules on constructors) and matching on
function symbols, which is not possible in the Calculus of Inductive
Constructions, we use a notion of ``constructor'' more general than
the usual one (see Subsection~\ref{sec-ind-typ}).

For ensuring the subject reduction property, that is, the preservation
of typing under reduction, we introduce new conditions more general
than the ones previously used. In particular, these conditions allow
us to get rid of many non-linearities due to typing, which makes
rewriting more efficient and confluence easier to prove.




\section{Outline of the thesis}

\hs[4.5mm]
{\bf Chapter~\ref{chap-tsm}~:} We study the basic properties of Pure
Type Systems whose type conversion relation is abstract. We call such
a system a Type System Modulo (TSM).

~

{\bf Chapter~\ref{chap-rts}~:} We study the properties of a particular
class of TSM's, those whose conversion relation is generated from a
reduction relation. We call such a system a Reduction Type System
(RTS). An essential problem in these systems is to make sure that the
reduction relation preserves typing (subject reduction property).

~

{\bf Chapter~\ref{chap-ats}~:} We give sufficient conditions for
ensuring the subject reduction property in RTS's whose reduction
relation is generated from rewrite rules. We call such a system an
Algebraic Type System (ATS).

~

{\bf Chapter~\ref{chap-conditions}~:} In this chapter and the
following ones, we consider a particular ATS, the Calculus of
Algebraic Constructions (CAC). We give sufficient conditions for
ensuring its strong normalization.

~

{\bf Chapter~\ref{chap-examples}~:} We give important examples of type
systems satisfying our strong normalization conditions. Among these
systems, we find a sub-system with strong elimination of the Calculus
of Inductive Constructions (CIC) which is the basis of the proof
assistant Coq \cite{coq01}. We also find Natural Deduction Modulo
(NDM) a large class of equational theories.

~

{\bf Chapter~\ref{chap-correctness}~:} We prove the correctness of our
strong normalization conditions and clearly indicate which conditions
are used. An index enables one to find where each conditions is used.

~

{\bf Chapter~\ref{chap-future}~:} We finish by enumerating several
directions for future research which could improve or extend our
strong normalization conditions.


%% file: prelim.tex



\chapter{Preliminaries}
\label{chap-prelim}

In this chapter, we define the syntax of the systems we will study and
recall a few elementary notions about $\la$-calculus, Pure Type
Systems (PTS) (see \cite{barendregt92book} for more details) and
relations. This syntax simply extends the syntax of PTS's by adding
symbols ($nat$, $0$, $+$, $\ge$, \ldots) which must be applied to as
many arguments as are required by their specified {\em arity\,} (see
Remark~\ref{rem-arity} for a discussion about this notion).


\renewcommand{\at}{\alpha}

\begin{dfn}[Sorted $\la$-systems]
\label{def-lambda-sys}
\index{sorted lambda-system@sorted $\la$-system|inddef[def-lambda-sys]}
\index{sort|inddef[def-lambda-sys]}
\index{variable|inddef[def-lambda-sys]}
\index{symbol|inddef[def-lambda-sys]}
\index{arity!symbol|inddef[def-lambda-sys]}
\index{S@$\cS$|inddef[def-lambda-sys]}
\index{X@$\cX,\cX^s$|inddef[def-lambda-sys]}
\index{F@$\cF,\cF^s,\cF_n,\cF^s_n$|inddef[def-lambda-sys]}

A {\em sorted $\la$-system\,} is given by~:

\begin{lst}{--}
\item a set of {\em sorts\,} $\cS$,
\item a family $\cF= (\cF^s_n)^{s\in\cS}_{n\ge 0}$ of sets of {\em
symbols\,},
\item a family $\cX= (\cX^s)^{s\in\cS}$ of infinite denumerable sets
of {\em variables\,},
\end{lst}

\noindent
such that all sets are disjoint. A symbol $f\in \cF^s_n$ is of {\em
arity\,} $\at_f= n$ and of {\em sort\,} $s$. We will denote the set of
symbols of sort $s$ by $\cF^s$ and the set of symbols of arity $n$ by
$\cF_n$.
\end{dfn}


\begin{dfn}[Terms]
\label{def-term}
\index{term|inddef[def-term]}
\index{$\lx{T}u$|inddef[def-term]}
\index{abstraction|inddef[def-term]}
\index{$\px{T}U$|inddef[def-term]}
\index{product!dependent|inddef[def-term]}
\index{dependent!product|inddef[def-term]}
\index{application|inddef[def-term]}
\index{$uv$|inddef[def-term]}
\index{$f(\vt)$|inddef[def-term]}
\index{T@$\cT$|inddef[def-term]}

The set $\cT$ of {\em terms\,} is the smallest set such that~:

\begin{lst}{--}
\item sorts and variables are terms;
\item if $x$ is a variable and $t$ and $u$ are terms then the {\em
dependent product\,} $\px{t}u$ and the {\em abstraction\,} $\lx{t}u$
are terms;
\item if $t$ and $u$ are terms then the {\em application\,} $tu$ is a
term;
\item if $f$ is a symbol of arity $n$ and $t_1, \ldots, t_n$ are terms
then $f(t_1, \ldots, t_n)$ is a term (some binary symbols like $+$,
$\times$, \ldots will sometimes be written infix).
\end{lst}
\end{dfn}


\newcommand{\FVB}{\FV^\B}

\newcommand{\vx}{\vec{x}}
\newcommand{\vt}{\vec{t}}
\newcommand{\vT}{\vec{T}}
\newcommand{\vy}{\vec{y}}
\newcommand{\vU}{\vec{U}}
\newcommand{\vl}{\vec{l}}
\newcommand{\vu}{\vec{u}}
\newcommand{\vz}{\vec{z}}
\newcommand{\vV}{\vec{V}}
\newcommand{\vv}{\vec{v}}
\newcommand{\va}{\vec{a}}
\newcommand{\vw}{\vec{w}}
\newcommand{\vS}{\vec{S}}
\newcommand{\vW}{\vec{W}}

\newcommand{\pvxT}{(\vx:\vT)}
\newcommand{\pvxTp}{(\vx:\vT')}
\newcommand{\pvxt}{(\vx:\vt)}
\newcommand{\pvxU}{(\vx:\vU)}
\newcommand{\pvxV}{(\vx:\vV)}
\newcommand{\pvyT}{(\vy:\vT)}
\newcommand{\pvyV}{(\vy:\vT)}
\newcommand{\pvyU}{(\vy:\vU)}
\newcommand{\pvzV}{(\vz:\vV)}

\newcommand{\xt}{\{x\to t\}}
\newcommand{\xup}{\{x\to u'\}}
\newcommand{\xv}{\{x\to v\}}
\newcommand{\yv}{\{y\to v\}}
\newcommand{\XU}{\{X\to U\}}
\newcommand{\XS}{\{X\to S\}}
\newcommand{\xS}{\{x\to S\}}

\newcommand{\vyu}{\{\vy\to\vu\}}
\newcommand{\vyt}{\{\vy\to\vt\}}
\newcommand{\vxt}{\{\vx\to\vt\}}
\newcommand{\vxtp}{\{\vx\to\vt'\}}
\newcommand{\vxl}{\{\vx\to\vl\}}
\newcommand{\vxv}{\{\vx\to\vv\}}
\newcommand{\vxS}{\{\vx\to\vS\}}
\newcommand{\vxu}{\{\vx\to\vu\}}
\newcommand{\vxup}{\{\vx\to\vu'\}}
\newcommand{\vzv}{\{\vz\to\vv\}}


\noindent{\bf Free and bound variables}
\index{variable!free|indchap[chap-prelim]}
\index{variable!bound|indchap[chap-prelim]}
\index{alpha-equivalence@$\alpha$-equivalence|indchap[chap-prelim]}
\index{term!closed|indchap[chap-prelim]}
\index{product!non-dependent|indchap[chap-prelim]}
\index{$T\a U$|indchap[chap-prelim]}
\index{FVt@$\FV(t),\FV^s(t)$|indchap[chap-prelim]}

A variable $x$ in the scope of an abstraction $\lx{T}$ or a product
$\px{T}$ is {\em bound\,}. As usual, it may be replaced by another
variable of the same sort. This is {\em $\alpha$-equivalence\,}. A
variable which is not bound is {\em free\,}. We denote by $\FV(t)$ the
set of free variables of a term $t$ and by $\FV^s(t)$ the set of free
variables of sort $s$. A term without free variables is {\em
closed\,}. We often denote by $U\a V$ a product $\px{U}V$ such that
$x\notin \FV(V)$ (non dependent product).


\newpage

\noindent{\bf Vectors}
\index{vector|indchap[chap-prelim]}
\index{$\vt$|indchap[chap-prelim]}
\index{$"|\vt"|$|indchap[chap-prelim]}

We will often use vectors ($\vt, \vu, \ldots$) for sequences of terms
(or anything else). The size of a vector $\vt$ is denoted by
$|\vt|$. For example, $[\vx:\vT] u$ denotes the term $\l{x_1}{T_1}
\ldots \l{x_n}{T_n} u$ where $n=|\vx|$.


\begin{dfn}[Positions]
\label{def-pos}
\index{position!term|inddef[def-pos]}
\index{post@$\pos(t),\pos(f,t),\pos(x,t)$|inddef[def-pos]}
\index{subterm|inddef[def-pos]}
\index{$t"|_p,t[u]_p$|inddef[def-pos]}
\index{$\tgt$|inddef[def-pos]}
\index{ordering!$\tgt$|inddef[def-pos]}
\index{ordering!subterm|inddef[def-pos]}

To designate a subterm of a term, we use a system of {\em
positions\,}. Formally, the set $\pos(t)$ of the positions in a term
$t$ is the smallest set of words over the alphabet of positive
integers such that~:

\begin{lst}{--}
\item $\pos(s)= \pos(x)= \{\vep\}$,
\item $\pos(\px{t}u)= \pos(\lx{t}u)= \pos(tu)= 1\cdot\pos(t)
  \,\cup\, 2\cdot\pos(u)$,
\item $\pos(f(\vt))= \{\vep\} \,\cup\, \bigcup \,\{ i\cdot\pos(t_i)
  ~|~ 1\le i\le \at_f \}$,
\end{lst}

\noindent
where $\vep$ denotes the empty word and $\cdot$ the concatenation. We
denote by $t|_p$ the subterm of $t$ at the position $p$ and by
$t[u]_p$ the term obtained by replacing $t|_p$ by $u$ in $t$. The
relation ``is subterm of'' is denoted by $\tlt$.

Let $t$ be a term and $f$ be a symbol. We denote by $\pos(f,t)$ the
set of the positions $p$ in $t$ where $t|_p$ is of the form
$f(\vt)$. If $x$ is a variable, we denote by $\pos(x,t)$ the set of
the positions $p$ in $t$ such that $t|_p$ is a free occurrence of $x$.
\end{dfn}


\begin{dfn}[Substitution]
\label{def-subs}
\index{substitution!term|inddef[def-subs]}
\index{domain!substitution|inddef[def-subs]}
\index{domtheta@$\dom(\t),\dom^s(\t)$|inddef[def-subs]}

A {\em substitution\,} $\t$ is an application from $\cX$ to $\cT$
whose {\em domain\,} $\dom(\t)= \{ x \IN \cX ~|~ x\t \neq x \}$ is
finite. Applying a substitution $\t$ to a term $t$ consists of
replacing all the free variables of $t$ by their image in $\t$ (to
avoid variable captures, bound variables must be distinct from free
variables). The result is denoted by $t\t$. We let $\dom^s(\t)=
\dom(\t) \cap \cX^s$. We denote by $\vxt$ the substitution which
associates $t_i$ to $x_i$ and by $\t\cup\xt$ the substitution which
associates $t$ to $x$ and $y\t$ to $y\neq x$.
\end{dfn}


\noindent{\bf Relations}
\index{stable by!context|indchap[chap-prelim]}
\index{stable by!substitution|indchap[chap-prelim]}
\index{$\a^*,\a^+,\al,\aa^*,\ad$|indchap[chap-prelim]}
\index{beta-reduction@$\b$-reduction|indchap[chap-prelim]}
\index{eta-reduction@$\eta$-reduction|indchap[chap-prelim]}
\index{$\ab,\ae$|indchap[chap-prelim]}
\index{reduction!beta@$\b$|indchap[chap-prelim]}
\index{reduction!eta@$\eta$|indchap[chap-prelim]}

We now recall a few elementary definitions on relations. Let $\a$ be a
relation on terms.

\begin{lst}{--}
\item $\al\,$ is the inverse of $\a$.
\item $\a^+$ is the smallest transitive relation containing $\a$.
\item $\a^*$ is the smallest reflexive and transitive relation
containing $\a$.
\item $\aa^*$ is the smallest reflexive, transitive and symmetric
relation containing $\a$.
\item $\ad\,$ is the relation $\a^*\als$ ~($t\ad u$ if there exists
$v$ such that $t\a^* v$ and $u \a^* v$).
\end{lst}

If $t\a t'$, we say that $t$ {\em rewrites\,} to $t'$. If $t\a^* t'$,
we say that $t$ {\em reduces\,} to $t'$.

The relation $\a$ is {\em stable by context\,} if $u\a u'$ implies
$t[u]_p \a t[u']_p$ for all term $t$ and position $p\in\pos(t)$.

The relation $\a$ is {\em stable by substitution\,} if $t\a t'$
implies $t\t\a t'\t$ for all substitution $\t$.

The {\em $\b$-reduction\,} (resp. {\em $\eta$-reduction\,}) relation
is the smallest relation stable by context and substitution containing
$\lx{U}t ~u \ab t\xu$ (resp. $\lx{U}tx \ae t$ if $x\notin \FV(t)$).  A
term of the form $\lx{U}t ~u$ (resp. $\lx{U}tx$ with $x\notin \FV(t)$)
is a {\em $\b$-redex\,} (resp. {\em $\eta$-redex\,}).


\newpage

\noindent{\bf Normalization}
\index{normalization!weak|indchap[chap-prelim]}
\index{normalization!strong|indchap[chap-prelim]}
\index{well-founded|indchap[chap-prelim]}
\index{n\oe{}therian|indchap[chap-prelim]}

The relation $\a$ is {\em weakly normalizing\,} if, for all term $t$,
there exists an irreducible term $t'$ to which $t$ reduces. We say
that $t'$ is a {\em normal form\,} of $t$. The relation $\a$ is {\em
strongly normalizing\,} (well-founded, n\oe{}therian) if, for all term
$t$, any reduction sequence issued from $t$ is finite.


~

\noindent{\bf Confluence}
\index{confluence|indchap[chap-prelim]}
\index{confluence!local|indchap[chap-prelim]}
\index{local confluence|indchap[chap-prelim]}
\index{normal form|indchap[chap-prelim]}

The relation $\a$ is {\em locally confluent\,} if, whenever a term $t$
rewrites to two distinct terms $u$ and $v$, then $u \ad v$. The
relation $\a$ is {\em confluent\,} if, whenever a term $t$ reduces to
two distinct terms $u$ and $v$, then $u \ad v$.

If $\a$ is locally confluent and strongly normalizing then $\a$ is
confluent \cite{newman42}. If $\a$ is confluent and weakly normalizing
then every term $t$ has a normal form denoted by $t\ad$.


~

\noindent{\bf Lexicographic and multiset orderings}
\index{multiset|indchap[chap-prelim]}
\index{ordering!multiset|indchap[chap-prelim]}
\index{lexicographic ordering|indchap[chap-prelim]}
\index{ordering!lexicographic|indchap[chap-prelim]}
\index{$>\mul$|indchap[chap-prelim]}
\index{ordering!$>\mul$|indchap[chap-prelim]}
\index{$(>_1,\ldots,>_n)\lex$|indchap[chap-prelim]}
\index{ordering!$(>_1,\ldots,>_n)\lex$|indchap[chap-prelim]}

Let $>_1,\ldots,>_n$ be orderings on $E_1,\ldots,E_n$ respectively. We
denote by $(>_1,\ldots,>_n)\lex$ the {\em lexicographic\,} ordering on
$E_1\times\ldots\times E_n$ from $>_1,\ldots,>_n$. For example, for
$n=2$, $(x,y) (>_1,>_2)\lex (x',y')$ if $x >_1 x'$ or, $x =_1 x'$ and
$y >_2 y'$.

Let $E$ be a set. A {\em multiset\,} $M$ on $E$ is a function from $E$
to $\mb{N}$ ($M(x)$ denotes the number of occurrences of $x$ in
$M$). We denote by $\cM(E)$ the set of finite multisets on $E$. Let
$>$ be an ordering on $E$, the {\em multiset extension\,} of $>$ is
the ordering $>\mul$ on $\cM(E)$ defined as follows~: $M >\mul N$ if
there exists $P,Q\in \cM(E)$ such that $P\neq\vide$, $P \sle M$, $N=
(M\moins P)\cup Q$ and, for all $y\in Q$, there exists $x\in P$ such
that $x>y$.

An important property of these extensions is that they preserve the
well-founded\-ness. For more details on these notions, one can consult
\cite{baader98book}.


%% file: tsm.tex



\chapter{Type Systems Modulo (TSM's)}
\markboth{CHAPTER \thechapter. TYPE SYSTEMS MODULO}{}
\label{chap-tsm}

In this chapter, we consider an extension of PTS's with function and
predicate symbols and a conversion rule (conv) where $\aa^*_\b$ is
replaced by an arbitrary conversion relation $\cC$.

~

There has already been different extension of PTS's, in particular~:

\begin{lst}{--}
\item In 1989, Z. Luo \cite{luo90thesis} studied an extension of the
Calculus of Constructions with a cumulative hierarchy of sorts ($\st
\prec \B= \B_0 \prec \B_1 \prec \ldots$), the Extended Calculus of
Constructions (ECC)~: $\cC$ is the smallest quasi-ordering including
$\aa^*_\b$, $\prec$ and which is compatible with the product structure
($U' ~\cC~ U$ and $V ~\cC~ V'$ implies $\px{U}V ~\cC~ \px{U'}V'$).

\item In 1993, H. Geuvers \cite{geuvers93thesis} studied the PTS's
with $\eta$-reduction~: $\cC=\, \aa^*_{\b\eta}$.
  
\item In 1993, M. Fern\'andez \cite{fernandez93thesis} studied an
extension of the Calculus of Constructions with higher-order rewriting
{\em \`a la\,} Jouannaud-Okada \cite{jouannaud91lics}, the $\la
R$-cube~: $\cC=\, \abr^* \cup ~\als[\b R]$.

\item In 1994, E. Poll and P. Severi \cite{poll94lfcs} studied the
PTS's with abbreviations ({\tt let x= ... in ...})~: $\cC=\, \aa^*_\b
\cup \aa^*_\d$ where $\a_\d$ is the replacement of an abbreviation by
its definition.
  
\item In 1994, B. Werner \cite{werner94thesis} studied an extension of
the Calculus of Constructions with inductive types, the Calculus of
Inductive Constructions (CIC), introduced by T. Coquand and C. Paulin
in 1988 \cite{coquand88colog}~: $\cC=\, \aa^*_{\b\eta\io}$ where $\ai$
is the reduction relation associated with the elimination schemas of
inductive types.
  
\item Between 1995 and 1998, G. Barthe and his co-authors
\cite{barthe95hoa,barthe96csl,barthe97alp,barthe98icalp} considered
different extensions of the Calculus of Constructions or of the PTS's
with conversion relations more or less abstract, often based on
rewriting {\em \`a la\,} Jouannaud-Okada \cite{jouannaud91lics}, hence
extending the work of M. Fern\'andez \cite{fernandez93thesis}.
\end{lst}

In all this work, basic properties well known in the case of PTS's
must be proved again since new constructions or a new conversion rule
$\cC$ is introduced. This is why it appears useful for us to study the
properties of PTS's equipped with an abstract conversion relation
$\cC$.

~

Such a need is not new since it has already been undertaken in formal
developments~:

\begin{lst}{--}
\item In 1994, R. Pollack \cite{pollack94thesis} formally proved in
LEGO \cite{lego92}, an implementation of ECC with inductive types,
that type checking in ECC (without inductive types) is decidable (by
assuming of course that the calculus is strongly normalizing). Type
checking is saying if, in some environment $\G$, a term $t$ is of type
$T$ ({\em i.e.\,} is a proof of $T$). To this end, he showed many
properties of PTS's in the case of a conversion relation $\cC=\,\le$
reflexive, transitive and stable by substitution and context.
  
\item In 1999, B. Barras \cite{barras99thesis} formally proved in Coq
\cite{coq99}, another implementation of ECC with inductive types, that
type checking of Coq (so, with inductive types) is decidable (again of
course by assuming that the calculus is strongly normalizing). To this
end, he also showed some properties of PTS's in the case of a
conversion relation $\cC=\,\le$ also reflexive, transitive and stable
by substitution and context. In fact, he considered an extension of
PTS's with a schema of typing rules for introducing new constructions
in a generic way (abbreviations, inductive types, elimination
schemas).
\end{lst}

~

Hence, on the one hand, we make fewer assumptions on the conversion
relation $\cC$ than R. Pollack or B. Barras.  This is justified by the
fact that, in the work of M. Fern\'andez \cite{fernandez93thesis} for
example, to prove that reduction preserves typing, they use the fact
that the conversion relation is not transitive. On the other hand, our
typing rule for function symbols is not as general as the one of
B. Barras.




\section{Definition}


\begin{dfn}[Environment]
\label{def-env}
\index{environment|inddef[def-env]}
\index{FV@$\FV(\G)$|inddef[def-env]}
\index{$\sle$|inddef[def-env]}
\index{dom@$\dom(\G)$|inddef[def-env]}
\index{domain!environment|inddef[def-env]}
\index{E@$\cE$|inddef[def-env]}
\index{$\vide$|inddef[def-env]}

An {\em environment\,} $\G$ is a list of pairs $x_i\!:\!T_i$ made of a
variable $x_i$ and a term $T_i$. We denote by $\vide$ the empty
environment, by $\cE$ the set of environments and by $x_i\G$ the term
$T_i$ associated to $x_i$ in $\G$. The set of {\em free variables\,}
of an environment $\G$ is $\FV(\G)= \bigcup \{\FV(x\G)~|~
x\in\dom(\G)\}$. The {\em domain\,} of an environment $\G$ is the set
of variables to which $\G$ associates a term. Given two environments
$\G$ and $\G'$, $\G$ is {\em included\,} in $\G'$, written
$\G\sle\G'$, if all the elements of $\G$ occur in $\G'$ in the same
ordering.
\end{dfn}


\newcommand{\tf}{\tau_f}
\newcommand{\tg}{\tau_g}
\newcommand{\tF}{\tau_F}
\newcommand{\tG}{\tau_G}
\newcommand{\tc}{\tau_c}
\newcommand{\td}{\tau_d}
\newcommand{\tC}{\tau_C}
\newcommand{\tD}{\tau_D}

\newcommand{\ti}{\tau^i}

\begin{dfn}[Type assignment]
\label{def-typ-assign}
\index{assignment!type|inddef[def-typ-assign]}
\index{type!assignment|inddef[def-typ-assign]}
\index{tau@$\tau,\tau_f$|inddef[def-typ-assign]}
\index{Gammaf@$\G_f$|inddef[def-typ-assign]}

A {\em type assignment\,} is a function $\tau$ from $\cF$ to $\cT$
which, to a symbol $f$ of arity $n$, associates a closed term $\tf$ of
the form $\pvxT U$ where $|\vx|=n$. We will denote by $\G_f$ the
environment $\vx:\vT$.
\end{dfn}


\begin{dfn}[TSM]
\label{def-tsm}
\index{TSM|inddef[def-tsm]}
\index{Type System!Modulo|inddef[def-tsm]}
\index{TSM!beta@$\b$|inddef[def-tsm]}
\index{TSM!eta@$\eta$|inddef[def-tsm]}
\index{axiom|inddef[def-tsm]}
\index{rule!product formation|inddef[def-tsm]}
\index{A@$\cA$|inddef[def-tsm]}
\index{B@$\cB$|inddef[def-tsm]}
\index{C@$\cC$|inddef[def-tsm]}

A {\em Type System Modulo\,} (TSM) is a sorted $\la$-system
$(\cS,\cF,\cX)$ with~:
\begin{lst}{--}
\item a set of {\em axioms\,} $\cA\sle\cS^2$,
\item a set of product formation {\em rules\,} $\cB\sle\cS^3$,
\item a type assignment $\tau$,
\item a {\em conversion relation\,} $\cC\sle\cT^2$.
\end{lst}

\noindent
A $\b$TSM (resp. $\eta$TSM) is a TSM such that $\ad_\b\sle\cC$
(resp. $\ad_\eta\sle\cC$).
\end{dfn}


\newcommand{\thw}{\th\!\!\!_w}
\newcommand{\ths}{\th\!\!\!_s}

\newcommand{\E}{\mb{E}}
\renewcommand{\T}{\mb{T}}
\newcommand{\C}{\mb{C}}
\newcommand{\K}{\mb{K}}
\renewcommand{\P}{\mb{P}}
\renewcommand{\O}{\mb{O}}

\begin{dfn}[Typing]
\label{def-th}
\label{def-ths}
\label{def-thw}
\index{$\th$|inddef[def-th]}
\index{type|inddef[def-th]}
\index{typing!TSM|inddef[def-th]}
\index{valid!environment|inddef[def-th]}
\index{T@$\T,\T^s_0,\T^s_1$|inddef[def-th]}
\index{predicate|inddef[def-th]}
\index{object|inddef[def-th]}
\index{E@$\E$|inddef[def-th]}
\index{$\G\th t:T$|inddef[def-th]}

The typing relation of a TSM is the smallest ternary relation
$\th\,\sle \cE\times \cT\times \cT$ defined by the inference rules of
Figure~\ref{fig-th}. Compared with PTS's, we have a new rule, (symb),
for typing the symbols and, in the conversion rule (conv), instead of
the $\b$-conversion, we have an abstract conversion relation $\cC$. A
term $t$ is {\em typable\,} if there exists an environment $\G$ and a
term $T$ such that $\G\th t:T$ ($T$ is a {\em type\,} of $t$ in
$\G$). An environment $\G$ is {\em valid\,} if there exists a term
typable in $\G$. In the rule (symb), the premise ``$\G$ valid'' is
therefore useful only if $f$ is of null arity ($n=0$).

\begin{lst}{--}
\item $\T= \{t\in\cT ~|~ \ex \G\in\cE, \ex T\in\cT,~ \G\th t:T\}$ is
the set of typable terms,
\item $\T^s_0= \{T\in\cT ~|~ \ex \G\in\cE,~ \G\th T:s\}$ is the set of
{\em predicates\,} of sort $s$,
\item $\T^s_1= \{t\in\cT ~|~ \ex \G\in\cE, \ex T\in\cT,~ \G\th t:T$
and $\G\th T:s\}$ is the set of {\em objects\,} of sort $s$,
\item $\E= \{\G\in\cE ~|~ \ex t,T\in\cT,~ \G\th t:T\}$ is the set of
valid environments.
\end{lst}
\end{dfn}


\begin{figure}[ht]
\centering
\vs[5mm]
\caption{TSM typing rules\label{fig-th}}
\index{typing!TSM|indfig[fig-th]}
\index{TSM|indfig[fig-th]}

\begin{tabular}{rcc}
\\ (ax) & $\cfrac{}{\th s_1:s_2}$ & $((s_1,s_2) \in \cA)$\\

\\ (symb) & $\cfrac{
\begin{array}{c}
\th \tf:s \quad \G \mbox{ valid}\\
\G\th t_1:T_1\g \quad \ldots \quad \G\th t_n:T_n\g\\
\end{array}
}{\G\th f(\vt):U\g}$ &
\hs[-3mm]
$\begin{array}{c}
(f\in \cF^s_n,\\
\tf = \pvxT U,\\
\g = \vxt)\\
\end{array}$\\

\\ (var) & $\cfrac{\G\th T:s}{\GxT \th x:T}$
& $(x\in \cX^s \moins \dom(\G))$ \\

\\ (weak) & $\cfrac{\G\th t:T \quad \G\th U:s}{\GxU \th t:T}$
& $(x \in \cX^s \moins \dom(\G))$ \\

\\ (prod) & $\cfrac{\G\th T:s_1 \quad \GxT \th U:s_2}
{\G\th \px{T}U:s_3}$
& $((s_1,s_2,s_3) \in \cB)$\\

\\ (abs) & $\cfrac{\GxT \th u:U \quad \G\th \px{T}U:s}
{\G\th \lx{T}u:\px{T}U}$\\

\\ (app) & $\cfrac{\G\th t:\px{U}V \quad \G\th u:U}
{\G\th tu:V\xu}$\\

\\ (conv) & $\cfrac{\G\th t:T \quad \G\th T':s'}{\G\th t:T'}$
& $(T ~\cC~ T')$\\

\alaligne
\end{tabular}
\end{figure}


\newpage

\begin{rem}[Conversion]\hfill
\label{rem-conv}
\index{C@$\cC_\G,\C_\G$|indrem[rem-conv]}
\index{$\ths$|indrem[rem-conv]}
\index{$\thw$|indrem[rem-conv]}
\index{(weak')|indrem[rem-conv]}
\index{(conv')|indrem[rem-conv]}

It might seem more natural to define (conv) in a symmetric way by
adding the premise $\G\th T:s$ or the premise $\G\th T:s'$. We have
chosen this definition for two reasons. First, it is defined in this
way in the reference papers on PTS's
\cite{geuvers91jfp,barendregt92book}. Second, from a practical point
of view, for type checking, this avoids an additional test. However,
we will see in Lemma~\ref{lem-equiv-conv} that, for many TSM's, we
have $\G\th T:s'$. We will denote by $\ths$ the typing relation
defined by the same inference rules as for $\th$ but with (conv)
replaced by~:

\begin{center}
  (conv') \quad $\cfrac{\G\th t:T \quad \G\th T:s \quad \G\th
    T':s}{\G\th t:T'}$ \quad $(T ~\cC~ T')$
\end{center}

\noindent
We will show the equivalence between $\ths$ and $\th$ in
Lemma~\ref{lem-equiv-ths}.

In the same way, in {\em full\,} TSM's ($\all s_1,s_2\IN\cS, \ex
s_3\IN\cS,\, (s_1,s_2,s_3)\in\cB$), the rule (abs) can be replaced
by~:

\begin{center}
  (abs') \quad $\cfrac{\GxT \th u:U} {\G\th \lx{T}u:\px{T}U}$ \quad
  $(U\notin\cS$ or $\ex s\IN\cS,\, (U,s)\in\cA)$
\end{center}
\end{rem}

Finally, we will denote by $\thw$ the typing relation defined by the
same inference rules as for $\th$ but with (weak) replaced by~:

\begin{center}
  (weak') \quad $\cfrac{\G\th t:T \quad \G\th U:s}{\GxU\th t:T}$
  \quad $(x\in \cX^s\moins\dom(\G),\, t\in \cX\cup\cS)$
\end{center}

\noindent
that is, where weakening is restricted to variables and sorts ($t\in
\cX\cup\cS$). We will show the equivalence between $\thw$ and $\th$ in
Lemma~\ref{lem-equiv-thw}.


\begin{rem}[Arity]\hfill
\label{rem-arity}
\index{arity!symbol|indrem[rem-arity]}

On can wonder why symbols are equipped with a fixed arity since, in
general, in $\la$-calculus, one is used to consider higher-order
constants. Of course, having an arity is not a restriction since, to a
symbol $f$ of arity $n$ and type $\pvxT U$, one can always associate a
curried version $f^c$ of null arity defined by the rewrite rule $f^c
\a [\vx:\vT]f(\vx)$. Furthermore, in practice, the existence of
arities can be masked by doing $\eta$-expansions if $f$ is not applied
to sufficient arguments, or by doing additional applications if $f$ is
applied to too many arguments.  However, without arities, we would
have a simpler presentation where the rule (symb) would be reduced
to~:

\begin{center}
(symb') \quad $\cfrac{\th \tf:s}{\th f:\tf}$ \quad $(f\in\cF^s)$
\end{center}

To our knowledge, except the works of Jouannaud and Okada
\cite{jouannaud97tcs} and of G. Barthe and his co-authors
\cite{barthe95hoa,barthe96csl,barthe97alp,barthe98icalp}, most of the
other works on the combination of typed $\la$-calculi and rewriting
\cite{breazu88lics,breazu89icalp,fernandez93thesis} do not use arities
for typing symbols. We have chosen to use arities for technical
reasons. With the method we use for proving strong normalization, we
need an application $uv$ not to be a rewriting redex (see
Chapter~\ref{chap-ats} for an explanation of these notions and
Lemma~\ref{lem-cor-schema-int}, case $T=\px{U}V$, (b), (R3) for the
use of this property).  The introduction of arities enable us to
syntactically distinguish between the application of the
$\la$-calculus and the application of a symbol. But one may wonder
whether this notion is really necessary.
\end{rem}


\begin{dfn}[Well-typed substitution]
\label{def-wt-sub}
\index{substitution!well-typed|inddef[def-wt-sub]}
\index{well-typed!substitution|inddef[def-wt-sub]}

Given two valid environments $\G$ and $\D$, a substitution $\t$ is
{\em well typed between $\G$ and $\D$\,}, $\t: \G\a\D$, if, for all
$x\in\dom(\G)$, $\D\th x\t:x\G\t$.
\end{dfn}

For example, in the rule (symb), we have $\g:\G_f\a\G$ where $\G_f=
\vx:\vT$.


%% file: tsm-prop.tex



\section{Properties}

In this section and the following one, we prove some properties of
TSM's that are well known for PTS's (except
Lemma~\ref{lem-conv-env}). Apart from the new case (symb) that we will
detail each time, proofs are identical to the ones for PTS's. The fact
that, in (conv), $\aa_\b^*$ is replaced by an arbitrary relation $\cC$
is not important. For more details, the reader is invited to look at
\cite{geuvers91jfp,barendregt92book,geuvers93thesis}.


\begin{lem}[Free variables]
\label{lem-fv}
\index{variable!free|indlem[lem-fv]}

Let $\G= \vx:\vT$ be an environment. If $\G\th t:T$ then~:
\begin{enumalphai}
\item the $x_i$'s are distinct from one another,
\item $\FV(t) \cup \FV(T) \sle \dom(\G)$,
\item for all $i$, $\FV(T_i) \sle \{x_1,\ldots,x_{i-1}\}$.
\end{enumalphai}
\end{lem}

\begin{prf}
  By induction on $\G\th t:T$. We only detail the new case (symb). (a)
  and (c) are true by induction hypothesis. Let us see (b) now. By
  induction hypothesis, $\FV(\tf)= \vide$ and, for all $i$, $\FV(t_i)
  \sle \dom(\G)$. Therefore, $\FV(f(\vt)) \sle \dom(\G)$. Hence,
  $\FV(U\g) \sle \dom(\G)$ since $\FV(U) \sle \{\vx\}$ and $\g=
  \vxt$.\cqfd
\end{prf}


\begin{lem}[Subterms]
\label{lem-subterm}
\index{subterm|indlem[lem-subterm]}

  If a term is typable then all its subterms are typable.
\end{lem}

\begin{prf}
  By induction on $\G\th t:T$. In the case of (symb), by induction
  hypothesis, for all $i$, all the subterms of $t_i$ are
  typable. Therefore, all the subterms of $f(\vt)$ are typable.\cqfd
\end{prf}


\begin{lem}[Environment]
\label{lem-env}
\index{environment|indlem[lem-env]}

  Let $\G= \vx:\vT$ be a valid environment.
\begin{enumalphai}
\item If $x_i$ is of sort $s$ then $x_1\!:\!T_1, \ldots,
  x_{i-1}\!:\!T_{i-1} \th T_i:s$.
\item For all $i$, $x_1\!:\!T_1, \ldots, x_i\!:\!T_i \th
  x_i\!:\!T_i$.
\end{enumalphai}
\end{lem}

\begin{prf}
  By (var), (b) is an immediate consequence of (a). We prove (a) by
  induction on $\G\th t:T$. In the case (symb), as $\G$ is valid,
  there exists $v$ and $V$ such that $\G\th v:V$. Therefore, by
  induction hypothesis, (a) is true.\cqfd
\end{prf}


~

The following lemma is a form of $\alpha$-equivalence on the variables
of an environment.

\begin{lem}[Replacement]
\label{lem-repl}
\index{replacement|indlem[lem-repl]}

If $\G,y\!:\!W,\G'\th t:T$, $y\in \cX^s$ and $z\in \cX^s\moins
\dom(\G,y\!:\!W,\G')$ then $\G,z\!:\!W,\G'\{y\to z\}\th t\{y\to z\}:
T\{y\to z\}$.
\end{lem}

\begin{prf}
  By induction on $\G,y\!:\!W,\G'\th t:T$. Let $\t=\{y\to z\}$, $\D=
  \G,y\!:\!W,\G'$ and $\D'= \G,z\!:\!W,\G'\t$. In the case (symb), by
  induction hypothesis, we have $\D'$ valid and, for all $i$, $\D'\th
  t_i\t:T_i\g\t$. Therefore, by (symb), $\D'\th f(\vt\t):U\g\t$.\cqfd
\end{prf}


\begin{lem}[Weakening]
\label{lem-weak}
\index{weakening|indlem[lem-weak]}

  If $\G\th t:T$ and $\G\sle\G'\in\E$ then $\G'\th t:T$.
\end{lem}

\begin{prf}
  By induction on $\G\th t:T$. In the case (symb), by induction
  hypothesis, we have $\G'$ valid and, for all $i$, $\G'\th
  t_i:T_i\g$. Therefore, by (symb), $\G'\th f(\vt):U\g$.\cqfd
\end{prf}


\begin{lem}[Transitivity]
\label{lem-trans}
\index{transitivity|indlem[lem-trans]}

  Let $\G$ and $\D$ be two valid environments. If $\G\th t:T$ and, for
  all $x\in\dom(\G)$, $\D\th x:x\G$, that we will denote by $\D\th
  \G$, then $\D\th t:T$.
\end{lem}

\begin{prf}
  By induction on $\G\th t:T$. In the case (symb), by induction
  hypothesis, we have $\D$ valid and, for all $i$, $\D\th
  t_i:T_i\g$. Therefore, by (symb), $\D\th f(\vt):U\g$.\cqfd
\end{prf}


\begin{lem}[Weak permutation]
\label{lem-weak-permut}
\index{permutation!weak|indlem[lem-weak-permut]}

If $\G,y\!:\!A,z\!:\!B,\G'\th t:T$ and $\G\th B:s$ then
$\G,z\!:\!B,y\!:\!A,\G'\th t:T$.
\end{lem}

\begin{prf}
  Let $\D= \G,y\!:\!A,z\!:\!B,\G'$ and $\D'= \G,z\!:\!B,y\!:\!A,\G'$.
  By transitivity, it suffices to prove that $\D'$ is valid and that
  $\D'\th\D$. And to this end, it suffices to prove that $\D'$ is
  valid. By the Environment Lemma, we have $\G\th A:s'$ and, by
  hypothesis, we have $\G\th B:s$. Therefore, by weakening,
  $\G,z\!:\!B,y\!:\!A$ is valid. Assume that $\G'= \vx:\vT$ and let
  $\D_i= \G,y\!:\!A,z\!:\!B,x_1\!:\!T_1,\ldots,x_i\!:\!T_i$ and
  $\D'_i= \G,z\!:\!B,y\!:\!A,x_1\!:\!T_1,\ldots,x_i\!:\!T_i$. We prove
  by induction on $i$ that $\D'_i$ is valid. We have already proved
  that $\D'_0$ is valid. Assume that $\D'_i$ is valid. By the
  Environment Lemma, $\D_i\th T_{i+1}:s_{i+1}$. As $\D'_i\th\D_i$,
  $\D'_i\th T_{i+1}:s_{i+1}$ and $\D'_{i+1}$ is valid. Therefore,
  $\D'$ is valid and $\D'\th t:T$.\cqfd
\end{prf}


\begin{lem}[Equivalence of $\thw$ and $\th$]
\label{lem-equiv-thw}
\index{$\thw$|indlem[lem-equiv-thw]}

~$\thw =\, \th$.
\end{lem}

\begin{prf}
  First of all, it is clear that $\thw \sle\, \th$. We prove the
  reverse by induction on $\G\th t:T$. The only difficult case is of
  course (weak)~: from $\G\th t:T$ and $\G\th U:s$, we get $\GxU\th
  t:T$. By induction hypothesis, we have $\G\thw~ t:T$ and $\G\thw~
  U:s$. One has to modify the proof of $\G\thw~ t:T$ by adding
  $x\!:\!U$ at the appropriate places in order to obtain a proof of
  $\GxU\thw~ t:T$. See Lemma 4.4.21 page 102 in \cite{geuvers93thesis}
  for more details.\cqfd
\end{prf}


~

Now, let us see what we can do about the derivations of $\G\th t:T$
and the form of $T$ with respect to $t$. To this end, we introduce
relations related to the rule (conv).

\newcommand{\cv}[1][]{~\cC_{#1}^*~}
\newcommand{\cvG}{\cv[\G]}
\newcommand{\CV}[1][]{~\C_{#1}^*~}
\newcommand{\CVG}{\CV[\G]}
\newcommand{\CVGp}{\CV[\G']}

\begin{dfn}[Conversion relations]
\label{def-rel-conv}
\index{C@$\cC_\G,\C_\G$|inddef[def-rel-conv]}
\index{C@$\C$|inddef[def-rel-conv]}

\hfill
\begin{lst}{--}
\item $T ~\cC_\G~ T'$ \,iff\, $T ~\cC~ T'$ and there exists $t$, $t'$
and $s'$ such that $\G \th t:T$, $\G \th t':T'$ and $\G \th T':s'$,
\item $T ~\C_\G~ T'$ \,iff\, $T ~\cC_\G~ T'$ and there exists $s$ such
that $\G \th T:s$,
\item $\G ~\C~ \G'$ \,iff\, $\G = \vx:\vT$, $\G' = \vx:\vT'$ and,
either $|\vx|=0$, or there exists $j$ such that $T_j ~\C_{x_1:T_1,
\ldots, x_{j-1}:T_{j-1}}~ T_j'$ and, for all $i\neq j$, $T_i = T_i'$.
\end{lst}
\end{dfn}

We have $\C_\G \sle\, \cC_\G$ but, as opposed to $\C_\G$, $\cC_\G$ is
not defined in a symmetric way. This comes from the asymmetry of the
rule (conv) which requires $\G\th T':s'$ but not $\G\th T:s$. However,
we will see in Lemma~\ref{lem-equiv-conv} that, for many TSM's, these
two relations are equal.


\begin{lem}[Inversion]
\label{lem-inv}
\index{inversion!TSM|indlem[lem-inv]}

Assume that $\G \th t:T$.
\begin{lst}{--}
\item If $t=s$ then there exists $s'$ such that $(s,s') \in \cA$ and
  $s'\cvG T$.
\item If $t=f(\vt)$, $f\in \cF^s$, $\tf = \pvxT U$ and $\g = \vxt$
  then $\th \tf:s$, $\g : (\vx:\vT) \a \G$ and $U\g \cvG T$.
\item If $t=x\in \cX^s$ then $\G \th x\G:s$ and $x\G \CVG T$.
\item If $t=\px{U}V$ then there exists $(s_1,s_2,s_3) \in \cP$ such that
  $\G \th U : s_1$, $\GxU \th V:s_2$ and $s_3 \cvG T$.
\item If $t=\lx{U}v$ then there exists $V$ such that $\GxU \th v:V$ and
  $\px{U}V \CVG T$.
\item If $t=uv$ then there exists $V$ and $W$ such that $\G \th
  u:\px{V}W$, $\G \th v:V$ and $W\xv \cvG T$.
\end{lst}
\end{lem}

\begin{prf}
  A typing derivation always finishes by a rule distinct from (weak)
  and (conv) followed by a possibly empty sequence of (weak)'s and
  (conv)'s. We hence get the term to which $T$ is convertible. For
  typing judgments, it suffices to do a weakening to express them in
  $\G$.\cqfd
\end{prf}


\begin{lem}[Environment conversion]
\label{lem-conv-env}
\index{environment conversion|indlem[lem-conv-env]}

If $\G\th t:T$ and $\G ~\C~ \G'$ then $\G' \th t:T$.
\end{lem}

\begin{prf}
  We have $\G= \vx:\vT$, $\G'= \vx:\vT'$ and there exists $j$ such
  that $T_j ~\C_\D~ T_j'$ with $\D= x_1:T_1, \ldots, x_{j-1}:T_{j-1}$
  and, for all $i\neq j$, $T_i= T_i'$. By transitivity, it suffices to
  prove that, for all $i$, $\G'\th x_i:T_i$. Let $n= |\vx|$, $\G_1=
  x_1\!:\!T_1, \ldots, x_{j-1}\!:\!T_{j-1}$ and $\G_2=
  x_{j+1}\!:\!T_{j+1}, \ldots, x_n\!:\!T_n$. We proceed by induction
  on the size of $\G_2$.
  
  If $\G_2$ is empty then $\G= \G_1, x_j\!:\!T_j$ and $\G'= \G_1,
  x_j\!:\!T_j'$. Since $\G$ is valid, $\G_1$ is valid and, for all
  $i<j$, $\G_1\th x_i:T_i$. Since $T_j ~\C_{\G_1}~ T_j'$, there exists
  $s$ and $s'$ such that $\G_1\th T_j:s$ and $\G_1\th T_j':s'$.  By
  (weak), we get, for all $i<j$, $\G'\th x_i:T_i$, and by (var), we
  get $\G'\th x_j:T_j'$. From $\G_1\th T_j:s$, by (weak), we also get
  $\G'\th T_j:s$. Therefore, by (conv), $\G' \th x_j:T_j$.
  
  assume now that $\G_2= \G_3, x_n\!:\!T_n$. Let $\D= \G_1,
  x_j\!:\!T_j, \G_3$ and $\D'= \G_1, x_j\!:\!T_j', \G_3$. By induction
  hypothesis, for all $i<n$, $\D'\th x_i:T_i$. Since $\G$ is valid,
  there exists $s$ such that $\D\th T_n:s$. By transitivity, we get
  $\D'\th T_n:s$. Therefore, by (var), we get $\G'\th x_n:T_n$, and by
  (weak), $\G'\th x_i:T_i$.\cqfd
\end{prf}




\section{TSM's stable by substitution}


\begin{dfn}[TSM stable by substitution]
\label{def-tsm-subs}
\index{TSM!stable by substitution|inddef[def-tsm-subs]}

A TSM is stable by substitution if its conversion relation $\cC$ is
stable by substitution.
\end{dfn}


\begin{lem}[Substitution]
\label{lem-subs}
\index{substitution|indlem[lem-subs]}

If $\cC$ is stable by substitution, $\G\th t:T$ and $\t: \G\a\D$
then $\D\th t\t:T\t$.
\end{lem}

\begin{prf}
  By induction on $\G\th t : T$. In the case (symb), by induction
  hypothesis, we have $\D\th t_i\t:T_i\g\t$. Therefore, by (symb),
  $\D\th f(\vt\t):U\g\t$.\cqfd
\end{prf}


\begin{cor}
  If $\cC$ is stable by substitution, $\GxU, \G'\th t:T$ and $\G\th
  u:U$ then $\G, \G'\xu\th t\xu:T\xu$.
\end{cor}

\begin{prf}
  We have to prove that $\t= \xu$ is a well-typed substitution from
  $\GxU,\G'$ to $\G,\G'\t$. We proceed by induction on the size of
  $\G'$. If $\G'$ is empty, this is immediate since $\G$ is valid and
  $\G\th u:U$. Assume now that $\G'= \G'',y\!:\!V$. Let $\D=
  \GxU,\G''$ and $\D'= \G,\G''\t$. By induction hypothesis, $\t:
  \D\a\D'$. Since $\D\th V:s$, by substitution, we get $\D'\th
  V\t:s$. Therefore, by (var), $\D', y\!:\!V\t\th y:V\t$. Now, let
  $z\in \dom(\D)$. As $\D\th z:z\D$, by substitution, $\D'\th
  z:z\D\t$. Then, by (weak), $\D', y\!:\!V\t\th z:z\D\t$.\cqfd
\end{prf}


\begin{cor}
  If $\cC$ is stable by substitution, $\t_1: \G_0\a\G_1$ and $\t_2:
  \G_1\a\G_2$ then $\t_1\t_2: \G_0\a\G_2$.
\end{cor}

\begin{prf}
  Let $x\in \dom(\G_0)$. Since $\t_1: \G_0\a\G_1$, by substitution, we
  get $\G_1\th x\t_1:x\G_0\t_1$, and since $\t_2: \G_1\a\G_2$, by
  substitution again, we get $\G_2\th x\t_1\t_2:x\G_0\t_1\t_2$.\cqfd
\end{prf}


\begin{dfn}[Maximal sort]
\label{def-sort-max}
\index{maximal!sort|inddef[def-sort-max]}

A sort $s$ is {\em maximal\,} if there does not exist any sort $s'$
such that $(s,s')\in \cA$.
\end{dfn}


\begin{lem}[Correctness of types]
\label{lem-cor-types}
\index{correctness of types|indlem[lem-cor-types]}

If $\cC$ is stable by substitution and $\G\th t:T$ then, either $T$ is
a maximal sort, or there exists a sort $s$ such that $\G\th T:s$. In
other words, $\T= \bigcup \{ \T^s_0 \cup \T^s_1 ~|~ s \in \cS \}$.
\end{lem}

\begin{prf}
  By induction on $\G\th t:T$. In the case (symb), we have $\th
  \tf:s$. By inversion, there exists $s'$ such that $\G_f\th U:s'$.
  As $\g: \G_f\a\G$, by substitution, $\G\th U\g:s'$.\cqfd
\end{prf}


\begin{lem}[Inversion for TSM's stable by substitution]
\label{lem-inv-subs}
\index{inversion!TSM stable by substitution|indlem[lem-inv-subs]}

Assume that $\G\th t:T$.
\begin{lst}{--}
\item If $t=s$ then there exists $s'$ such that $(s,s') \in \cA$ and
  $s'\cvG T$.
\item If $t=f(\vt)$, $f\in \cF^s$, $\tf= \pvxT U$ and $\g= \vxt$
  then $\th \tf:s$, $\g : (\vx:\vT) \a \G$ and $U\g \CVG T$.
\item If $t=x\in \cX^s$ then $\G\th x\G:s$ and $x\G \CVG T$.
\item If $t=\px{U}V$ then there exists $(s_1,s_2,s_3) \in \cP$ such that
  $\G\th U:s_1$, $\GxU\th V:s_2$ and $s_3 \cvG T$.
\item If $t=\lx{U}v$ then there exists $V$ such that $\GxU\th v:V$ and
  $\px{U}V \CVG T$.
\item If $t=uv$ then there exists $V$ and $W$ such that $\G\th
  u:\px{V}W$, $\G\th v:V$ and $W\xv \CVG T$.
\end{lst}
\end{lem}

\begin{prf}
Only the cases $t=f(\vt)$ and $t=uv$ have been modified.
\begin{lst}{--}
\item $t=f(\vt)$. By inversion, $\th \tf:s$, $\g: (\vx:\vT)\a\G$ and
$U\g \cvG T$. By inversion again, there exists $s'$ such that
$\vx:\vT\th U:s'$. Therefore, by substitution, $\G\th U\g:s'$ and $U\g
\CVG T$.
\item $t=uv$. By inversion, there exists $V$ and $W$ such that $\G\th
u:\px{V}W$, $\G\th v:V$ and $W\xv \cvG T$. By correctness of types,
there exists $s$ such that $\G\th \px{V}W : s$. By inversion, there
exists $s'$ such that $\GxV\th W:s'$. Therefore, by substitution,
$\G\th W\xv:s'$ and $W\xv \CVG T$.\cqfd
\end{lst}
\end{prf}




\section{Logical TSM's}


We now introduce an important class of TSM's for which $\b$-reduction
preserves typing.

\begin{dfn}[Logical TSM]
\label{def-logic}
\index{logical TSM|inddef[def-logic]}
\index{TSM!logical|inddef[def-logic]}

A TSM is {\em logical\,} if its conversion relation is {\em product
compatible\,}~:
\begin{center}
  $\px{T}U \CVG \p{x'}{T'}U'$ implies $T \CVG T'$ and $U \CV[\G,x:T]
  U'\{x'\to x\}$.
\end{center}
\end{dfn}

~

$T \CVG T'$ means that there exists a sequence of terms $\vT$ such
that $T_0= T ~\C_\G~ T_1$ \ldots $T_{n-1} ~\C_\G~ T_n= T'$. So, there
is no reason {\em a priori\,} to take $U \CV[\G,x:T] U'\{x'\to x\}$
instead of $U \CV[\G,x:T_i] U'\{x'\to x\}$ with $i\neq 0$. However, as
$T \CVG T'$, by environment conversion, this choice is not important.

Product compatibility is not a new condition and appears in all the
previously cited works but, to our knowledge, it has never received
any special name.

All the TSM's cited at the beginning of this chapter are logical. In
the case where $\cC= \aa^*$ with $\a$ a confluent reduction relation,
it is clear that $\cC$ is product compatible. To prove this property
without using confluence is more delicate. This is the case of PTS's
with $\eta$-reduction \cite{geuvers93thesis} or of the $\la R$-cube
\cite{fernandez93thesis}, an extension of the Calculus of
Constructions with higher-order rewriting {\em \`a la\,}
Jouannaud-Okada at the object-level.


\begin{lem}[Subject reduction for $\b$]
\label{lem-cor-ab}
\index{subject reduction!beta@$\b$|indlem[lem-cor-ab]}

  In a logical $\b$TSM, if $\G\th t:T$ and $t\ab t'$ then $\G\th
  t':T$.
\end{lem}

\begin{prf}
  We will say that an environment $\vx:\vT$ $\b$-rewrites to an
  environment $\vx':\vT'$, written $\vx:\vT \ab \vx':\vT'$, if $\vx=
  \vx'$ and there exists $j$ such that $T_j \ab T'_j$ and, for all
  $i\neq j$, $T_i= T'_i$. We simultaneously prove that~:

\begin{enumalphaii}
\item if $t\ab t'$ then $\G\th t':T$,
\item if $\G\ab \G'$ then $\G'\th t:T$,
\end{enumalphaii}

\noindent
by induction on $\G\th t:T$.

\begin{lst}{}


\item [\bf(ax)] $\th s_1:s_2$ \quad $((s_1,s_2)\in\cA)$
  
  No $\b$-reduction is possible in $s_1$ or in $\G=\vide$.

\item [\bf(symb)] $\cfrac{\th \tf:s \quad \G \mbox{ valid} \quad \G\th
    t_1:T_1\g \ldots \G\th t_n:T_n\g}{\G\th f(\vt):U\g}$
  \quad $\begin{array}{c}
(f\in\cF^s_n,\\
\tf= \pvxT U,\\
\g= \vxt)\\
\end{array}$

\begin{enumalphai}
\item If $f(\vt)\ab t'$ then $t'= f(\vt')$ with $j$ such that $t_j\ab
t_j'$ and, for all $i\neq j$, $t_i=t_i'$. By induction hypothesis, we
have, for all $i$, $\G\th t_i':T_i\g$. Let $\g'= \{\vx\to\vt'\}$. We
have $U\g' \als\b U\g$ and, for all $i$, $T_i\g \ab^* T_i\g'$. As
$\ad_\b \sle \cC$, $U\g' ~\cC~ U\g$ and $T_i\g ~\cC~ T_i\g'$. If we
prove that every $T_i\g'$ is typable in $\G$ by a sort then, by
(conv), we have $\G \th t_i':T_i\g'$ and, by (symb), $\G\th
t':U\g'$. It then suffices to prove that $U\g$ is typable by a sort in
$\G$ to apply again (conv) and conclude that $\G\th t':U\g$.
  
  \hs Let us begin by verifying that $U\g$ is typable by a sort. We
  have $\th \tf:s$. By inversion, $\g: (\vx:\vT)\a\G$ and there exists
  $s'$ such that $\vx:\vT\th U:s'$. By substitution, we therefore get
  $\G\th U\g:s'$.
  
  \hs Now, we are going to prove that every $T_i\g'$ is typable by a
  sort. To this end, it suffices to prove that $\g':
  (\vx:\vT)\a\G$. Indeed, since $\th \tf:s$, by inversion, every $T_i$
  is typable by a sort in $\G_{i-1}= x_1\!:\!T_1, \ldots,
  x_{i-1}\!:\!T_{i-1}$. Let us prove by induction on $i$ that $\g':
  \G_i\a\G$.
  
  \hs For $i=0$, there is nothing to prove. Assume therefore that
  $\g': \G_i\a\G$. Then $\g': \G_{i+1}\a\G$ if $\G\th
  t_{i+1}':T_{i+1}\g'$. We know that $\G\th t_{i+1}':T_{i+1}\g$,
  $T_{i+1}\g \ab^* T_{i+1}\g'$ and that there exists $s$ such that
  $\G_i \th T_{i+1}:s$. Therefore, by substitution, $\G\th
  T_{i+1}\g':s$ and, by (conv), $\G\th t_{i+1}':T_{i+1}\g'$.
  
\item If $\G\ab \G'$ then, by induction hypothesis, $\G'$ is valid and
$\G'\th t_i:T_i\g$. Therefore, by (symb), $\G'\th f(\vt):U\g$.
\end{enumalphai}


\item [\bf(var)] $\cfrac{\G\th T:s}{\GxT\th x:T}$

\begin{enumalphai}
\item No $\b$-reduction is possible in $x$.

\item There is two cases, depending on where takes place the
$\b$-reduction~:

\begin{lst}{--}
\item $\G\ab \G'$. By induction hypothesis, $\G'\th T:s$.
  Therefore, by (var), $\GpxT\th x:T$.
  
\item $T\ab T'$. By induction hypothesis, $\G\th T':s$. Therefore,
  by (var), $\GxT'\th x:T'$. As $\ad_\b \sle \cC$, $T' ~\cC~ T$.
  As $\G\th T:s$, by (conv), $\GxT'\th x:T$.
\end{lst}
\end{enumalphai}


\item [\bf(weak)] $\cfrac{\G\th t:T \quad \G\th U:s}{\GxU\th t:T}$

\begin{enumalphai}
\item If $t\ab t'$ then, by induction hypothesis, $\G\th
  t':T$. As $\G\th U:s$, by (weak), $\GxU\th t':T$.

\item There is two cases, depending on where takes place the
$\b$-reduction~:

\begin{lst}{--}
\item $\G\ab \G'$. By induction hypothesis, $\G'\th t:T$ and $\G'\th
U:s$. Therefore, by (weak), $\GpxU\th t:T$.
  
\item $U\ab U'$. By induction hypothesis, $\G\th U':s$. Therefore, by
(weak), $\G,x\!:\!U'\th t:T$.
\end{lst}
\end{enumalphai}


\item [\bf(prod)] $\cfrac{\G\th T:s_1 \quad \GxT\th U:s_2}{\G\th
    \px{T}U:s_3}$ \quad $((s_1,s_2,s_3)\in\cB)$

\begin{enumalphai}
\item There is two cases, depending on where takes place the
$\b$-reduction~:

\begin{lst}{--}
\item $T \ab T'$. By induction hypothesis, $\G\th T':s_1$ and
$\GxT'\th U:s_2$. Therefore, by (prod), we get $\G\th \px{T'}U:s_3$.

\item $U \ab U'$. By induction hypothesis, $\GxT\th U':s_2$.
Therefore, by (prod), $\G\th \px{T}U':s_3$.
\end{lst}

\item If $\G\ab \G'$ then, by induction hypothesis, $\G'\th T:s_1$ and
$\GpxT\th U:s_2$. Therefore, by (prod), $\G'\th \px{T}U:s_3$.
\end{enumalphai}


\item [\bf(abs)] $\cfrac{\GxT\th u:U \quad \G\th \px{T}U:s}{\G\th
    \lx{T}u:\px{T}U}$

\begin{enumalphai}
\item There is two cases, depending on where takes place the
$\b$-reduction~:

\begin{lst}{--}
\item $T \ab T'$. By induction hypothesis, $\GxT'\th u:U$ and $\G\th
\px{T'}U:s$. By (abs), $\G\th \lx{T'}u:\px{T'}U$.  As $\px{T'}U \al\b
\px{T}U$ and $\ad_\b \sle \cC$, $\px{T'}U ~\cC~ \px{T}U$. As $\G\th
\px{T}U:s$, by (conv), $\G\th \lx{T'}u:\px{T}U$.

\item $u\ab u'$. By induction hypothesis, $\GxT\th u':U$. As $\G\th
\px{T}U:s$, by (abs), $\G\th \lx{T}u':\px{T}U$.
\end{lst}

\item If $\G\ab \G'$ then, by induction hypothesis, $\GpxT\th u:U$ and
$\G'\th \px{T}U:s$. Therefore, by (abs), $\G'\th \lx{T}u:\px{T}U$.
\end{enumalphai}


\item [\bf(app)] $\cfrac{\G\th t:\px{U}V \quad \G\th u:U}{\G\th tu:V\xu}$

\begin{enumalphai}
\item There is three cases, depending on where takes place the
$\b$-reduction~:

\begin{lst}{--}
\item $t \ab t'$. By induction hypothesis, $\G\th t':\px{U}V$. As
$\G\th u:U$, by (app), $\G\th t'u:V\xu$.
  
\item $u \ab u'$. By induction hypothesis, $\G\th u':U$.  By (app),
$\G\th tu':V\xup$. As $V\xup \als\b V\xu$ and $\ad_\b \sle \cC$,
$V\xup ~\cC~ V\xu$. By inversion, there exists $s$ such that $\G\th
V\xu:s$. Therefore, by (conv), $\G\th tu':V\xu$.
  
\item $t= \lx{U'}v$ and $tu \ab v\xu$. By inversion, there exists $V'$
such that $\GxU'\th v:V'$ and $\px{U'}V' \CVG \px{U}V$. By product
compatibility, $U' \CVG U$ and $V' \CV[\G,x:U] V$.  By environment
conversion, $\GxU\th v:V'$ and, by (conv), $\GxU\th v:V$.
\end{lst}

\item If $\G\ab \G'$ then, by induction hypothesis, $\G'\th t:\px{U}V
\quad \G'\th u:U$. Therefore, by (app), $\G'\th tu:V\xu$.
\end{enumalphai}


\item [\bf(conv)] $\cfrac{\G\th t:T \quad T ~\cC~ T' \quad \G\th
    T':s'}{\G\th t:T'}$

\begin{enumalphai}
\item If $t\ab t'$ then, by induction hypothesis, $\G\th t':T$. As $T
~\cC~ T'$ and $\G\th T':s'$, by (conv), $\G\th t':T'$.
  
\item If $\G\ab \G'$ then, by induction hypothesis, $\G'\th t:T$ and
$\G'\th T':s'$. Therefore, by (conv), $\G'\th t:T'$.\cqfd
\end{enumalphai}

\end{lst}
\end{prf}


%% file: rts.tex



\chapter{Reduction Type Systems (RTS's)}
\markboth{CHAPTER \thechapter. REDUCTION TYPE SYSTEMS}{}
\label{chap-rts}

Now, we are going to study the case of TSM's whose conversion relation
$\cC$ is of the form $\ad$ with $\a$ a reduction relation. We shall
call such systems Reduction Type Systems (RTS). Except for ECC
\cite{luo90thesis} which uses a notion of subtyping, all the systems
previously mentioned at the beginning of Chapter~\ref{chap-tsm} are
RTS's, either because they are defined in this way
\cite{fernandez93thesis,barthe95hoa}, or because they are defined with
$\cC=\aa^*$ and $\a$ confluent, which is equivalent
\cite{paulin93tlca,geuvers93thesis,werner94thesis,poll94lfcs}. The
general study of such systems is justified by the fact that, in
\cite{fernandez93thesis}, the proof that reduction preserves typing
uses the fact that $\cC$ is of the form $\ad$.

The proofs of the Lemmas \ref{lem-conv-typ}, \ref{lem-sep},
\ref{lem-dep-var} and \ref{lem-strong-perm} are widely inspired from
the ones given by H. Geuvers and M.-J. Nederhof \cite{geuvers91jfp} or
H. Geuvers \cite{geuvers93thesis}.




\section{Definition}


\begin{dfn}[RTS]
\label{def-rts}
\index{pre-RTS|inddef[def-rts]}
\index{pre-RTS!admissible|inddef[def-rts]}
\index{admissible!pre-RTS|inddef[def-rts]}
\index{RTS|inddef[def-rts]}
\index{Type System!Reduction|inddef[def-rts]}
\index{subject reduction|inddef[def-rts]}

A {\em pre-RTS\,} is a TSM whose conversion relation $\cC$ is of the
form $\ad$ with $\a$ a relation stable by substitution and
context. The relation $\a$ is called the {\em reduction relation\,} of
the pre-RTS. A pre-RTS is {\em confluent\,} if its reduction relation
is confluent. An {\em RTS\,} is a pre-RTS which is {\em admissible\,},
that is, whose reduction relation {\em preserves typing\,}~: $\G\th
t:T$ and $t\a t'$ imply $\G\th t':T$, a property often called {\em
subject reduction\,}.
\end{dfn}

Any pre-RTS satisfies the following elementary properties~:


\begin{lem}
\label{lem-prop-elem}
\index{symmetry|indlem[lem-prop-elem]}
\index{stable by!substitution|indlem[lem-prop-elem]}
\index{stable by!context|indlem[lem-prop-elem]}
\index{preservation of sorts|indlem[lem-prop-elem]}
\index{sort!preservation|indlem[lem-prop-elem]}

The relation $\cC=\ad$ is~:
\begin{lst}{--}
\item symmetric~: $T ~\cC~ T'$ implies $T' ~\cC~ T$.
\item stable by substitution~: $T ~\cC~ T'$ implies $T\t ~\cC~ T'\t$.
\item stable by context~: $T ~\cC~ T'$ implies $C[T]_p ~\cC~ C[T']_p$.
\item preserves sorts~: $s ~\cC~ s'$ implies $s=s'$.
\end{lst}
\end{lem}

In ECC, the conversion relation $\cC$ is not symmetric and does not
preserve sorts. It would be interesting to try to formulate some
properties below in the more general framework of Cumulative pure Type
Systems (CTS) to which belongs ECC. To this end, the reader is refered
to the works of Z. Luo \cite{luo90thesis}, R. Pollack
\cite{pollack94thesis} and Barras \cite{barras99thesis}.


~

Subject reduction can be extended to types, environments and
substitutions~:

\begin{dfn}
  A substitution $\t$ {\em rewrites\,} to a substitution $\t'$, $\t\a
  \t'$, if there exists $x$ such that $x\t\a x\t'$ and, for all $y\neq
  x$, $y\t=y\t'$. An environment $\G=\vx:\vT$ {\em rewrites\,} to an
  environment $\G'$, $\G\a\G'$, if $\G'=\vx:\vT'$ and there exists $i$
  such that $T_i\a T_i'$ and, for all $j\neq i$, $T_j=T_j'$.
\end{dfn}


\begin{lem}
\label{lem-sr-typ}
\label{lem-sr-subs}
\label{lem-sr-env}
\index{subject reduction|indlem[lem-sr-typ]}

In an RTS~:

\begin{enumalphai}
\item if $\G\th t:T$ and $T\a T'$ then $\G\th t:T'$,
\item if $\t: \G\a\D$ and $\t\a \t'$ then $\t': \G\a\D$,
\item if $\G\th t:T$ and $\G\a\G'$ then $\G'\th t:T$.
\end{enumalphai}
\end{lem}

\begin{prf}
\begin{enumalphai}
\item By correctness of types, either $T=s$ or $\G\th T:s$. The case
$T=s$ is not possible since $s$ is not reducible. Therefore, $\G\th
T:s$ and, by subject reduction, $\G\th T':s$. Hence, by (conv), $\G\th
t:T'$.

\item By induction on the size of $\G$. If $\G$ is empty, this is
immediate. Assume then that $\G= \G', x\!:\!T$. Since $\t: \G'\a\D$,
by induction hypothesis, $\t': \G'\a\D$. Then it suffices to prove
that $\D\th x\t':T\t'$. As $\t: \G\a\D$, we have $\D\th x\t:T\t$. By
subject reduction, $\D\th x\t':T\t$. After the Environment Lemma,
there exists $s$ such that $\G\th T:s$. By substitution, $\D\th
T\t:s$. Since $T\t \a^* T\t'$, $T\t ~\cC~ T\t'$ and, by subject
reduction, $\D\th T\t':s$. Therefore, by (conv), $\D\th x\t':T\t'$.
  
\item Assume that $\G= \G_1,x\!:\!T,\G_2$ and $\G'=
\G_1,x\!:\!T',\G_2$. By the Environment Lemma, $\G_1\th T:s$.  By
subject reduction, $\G_1\th T':s$. Therefore, $\G ~\C~ \G'$ and, by
the Environment conversion Lemma, $\G'\th t:T$.\cqfd
\end{enumalphai}
\end{prf}


\begin{lem}[Inconvertibility of maximal sorts]
\label{lem-inconv-sort-max}
\index{inconvertibility of maximal sorts|indlem[lem-inconv-sort-max]}
\index{sort!maximal|indlem[lem-inconv-sort-max]}
\index{maximal!sort|indlem[lem-inconv-sort-max]}

  \hfill In an RTS, if $s \cvG T$ then\\ $s \CVG T$. Therefore, if $s$
  is maximal then $T=s$.
\end{lem}

\begin{prf}
  By case on the number of conversions between $s$ and $T$. If $s=T$,
  this is immediate. Assume then that $s ~\cC_\G~ T' \cvG T$. By
  definition of $\cC_\G$, there exists $s'$ such that $\G\th T':s'$.
  As $\cC=\ad$ and $s$ is irreducible, $T' \a^* s$. By subject
  reduction, $\G\th s:s'$ and $s \CVG T$.\cqfd
\end{prf}

~

Hence we get the equivalence of the two relations $\cC_\G$ and
$\C_\G$, and a refinement of the Inversion Lemma for RTS's.


\begin{lem}[Equivalence of $\cC_\G$ and $\C_\G$]
\label{lem-equiv-conv}
\index{C@$\cC_\G,\C_\G$|indlem[lem-equiv-conv]}

In an RTS, $\cC_\G = \C_\G$.
\end{lem}

\begin{prf}
  First of all, we have $\C_\G \sle \cC_\G$. We prove the
  reverse. Assume that $T ~\cC_\G~ T'$. As there exists $t$ such that
  $\G\th t:T$, by correctness of types, either $T$ is a maximal sort,
  or there exists $s$ such that $\G\th T:s$. After the previous lemma,
  $T$ cannot be a maximal sort. Therefore, there exists $s$ such that
  $\G\th T:s$ and $T ~\C_\G~ T'$.\cqfdd
\end{prf}


\begin{dfn}[Regular sort]
\label{def-sort-reg}
\index{regular!sort|inddef[def-sort-reg]}
\index{sort!regular|inddef[def-sort-reg]}
\index{TSM!regular|inddef[def-sort-reg]}
\index{regular!TSM|inddef[def-sort-reg]}

A sort $s$ is {\em regular\,} if, for all $(s_1,s_2,s)\in \cB$,
$s_2=s$. A TSM is {\em regular\,} if all its sorts are regular.
\end{dfn}

Most of the PTS's that one can find in the literature are regular. For
these systems, it is often made use of the abbreviation
$(s_1,s_2)\in\cB$ for $(s_1,s_2,s_2)\in\cB$
\cite{geuvers91jfp,barendregt92book}.


\begin{lem}[Inversion for RTS's]
\label{lem-inv-rts}
\index{inversion!RTS|indlem[lem-inv-rts]}

Assume that $\G\th t:T$.
\begin{lst}{--}
\item If $t=s$ then there exists $s'$ such that $(s,s')\in \cA$ and
$s'\CVG T$.
\item If $t=f(\vt)$, $f\in \cF^s$, $\tf= \pvxT U$ and $\g= \vxt$ then
$\th \tf:s$, $\g: (\vx:\vT)\a\G$ and $U\g \CVG T$. Moreover, if $s$ is
regular then $\G\th U\g:s$.
\item If $t=x\in \cX^s$ then $\G\th x\G:s$ and $x\G \CVG T$.
\item If $t= \px{U}V$ then there exists $(s_1,s_2,s_3)\in \cP$ such
that $\G\th U:s_1$, $\GxU\th V:s_2$ and $s_3 \CVG T$.
\item If $t=\lx{U}v$ then there exists $V$ such that $\GxU\th v:V$ and
$\px{U}V \CVG T$.
\item If $t=uv$ then there exists $V$ and $W$ such that $\G\th
u:\px{V}W$, $\G\th v:V$ and $W\xv \CVG T$. Moreover, if $\G\th
\px{V}W:s$ and $s$ is regular then $\G\th W\xv:s$.
\end{lst}
\end{lem}

\begin{prf}
  The modifications of the cases $t=s$ and $t=\px{U}V$ are a
  consequence of the inconvertibility of maximal sorts. Hence, we are
  left to prove the additional properties in case of regular
  sorts. The property for $t=f(\vt)$ can be obtained by iteration from
  the one for $t=uv$.
  
  Assume that $\G\th \px{V}W:s$. By inversion, there exists
  $(s_1,s_2,s_3)\in \cP$ such that $\GxV\th W:s_2$ and $s_3 \CVG s$.
  By preservation of sorts, $s_3=s$. By regularity,
  $s_2=s_3$. Therefore, $\GxV\th W:s$ and, by substitution, $\G\th
  W\xv:s$.\cqfd
\end{prf}




\section{Logical and functional RTS's}


\begin{dfn}[Functional TSM]
\label{def-tsm-func}
\index{TSM!functional|inddef[def-tsm-func]}
\index{functional TSM|inddef[def-tsm-func]}

A set of rules $\cB$ is {\em functional\,} if $(s_1,s_2,s_3) \in \cB$
and $(s_1,s_2,s_3') \in \cB$ imply $s_3=s_3'$. A TSM is {\em
functional\,} if $\cA$ is a functional relation and $\cB$ is
functional.
\end{dfn}

Most of the PTS's one can encounter in the literature are functional.

In a regular TSM, $\cB$ is functional. Therefore, for a regular TSM to
be functional, it suffices that $\cA$ is a functional relation.


\begin{lem}[Convertibility of types]
\label{lem-conv-typ}
\index{convertibility of types|indlem[lem-conv-typ]}

  In a logical and functional RTS, if $\G\th t:T$ and $\G\th t:T'$
  then $T \CVG T'$.
\end{lem}

\begin{prf}
  By induction on $t$. We follow the notations of the Inversion Lemma.

\begin{lst}{--}
\item $t=s$. By inversion, there exists $s'_1$ and $s'_2$ such that
$(s,s'_1)\in \cA$, $(s,s'_2)\in \cA$, $s'_1 \CVG T$ and $s'_2 \CVG
T'$. By functionality, $s'_1=s'_2$. Therefore, by symmetry, $T \CVG
T'$.
  
\item $t=f(\vt)$. By inversion, $U\g \CVG T$ and $U\g \CVG T'$.
Therefore, by symmetry, $T \CVG T'$.
  
\item $t=x$. By inversion, $x\G \CVG T$ and $x\G \CVG T'$. Therefore,
by symmetry, $T \CVG T'$.
  
\item $t=\px{U}V$. By inversion, there exists $(s_1,s_2,s_3)\in \cP$
and $(s_1',s_2',s_3')\in \cP$ such that $\G\th U:s_1$, $\G\th U:s_1'$,
$\GxU\th V:s_2$, $\GxU\th V:s_2'$, $s_3 \CVG T$ and $s_3' \CVG T'$. By
induction hypothesis, $s_1 \CVG s_1'$ and $s_2 \CV[\G,x:U] s_2'$. By
preservation of sorts, $s_1=s_1'$ and $s_2=s_2'$. Therefore, by
functionality, $s_3=s_3'$ and, by symmetry, $T \CVG T'$.
  
\item $t=\lx{U}v$. By inversion, there exists $V$ and $V'$ such that
$\GxU\th v:V$, $\GxU\th v:V'$, $\px{U}V \CVG T$ and $\px{U}V' \CVG
T'$.  By induction hypothesis, $V \CV[\G,x:U] V'$. By stability by
context, $\px{U}V \CVG \px{U}V'$. Therefore, by symmetry, $T \CVG T'$.
  
\item $t=uv$. By inversion, there exists $V$, $V'$, $W$ and $W'$ such
that $\G\th u:\px{V}W$, $\G\th u:\px{V'}W'$, $W\xv \CVG T$ and $W'\xv
\CVG T'$. By induction hypothesis, $\px{V}W \CVG \px{V'}W'$. By
product compatibility, $W \CV[\G,x:V] W'$. By substitution and
stability by substitution, $W\xv \CVG W'\xv$. Therefore, by symmetry,
$T \CVG T'$.\cqfd
\end{lst}
\end{prf}


\begin{lem}[Conversion correctness]
\label{lem-cor-conv}
\index{conversion correctness|indlem[lem-cor-conv]}

In a logical and functional RTS, if $\G\th T:s$ and $T ~\C_\G~ T'$
then $\G\th T':s$.
\end{lem}

\begin{prf}
  By definition of $\C_\G$, there exists $s'$ such that $\G\th T':s'$.
  As $\cC=\,\ad$, there exists $U$ such that $T \a^* U$ and $T' \a^*
  U$.  By subject reduction, $\G\th U:s$ and $\G\th U:s'$. By
  convertibility of types and preservation of sorts, $s=s'$ and $\G\th
  T':s$.\cqfdd
\end{prf}


\begin{lem}[Equivalence of $\ths$ and $\th$]
\label{lem-equiv-ths}
\index{$\ths$|indlem[lem-equiv-ths]}

In a logical and functional RTS,\, $\ths=\,\th$.
\end{lem}

\begin{prf}
  First of all, it is immediate that $\ths\,\sle\,\th$. We show the
  inverse by induction on $\G\th t:T$. The only difficult case is of
  course (conv). By induction hypothesis, we have $\G\ths t:T$ and
  $\G\ths T':s'$. It is easy to verify that the Substitution Lemma and
  the correctness of types are also valid for $\ths$. Hence, by
  correctness of types, either $T$ is a maximal sort, or there exists
  $s$ such that $\G\ths T:s$. If $T$ is a maximal sort $s$ then $T'
  \a^* s$ and $s$ is typable, which is excluded. Therefore, $\G\ths
  T:s$. By convertibility of types and preservation of sorts, $s=s'$
  and, by (conv'), $\G\ths t:T'$.\cqfd
\end{prf}


\begin{lem}[$\alpha$-equivalence]
\label{lem-alpha-equiv}
\index{alpha-equivalence@$\alpha$-equivalence|indlem[lem-alpha-equiv]}

In a logical and functional RTS, if $\px{T}U ~\C_\G~ \p{x'}{T'}U'$
then $x$ and $x'$ are of the same sort and $\p{x'}{T'}U'$ is
$\alpha$-equivalent to $\px{T'}U'\{x'\to x\}$.
\end{lem}

\begin{prf}
  Assume that $x$ is of sort $s$ and $x'$ is of sort $s'$. By
  definition of $\C_\G$, we have $\G\th \px{T}U:s_3$ and $\G\th
  \p{x'}{T'}U':s_3'$. By inversion, we have $\GxT\th U:s_1$ and
  $\G,x'\!:\!T'\th U':s_1'$. By the Environment Lemma, we have $\G\th
  T:s$ and $\G\th T':s'$. By conversion correctness and preservation
  of sorts, $s=s'$. Therefore $x$ and $x'$ are of the same sort and
  $\p{x'}{T'}U'$ is $\alpha$-equivalent to $\px{T'}U'\{x'\to
  x\}$.\cqfd
\end{prf}


\begin{lem}[Maximal sort]
\label{lem-sort-max}
\index{maximal!sort|indlem[lem-sort-max]}
\index{sort!maximal|indlem[lem-sort-max]}

  In a logical and functional RTS, if $s$ is a maximal sort and $\G\th
  t:s$ then $t$ is of the form $\pvxt s'$. Moreover, if $t \CVG t'$
  then $t'$ is of the form $(\vy:\vt') s'$ with $|\vy|=|\vx|$.
\end{lem}

\begin{prf}
  We prove the first assertion by case on $t$. Note first of all that
  there is no $s'$ such that $\G\th s:s'$. Otherwise, by inversion,
  there would exist $s''$ such that $(s,s'')\in \cA$ and $s'' \CVG
  s'$, which is excluded since $s$ is maximal.

\begin{lst}{--}
\item $t=f(\vt)$. Let $\tf=\pvxT U$ and $\g=\vxt$. By inversion, there
exists $s'$ such that $\G\th U\g:s'$ and $U\g \CVG s$. By conversion
correctness, $\G\th s:s'$. This case is therefore impossible.

\item $t=x\in \cX^{s'}$. By inversion, $\G\th x\G:s'$ and $x\G \CVG
s$. By conversion correctness, $\G\th s:s'$. This case is therefore
impossible.
  
\item $t=\lx{U}v$. By inversion, there exists $V$ and $s'$ such that
$\GxU\th v:V$, $\G\th \px{U}V:s'$ and $\px{U}V \CVG s$. By conversion
correctness, $\G\th s:s'$. This case is therefore impossible.
  
\item $t=uv$. By inversion, there exists $V$, $W$ and $s'$ such that
$\G\th W\xu:s'$ and $W\xu \CVG s$. By conversion correctness, $\G\th
s:s'$. This case is therefore impossible.
\end{lst}

We are left with the cases $t=\px{U}V$ and $t=s'$. Therefore $t$ must
be of the form $\pvxt s'$.

We now show the second assertion. By conversion correctness, $\G\th
t':s$. After the first assertion, $t'$ is of the form
$(\vy:\vt')s''$. By product compatibility and $\alpha$-equivalence,
and by exchanging the roles of $t$ and $t'$, we can assume that $\vy=
\vx\vz$ and $\vt'= \vt\vu$.  Hence, $s' \CV[\G'] (\vz:\vu)s''$ where
$\G'= \vx:\vt$. We then prove by induction on the number of
conversions between $s'$ and $(\vz:\vu)s''$ that $|\vz|=0$ and
$s'=s''$. If $s'= (\vz:\vu)s''$, this is immediate.  Assume then that
$s' ~\C_\G'~ v \CV[\G'] (\vz:\vu)s''$. By conversion correctness,
$\G'\th v:s$. Therefore, after the first assertion, $v$ is of the form
$(\vz':\vu')s'''$. As $\cC=\ad$ and $s'$ is irreducible, $v \a^*
s'$. Therefore $|\vz'|=0$ and $s'''=s'$. So, $v=s'$ and, by induction
hypothesis, $|\vz|=0$ et $s''=s'$.\cqfd
\end{prf}




\section{Logical and injective RTS's}


\begin{dfn}[Injective TSM]
\label{def-tsm-inj}
\index{injective TSM|inddef[def-tsm-inj]}
\index{TSM!injective|inddef[def-tsm-inj]}

A set of rules $\cB$ is {\em injective\,} if $(s_1,s_2,s_3) \in \cB$
and $(s_1,s_2',s_3) \in \cB$ imply $s_2=s_2'$. A TSM is {\em
injective\,} if $\cA$ is a functional and injective relation and $\cB$
is functional and injective.
\end{dfn}

In a regular TSM, $\cB$ is functional and injective. Therefore, for a
regular TSM to be injective, it suffices that $\cA$ is a functional
and injective relation.


\newcommand{\CVj}{\CV[\G_j]}

\begin{lem}[Separation]
\label{lem-sep}
\index{separation|indlem[lem-sep]}

  In a logical and injective RTS, if $s_1\neq s_2$ then, for all
  $i\in\{0,1\}$, $\T_i^{s_1} \cap \T_i^{s_2}= \vide$.
\end{lem}

\begin{prf}
  We show that $t \in \T_i^{s_1} \cap \T_i^{s_2}$ implies $s_1=s_2$,
  by induction on $t$.
  
\begin{lst}{}
  
\item \u{\bf Case $\mg{i=0}$}. There exists $\G_j$ such that $\G_j\th
t:s_j$.

\begin{lst}{--}
\item $t=s$. By inversion, there exists $s_j'$ such that $(s,s_j') \in
\cA$ and $s_j' \CVj s_j$. By functionality, $s_1'=s_2'$.  Let $s'$ be
the sort $s_1'=s_2'$. Then, $s' \CVj s_j$. Therefore, by preservation
of sorts, $s_1=s_2=s'$.
  
\item $t=f(\vt)$, $f\in \cF^s$ and $\tf= \pvxT U$. Let $\G= \vx:\vT$
and $\g= \vxt$. By inversion, $\th \tf:s$, $\g: \G\a\G_j$ and $U\g
\CVj s_j$. By inversion again, there exists $s'$ such that $\G\th
U:s'$. By substitution, $\G_j\th U\g:s'$. By conversion correctness,
$\G_j\th s_j:s'$. By inversion and preservation of sorts, $(s_j,s')
\in \cA$. Therefore, by injectivity, $s_1=s_2$.
  
\item $t=x\in \cX^s$. By inversion, $\G_j\th x\G_j:s$ and $x\G_j \CVj
s_j$. By conversion correctness, $\G_j\th s_j:s$.  By inversion and
preservation of sorts, $(s_j,s) \in \cA$.  Therefore, by injectivity,
$s_1=s_2$.
  
\item $t=\px{U}V$. By inversion, there exists $(s_j^1,s_j^2,s_j^3) \in
\cP$ such that $\G_j\th U:s_j^1$, $\G_j, x\!:\!U\th V:s_j^2$ and
$s_j^3 \CVj s_j$. By induction hypothesis, $s_1^1=s_2^1$ and
$s_1^2=s_2^2$. Therefore, by functionality, $s_1^3=s_2^3$. Let $s$ be
the sort $s_1^3=s_2^3$. Then, $s \CVj s_j$ and, by preservation of
sorts, $s_1=s_2=s$.
  
\item $t=\lx{U}v$. By inversion, there exists $V_j$ and $s_j^4$ such
that $\G_j, x\!:\!U\th v:V_j$, $\G_j\th \px{U}V_j:s_j^4$ and
$\px{U}V_j \CVj s_j$. By inversion again, there exists
$(s_j^1,s_j^2,s_j^3) \in \cP$ such that $\G_j\th U:s_j^1$, $\G_j,
x\!:\!U\th V_j:s_j^2$ and $s_j^3 \CVj s_j^4$. By preservation of
sorts, $s_j^3=s_j^4$. By induction hypothesis, $s_1^1=s_2^1$ and
$s_1^2=s_2^2$. Therefore, by functionality, $s_1^3=s_2^3$. Let $s$ be
the sort $s_1^3=s_2^3=s_1^4=s_2^4$. By conversion correctness, $\G\th
s_j:s$. By inversion and preservation of sorts, $(s_j,s) \in
\cA$. Therefore, by injectivity, $s_1=s_2$.
  
\item $t=uv$. By inversion, there exists $V_j$ and $W_j$ such that
$\G_j\th u:\p{x_j}{V_j}W_j$, $\G_j\th v:V_j$ and $W_j\{x_j\to v\} \CVj
s_j$. Let $\G_j'= \G_j, x_j\!:\!V_j$ and $\t_j= \{x_j\to v\}$. By
correctness of types, there exists $s_j'$ such that $\G_j\th
\p{x_j}{V_j}W_j:s_j'$. By induction hypothesis on $u$,
$s_1'=s_2'$. Let $s'=s_1'=s_2'$. By inversion, there exists
$(s_j^1,s_j^2,s_j^3) \in \cP$ such that $\G_j\th V_j:s_j^1$, $\G_j'
\th W_j:s_j^2$ and $s_j^3 \CVj s'$. By preservation of sorts,
$s_j^3=s'$. By induction hypothesis on $v$, $s_1^1=s_2^1$.  Therefore,
by injectivity, $s_1^2=s_2^2$. Let $s''=s_1^2=s_2^2$.  As $\t_j:\G_j'
\to \G_j$, by substitution, $\G_j\th W_j\t_j:s''$. By conversion
correctness, $\G_j\th s_j:s''$.  By inversion and preservation of
sorts, $(s_j,s'') \in \cA$. Therefore, by injectivity, $s_1=s_2$.
\end{lst}

\item \u{\bf Case $i=1$}. There exists $\G_j$ and $T_j$ such that
$\G_j \th t:T_j$ and $\G_j\th T_j:s_j$.

\begin{lst}{--}
\item $t=s$. After case $i=0$, there exists $s'$ such that $s' \CVj
T_j$. By conversion correctness, $\G_j\th s':s_j$. By inversion and
preservation of sorts, $(s',s_j) \in \cA$.  Therefore, by
functionality, $s_1=s_2$.
  
\item $t=f(\vt)$. After case $i=0$, there exists $s'$ such that $\G_j
\th T_j:s'$. By convertibility of types and preservation of sorts,
$s_1=s_2=s'$.
  
\item $t=x\in \cX^s$. After case $i=0$, $\G_j\th T_j:s$. Therefore, by
convertibility of types and preservation of sorts, $s_1=s_2=s$.
  
\item $t=\px{U}V$. After case $i=0$, there exists $s$ such that $s
\CVj T_j$. By conversion correctness, $\G_j\th s:s_j$.  By inversion
and preservation of sorts, $(s,s_j) \in \cA$.  Therefore, by
functionality, $s_1=s_2$.
  
\item $t=\lx{U}v$. After case $i=0$, there exists $s$ such that $\G_j
\th T_j:s$. By convertibility of types and preservation of sorts,
$s_1=s_2=s$.
  
\item $t=uv$. After case $i=0$, there exists $s''$ such that $\G_j \th
T_j:s''$. By convertibility of types and preservation of sorts,
$s_1=s_2=s'$.\cqfd
\end{lst}
\end{lst}
\end{prf}


\newpage

\begin{lem}[Classification]
\label{lem-class}
\index{classification|indlem[lem-class]}

  In a logical and injective RTS, either $(s_1,s_2) \in \cA$ and
  $\T_0^{s_1} \sle \T_1^{s_2}$, or $(s_1,s_2) \notin \cA$ and
  $\T_0^{s_1} \cap \T_1^{s_2}= \vide$.
\end{lem}

\begin{prf}
  If $(s_1,s_2) \in \cA$, it is clear that $\T_0^{s_1} \sle
  \T_1^{s_2}$. We then prove that $t \in \T_0^{s_1} \cap \T_1^{s_2}$
  implies $(s_1,s_2) \in \cA$, by induction on $t$. Let $\G$, $\G'$
  and $T$ such that $\G\th t:s_1$, $\G'\th t:T$ and $\G'\th T:s_2$.

\begin{lst}{--}
\item $t=s$. By inversion, there exists $s_1'$ and $s_2'$ such that
$(s,s_1') \in \cA$, $s_1' \CVG s_1$, $(s,s_2') \in \cA$ and $s_2'
\CVGp T$. By preservation of sorts, $s_1'=s_1$. By functionality,
$s_1'=s_2'$. By conversion correctness, $\G \th s_2':s_2$. Therefore,
by inversion and preservation of sorts, $(s_2',s_2) \in \cA$ and
$(s_1,s_2) \in \cA$.
  
\item $t=f(\vt)$. Let $\tf=\pvxT U$ and $\g=\vxt$. By inversion, there
exists $s_1'$ and $s_2'$ such that $\G\th U\g:s_1'$, $U\g \CVG s_1$,
$\G'\th U\g:s_2'$ and $U\g \CVGp T$. By conversion correctness, $\G\th
s_1:s_1'$. Therefore, by inversion and preservation of sorts,
$(s_1,s_1') \in \cA$. By separation, $s_1'=s_2'$. By conversion
correctness, $\G'\th T:s_2'$.  Therefore, by separation, $s_2'=s_2$
and $(s_1,s_2) \in \cA$.
  
\item $t=x\in \cX^s$. By inversion, $\G\th x\G:s$, $x\G \CVG s_1$,
$\G'\th x\G':s$ and $x\G' \CVGp T$. By conversion correctness, $\G\th
s_1:s$. By inversion and preservation of sorts, $(s_1,s) \in \cA$. By
conversion correctness, $\G'\th T:s$. Therefore, by separation,
$s=s_2$ and $(s_1,s_2) \in \cA$.
  
\item $t=\px{U}V$. By inversion, there exists $(s_a,s_b,s_c)$ and
$(s_a',s_b',s_c') \in \cP$ such that $\G\th U:s_a$, $\GxU\th V:s_b$,
$s_c \CVG s_1$, $\G'\th U:s_a'$, $\GpxU\th V:s_b'$ and $s_c' \CVGp
T$. By preservation of sorts, $s_c=s_1$. By conversion correctness,
$\G'\th s_c':s_2$. By inversion and preservation of sorts, $(s_c',s_2)
\in \cA$. By separation, $s_a=s_a'$ and $s_b=s_b'$. Therefore, by
functionality, $s_c=s_c'$ and $(s_1,s_2)\in \cA$.

\item $t=\lx{U}v$. By inversion, there exists $V$, $s$, $V'$ and $s'$
such that $\GxU\th v:V$, $\G\th \px{U}V:s$, $\px{U}V \CVG s_1$,
$\GpxU\th v:V'$, $\G'\th \px{U}V':s'$ and $\px{U}V' \CVGp T$. By
conversion correctness, $\G\th s_1:s$ and $\G'\th T:s'$. By inversion
and preservation of sorts, $(s_1,s) \in \cA$. By separation
$s'=s_2$. By inversion again, there exists $(s_a,s_b,s_c)$ and
$(s_a',s_b',s_c') \in \cP$ such that $\G \th U:s_a$, $\GxU\th V:s_b$,
$s_c \CVG s$, $\G'\th U:s_a'$ and $\GpxU\th V':s_b'$ and $s_c' \CVGp
s'$. By preservation of sorts, $s_c=s$ and $s_c'=s'$. By separation,
$s_a=s_a'$ and $s_b=s_b'$. Therefore, by functionality, $s_c=s_c'$ and
$(s_1,s_2) \in \cA$.
  
\item $t=uv$. By inversion, there exists $V$, $W$, $s$, $V'$ and $W'$
and $s'$ such that $\G\th u:\px{V}W$, $\G\th W\xv:s$, $W\xv \CVG s_1$,
$\G'\th u:\px{V'}W'$, $\G'\th W'\xv:s'$ and $W'\xv \CVGp T$. By
conversion correctness, $\G\th s_1:s$ and $\G' \th T:s'$. By inversion
and preservation of sorts, $(s_1,s) \in \cA$. By separation, $s=s'$
and $s'=s_2$. Therefore, $(s_1,s_2) \in \cA$.\cqfd
\end{lst}
\end{prf}


\begin{rem}[Typing classes]\hfill
\label{rem-class-typ}
\index{typing!class|indrem[rem-class-typ]}

With the correctness of types, we have seen that a typable term is
necessarily in one of the sets $\T_i^s$ where $i\in\{0,1\}$ and $s\in
\cS$. With the Separation and Classification Lemmas, we can describe
the relations between these sets more precisely.

In an injective TSM, the set of axioms $\cA$ is necessarily an union
of disjoint maximal ``chains'', that is, sets $\cA'$ such that~:

\newpage

\begin{lst}{--}
\item if $(s_1,s_2)\in\cA'$ and $(s_2,s_3)\in\cA$ then
$(s_2,s_3)\in\cA'$,
\item if $(s_2,s_3)\in\cA'$ and $(s_1,s_2)\in\cA$ then
$(s_1,s_2)\in\cA'$.
\end{lst}

For example, in the case where $\cA$ is finite, a maximal chain is of
the form $\{(s_1,s_2)$, $(s_2,s_3)$, \ldots, $(s_n,s_{n+1})\}$ with
$s_1, \ldots, s_n$ distinct from one another. For such a chain, we
obtain $n$ classes $\T_1^{s_1}$, $\T_1^{s_2}$, \ldots, $\T_1^{s_n}$,
plus two classes $\T_1^{s_{n+1}}$ and $\T_0^{s_{n+1}}$ if $s_{n+1}$ is
distinct from the other $s_i$'s. Finally, a sort $s$ which does not
belong to any axiom gives two classes, $\T_1^s$ and $\T_0^s$.
\end{rem}




\section{Confluent RTS's}

In the following, we prove results about the dependence of types with
respect to variables and symbols. The first result, dependence with
respect to variables, is better known under the name of
``Strengthening Lemma''. We give a proof of this lemma in the case of
a functional (and confluent) RTS inspired from the one of H. Geuvers
and M.-J. Nederhof \cite{geuvers91jfp}. L. S. van Benthem Jutting
\cite{jutting93ic} proved the same lemma for all PTS's. It would be
interesting to examine his proof to adapt it to the case of RTS's.


\begin{lem}[Dependence w.r.t. variables]
\label{lem-dep-var}
\index{strengthening|indlem[lem-dep-var]}
\index{dependence w.r.t.!variable|indlem[lem-dep-var]}

In a confluent and functional RTS, if $\D,z\!:\!V,\D'\th t:T$ and
$z\notin \FV(\D',t)$ then there exists $T'$ such that $T \a^* T'$ and
$\D,\D'\th t:T'$.
\end{lem}

\begin{prf}
By induction on $\D,z\!:\!V,\D'\th t:T$.

\begin{lst}{}
\item [\bf(ax)] Impossible.
  
\item [\bf(symb)] Let $\G= \D,z\!:\!V,\D'$. We prove the property for
$U\g$ itself. By induction hypothesis, for all $i$, there exists
$T_i'$ such that $T_i\g \a^* T_i'$ and $\D,\D'\th t_i:T_i'$. We prove
that $\g: \G_f\a\D,\D'$. Let $\g_i= \{x_1\to t_1, \ldots, x_i\to
t_i\}$ and $\G_i= x_1\!:\!T_1, \ldots, x_i\!:\!T_i$. We prove that
$\g_i: \G_i\a\D,\D'$ by induction on $i$. For $i=0$, there is nothing
to prove. Assume then that $\g_i: \G_i\a\D,\D'$ and let us prove that
$\g_{i+1}: \G_{i+1}\a\D,\D'$.  As $\th \tf:s$, by inversion, for all
$j$, there exists $s_j$ such that $\G_{j-1}\th T_j:s_j$. Hence, by the
Environment Lemma, $\FV(T_j)\sle \{x_1,\ldots,x_{j-1}\}$. Therefore,
for all $j\le i+1$, $T_j\g_{i+1}= T_j\g_i$. So, $\g_{i+1}:
\G_i\a\D,\D'$ and we are left to prove that $\D,\D'\th
t_{i+1}:T_{i+1}\g_i$. We have $\D,\D'\th t_{i+1}:T_{i+1}'$. As
$\G_i\th T_{i+1}:s_{i+1}$ and $\g_i: \G_i\a\D,\D'$, by substitution,
$\D,\D'\th T_{i+1}\g_i:s_{i+1}$. Therefore, by (conv), $\D,\D'\th
t_{i+1}:T_{i+1}\g_i$ and $\g_{i+1}: \G_{i+1}\a\D,\D'$. Finally,
$\g=\g_n: \G_f\a\D,\D'$. As $\G_f\th U:s$, by substitution, $\D,\D'\th
U\g:s$.
  
\item [\bf(var)] Let $\G= \D,z\!:\!V,\D'$. By induction hypothesis,
$\D,\D'\th T:s$. Therefore, by (var), $\D,\D',x:T\th x:T$.

\item [\bf(weak)] If $z=x$, $T$ itself satisfies the property since
$\G\th t:T$. Otherwise, let $\G= \D,z\!:\!V,\D'$. By induction
hypothesis, there exists $T'$ such that $T \a^* T'$, $\D,\D'\th t:T'$
and $\D,\D'\th U:s$. Therefore, by (weak), $\D,\D',x\!:\!U\th t:T'$.

\item [\bf(prod)] Let $\G= \D,z\!:\!V,\D'$. By induction hypothesis,
$\D,\D'\th T:s_1$ and $\D,\D',x\!:\!T\th U:s_2$. Therefore, by (prod),
$\D,\D'\th \px{T}U:s_3$.
  
\item [\bf(abs)] Let $\G= \D,z\!:\!V,\D'$. By induction hypothesis,
there exists $U'$ such that $U \a^* U'$ and $\D,\D',x\!:\!T\th u:U'$.
We prove that $\D,\D'\th \px{T}U':s$. Then, by (abs), $\D,\D'\th
\lx{T}u:\px{T}U'$. As $\G\th \px{T}U:s$, by inversion, there exists
$(s_1,s_2,s)\in\cP$ such that $\G\th T:s_1$ and $\G,x\!:\!T\th
U:s_2$. As $z\notin \FV(T)$, by induction hypothesis, $\D,\D'\th
T:s_1$. As $\D,\D',x\!:\!T\th u:U'$, by correctness of types, either
$U'=s'$ or $\D,\D',x\!:\!T\th U':s'$. Assume that $U'=s'$. As
$\G,x\!:\!T\th U:s_2$, by subject reduction, $\G,x\!:\!T\th
U':s_2$. Therefore $(s',s_2)\in\cA$ and $\D,\D',x\!:\!T\th U':s_2$. If
now $\D,\D',x\!:\!T\th U':s'$ then, by convertibility of types,
$s'=s_2$. Hence, in all cases, $\G,x\!:\!T\th U':s_2$. Therefore, by
(prod), $\D,\D'\th \px{T}U':s$.
  
\item [\bf(app)] Let $\G= \D,z\!:\!V,\D'$. By induction hypothesis,
there exists $U'_1$, $U'_2$ and $V$ such that $U \a^* U'_1$, $U \a^*
U'_2$, $V \a^* V'$, $\D,\D'\th t:\px{U'_1}V'$ and $\D,\D'\th u:U'_2$.
By confluence, there exists $U''$ such that $U'_1 \a^* U''$ and $U'_2
\a^* U''$. By subject reduction, $\D,\D'\th t:\px{U''}V'$ and
$\D,\D'\th u:U''$. Therefore, by (app), $\D,\D'\th tu:V'\xu$.
  
\item [\bf(conv)] By induction hypothesis, there exists $T''$ such
that $T \a^* T''$ and $\D,\D'\th t:T''$. By confluence, there exists
$T'''$ such that $T'' \a^* T'''$ and $T' \a^* T'''$. By subject
reduction, $\D,\D'\th t:T'''$.\cqfd
\end{lst}
\end{prf}


\begin{cor}
  In a confluent and functional RTS, if $\D,z\!:\!V,\D'\th t:T$ and
  $z\notin \FV(\D',t,T)$ then $\D,\D'\th t:T$.
\end{cor}

\begin{prf}
  After the lemma, there exists $T'$ such that $T \a^* T'$ and
  $\D,\D'\th t:T'$. By correctness of types, either $T$ is a maximal
  sort and $T'=T$, or $\D,z\!:\!V,\D'\th T:s$. Then, after the lemma,
  $\D,\D'\th T:s$. Therefore, by (conv), $\D,\D'\th t:T$.\cqfd
\end{prf}


\begin{lem}[Strong permutation]
\label{lem-strong-perm}
\index{permutation!strong|indlem[lem-dep-var]}
  If $\G,y\!:\!A,z\!:\!B,\G'\th t:T$ and $y \notin \FV(B)$ then
  $\G,z\!:\!B,y\!:\!A,\G'\th t:T$.
\end{lem}

\begin{prf}
  Let $\D= \G,y\!:\!A,z\!:\!B,\G'$ and $\D'= \G,z\!:\!B,y\!:\!A,\G'$.
  By transitivity, it suffices to prove that $\D'$ is valid and that
  $\D'\th\D$. To this end, it suffices to prove that $\D'$ is
  valid. By the Environment Lemma, we have $\G\th A:s$ and
  $\G,y\!:\!A\th B:s'$. After the previous lemma, $\G\th
  B:s'$. Therefore, $\G,z\!:\!B$ is valid and, by weakening,
  $\G,z\!:\!B,y\!:\!A$ too. Assume that $\G'= \vx:\vT$ and let $\D_i=
  \G,y\!:\!A,z\!:\!B,x_1\!:\!T_1,\ldots,x_i\!:\!T_i$ and $\D'_i=
  \G,z\!:\!B,y\!:\!A,x_1\!:\!T_1,\ldots,x_i\!:\!T_i$. We prove by
  induction on $i$ that $\D'_i$ is valid. We have already proved that
  $\D'_0$ is valid. Assume that $\D'_i$ is valid. By the Environment
  Lemma, $\D_i\th T_{i+1}:s_{i+1}$. As $\D'_i\th\D_i$, $\D'_i\th
  T_{i+1}:s_{i+1}$ and $\D'_{i+1}$ is valid. Therefore, $\D'$ is valid
  and $\D'\th t:T$.\cqfd
\end{prf}


\begin{dfn}[Compatibility w.r.t. a quasi-ordering]
\label{def-comp}
\index{compatible with!ordering!reduction|inddef[def-comp]}
\index{compatible with!ordering!type assignment|inddef[def-comp]}
\index{$\th_f$|inddef[def-comp]}

\hfill Let $\ge$ be a quasi-\\ordering on $\cF$. Given a symbol $g$,
we will denote by $\th_g$ the typing relation of the RTS whose symbols
are strictly smaller than $g$.

\begin{lst}{--}
\item $\a$ is {\em compatible\,} with $\ge$ if, for all symbol $g$ and
all term $t,t'$, if all the symbols in $t$ are strictly smaller than
$g$ and $t\a t'$ then the symbols in $t'$ are strictly smaller than
$g$.
  
\item $\tau$ is {\em compatible\,} with $\ge$ if, for all symbol $g$,
all the symbols in $\tg$ are smaller than $g$.
\end{lst}
\end{dfn}


\begin{lem}[Dependence w.r.t. symbols]
\label{lem-dep-symb}
\index{dependence w.r.t.!symbol|indlem[lem-dep-symb]}

Consider a confluent and functional RTS and $\ge$ a quasi-ordering on
$\cF$ such that $\a$ and $\tau$ are compatible with $\ge$. If $\G\th
t:T$ and the symbols in $\G$ and $t$ are strictly smaller than $g$ then
there exists $T'$ such that $T \a^* T'$ and $\G\th_g t:T'$.
\end{lem}

\begin{prf}
By induction on $\G\th t:T$.

\begin{lst}{}
\item [\bf(ax)] Immediate.
  
\item [\bf(symb)] We prove that $\G\th_g f(\vt):U\g$. By induction
hypothesis, for all $i$, there exists $T_i'$ such that $T_i\g \a^*
T_i'$ and $\G\th_g t_i:T_i'$. We prove that $\g: \G_f\a\G$ in
$\th_g$. Let $\g_i= \{x_1\to t_1, \ldots, x_i\to t_i\}$ and $\G_i=
x_1\!:\!T_1, \ldots, x_i\!:\!T_i$. We prove that $\g_i: \G_i\a\G$ by
induction on $i$. For $i=0$, there is nothing to prove.  Assume then
that $\g_i: \G_i\a\G$ and let us prove that $\g_{i+1}: \G_{i+1}\a\G$.
As $\tau$ is compatible with $\ge$, by induction hypothesis, $\th_g
\tf:s$. By inversion, for all $j$, there exists $s_j$ such that
$\G_{j-1}\th_g T_j:s_j$. Hence, by the Free variables Lemma,
$\FV(T_j)\sle \{x_1,\ldots,x_{j-1}\}$.  Therefore, for all $j\le i+1$,
$T_j\g_{i+1}= T_j\g_i$. So, $\g_{i+1}: \G_i\a\G$ and we are left to
prove that $\G\th_g t_{i+1}:T_{i+1}\g_i$. We have $\G\th_g
t_{i+1}:T_{i+1}'$. As $\G_i\th_g T_{i+1}:s_{i+1}$ and $\g_i:
\G_i\a\G$, by substitution, $\G\th_g T_{i+1}\g_i:s_{i+1}$. Therefore,
by (conv), $\G\th_g t_{i+1}:T_{i+1}\g_i$ and $\g_{i+1}:
\G_{i+1}\a\G$. Finally, $\g=\g_n: \G_f\a\G$. As $\G_f\th_g U:s$, by
substitution, $\G\th_g U\g:s$.
  
\item [\bf(var)] By induction hypothesis, $\G\th_g T:s$. Therefore, by
(var), $\GxT\th_g x:T$.
  
\item [\bf(weak)] By induction hypothesis, there exists $T'$ such that
$T \a^* T'$, $\G\th_g t:T'$ and $\G\th_g U:s$. Therefore, by (weak),
$\GxU\th_g t:T'$.
  
\item [\bf(prod)] By induction hypothesis, $\G\th_g T:s_1$ and
$\GxT\th_g U:s_2$. Therefore, by (prod), $\G\th_g \px{T}U:s_3$.
  
\item [\bf(abs)] By induction hypothesis, there exists $U'$ such that
$U \a^* U'$ and $\GxT\th_g u:U'$. We prove that $\G\th_g \px{T}U':s$.
Then, by (abs), $\G\th_g \lx{T}u:\px{T}U'$. As $\G\th \px{T}U:s$, by
inversion, there exists $(s_1,s_2,s)\in\cP$ such that $\G\th T:s_1$
and $\GxT\th U:s_2$. By induction hypothesis, $\G\th_g T:s_1$. As
$\GxT\th_g u:U'$, by correctness of types, either $U'=s'$ or
$\GxT\th_g U':s'$. Assume that $U'=s'$. As $\GxT\th U:s_2$, by subject
reduction, $\GxT\th U':s_2$.  Therefore $(s',s_2)\in\cA$ and
$\GxT\th_g U':s_2$. If now $\GxT\th_g U':s'$ then, by convertibility
of types, $s'=s_2$.  Hence, in all cases, $\GxT\th_g
U':s_2$. Therefore, by (prod), $\G\th_g \px{T}U':s$.
  
\item [\bf(app)] By induction hypothesis, there exists $U'_1$, $U'_2$
and $V$ such that $U \a^* U'_1$, $U \a^* U'_2$, $V \a^* V'$, $\G\th_g
t:\px{U'_1}V'$ and $\G\th_g u:U'_2$. By confluence, there exists $U''$
such that $U'_1 \a^* U''$ and $U'_2 \a^* U''$. By subject reduction,
$\G\th_g t:\px{U''}V'$ and $\G\th_g u:U''$. Therefore, by (app),
$\G\th_g tu:V'\xu$.
  
\item [\bf(conv)] By induction hypothesis, there exists $T''$ such
that $T \a^* T''$ and $\G\th_g t:T''$. By confluence, there exists
$T'''$ such that $T'' \a^* T'''$ and $T' \a^* T'''$. By subject
reduction, $\G\th_g t:T'''$. As $\a$ is compatible with $\ge$, the
symbols in $T'''$ are strictly smaller than $g$.\cqfd
\end{lst}
\end{prf}


%% file: ats.tex



\chapter{Algebraic Type Systems (ATS's)}
\markboth{CHAPTER \thechapter. ALGEBRAIC TYPE SYSTEMS}{}
\label{chap-ats}

Now, we are going to study the case of RTS's whose reduction relation
is made of $\b$-reduction and rewrite rules. But, before, we must
properly define what rewriting means in a strongly typed higher-order
framework.

~

In first-order frameworks, that is, within a first-order term algebra,
a rewrite rule is generally defined as a pair $l\a r$ of terms such
that $l$ is not a variable and the variables occurring in $r$ also
occur in $l$ (otherwise, rewriting does not terminate). Then, one says
that a term $t$ rewrites to a term $t'$ at position $p$ if there
exists a substitution $\s$ such that $t|_p=l\s$ (one says that $t|_p$
matches $l$) and $t'=t[r\s]_p$ (the subterm of $t$ at position $p$,
$l\s$, is replaced by $r\s$). The reader is invited to look at, for
example, \cite{dershowitz90book,baader98book} to get more details on
first-order rewriting.

Here, we are going to consider a very similar rewriting mechanism by
restricting left-hand sides of rules to belong to the first-order-like
term algebra generated from $\cF$ and $\cX$.  On the other hand,
right-hand sides can be arbitrary. This is a particular the case of
{\em Combinatory Reduction System\,} (CRS)
\footnote{To see this, it suffices to translate $\lx{T}u$ by
$\L(T,[x]u)$, $\px{T}U$ by $\Pi(T,[x]U)$ and $uv$ by $@(u,v)$, where
$\L$, $\Pi$ and $@$ are symbols of arity 2 and $[\_]\_$ is the
abstraction operator of CRS's.} of W. Klop \cite{klop93tcs} for which
it is not necessary to use {\em higher-order pattern matching\,} {\em
\`a la\,} Klop or {\em \`a la\,} Miller
\cite{miller89elp,nipkow91lics}.

However, we proved in \cite{blanqui00rta} that a weaker version of the
termination criteria that we are going to present in next chapter can
be adapted, in case of simply-typed $\la$-calculus, to rewriting with
higher-order matching {\em \`a la\,} Klop or {\em \`a la\,} Miller. It
would therefore be interesting to try to define a notion of rewriting
with higher-order matching in case of polymorphic and dependent types,
and to study if our termination criteria can also be adapted to this
notion of rewriting.


\begin{dfn}[Algebraic terms]
\label{def-alg-term}
\index{term!algebraic|inddef[def-alg-term]}
\index{algebraic!term|inddef[def-alg-term]}
\index{TFX@$\cT(\cF,\cX),\T(\cF,\cX)$|inddef[def-alg-term]}

  A term is {\em algebraic\,} if it is a variable or of the form
  $f(\vt)$ with all the $t_i$'s themselves algebraic.  We denote by
  $\cT(\cF,\cX)$ the set of algebraic terms built from $\cF$ and
  $\cX$, and by $\T(\cF,\cX)$ the set of typable algebraic terms.
\end{dfn}


\begin{dfn}[Rewriting]
\label{def-rew}
\index{rewriting|inddef[def-rew]}
\index{rewrite rule|inddef[def-rew]}
\index{rule!rewrite|inddef[def-rew]}
\index{left-linear rule|inddef[def-rew]}
\index{R@$\cR,\cR_\cG$|inddef[def-rew]}
\index{constant symbol|inddef[def-rew]}
\index{symbol!constant|inddef[def-rew]}
\index{defined symbol|inddef[def-rew]}
\index{symbol!defined|inddef[def-rew]}
\index{$l\a r$|inddef[def-rew]}
\index{CF@$\cC\cF,\cC\cF^s$|inddef[def-rew]}
\index{DF@$\cD\cF,\cD\cF^s$|inddef[def-rew]}
\index{R-reduction@$\cR$-reduction|inddef[def-rew]}
\index{reduction!R@$\cR$|inddef[def-rew]}

  A {\em rewrite rule\,} is a pair of terms $l\a r$ such that $l$ is
  an algebraic term distinct from a variable and $\FV(r) \sle
  \FV(l)$. A rule $l\a r$ is {\em left-linear\,} if no variable occurs
  more than once in $l$.
  
  Given a set of rewrite rules $\cR$, {\em $\cR$-reduction\,} $\ar$ is
  the smallest relation containing $\cR$ and stable by substitution
  and context. A term of the form $l\s$ with $l\a r\in \cR$ and $\s$ a
  substitution is an {\em $\cR$-redex\,}.
  
  Given a set of symbols $\cG$, we denote by $\cR_\cG$ the set of
  rules which {\em define\,} a symbol in $\cG$, that is, the set of
  rules such that the head symbol of the left-hand side is a symbol of
  $\cG$.
  
  A symbol $f$ is {\em constant\,} if $\cR_{\{f\}} = \vide$, otherwise
  it is (partially) {\em defined\,}. We denote by $\cC\cF$ the set of
  constant symbols and by $\cD\cF$ the set of defined symbols.
\end{dfn}


\begin{dfn}[ATS]
\label{def-ats}
\index{ATS|inddef[def-ats]}
\index{Type System!Algebraic|inddef[def-ats]}
\index{algebraic!type system|inddef[def-ats]}

  An {\em ATS\,} is a pre-RTS whose reduction relation $\a$ is of the
  form $\ar\cup\ab$ with $\cR$ a set of rewrite rules.
\end{dfn}


Now that we have introduced our notion of rewriting, we can wonder
under which conditions it has the subject reduction property.

With first-order rewriting in sorted algebras, for rewriting to
preserve the sort of terms, it suffices that, for all rules, both
sides of the rule have the same sort.

Carried over to type systems, this condition gives~: there exists an
environment $\G$ and a type $T$ such that $\G\th l:T$ and $\G\th
r:T$. This condition is the one which has been taken in all previous
work combining typed $\la$-calculus and rewriting.

However, this condition has an important drawback. With polymorphic or
dependent types, it leads to strongly non-left-linear rules. This has
two important consequences. First, rewriting is strongly slowed down
because of the necessary equality tests. Second, it is more difficult
to prove confluence with non-left-linear rules.

~

Let us take the example of the concatenation of two polymorphic lists
in the Calculus of Constructions ($\cS=\{\st,\B\}$, $\cA=\{(\st,\B)\}$
and $\cB= \{(s_1,s_2,s_3) \in\cS^3~|~ s_2=s_3\}$)~:

\newcommand{\FB}{\cF^\B}
\newcommand{\Fs}{\cF^\st}

\begin{lst}{--}
\item $list\in\FB_1$ with $\tau_{list}= \st\a\st$ the type of
polymorphic lists,
\item $nil\in\Fs_1$ with $\tau_{nil}= \p{A}\st list(A)$ the empty
list,
\item $cons\in\Fs_3$ with $\tau_{cons}= \p{A}\st A\a list(A)\a
list(A)$ the function adding an element at the head of a list,
\item $app\in\Fs_3$ with $\tau_{app}= \p{A}\st list(A)\a list(A)\a
list(A)$ the concatenation function.
\end{lst}

A usual definition for $app$ is~:

\begin{lst}{--}
\item $app(A,nil(A),\ell') \a \ell'$
\item $app(A,cons(A,x,\ell),\ell') \a cons(A,x,app(A,\ell,\ell'))$
\end{lst}

This definition satisfies the usual condition; it suffices to take
$\G= A\!:\!\st, x\!:\!A, \ell\!:\!list(A), \ell'\!:\!list(A)$ and $T=
list(A)$. But one may wonder whether it is really necessary to do an
equality test between the first argument of $app$ and the first
argument of $cons$ when one wants to apply the second rule. Indeed, if
$app(A,cons(A',x,\ell),\ell')$ is well typed then, by inversion,
$cons(A',x,\ell)$ is of type $list(A)$ and, by inversion again,
$list(A')$ is convertible to $list(A)$. Hence, to allow the reduction
even though $A'$ is different from $A$ does not seem to be harmful
since $list(A')$ is convertible to $list(A)$.

~

In fact, what is important is not that the left-hand side of a rule is
typable, but that, if an instance of the left-hand side of a rule is
typable, then the corresponding instance of the right-hand side has
the same type. We express this by requiring that there exists an
environment $\G$ in which the right-hand side is typable, and a
substitution $\r$ which replaces the variables of the left-hand side
not belonging to $\G$ by terms typable typable in $\G$. Hence, one can
consider the following rules~:

\begin{lst}{--}
\item $app(A,nil(A'),\ell') \a \ell'$
\item $app(A,cons(A',x,\ell),\ell') \a cons(A,x,app(A,\ell,\ell'))$
\end{lst}

\noindent
by taking $\G= A\!:\!\st, x\!:\!A, \ell\!:\!list(A),
\ell'\!:\!list(A)$ and $\r=\{A'\to A\}$. In \cite{blanqui01lics}, we
give 5 conditions, (S1) to (S5), which must be satisfied by the rule
$l\a r$, $\G$ and $\r$. Assume that $l= f(\vl)$, $\tf= \pvxT U$ and
$\g= \vxl$. Then, (S1) is $\dom(\r)\sle \FV(l)\moins\dom(\G)$ and (S2)
is $\G\th l\r:U\g\r$. Although these two first conditions are often
true, they are not necessary for proving the subject reduction
property. This is why, in the following definition, they are not
written. However, we will see that (S2) is necessary for proving the
strong normalization property (see Definition~\ref{def-wf-rule}).


\begin{dfn}[Well-typed rule]
\label{def-wt-rule}
\index{well-typed!rule|inddef[def-wt-rule]}
\index{rule!well-typed|inddef[def-wt-rule]}
\index{rho@$\r$|inddef[def-wt-rule]}
\index{$(l\a r,\G,\r)$|inddef[def-wt-rule]}

A rule $l\a r$ is {\em well-typed\,} if there exists an environment
$\G$ and a substitution $\r$ such that, if $l= f(\vl)$, $\tf= \pvxT U$
and $\g= \vxl$ then~:

\index{S3@\textbf{S3}|inddef[def-wt-rule]}
\index{S4@\textbf{S4}|inddef[def-wt-rule]}
\index{S5@\textbf{S5}|inddef[def-wt-rule]}

\begin{bfenumi}{S}
\setcounter{enumi}{2}
\item $\G\th r:U\g\r$,
\item for all $\D$, $\s$ and $T$, if $\D\th l\s:T$ then $\s: \G\a\D$,
\item for all $\D$, $\s$ and $T$, if $\D\th l\s:T$ then, for all $x$,
$x\s \ad x\r\s$.
\end{bfenumi}

\noindent
In the following, we will write $(l\a r,\G,\r) \in \cR$ when the
previous conditions are satisfied.
\end{dfn}


An example using a dependent type is given by the concatenation of two
lists of given length and the function $map$ which, to a function $f$
and a list $a_1\ldots a_n$, associates the list $f(a_1)\ldots
f(a_n)$~:

\begin{lst}{--}
\item $T\in\FB_0$ with $\tau_T=\st$ a type constant,
\item $nat\in\FB_0$ with $\tau_{nat}=\st$ the type of natural numbers,
\item $0\in\Fs_0$ with $\tau_0=nat$ zero,
\item $s\in\Fs_1$ with $\tau_s=nat\a nat$ the successor function,
\item $+\in\Fs_2$ with $\tau_+=nat\a nat\a nat$ the addition on $nat$,
\item $listn\in\FB_1$ with $\tau_{listn}= nat\a\st$ the type of lists
of given length,
\item $niln\in\Fs_0$ with $\tau_{niln}= listn(0)$ the empty list,
\item $consn\in\Fs_3$ with $\tau_{consn}= T\a \p{n}{nat} listn(n)\a
listn(s(n))$ the function adding an element at the head of a list,
\item $appn\in\Fs_4$ with $\tau_{appn}= \p{n}{nat} listn(n)\a
  \p{n'}{nat} listn(n')\a listn(n+n')$ the concatenation function,
\item $mapn\in\Fs_3$ with $\tau_{mapn}= (T\a T)\a \p{n}{nat}
listn(n)\a listn(n)$ the function which, to a function $f:T\a T$ and a
list $a_1\ldots a_n$, associates the list $f(a_1)\ldots f(a_n)$,
\end{lst}

\noindent
where $+$, $appn$ and $mapn$ are defined by~:

\begin{lst}{--}
\item $+(0,n') \a n'$
\item $+(s(n),n') \a s(n+n')$
\item $appn(0,\ell,n',\ell') \a \ell'$
\item $appn(p,consn(x,n,\ell),n',\ell') \a
  consn(x,n+n',appn(n,\ell,n',\ell'))$
\item $mapn(f,0,\ell) \a \ell$
\item $mapn(f,p,consn(x,n,\ell)) \a consn(fx,n,mapn(f,n,\ell))$
\item $mapn(f,p,appn(n,\ell,n',\ell')) \a
  appn(n,mapn(f,n,\ell),n',mapn(f,n',\ell'))$
\end{lst}

For the second rule of $appn$, we take $\G= x\!:\!T, n\!:\!nat,
\ell\!:\!listn(n), n'\!:\!nat, \ell'\!:\!listn(n')$ and $\r= \{p\to
s(n)\}$. This avoids checking that $p$ is convertible to $s(n)$.

For the third rule of $mapn$, we take $\G= f\!:\!T\a T, n\!:\!nat,
\ell\!:\!listn(n), n'\!:\!nat, \ell'\!:\!listn(n')$ and $\r= \{p\to
n+n'\}$. This avoids checking that $p$ is convertible to $n+n'$.

The reader will find other examples in Section~\ref{sec-cicr}.


\begin{lem}[Subject reduction for rewriting]
\label{lem-cor-ar}
\index{subject reduction!rewriting|indlem[lem-cor-ar]}

If $\cR$ is a set of well-typed rules then $\ar$ preserves typing.
\end{lem}

\begin{prf}
  We proceed as for the correctness of $\ab$ and only consider case
  (symb)~:
\begin{lst}{}
\item [\bf(symb)] $\cfrac{\th \tf:s \quad \G \mbox{ valid} \quad \G\th
    t_1:T_1\g \ldots \G\th t_n:T_n\g}{\G\th f(\vt):U\g}$
  \quad $\begin{array}{c}
(f\in\cF^s_n,\\
\tf= \pvxT U,\\
\g= \vxt)\\
\end{array}$

  Let $(l\a r,$ $\G_0,\r) \in \cR$ with $l=f(\vl)$, $\tf= \pvxT U$ and
  $\g_0= \vxl$. Assume that $t=l\s$. We prove that $\G\th r\s:U\g$. By
  {\bf(S4)}, \index{S4@\textbf{S4}|indthm[thm-admis]} $\s:
  \G_0\a\G$. By {\bf(S3)}, \index{S3@\textbf{S3}|indthm[thm-admis]}
  $\G_0\th r:U\g_0\r$. Therefore, by substitution, $\G\th
  r\s:U\g_0\r\s$. By {\bf(S5)},
  \index{S5@\textbf{S5}|indthm[thm-admis]} for all $x$, $x\r\s$ and
  $x\s$ have a common reduct that we will call $t_x$. Therefore, by
  successively reducing in $U\g_0\r\s$ each $x\r\s$ to $t_x$, and in
  $U\g_0\s$ each $x\s$ to $t_x$, we obtain $U\g_0\r\s \ad
  U\g_0\s$. But $U\g_0\s= U\g$ and, by inversion, there exists $s'$
  such that $\G\th U\g:s'$. Therefore, by (conv), $\G\th
  r\s:U\g_0\s$.\cqfd
\end{lst}
\end{prf}


\begin{thm}[Admissibility]
\label{thm-admis}
\index{admissible!ATS|indthm[thm-admis]}
\index{subject reduction|indthm[thm-admis]}
\index{RTS|indthm[thm-admis]}

  A logical ATS whose rules are well-typed is an RTS, {\em i.e.\,} its
  reduction relation preserves typing.
\end{thm}

\begin{prf}
  It is true for $\ab$ since we assume that the ATS is logical. For
  $\ar$, this comes from the correctness of rewriting.\cqfd
\end{prf}


~

How to check the conditions (S3), (S4) and (S5) ?
\index{decidability!well-typed rule|indpage} In all their generality,
they are certainly undecidable. On the one hand, we do not know
whether $\th$ and $\ad$ are decidable and, on the other hand, in (S4)
and (S5), we arbitrarily quantify on $\D$, $\s$ and $T$. It is
therefore necessary to make additional hypothesis. In the following,
we successively consider the three conditions.


~

Let us look at (S3). \index{S3@\textbf{S3}|indpage} In practice, the
symbols and their defining rules are often added one after another (or
by groups but the following argument can be generalized). Let
$(\cF,\cR)$ be a system in which $\th$ is decidable (for example, a
functional, confluent and strongly normalizing system), $f\notin\cF$
and $\cR_f$ a set of rules defining $f$ and whose symbols belong to
$\cF'=\cF\cup\{f\}$. Then, in $(\cF',\cR)$, $\th$ is still
decidable. One can therefore try to check (S3) in this system. This
does not seem an important restriction~: it would be surprising if the
typing of a rule required the use of the rule itself !


~

We now consider (S4).


\begin{dfn}[Canonical and derived types]
\label{def-der-typ}
\label{def-can-typ}
\index{canonical!type|inddef[def-der-typ]}
\index{type!canonical|inddef[def-der-typ]}
\index{derived type|inddef[def-der-typ]}
\index{type!derived|inddef[def-der-typ]}
\index{tautp@$\tau(t,p)$|inddef[def-der-typ]}

Let $t$ be a term of the form $l\s$ with $l=f(\vl)$ algebraic,
$\tf=\pvxT U$ and $\g=\vxl$. The term $U\g\s$ will be called the {\em
canonical type\,} of $t$.
  
Let $p\in \pos(l)$ with $p\neq\vep$. We define the {\em type of $t|_p$
derived from $t$\,}, $\tau(t,p)$, as follows~:

\begin{lst}{--}
\item if $p=i$ then $\tau(t,p)= T_i\g\s$,
\item if $p=iq$ and $q\neq \vep$ then $\tau(t,p)= \tau(t_i,q)$.
\end{lst}
\end{dfn}

In fact, the type of $t|_p$ derived from $t$ only depends on the term
just above $t|_p$ in $t$.

The following lemma shows that the canonical type of $t$ and the type
of $t|_p$ derived from $t$ are indeed types for $t$ and $t|_p$
respectively.


\begin{lem}
\label{lem-der-typ-prop}
\index{canonical!type|indlem[lem-der-typ-prop]}
\index{type!canonical|indlem[lem-der-typ-prop]}
\index{derived type|indlem[lem-der-typ-prop]}
\index{type!derived|indlem[lem-der-typ-prop]}

  Let $t$ be a term of the form $l\s$ with $l=f(\vl)$ algebraic and
  $\G\th t:T$, $V$ the canonical type of $t$ and $p\in \pos(l)$ with
  $p\neq\vep$. In any TSM, $\G\th t:V$ and $\G\th t|_p:\tau(t,p)$.
\end{lem}

\begin{prf}
  From $\G\th t:T$, by inversion, we immediately obtain $\G\th
  t:V$. Let us consider $\G\th t|_p:\tau(t,p)$ now. As $p\neq\vep$, we
  have $p=qi$, $t|_q$ of the form $g(\vec{k}\s)$ with $g(\vec{k})$
  algebraic and $t|_q$ typable in $\G$. Assume that $\tg= \pvxT U$ and
  $\g= \{\vx\to\vec{k}\}$. Then, $\tau(t,p)= T_i\g\s$ and, by
  inversion, $\G\th t|_p:T_i\g\s$.\cqfd
\end{prf}


\begin{lem}[S4]
\label{lem-S4}
\index{S4@\textbf{S4}|indlem[lem-S4]}

  Let $l\a r$ be a rule and $\G$ be an environment. If, for all $x\in
  \dom(\G)$, there exists $p_x\in \pos(x,l)$ such that $\tau(l,p_x)=
  x\G$, then (S4) is satisfied.
\end{lem}

\begin{prf}
  Assume that $\D\th l\s:T$. As $l$ is algebraic, by inversion, $\D\th
  x\s:\tau(l\s,p_x)= \tau(l,p_x)\s= x\G\s$.\cqfd
\end{prf}


~

On the other hand, for (S5), \index{S5@\textbf{S5}|indpage} we have no
general result. By inversion, (S5) can be seen as a problem of
unification modulo $\ad^*$. The confluence of $\a$ (which implies
$\ad^*=\ad$) can therefore be very useful. Unfortunately, there are
very few general results on the confluence of $\ar\cup\ab$ (see the
discussion after Definition~\ref{def-cond}). On the other hand, one
can easily prove that the local confluence is preserved.


\begin{lem}[Local confluence]
\label{lem-loc-confl}
\index{local confluence|indlem[lem-loc-confl]}
\index{confluence!local|indlem[lem-loc-confl]}

If $\ar$ is locally confluent on $\cT(\cF,\cX)$ then $\a=$
$\ar\cup\ab$ is locally confluent on $\cT$.
\end{lem}

\begin{prf}
  We write $t\a^p t'$ when there exists $u$ such that $t|_p \a u$ and
  $t'= t[u']_p$ (reduction at position $p$). Assume that $t\a^p t_1$
  and $t\a^q t_2$. We prove by induction on $t$ that there exists $t'$
  such that $t_1\a^* t'$ and $t_2\a^* t'$. There is three cases~:

\begin{lst}{\bu}
\item $p$ and $q$ have no common prefix. The reductions at $p$ and $q$
can be done in parallel~: $t_1\a^q t_1'$, $t_2\a^p t_2'$ and
$t_1'=t_2'$.

\item $p=ip'$ and $q=iq'$. We can conclude by induction hypothesis on
$t|_i$.
  
\item $p=\vep$ or $q=\vep$. By exchanging the roles of $p$ and $q$, we
can assume that $p=\vep$. Then, there is two cases~:

\begin{lst}{--}
\item $t= \lx{V}u ~v$ and $t_1= u\xv$. One can distinguish three
sub-cases~:

\begin{lst}{$\circ$}
\item $q=11q'$ and $V\a^{q'} V'$. Then $t'= t_1$ works.
\item $q=12q'$ and $u\a^{q'} u'$. Then $t'= u'\xv$ works.
\item $q=2q'$ and $v\a^{q'} v'$. Then $t'= u\{x\to v'\}$ works.
\end{lst}

\item $t=l\s$, $l\a r\in\cR$ and $t_1=r\s$. There exists an algebraic
term $u$ of maximal size and a substitution $\t$ such that $t=u\t$ and
$x\t=y\t$ implies $x=y$ ($u$ and $\t$ are unique up to the choice of
variables and $u$ has the same non-linearities than $t$). As the
left-hand sides of rules are algebraic, $u=l\s'$ and $\s=\s'\t$. Now,
one can distinguish two sub-cases~:

\begin{lst}{$\circ$}
\item $q\in\pos(u)$. As the left-hand sides of rules are algebraic, we
have $u\ar r\s'$ and $u\ar v$. By local confluence of $\ar$ on
$\cT(\cF,\cX)$, there exists $u'$ such that $r\s' \a^* u'$ and $v \a^*
u'$. Hence, $t_1= r\s'\t \a^* u'\t$ and $t_2= v\t \a^* u'\t$.

\item $q=q_1q'$ and $u|_{q_1}= x$. Let $q_2, \ldots, q_n$ be the
positions of the other occurrences of $x$ in $u$. If one reduces $t_2$
at each position $q_iq'$, one obtains a term of the form $l\s'\t'$
where $\t'$ is the substitution equal to $\t$ but for $x$ where it is
equal to the reduct of $x\t$. Then , it suffices to take $t'=
r\s'\t'$.\cqfd
\end{lst}
\end{lst}
\end{lst}
\end{prf}


%% file: conditions.tex



\chapter{Conditions of Strong Normalization}
\label{chap-conditions}

In this chapter, we are going to give strong normalization conditions
for ATS's based on the Calculus of Constructions.


\begin{dfn}[CAC]
\label{def-cac}
\index{CAC|inddef[def-cac]}
\index{Calculus of Constructions!Algebraic|inddef[def-cac]}
\index{$\st,\B$|inddef[def-cac]}
\index{typing!CAC|inddef[def-cac]}

A {\em Calculus of Algebraic Constructions\,} (CAC) is an ATS
$(\cS,\cF,\cX,\cA,\cB,\tau,\cR)$ such that $\cS= \{\st,\B\}$, $\cA=
\{(\st,\B)\}$ and $\cB= \{(s_1,s_2,s_3) \in \cS^3 ~|~ s_2=s_3\}$.
\end{dfn}

A CAC is injective and regular.




\section{Term classes}


After the Separation and Classification Lemmas, typable terms can be
divided into three disjoint classes~: $\T_0^\B$, $\T_1^\B$ and
$\T_1^\st$. For denoting them, we introduce more explicit notations.

\begin{dfn}[Typing classes]
\label{def-typ-class}
\index{class!typing|inddef[def-typ-class]}
\index{typing!class|inddef[def-typ-class]}
\index{K@$\K$|inddef[def-typ-class]}
\index{P@$\P$|inddef[def-typ-class]}
\index{O@$\O$|inddef[def-typ-class]}
\index{predicate!type|inddef[def-typ-class]}
\index{predicate|inddef[def-typ-class]}
\index{object|inddef[def-typ-class]}

\hfill
\begin{lst}{--}
\item Let $\K= \T_0^\B$ be the class of {\em predicate types\,}.
\item Let $\P= \T_1^\B$ be the class of {\em predicates\,}.
\item Let $\O= \T_1^\st$ be the class of {\em objects\,}.
\end{lst}
\end{dfn}


\newcommand{\XB}{\cX^\B}
\newcommand{\Xs}{\cX^\st}

That a well-typed term belongs to one of these classes can be easily
decided by introducing the following syntactic classes~:

\begin{dfn}[Syntactic classes]
\label{def-syn-class}
\index{class!syntactic|inddef[def-syn-class]}
\index{syntactic class|inddef[def-syn-class]}
\index{K@$\cK$|inddef[def-syn-class]}
\index{P@$\cP$|inddef[def-syn-class]}
\index{O@$\cO$|inddef[def-syn-class]}

\hfill
\begin{lst}{\bu}
\item The syntactic class $\cK$ of {\em predicate types\,}~:
\begin{lst}{--}
\item $\st\in \cK$,
\item if $x\in \cX$, $T\in \cT$ and $K\in \cK$ then $\px{T}K \in \cK$.
\end{lst}

\item The syntactic class $\cP$ of {\em predicates\,}~:
\begin{lst}{--}
\item $\XB \sle \cP$,
\item if $x\in \cX$, $T\in \cT$ and $P\in \cP$ then $\px{T}P \in \cP$
  and $\lx{T}P \in \cP$,
\item if $P\in \cP$ and $t\in \cT$ then $Pt \in \cP$,
\item if $F\in\FB_n$ and $t_1,\ldots,t_n\in \cT$ then $F(\vt)\in \cP$.
\end{lst}

\item The syntactic class $\cO$ of {\em objects\,}~:
\begin{lst}{--}
\item $\Xs \sle \cO$,
\item if $x\in \cX$, $T\in \cT$ and $u\in \cO$ then $\lx{T}u \in \cO$,
\item if $u\in \cO$ and $t\in \cT$ then $ut \in \cO$,
\item if $f\in\Fs_n$ and $t_1,\ldots,t_n\in \cT$ then $f(\vt)\in \cO$.
\end{lst}
\end{lst}
\end{dfn}


\begin{lem}
  Syntactic classes are disjoint from one another and each typing
  class is included in its corresponding syntactic class~: $\K \sle
  \cK$, $\P \sle \cP$ and $\O \sle \cO$.
\end{lem}

\begin{prf}
  That the syntactic classes are disjoint from one another comes from
  their definition. We prove that if $\G \th t:T$ then $t$ belongs to
  the syntactic class corresponding to its typing class by induction
  on $\G \th t:T$. We follow the notations used in the typing rules.

\begin{lst}{}
\item [\bf(ax)] As $\cA= \{(\st,\B)\}$, we necessarily have $s_1=\st$
and $s_2=\B$. But, $\st\in \K\cap\cK$.
  
\item [\bf(symb)] By inversion and regularity, $\G\th
U\g:s$. Therefore, if $f\in \Fs$ then $f(\vt)\in \O\cap\cO$, and if
$f\in \FB$ then $f(\vt)\in \P\cap\cP$.
  
\item [\bf(var)] If $x\in \Xs$ then $x\in \O\cap\cO$, and if $x\in
\XB$ then $x\in \P\cap\cP$.
  
\item [\bf(weak)] By induction hypothesis.
  
\item [\bf(prod)] By regularity, $U$ and $\px{T}U$ have the same
type. We can therefore conclude by induction hypothesis on $U$.
  
\item [\bf(abs)] By inversion and regularity, $\px{T}U$ and $U$ have
the same type. We can therefore conclude by induction hypothesis on
$u$.
  
\item [\bf(app)] By inversion and regularity, $V\xu$ and $\px{U}V$
have the same type. We can therefore conclude by induction hypothesis
on $t$.
  
\item [\bf(conv)] By conversion correctness, $T$ and $T'$ have the
same type. We can therefore conclude by induction hypothesis.\cqfd
\end{lst}
\end{prf}




\section{Inductive types and constructors}
\label{sec-ind-typ}
\index{inductive!type|indsec[sec-ind-typ]}

Until now we made few hypothesis on symbols and rewrite rules. However
N. P. Mendler \cite{mendler87thesis} showed that the extension of the
simply-typed $\la$-calculus with recursion on inductive types is
strongly normalizing if and only if the inductive types satisfy some
positivity condition.

A base type $T$ occurs positively in a type $U$ if all the occurrences
of $T$ in $U$ are on the left of a even number of $\a$. A type $T$ is
positive if $T$ occurs positively in the type of the arguments of its
constructors. Usual inductive types like natural numbers and lists of
natural numbers are positive.

~

Now let us see an example of a non-positive type $T$. Let $U$ be a
base type. Assume that $T$ has constructor $c$ of type $(T\a U)\a T$.
$T$ is not positive because $T$ occurs at a negative position in $T\a
U$.  Consider now the function $p$ of type $T\a (T\a U)$ defined by
the rule $p(c(x)) \a x$. Let $\w= \la x. p(x)x$ of type $T\a U$. Then
the term $\w c(\w)$ of type $U$ is not normalizable~:

\begin{center}
$\w c(\w) \,\ab\, p(c(\w))c(\w) \,\ar\, \w c(\w) \,\ab\, \ldots$
\end{center}

In the case where $U=\st$, we can interpret this as Cantor's Theorem~:
there is no surjection from a set $T$ to the set of its subsets
$T\a\st$. In this interpretation, $p$ is the natural injection between
$T$ and $T\a\st$. Saying that $p$ is surjective is equivalent to
saying (with the Axiom of Choice) that there exists $c$ such that $p
\circ c$ is the identity, that is, such that $p(c(x)) \a x$. In
\cite{dowek99hab}, G.  Dowek shows that such an hypothesis is
incoherent. Here, we show that this is related to the
non-normalization of non-positive inductive types.

N. P. Mendler also gives a condition, strong positivity, in the case
of dependent and polymorphic types. A similar notion, but more
restrictive, strict positivity, is used by T. Coquand and C.  Paulin
in the Calculus of Inductive Constructions \cite{coquand88colog}.

~

Hereafter we introduce the more general notion of {\em structure
  inductive admissible\,}. In particular, we do not consider that a
constructor must be constant~: it will be possible to have rewrite
rules on constructors. This will allow us to formalize quotient types
as the type $int$ of integers~:

\begin{lst}{--}
\item $int\in\FB_0$ with $\tau_{int}=\st$ the type of integers,
\item $0\in\Fs_0$ with $\tau_0=int$ the constant zero,
\item $s\in\Fs_1$ with $\tau_s=int\a int$ the successor function,
\item $p\in\Fs_1$ with $\tau_p=int\a int$ the predecessor function,
\end{lst}

\noindent
where $s$ and $p$ are defined by~:

\begin{lst}{--}
\item $s(p(x)) \a x$
\item $p(s(x)) \a x$
\end{lst}


\newcommand{\FC}{\cC\FB}
\newcommand{\FD}{{\cD\FB}}
\newcommand{\co}{\cC o}

\newcommand{\ind}{\mr{Ind}}
\newcommand{\acc}{\mr{Acc}}

\begin{dfn}[Constructors]
\label{def-cons}
\index{constructor|inddef[def-cons]}

Let $C$ be a constant predicate symbol. A symbol $f$ is a {\em
  constructor\,} of $C$ if $\tf$ is of the form $\pvyU C(\vv)$ with
$\at_f=|\vy|$.
\end{dfn}

Our notion of constructor not only includes the usual (constant) constructors
but also any symbol producing terms of type $C$. For example~:

\begin{lst}{--}
\item $+\in\Fs_2$ with $\tau_+=int\a int\a int$ the addition on
  integers,
\item $\times\in\Fs_2$ with $\tau_\times=int\a int\a int$ the
  multiplication on integers,
\end{lst}

\noindent
or with polymorphic lists~:

\begin{lst}{--}
\item $app\in\Fs_3$ with $\tau_{app}= \p{A}\st list(A)\a list(A)\a
  list(A)$ the concatenation function.
\end{lst}


A constant predicate symbol having some constructors cannot have any
arity~:

\begin{dfn}[Maximal arity]
\label{def-max-arity}
\index{maximal!arity|inddef[def-max-arity]}
\index{arity!maximal|inddef[def-max-arity]}

A predicate symbol $F$ is of {\em maximal arity\,} if $\tF=\pvxT\st$
and $\at_F=|\vx|$.
\end{dfn}


\begin{lem}
\label{lem-max-arity}
\index{maximal!arity|indlem[lem-max-arity]}
\index{arity!maximal|indlem[lem-max-arity]}

Let $C$ be a constant predicate symbol and $c$ a constructor of $C$.
In a logical CAC, if $\th\tC:\B$ and $\th\tc:s$ then $s=\st$ and $C$
is of maximal arity.
\end{lem}

\begin{prf}
  Assume that $\tC= \pvxV W$ and $\tc= \pvyU C(\vv)$. Let $\g=\vxv$.
  As $\B$ is a maximal sort and $\th\tC:\B$, after the lemma on
  maximal sorts, $W$ is of the form $(\vx':\vV')\st$. Now, from
  $\th\tc:s$, by inversion and regularity, on deduce that $\G_c\th
  C(\vv):s$, $\G_c\th C(\vv):W\g$ and $W\g \CV[\G_c] s$. As $\G_c\th
  W\g:\B$, by conversion correctness, $\G_c\th s:\B$ and, by
  inversion, $s=\st$. Therefore, after the lemma on maximal sorts,
  $|\vx'|=0$ and $W=\st$.\cqfd
\end{prf}


\begin{dfn}[Inductive structure]
\label{def-ind-str}
\index{inductive!structure|inddef[def-ind-str]}
\index{$>_\cC,\ge_\cC,=_\cC$|inddef[def-ind-str]}
\index{ordering!$>_\cC$|inddef[def-ind-str]}
\index{IndC@$\ind(C)$|inddef[def-ind-str]}
\index{Accc@$\acc(c)$|inddef[def-ind-str]}
\index{inductive!position|inddef[def-ind-str]}
\index{position!inductive|inddef[def-ind-str]}
\index{accessible!position|inddef[def-ind-str]}
\index{position!accessible|inddef[def-ind-str]}

  An {\em inductive structure\,} is given by~:
\begin{lst}{\bu}
\item a quasi-ordering $\ge_\cC$ on $\FC$ whose strict part $>_\cC$ is
well-founded;
\item for every constant predicate symbol $C$ of type $\pvxT\st$, a
set $\ind(C) \sle \{i\le\at_C ~|~ x_i\in\XB\}$ for the {\em inductive
positions\,} of $C$;
\item for every constructor $c$, a set $\acc(c) \sle
\{1,\ldots,\at_c\}$ for the {\em accessible positions\,} of $c$.
\end{lst}
\end{dfn}

The accessible positions denote the arguments that one wants to use in
the right hand-sides of rules. The inductive positions denote the
parameters in which constructors must be monotone.


\newcommand{\pp}{\pos^+}
\renewcommand{\pm}{\pos^-}
\renewcommand{\pz}{\pos^0}
\newcommand{\pnz}{\pos^{\neq 0}}
\newcommand{\pd}{\pos^\d}

\begin{dfn}[Positive and negative positions]
\label{def-neg-pos}
\index{position!positive|inddef[def-neg-pos]}
\index{positive!position|inddef[def-neg-pos]}
\index{position!negative|inddef[def-neg-pos]}
\index{negative position|inddef[def-neg-pos]}
\index{position!neutral|inddef[def-neg-pos]}
\index{neutral!position|inddef[def-neg-pos]}
\index{Posplus@$\pp,\pm,\pd,\pz,\pnz$|inddef[def-neg-pos]}

Let $T\in \cT\moins\cO$. The set of {\em positive positions\,} in $T$,
$\pp(T)$, and the set of {\em negative positions\,} in $T$, $\pm(T)$,
are simultaneously defined by induction on the structure of $T$~:

\begin{lst}{--}
\item $\pp(s) = \pp(F(\vt)) = \pp(X) = \vep$,
\item $\pm(s) = \pm(F(\vt)) = \pm(X) = \vide$,
\item $\pd(\px{V}W) = 1.\pos^{-\d}(V) \cup 2.\pd(W)$,
\item $\pd(\lx{V}W) = 1.\pos(V) \cup 2.\pd(W)$,
\item $\pd(Vu) = 1.\pd(V) \cup 2.\pos(u)$,
\item $\pd(VU) = 1.\pd(V)$,
\item $\pp(C(\vt)) = \{\vep\} \cup \bigcup\, \{i.\pp(t_i) ~|~
  i\in\ind(C)\}$,
\item $\pm(C(\vt)) = \bigcup\, \{i.\pm(t_i) ~|~ i\in\ind(C)\}$,
\end{lst}

\noindent
where $\d\in\{-,+\}$, $-+=-$, $--=+$ (usual rule of signs).  The set
of {\em neutral positions\,} in $T$ is $\pz(T)= \pp(T) \cap \pm(T)$.
The set of {\em non-neutral positions\,} in $T$ is $\pnz(T) = (\pp(T)
\cup \pm(T)) \moins \pz(T)$.
\end{dfn}

The positive and negative positions do not form two disjoint
sets. Their intersection forms the neutral positions. For example, all
the positions of $u$ in $Vu$ or all the positions of $V$ in $\lx{V}W$
are neutral. We will see in Section~\ref{sec-schema-int} that these
subterms are not taken into account into the interpretation of a type.


~

In \cite{blanqui01lics}, we give 6 conditions, (I1) to (I6), for
defining what is an admissible inductive structure. But we found that
(I1) can be eliminated if we modify (I2) a little bit. That is why, in
the following definition, there is no (I1) and (I2) is placed after
(I6).

\begin{dfn}[Admissible inductive structures]
\label{def-adm-ind-str}
\index{admissible!inductive structure|inddef[def-adm-ind-str]}
\index{inductive!structure!admissible|inddef[def-adm-ind-str]}

\hfill An inductive structure is\\ {\em admissible\,} if for all
constant predicate symbol $C$, for all constructor $c$ of type $\pvyU$
$C(\vv)$ and for all $j \in \acc(c)$~:

\begin{bfenumi}{I}
\setcounter{enumi}{2}
\index{I3@\textbf{I3}|inddef[def-adm-ind-str]}
\item $\all D\in\FC, D =_\cC C \A \pos(D,U_j) \sle \pp(U_j)$ (symbols
equivalent to $C$ must be at positive positions),

\index{I4@\textbf{I4}|inddef[def-adm-ind-str]}
\item $\all D\in\FC, D >_\cC C \A \pos(D,U_j) \sle \pz(U_j)$ (symbols
greater than $C$ must be at neutral positions),

\index{I5@\textbf{I5}|inddef[def-adm-ind-str]}
\item $\all F\in\FD, \pos(F,U_j) \sle \pz(U_j)$ (defined symbols must
be at neutral positions),

\index{I6@\textbf{I6}|inddef[def-adm-ind-str]}
\index{iotaY@$\io_Y$|inddef[def-adm-ind-str]}
\item $\all\, Y\!\in\FVB(U_j), \ex\, \io_Y\le\at_C, v_{\io_Y}= Y$
(every predicate variable in $U_j$ must be a parameter of $C$),

\index{I2@\textbf{I2}|inddef[def-adm-ind-str]}
\setcounter{enumi}{1}
\item $\all\, Y\!\in\FVB(U_j), \io_Y\in\ind(C) \A \pos(Y,U_j) \sle
\pp(U_j)$ (every predicate variable in $U_j$ which is an inductive
parameter of $C$ must be at a positive position).
\end{bfenumi}
\end{dfn}

For example, $\ind(list)= \{1\}$, $\acc(nil)= \{1\}$ and $\acc(cons)=
\{1,2,3\}$ is an admissible inductive structure. Assume we add~:

\begin{lst}{--}
\item $tree\in\FB_0$ with $\tau_{tree}=\st$ the type of finite
  branching trees,
\item $node\in\Fs_1$ with $\tau_{node}=list(tree)\a tree$ its
  constructor.
\end{lst}

Since $1\in\ind(list)$, if $\ind(tree)=\vide$ and $\acc(node)=\{1\}$
then we still have an admissible structure.

~

To allow greater or defined symbols does not matter if these symbols
are at neutral positions since neutral subterms are not taken into
account into the interpretation of a type.

~

The condition (I6) means that the predicate arguments of a constructor
must be parameters of their type. A similar condition appears in the
works of M. Stefanova \cite{stefanova98thesis} (``safeness'') and D.
Walukiewicz \cite{walukiewicz00lfm} (``$\st$-dependency''). On the
other hand, in the Calculus of Inductive Constructions (CIC)
\cite{paulin93tlca}, there is no such restriction. However, because of
the typing rules of the elimination scheme, no very interesting
function seems to be definable on a type not satisfying this
condition.

For example, let us take the type of heterogeneous non empty lists (in
the CIC syntax)~:

\begin{lst}{--}
\item $listh= Ind(X:\st)\{C_1|C_2\}$ where $C_1= \p{A}\st \px{A} X$ and
  $C_2= \p{A}\st \px{A} X\a X$,
\item $endh= Constr(1,listh)$,
\item $consh= Constr(2,listh)$.
\end{lst}

The typing rule of the non dependent elimination scheme
(Nodep$_{\st,\st}$) is~:

\begin{center}
  $\cfrac{\G\th \ell:listh \quad \G\th Q:\st \quad \G\th
    f_1:C_1\{listh,Q\} \quad \G\th f_2:C_2\{listh,Q\}}{\G\th
    Elim(\ell,Q)\{f_1|f_2\} : Q}$
\end{center}

\noindent
where $C_1\{listh,Q\}= \p{A}\st \px{A} Q$ and $C_2\{listh,Q\}=
\p{A}\st \px{A} listh\a Q\a Q$. So, $Q$, $f_1$ and $f_2$ must be
typable in $\G$. The result of $f_1$ or $f_2$ cannot depend on $A$ or
$x$. This means that, for example, it is possible to compute the
length of such a list but one cannot extract an element of such a
list. We think that the length of such a list is an information that
can surely be obtained without using such a data type.


~

We can distinguish several kinds of inductive types.

\begin{dfn}[Primitive, basic and strictly positive predicates]\hfill
\label{def-pred}
\index{predicate!primitive|inddef[def-pred]}
\index{predicate!basic|inddef[def-pred]}
\index{predicate!strictly positive|inddef[def-pred]}
\index{primitive!predicate|inddef[def-pred]}
\index{basic predicate|inddef[def-pred]}
\index{strictly positive!predicate|inddef[def-pred]}

A constant predicate symbol $C$ is~:

\begin{lst}{--}
\item {\em primitive\,} if for all $D =_\cC C$, for all constructor
$d$ of type $\pvyU D(\vw)$ and for all $j\in \acc(d)$, $U_j$ is either
of the form $E(\vt)$ with $E <_\cC D$ and $E$ primitive, or of the
form $E(\vt)$ with $E =_\cC D$.
  
\item {\em basic\,} if for all $D =_\cC C$, for all constructor $d$ of
type $\pvyU D(\vw)$ and for all $j\in \acc(d)$, if $E =_\cC D$ occurs
in $U_j$ then $U_j$ is of the form $E(\vt)$.
  
\item {\em strictly positive\,} if for all $D =_\cC C$, for all
constructor $d$ of type $\pvyU D(\vw)$ and for all $j\in \acc(d)$, if
$E =_\cC D$ occurs in $U_j$ then $U_j$ is of the form $\pvzV E(\vt)$
and no $D' =_\cC D$ occurs in the $V_i$'s.
\end{lst}
\end{dfn}

It is easy to see that a primitive predicate if basic and that a basic
predicate is strictly positive. The type $listn$ of lists of length
$n$ is primitive. The type $list$ of polymorphic lists is basic and
not primitive.

~

The strictly positive predicates are the predicates allowed in the
Calculus of Inductive Constructions (CIC). For example, the type of
well-founded trees or Brouwer's ordinals~:

\begin{lst}{--}
\item $ord\in\FB_0$ with $\tau_{ord}=\st$ the type of Brouwer's
  ordinals,
\item $0\in\Fs_0$ with $\tau_0=ord$ the ordinal zero,
\item $s\in\Fs_1$ with $\tau_s=ord\a ord$ the successor ordinal,
\item $lim\in\Fs_1$ with $\tau_{lim}=(nat\a ord)\a ord$ the limit
  ordinal.\footnote{A term of type $ord$ does not necessary
    correspond to a true ordinal. However, if one carefully chooses
    the functions $f$ for the limit ordinals then one can represent an
    initial enumerable segment of the true ordinals.}
\end{lst}

~

Another example is given by the following process algebra which uses a
choice operator $\S$ other some data type $data$
\cite{sellink93ssl}~:

\begin{lst}{--}
\item $data\in\FB_0$ with $\tau_{data}=\st$ a data type,
\item $proc\in\FB_0$ with $\tau_{proc}=\st$ the type of processes,
\item $\circ\in\Fs_2$ with $\tau_\circ=proc\a proc\a proc$ the
  sequence,
\item $+\in\Fs_2$ with $\tau_+=proc\a proc\a proc$ the
  parallelization,
\item $\d\in\Fs_0$ with $\tau_\d=proc$ the deadlock,
\item $\S\in\Fs_1$ with $\tau_\S=(data\a proc)\a proc$ the choice
  operator.
\end{lst}

~

A last example is given by the first-order predicate calculus~:

\begin{lst}{--}
\item $term\in\FB_0$ with $\tau_{term}=\st$ the type of terms,
\item $form\in\FB_0$ with $\tau_{form}=\st$ the type of formulas,
\item $\ou\in\Fs_2$ with $\tau_\ou=form\a form\a form$ the ``or'',
\item $\neg\in\Fs_1$ with $\tau_\neg=form\a form$ the ``not'',
\item $\all\in\Fs_1$ with $\tau_\all=(term\a form)\a form$ the
  universal quantification.
\end{lst}

~

For the moment, we do not forbid non-strictly positive predicates but
the conditions we will describe in the next section do not allow one
to define functions by recursion on such predicates.

Yet these predicates can be useful as shown by the following
breadth-first label listing function of binary trees defined with the
use of continuations \cite{matthes00}~:

\begin{lst}{--}
\item $tree\in\FB_0$ with $\tau_{tree}=\st$ the type of binary
labeled trees,
\item $L\in\Fs_1$ with $\tau_L= nat\a tree$ the leaf constructor,
\item $N\in\Fs_3$ with $\tau_N= nat\a tree\a tree\a tree$ the node
constructor,
\end{lst}

\vs[.5mm]
\begin{lst}{--}
\item $cont\in\FB_0$ with $\tau_{cont}=\st$ the type of continuations,
\item $D\in\Fs_0$ with $\tau_D=cont$,
\item $C\in\Fs_1$ with $\tau_C=((cont\a list)\a list)\a cont$ its
constructors,
\end{lst}

\vs[.5mm]
\begin{lst}{--}
\item $@\in\Fs_2$ with $\tau_@= cont\a (cont\a list)\a list$
  the application on continuations defined by~:
\item $@(D,g)\a g~D$
\item $@(C(f),g) \a f~g$
\end{lst}

\vs[.5mm]
\begin{lst}{--}
\item $ex\in\Fs_1$ with $\tau_{ex}= cont\a list$ the iterator on
continuations defined by~:
\item $ex(D)\a nil$
\item $ex(C(f))\a f~ \l{k}{cont}ex(k)$
\end{lst}

\newcommand{\brfst}{{\it br\_fst}}

\vs[.5mm]
\begin{lst}{--}
\item $br\in\Fs_2$ with $\tau_{br}= tree\a cont\a cont$,
\item $\brfst\in\Fs_1$ with $\tau_\brfst= tree\a list$ the
breadth-first label listing function defined by~:
\item $\brfst(t)\a ex(br(t,D))$
\item $br(L(x),k)\a C(\l{g}{cont\a list} cons(x,@(k,g)))$
\item $br(N(x,s,t),k)\a C(\l{g}{cont\a list} cons(x,@(k,\l{m}{cont} g~
  br(s,br(t,m))))$
\end{lst}

This function is strongly normalizable since it can be encoded into
the polymorphic $\la$-calculus
\cite{matthes98csl,matthes98thesis}. However, it is not clear how to
define a syntactic condition, a schema, ensuring the strong
normalization of such definitions. Indeed, in the right hand-side of
the second rule defining $ex$, $ex$ is explicitly applied to no
argument smaller than $f$. However $ex$ can only be applied to
subterms of reducts of $f$. But not all the subterms of a computable
term are {\em a priori\,} computable (see
Subsection~\ref{subsec-schema}).




\section{General Schema}




\subsection{Higher-order rewriting}

Which conditions on rewrite rules would ensure the strong
normalization of $\a= \ar\cup \ab$ ? Since the works of V.
Breazu-Tannen and J. Gallier \cite{breazu89icalp} and M. Okada
\cite{okada89issac} on the simply-typed $\la$-calculus or the
polymorphic $\la$-calculus, and later the works of F. Barbanera
\cite{barbanera90ctrs} on the Calculus of Constructions and of D.
Dougherty \cite{dougherty91rta} on the untyped $\la$-calculus, it is
well known that adding first-order rewriting to typed $\la$-calculi
preserves strong normalization. This comes from the fact that
first-order rewriting cannot create new $\b$-redexes. We will prove
that this result can be extended to predicate-level rewriting if some
conditions are fulfilled.

~

However, there are also many useful functions whose definition do not
enter the first-order framework, either because some arguments are not
primitive (the concatenation function $app$ on polymorphic lists), or
because their definition uses higher-order features like the function
$map$ which, to a function $f$ and a list $a_1\ldots a_n$ of elements,
associates the list $f(a_1)\ldots f(a_n)$~:

\begin{lst}{--}
\item $map\in\Fs_4$ with $\tau_{map}= \p{A}\st \p{B}\st (A\a B)\a
  list(A)\a list(B)$
\item $map(A,B,f,nil(A')) \a nil(B)$
\item $map(A,B,f,cons(A',x,\ell)) \a cons(B,fx,map(A,B,f,\ell))$
\item $map(A,B,f,app(A',\ell,\ell')) \a
  app(B,map(A,f,\ell),map(A,f,\ell'))$
\end{lst}

~

This is also the case of recursors~:

\begin{lst}{--}
\item $natrec\in\Fs_4$ with $\tau_{natrec}= \p{A}\st A\a (nat\a A\a A)\a
  nat\a A$ the recursor on natural numbers
\item $natrec(A,x,f,0) \a x$
\item $natrec(A,x,f,s(n)) \a f ~n ~natrec(A,x,f,n)$
\item $plus\in\Fs_0$ with $\tau_{plus}= nat\a nat\a nat$ the addition
  on natural numbers
\item $plus \a \l{p}{nat} \l{q}{nat} natrec(nat,p, \l{q'}{nat}
  \l{r}{nat} s(r), q)$
\end{lst}

\noindent
and of induction principles (recursors are just non-dependent
versions of the corresponding induction principles)~:

\begin{lst}{--}
\item $natind\in\Fs_4$ with $\tau_{natind}= \p{P}{nat\a\st} P0\a
  (\p{n}{nat} Pn\a Ps(n))\a \p{n}{nat} Pn$
\item $natind(P,h_0,h_s,0) \a h_0$
\item $natrec(P,h_0,h_s,s(n)) \a h_s \,n ~natind(P,h_0,h_s,n)$
\end{lst}

~

The methods used by V. Breazu-Tannen and J. Gallier
\cite{breazu89icalp} or D. Dougherty \cite{dougherty91rta} cannot be
applied to our calculus since, on the one hand, in contrast with
first-order rewriting, higher-order rewriting can create $\b$-redexes
and, on the other hand, rewriting is included in the type conversion
rule (conv), hence more terms are typable. But there exists other
methods, available in the simply-typed $\la$-calculus only or in
richer type systems, for proving the termination of this kind of
definitions~:

\begin{lst}{\bu}
\item The {\em General Schema\,}, \index{General Schema|indpage}
initially introduced by J.-P.  Jouannaud and M. Okada
\cite{jouannaud91lics} for the polymorphic $\la$-calculus and extended
to the Calculus of Constructions by F.  Barbanera, M. Fern\'andez and
H. Geuvers \cite{barbanera94lics}, is basically an extension of the
primitive recursion schema~: in the right hand-side of a rule
$f(\vl)\a r$, the recursive calls to $f$ must be done on strict
subterms of $\vl$. It can treat object-level and simply-typed symbols
defined on primitive types. It has been reformulated and extended to
strictly positive simple types by J.-P.  Jouannaud, M. Okada and
myself for the simply-typed $\la$-calculus \cite{blanqui02tcs} and the
Calculus of Constructions \cite{blanqui99rta}.

\item The {\em Higher-Order Recursive Path Ordering\,} (HORPO)
\index{HORPO|indpage} of J.-P. Jouannaud and A. Rubio
\cite{jouannaud99lics} \footnote{This generalizes the previous works
of C. Loria-Saenz and J. Steinbach \cite{loria92ctrs}, O. Lysne and
J. Piris \cite{lysne95rta} and J.-P. Jouannaud and A. Rubio
\cite{jouannaud96rta}.} is an extension of the RPO
\cite{plaisted78tr,dershowitz82tcs} for first-order terms to the terms
of the simply-typed $\la$-calculus.  It has been recently extended by
D. Walukiewicz \cite{walukiewicz00lfm} to the Calculus of
Constructions with polymorphic and dependent symbols at the
object-level and with basic types. The General Schema can be seen as a
non-recursive version of HORPO.

\item It is also possible to look for an interpretation of the symbols
such that the interpretation of a term strictly decreases when a rule
is applied. This method, introduced by J. van de Pol for the
simply-typed $\la$-calculus \cite{vandepol96thesis} extends to the
higher-order the method of the interpretations known for the
first-order framework. This is a very powerful method but difficult to
use in practice because the interpretations are themselves of
higher-order and also because it is not modular~: adding new rules or
new symbols may require finding new interpretations.
\end{lst}

~

For dealing with higher-order rewriting at the predicate-level
together with polymorphic and dependent symbols and strictly positive
predicates, we have chosen to extend the method of the General
Schema. For first-order symbols, we will use other conditions like in
\cite{jouannaud91lics,barbanera94lics}.




\subsection{Definition of the schema}
\label{subsec-schema}

This method is based on Tait and Girard's method of reductibility
candidates \cite{tait67jsl,girard88book} for proving the strong
normalization of the simply-typed and polymorphic $\la$-calculi. This
method consists of defining a subset of the strongly normalizable
terms, the {\em computable\,} terms, and in proving that each
well-typed term is computable. Indeed, a direct proof of strong
normalization by induction on the structure of terms does not go
through because of the application case~: if $u$ and $v$ are strongly
normalizable then it is not clear how to prove that $uv$ is also
strongly normalizable.

The idea of the General Schema is then, from a left hand-side $f(\vl)$
of a rule, to define a set of terms, the {\em computable closure\,},
whose elements are computable whenever the $l_i$'s are computable.
Hence, to prove the termination of a definition, it suffices to check
that, for each rule, the right hand-side belongs to the computable
closure of the left hand-side.

To build the computable closure, we first define a subset of the
subterms of the $l_i$'s that are computable whenever the $l_i$'s are
computable~: the {\em accessible\,} subterms of the $l_i$'s ({\em a
priori\,} not all the subterms of a computable term are computable).
Then we build the computable closure by closing the set of the
accessible variables of the left hand-side with
computability-preserving operations.

~

For most interesting functions, we must be able to accept recursive
calls. And to preserve strong normalization, recursive calls must
decrease in a well-founded ordering. The strict subterm relation
$\tgt$ (in fact, restricted to accessible subterms for preserving
computability) is sufficient for dealing with definition on basic
predicates. In the example of $map$, $\ell$ and $\ell'$ are strict
accessible subterms of $app(A',\ell,\ell')$. But, for non-basic
predicates, it is not sufficient as examplified by the following
``addition'' on Brouwer's ordinals~:

\begin{lst}{--}
\item $+\in\Fs_2$ with $\tau_+=ord\a ord\a ord$,
\item $+(x,0) \a x$
\item $+(x,s(y)) \a s(+(x,y))$
\item $+(x,lim(f)) \a lim(\l{n}{nat}+(x,fn))$
\end{lst}

~

Another example is given the following simplification rules on the
process algebra $proc$ \cite{sellink93ssl}~:

\begin{lst}{--}
\item $+(p,p) \a p$
\item $+(p,\d) \a p$
\item \ldots
\item $\circ(\S(f),p) \a \S(\l{d}{data}\circ(fd,p))$
\end{lst}

~

This is why, in our conditions, we will use two distinct
orderings. The first one, $>_1$, will be used for the arguments of
basic type and the second one, $>_2$, will be used for the arguments
of strictly-positive type.

Finally, for finer control over comparing the arguments, to each
symbol we will associate a {\em status\,} describing how to compare
two sequences of arguments by a simple combination of lexicographic
and multiset comparisons \cite{jouannaud97tcs}.


\begin{dfn}[Accessibility relations]
\label{def-acc}
\index{accessible!term|inddef[def-acc]}
\index{$\tgt_1,\tgt_2$|inddef[def-acc]}
\index{ordering!$\tgt_1,\tgt_2$|inddef[def-acc]}
\index{ordering!accessibility|inddef[def-acc]}
\index{Accc@$\acc(c)$|inddef[def-acc]}

  Let $c$ be a constructor of type $\pvyU$ $C(\vv)$, $\vu$ be
  arguments of $c$, $\g=\vyu$ and $j\in\acc(c)$ be an accessible
  position of $c$. Then~:

\begin{lst}{\bu}
\item $u_j\!:\!U$ is {\em weakly accessible modulo $\r$\,} in
$c(\vu)\!:\!T$, $c(\vu)\!:\!T ~\tgt_1^\r~ u_j\!:\!U$, if $T\r=
C(\vv)\g\r$ and $U\r= U_j\g\r$.
\item $u_j\!:\!U$ is {\em strongly accessible modulo $\r$\,} in
$c(\vu)\!:\!T$, $c(\vu)\!:\!T ~\tgt_2^\r~ u_j\!:\!U$, if $T\r=
C(\vv)\g\r$, $U\r= U_j\g\r$ and $U_j$ is of the form $\pvxT D(\vw)$.
\end{lst}
\end{dfn}

We can use these relations to define the orderings $>_1$ and $>_2$. We
have $\tgt_2^\r \sle \tgt_1^\r$. For technical reasons, we take into
account not only the terms themselves but also their types. This comes
from the fact that we are able to prove that two convertible types
have the same interpretation only if these two types are
computable. This may imply some restrictions on the types of the
symbols.

Indeed, accessibility requires the equality (modulo the application of
$\r$) between canonical types and derived types (see
Definition~\ref{def-can-typ}). More precisely, for having $t:T \tgt_1
u:U$, $T$ must be equal (modulo $\r$) to the canonical type of $t$ and
$U$ must be equal (modulo $\r$) to the type of $u$ derived from
$t$. In addition, if $u:U \tgt_1 v:V$ then $U$ must also be equal
(modulo $\r$) to the canonical type of $u$.


\begin{dfn}[Precedence]
\label{def-prec}
\index{precedence|inddef[def-prec]}
\index{$>_\cF,\ge_\cF,=_\cF$|inddef[def-prec]}
\index{ordering!precedence|inddef[def-prec]}
\index{ordering!$>_\cF$|inddef[def-prec]}

  A {\em precedence\,} is a quasi-ordering $\ge_\cF$ on $\cF$ whose
  strict part $>_\cF$ is well-founded. We will denote by $=_\cF$ its
  associated equivalence relation.
\end{dfn}


\renewcommand{\sf}{_{stat_f}}
\newcommand{\sF}{_{stat_F}}
\newcommand{\sg}{_{stat_g}}
\newcommand{\vm}{\vec{m}}

\begin{dfn}[Status]
\label{def-stat}
\index{status|inddef[def-stat]}
\index{status!assignment|inddef[def-stat]}
\index{stat@$stat,stat_f$|inddef[def-stat]}
\index{xi@$x_i$|inddef[def-stat]}
\index{lexm@$lex(\vm)$|inddef[def-stat]}
\index{mulx@$mul(\vx)$|inddef[def-stat]}
\index{arity!status|inddef[def-stat]}
\index{$>\sf$|inddef[def-stat]}
\index{ordering!$>\sf$|inddef[def-stat]}
\index{ordering!status|inddef[def-stat]}
\index{assignment!status|inddef[def-stat]}
\index{SPf@$SP(f)$|inddef[def-stat]}
\index{position!strictly positive|inddef[def-stat]}
\index{strictly positive!position|inddef[def-stat]}
\index{Cfiu@$C_f^i(\vu)$|inddef[def-stat]}
\index{compatible with!ordering!status assignment|inddef[def-stat]}

Let $(x_i)_{i \ge 1}$ be an indexed family of variables.

\begin{lst}{}
\item [\bf Status.] A {\em status\,} is a \u{linear} term of the form
$lex(m_1,\ldots,m_k)$ with $k\ge 1$ and each $m_i$ of the form
$mul(x_{k_1}, \ldots, x_{k_p})$ with $p \ge 1$. The {\em arity\,} of a
status $stat$ is the greatest indice $i$ such that $x_i$ occurs in
$stat$.
  
\item [\bf Status assignment.] A {\em status assignment\,} is an
application $stat$ which, to each symbol $f$ of arity $n>0$ and type
$\pvxT U$, associates a status $stat_f= lex(\vm)$ of arity smaller
than or equal to $n$ such that~:

\begin{lst}{--}
\item if $x_i\in \FV(stat_f)$ then $T_i$ is of the form $C_f^i(\vu)$
with $C_f^i$ a constant predicate symbol,
\item if $m_i= mul(x_{k_1},\ldots,x_{k_p})$ then $C_f^{k_1} =_\cC
\ldots =_\cC C_f^{k_p}$.
\end{lst}

\item [\bf Strictly positive positions.] Let $f$ be a symbol of status
$lex(\vm)$. We will denote by $SP(f)$ the set of indices $i$ such that
if $m_i= mul(x_{k_1},\ldots,x_{k_p})$ then $C_f^{k_1}$ (therefore also
$C_f^{k_2}, \ldots, C_f^{k_p}$) is strictly positive.

\item [\bf Assignment compatibility.] A status assignment $stat$ is
{\em compatible\,} with a precedence $\ge_\cF$ if~:
\begin{lst}{--}
\item $f =_\cF g$ implies $stat_f= stat_g$ and, for all $i$, $C_f^i
=_\cF C_g^i$.
\end{lst}

\item [\bf Status ordering.] Let $>$ be an ordering on terms and
$stat= lex(\vm)$ be a status of arity $n$. The {\em extension\,} by
$stat$ of $>$ to the sequences of terms of length at least $n$ is the
ordering $>_{stat}$ defined as follows~:
\begin{lst}{--}
\item $\vu >_{stat} \vv$ ~if~ $\vm\vxu ~(>^m)\lex~ \vm\vxv$,
\item $mul(\vu) >^m mul(\vv)$ ~if~ $\{\vu\} >\mul \{\vv\}$.
\end{lst}

\end{lst}
\end{dfn}

For example, if $stat= lex(mul(x_2),mul(x_1,x_3))$ then $\vu >_{stat}
\vv$ if $(\{u_2\},\{u_1,u_3\})$ $(>\mul)\lex~
(\{v_2\},\{v_1,v_3\})$. An important property of $>_{stat}$ is that it
is a well-founded ordering whenever $>$ so is.

We now define the orderings $>_1$ and $>_2$.


\begin{dfn}[Ordering on the arguments of a symbol]
\label{def-arg-ord}
\index{ordering!arguments|inddef[def-arg-ord]}
\index{$>_1,>_2,>$|inddef[def-arg-ord]}
\index{ordering!$>_1,>_2,>$|inddef[def-arg-ord]}

  Let $R= (l\a r$, $\G_0,\r)$ be a rule with $l=f(\vl)$, $\tf= \pvxT
  U$, $\g_0=\vxl$ and $stat_f=lex(\vm)$.

\begin{lst}{\bu}
\item $t:T >_1 u:U$ if $t:T ~(\tgt_1^\r)^+~ u:U$.
\item $t:T >_2 u:U$ if~:
\begin{lst}{--}
\item $t$ is of the form $c(\vt)$ with $c$ a constructor of type $\pvxT
C(\vv)$,
\item $u$ is of the form $x\vu$ with $x\in\dom(\G_0)$, $x\G_0$ of the
form $\pvyU D(\vw)$ and $D =_\cC C$,
\item $t:T ~(\tgt_2^\r)^+~ x:V$ with $V\r= x\G_0$.
\end{lst}
\end{lst}

\noindent
Now, we define the ordering $>$ on the arguments of $f$ (in fact the
pairs argument-type $u:U$). This is an adaptation of $>\sf$ where the
ordering $>$ is $>_1$ or $>_2$ depending on the type (basic or
strictly positive) of the argument. Assume that $stat_f=
lex(m_1,\ldots,m_k)$. Then~:

\begin{lst}{\bu}
\item $\vu > \vv$ ~if~ $\vm\vxu ~(>^1,\ldots,>^k)\lex~ \vm\vxv$,
\item $mul(\vu) >^i mul(\vv)$ ~if~ $\{\vu\} ~(>_{\phi(i)})\mul~
\{\vv\}$ with $>_{\phi(i)}=>_2$ if $i\in SP(f)$ and $>_{\phi(i)}=>_1$
otherwise.
\end{lst}
\end{dfn}


One can easily check that, in the third rule of the addition on
ordinals, $lim(f):ord >_2 fn:ord$. Indeed, for this rule, one can take
$\G_0= x\!:\!ord, f\!:\!nat\a ord$ and the identity for $\r$. Then,
$f\in \dom(\G_0)$, $\tau(lim(f),1)\r= f\G_0= nat\a ord$ and
$lim(f):ord \tgt_2^\r f:nat\a ord$.


\vs[1cm]

\newcommand{\thc}{\th_\mr{\!\!c}}

\begin{figure}[ht]
\centering
\caption{computable closure of $(f(\vl)\a r,\G_0,\r)$\label{fig-thc}}
\index{computable closure|indfig[fig-thc]}

\begin{tabular}{rcc}
\\ (ax) & $\cfrac{}{\G_0\thc \st:\B}$\\

\\ (symb$^<$) & $\cfrac{
\begin{array}{c}
\G_0\thc \tg:s \quad \G \mbox{ valid pour } \thc\\
\G\thc u_1:U_1\g \quad \ldots \quad \G\thc u_n:U_n\g\\
\end{array}}
{\G\thc g(\vu):V\g}$ &
$\begin{array}{c}
(g\in \cF^s_n,\, g <_\cF f,\\
\tg=\pvyU V,\\
\g=\vyu,\, \th \tg:s)\\
\end{array}$\\

\\ (symb$^=$) & $\cfrac{
\begin{array}{c}
\G_0\thc \tg:s\\
\G\thc u_1:U_1\g \quad \ldots \quad \G\thc u_n:U_n\g\\
\end{array}}
{\G\thc g(\vu):V\g}$ &
$\begin{array}{c}
(g\in \cF^s_n,\, g =_\cF f,\\
\tg=\pvyU V,\\
\g=\vyu,\, \th \tg:s,\\
\vl:\vT\g_0 > \vu:\vU\g)\\
\end{array}$\\

\\ (acc) & $\cfrac{\G_0\thc x\G_0:s}{\G_0\thc x:x\G_0}$
& $(x\in\dom^s(\G_0))$ \\

\\ (var) & $\cfrac{\G\thc T:s}{\GxT\thc x:T}$
& $\begin{array}{c}
(x\in \cX^s \moins \dom(\G)\\
\cup\, \FV(l))\\
\end{array}$ \\

\\ (weak) & $\cfrac{\G\thc t:T \quad \G\thc U:s}{\GxU\thc t:T}$
& $\begin{array}{c}
(x \in \cX^s \moins \dom(\G)\\
\cup\, \FV(l))\\
\end{array}$ \\

\\ (prod) & $\cfrac{\GxT\thc U:s}{\G\thc \px{T}U:s}$\\

\\ (abs) & $\cfrac{\GxT\thc u:U \quad \G\thc \px{T}U:s}
{\G\thc \lx{T}u : \px{T}U}$\\

\\ (app) & $\cfrac{\G\thc t:\px{U}V \quad \G\thc u:U}{\G\thc tu:V\xu}$\\

\\ (conv) & $\cfrac{\G\thc t:T \quad \G\thc T:s \quad \G\thc
  T':s}{\G\thc t:T'}$ & $(T \ad T')$\\
\end{tabular}
\end{figure}


\newpage

\begin{dfn}[Computable closure]
\label{def-thc}
\index{computable closure|inddef[def-thc]}
\index{$\thc$|inddef[def-thc]}

Let $R= (l\a r,\G_0,\r)$ be a rule with $l=f(\vl)$, $\tf= \pvxT U$ and
$\g_0=\vxl$. The {\em computable closure\,} of $R$ w.r.t. a precedence
$\ge_\cF$ and a status assignment $stat$ compatible with $\ge_\cF$ is
the smallest relation $\thc\,\sle \cE\times\cT\times\cT$ defined by
the inference rules of Figure~\ref{fig-thc}. We will denote by
$\thc^<$ the restriction of $\thc$ to the rules distinct from
(symb$^=$)
\end{dfn}

One can easily check that if $\G\thc t:T$ then $\G=\G_0,\G'$. And also
that $\thc\,\sle\,\ths$ and $\thc^<\,\sle\,\th_f$.

It is important to note that the computable closure can easily be
extended by adding new inference rules. For preserving the strong
normalization, it suffices to complete the proof of
Theorem~\ref{thm-cor-thc} where we prove that the rules of the
computable closure indeed preserve the computability.


\begin{dfn}[Well-formed rule]
\label{def-wf-rule}
\index{well-formed rule|inddef[def-wf-rule]}
\index{rule!well-formed|inddef[def-wf-rule]}

Let $R= (l\a r,\G_0,\r)$ be a rule with $l=f(\vl)$, $\tf= \pvxT U$ and
$\g_0=\vxl$. The rule $R$ is {\em well-formed\,} if~:
\begin{lst}{--}
\item $\G_0\th l\r:U\g_0\r$,
\item for all $x\in \dom(\G_0)$, there exists $i$ such that
$l_i:T_i\g_0 ~(\tgt_1^\r)^*~ x:x\G_0$,
\item $\dom(\r)\cap \dom(\G_0)= \vide$.
\end{lst}
\end{dfn}

For example, consider the rule~:

\begin{center}
  $appn(p,consn(x,n,\ell),n',\ell') \a
  consn(x,n+n',appn(n,\ell,n',\ell'))$
\end{center}

\noindent
with $\G_0= x\!:\!T, n\!:\!nat, \ell\!:\!listn(n), n'\!:\!nat,
\ell'\!:\!listn(n')$ and $\r=\{p\to s(n)\}$. We have $\G_0\th
l\r:listn(p+n')\r$. For $x$, we have $consn(x,n,\ell):listn(p)
\tgt_1^\r x:T$. One can easily check that the conditions are also
satisfied for the other variables.


\begin{dfn}[Recursive system]
\label{def-rec-sys}
\index{recursive rewrite system|inddef[def-rec-sys]}
\index{rewrite system!recursive|inddef[def-rec-sys]}
\index{General Schema|inddef[def-rec-sys]}

  Let $R= (l\a r,\G_0,\r)$ be a rule with $l=f(\vl)$, $\tf= \pvxT U$
  and $\g_0=\vxl$. The rule $R$ satisfies the {\em General Schema\,}
  w.r.t. a precedence $\ge_\cF$ and a status assignment $stat$
  compatible with $\ge_\cF$ if $R$ is well-formed and $\G_0\thc
  r:U\g_0\r$.

  A set of rules $\cR$ is {\em recursive\,} if there exists a
  precedence $\ge_\cF$ and a status assignment $stat$ compatible with
  $\ge_\cF$ for which every rule of $\cR$ satisfies the General Schema
  w.r.t. $\ge_\cF$ and $stat$.
\end{dfn}


\begin{rem}[Decidability]
\label{rem-thc}
\index{computable closure|indrem[rem-thc]}
\index{decidability!computable closure|indrem[rem-thc]}

{\em A priori\,}, because of the rule (conv) and of the condition
$\th\tg:s$ for the rules (symb$^<$) and (symb$^=$), the relation
$\thc$ may be undecidable. On the other hand, if we assume $\th\tg:s$
and restrict the rule (conv) to a confluent and strongly normalizing
fragment of $\a$ then $\thc$ becomes decidable (with an algorithm
similar to the one for $\th$). In practice, the symbols and the rules
are often added one after the other (or by groups, but the argument
can be generalized).

Let $(\cF,\cR)$ be a confluent and strongly normalizing system,
$f\notin\cF$ and $\cR_f$ a set of rules defining $f$ and whose symbols
belong to $\cF'=\cF\cup\{f\}$. Then $(\cF',\cR)$ is also confluent and
strongly normalizing. Thus we can check that the rules of $\cR_f$
satisfy the General Schema with the rule (conv) restricted to the case
where $T \ad_{\b R} T'$. This does not seem a big restriction~: it
would be surprising that the typing of a rule requires the use of the
rule itself!
\end{rem}


Before to give a detailed example, we are going to show several
properties of $\thc$. Indeed, to show $\G_0\thc r:U\g_0\r$, knowing
that by (S3) we have $\G_0\th r:U\g_0\r$, one can wonder whether it is
possible to transform a derivation of $\G_0\th r:U\g_0\r$ into a
derivation of $\G_0\thc r:U\g_0\r$. The best thing would be that it is
sufficient that the symbols in $r$ are smaller than $f$ and that the
recursive calls are made on smaller arguments. We prove hereafter that
it is sufficient when there is no recursive call.


\begin{dfn}
\label{def-comp-rule}
\index{rule!compatible|indrem[def-comp-rule]}
\index{compatible with!ordering!rule|indrem[def-comp-rule]}

A rule $l\a r$ is {\em compatible\,} with $\ge_\cF$ if all the symbols
in $r$ are smaller than or equivalent to $l$. A set of rule $\cR$ is
{\em compatible\,} with $\ge_\cF$ if every rule of $\cR$ is compatible
with $\ge_\cF$.
\end{dfn}


\begin{lem}
\label{lem-thf-thc}
\index{computable closure|indlem[lem-thf-thc]}

Assume that $\cR$ and $\tau$ are compatible with $\ge_\cF$. Let $\G$
be an environment with variables distinct from those of $\G_0$. If
$\G\th_f t:T$ then $\G_0,\G\thc^< t:T$.
\end{lem}

\begin{prf}
  By induction on $\G\th_f t:T$.\cqfd
\end{prf}


\begin{lem}[Substitution for $\thc$]
\label{lem-thc-subs}
\index{computable closure|indrem[lem-thc-subs]}

  If $\G_0,\G\thc^< t:T$ and $\t: \G_0,\G\a\G_0,\G'$ in $\thc$ (resp.
  in $\thc^<$) then $\G_0,\G'\thc t\t:T\t$ (resp. $\G_0,\G'\thc^<
  t\t:T\t$).
\end{lem}

\begin{prf}
  By induction on $\G_0,\G\thc^< t:T$.\cqfd
\end{prf}


\begin{lem}
\label{lem-thc-acc}
\index{computable closure|indlem[lem-thc-acc]}

If $\cR$ and $\tau$ are compatible with $\ge_\cF$, $\a$ is confluent
and the symbols of $\G_0$ are strictly smaller than $f$ then, for all
$x\in\dom^s(\G_0)$, $\G_0\thc^< x\G_0:s$.
\end{lem}

\begin{prf}
  Assume that $\G_0= \vy:\vU$ and that $y_i$ is of sort $s_i$. We show
  by induction on $i$ that, for all $j\le i$, $\G_0\thc^< U_j:s_j$. If
  $i=0$, this is immediate. So, assume that $i>0$. By induction
  hypothesis, for all $j<i$, $\G_0\thc^< U_j:s_j$. So, we are left to
  show that $\G_0\thc^< U_i:s_i$.
  
  Let $\G= y_1\!:\!U_1,\ldots,y_{i-1}:U_{i-1}$, $\vz$ be $i-1$ fresh
  variables, $\t= \{\vy\to\vz\}$, $\t'= \{\vz\to\vy\}$ and $\G'=
  \vz:\vU\t$. By {\bf(S3)}, \index{S3@\textbf{S3}|indlem[lem-thc-acc]}
  $\G_0$ is valid. So, by the Environment Lemma, $\G\th
  U_i:s_i$. Since $\a$ is confluent and the symbols in $\G$ are
  strictly smaller than $f$, by Lemma~\ref{lem-dep-symb},
  \index{dependence w.r.t.!symbol|indlem[lem-thc-acc]} $\G\th_f
  U_i:s_i$. By Replacement, $\G'\th_f U_i\t:s_i$. Therefore, by
  Lemma~\ref{lem-thf-thc}, $\G_0,\G'\thc^< U_i\t:s_i$.

  Now we show that $\t': \G_0,\G'\a\G_0$ in $\thc^<$. For this, it is
  sufficient to show that, for all $j<i$, $\G_0\thc^<
  z_j\t':U_j\t\t'$, that is, $\G_0\thc^< y_j:U_j$. By induction
  hypothesis, for all $j<i$, $\G_0\thc^< U_j:s_j$. Thus, by (acc),
  $\G_0\thc^< y_j:U_j$. So, $\t': \G'\a\G_0$ in $\thc^<$ and, by
  Lemma~\ref{lem-thc-subs}, $\G_0\thc^< U_i:s_i$.\cqfd
\end{prf}


\begin{lem}
\label{lem-thf-thc-plus}
\index{computable closure|indlem[lem-thf-thc-plus]}

If $\cR$ and $\tau$ are compatible with $\ge_\cF$, $\a$ is confluent
and $\G_0,\G\th_f t:T$ then $\G_0,\G\thc^< t:T$.
\end{lem}

\begin{prf}
  By induction on the size of $\G$. Assume that $\G_0,\G=
  \vy:\vU$. Let $\vz$ be $|\vy|$ fresh variables, $\t= \{\vy\to\vz\}$,
  $\t'= \{\vz\to\vy\}$ and $\D= \vz:\vU\t$. By Replacement, $\D\th_f
  t\t:T\t$. By Lemma~\ref{lem-thf-thc}, $\G_0,\D\thc^< t\t:T\t$. Now
  we show that $\t': \G_0,\D\a\G_0,\G$ in $\thc^<$ in order to
  conclude with Lemma~\ref{lem-thc-subs}.
  
  We must show that, for all $x\in\dom(\G_0,\D)$, $\G_0,\G\thc^<
  x\t':x(\G_0,\D)$. If $x\in\dom(\G_0)$ then we have to show that
  $\G_0,\G\thc^< x:x\G_0$, and if $x=z_i\in\dom(\D)$ then we have to
  show that $\G_0,\G\thc^< y_i:U_i$. So, it is sufficient to show that
  $\G_0,\G$ is valid in $\thc^<$. If $\G$ is empty, this is immediate
  since, by Lemma~\ref{lem-thc-acc}, $\G_0$ is valid in
  $\thc^<$. Assume now that $\G= \G',y\!:\!U$. Then, $\D=
  \D',z\!:\!U\t$. By the Environment Lemma, $\G_0,\G'\th_f U:s$. By
  induction hypothesis, $\G_0,\G'\thc^< U:s$. Therefore, by (var),
  $\G_0,\G\thc^< y:U$ and $\G_0,\G$ is valid in $\thc^<$.\cqfd
\end{prf}


A particular but useful case is~:

\begin{cor}
\label{cor-thc-tau}
\index{computable closure|indcor[cor-thc-tau]}

If $\cR$ and $\tau$ are compatible with $\ge_\cF$ and $\a$ is
confluent then, for all $g \le_\cF f$, if $\th \tg:s$ then $\G_0\thc^<
\tg:s$.
\end{cor}

\begin{prf}
  Since $\tau$ is compatible with $\ge_\cF$ and $\a$ is confluent, by
  Lemma~\ref{lem-dep-symb},
  \index{dependence w.r.t.!symbol|indlem[lem-thc-acc]} $\th_f
  \tg:s$. Thus, by Lemma~\ref{lem-thf-thc}, $\G_0\thc^< \tg:s$.\cqfd
\end{prf}


~

Now, we can detail an example. Let us consider the rule~:

\begin{center}
  $appn(p,consn(x,n,\ell),n',\ell') \a
  consn(x,n+n',appn(n,\ell,n',\ell'))$
\end{center}

\noindent
with $\G_0= x\!:\!T, n\!:\!nat, \ell\!:\!listn(n), n'\!:\!nat,
\ell'\!:\!listn(n')$ and $\r=\{p\to s(n)\}$. We take $stat_{appn}=
lex(mul(x_2))$; $appn >_\cF consn, +$; $consn >_\cF T$ and $+ >_\cF
s,0 >_\cF nat$. We have already seen that this rule is well
formed. Let us show that $\G_0\thc r:listn(s(n))$. We have $\cR$ and
$\tau$ compatible with $\ge_\cF$.

For applying (symb$^<$), we must show $\th \tau_{consn}:\st$,
$\G_0\thc \tau_{consn}:\st$, $\G_0\thc x:T$, $\G_0\thc n+n':nat$ and
$\G_0\thc appn(n,\ell,n',\ell'):listn(n+n')$. It is easy to check that
$\th \tau_{consn}:\st$. Then, by Lemma~\ref{cor-thc-tau}, we deduce
that $\G_0\thc \tau_{consn}:\st$. The assertions $\G_0\thc x:T$ and
$\G_0\thc n+n':nat$ come from Lemma~\ref{lem-thf-thc-plus}. We are
left to show that $\G_0\thc appn(n,\ell,n',\ell'):listn(n+n')$.

For applying (symb$^=$), we must show that $\th \tau_{appn}:\st$,
$\G_0\thc \tau_{appn}:\st$, $\G_0\thc n:nat$, $\G_0\thc
\ell:listn(n)$, $\G_0\thc n':nat$, $\G_0\thc \ell':listn(n')$ and
$consn(x,n,\ell):listn(s(n)) >_1 \ell:listn(n)$. It is easy to check
that $\th \tau_{appn}:\st$. Then, by Lemma~\ref{cor-thc-tau}, we
deduce that $\G_0\thc \tau_{appn}:\st$. The assertion
$consn(x,n,\ell):listn(p) >_1 \ell:listn(n)$ has already be shown for
proving that the rule is well formed. The other assertions come from
Lemma~\ref{lem-thf-thc-plus}.




\section{Strong normalization conditions}


\newcommand{\domB}{\dom^\B}

\begin{dfn}[Rewrite systems]
\label{def-rew-sys}
\index{rewrite system|inddef[def-rew-sys]}
\index{$(\cG,\cR_\cG)$|inddef[def-rew-sys]}

  Let $\cG$ be a set of symbols. The {\em rewrite system\,}
  $(\cG,\cR_\cG)$ is~:

\begin{lst}{\bu}
\index{first-order!rewrite system|inddef[def-rew-sys]}
\index{rewrite system!first-order|inddef[def-rew-sys]}
\item {\em of first-order\,} if~:
\begin{lst}{--}
\item $\cG$ is made of predicate symbols of maximal arity or of
constructors of primitive predicates,
\item the rules of $\cR_\cG$ have an algebraic right hand-side;
\end{lst}

\index{non-duplicating rewrite system|inddef[def-rew-sys]}
\index{rewrite system!non-duplicating|inddef[def-rew-sys]}
\item {\em non-duplicating\,} if, for all rule of $\cR_\cG$, no
variable has more occurrences in the right hand-side than in the left
hand-side;

\index{primitive!rewrite system|inddef[def-rew-sys]}
\index{rewrite system!primitive|inddef[def-rew-sys]}
\item {\em primitive\,} if all the rules of $\cR_\cG$ have a right
hand-side of the form $[\vx:\vT]g(\vu)\vv$ with $g$ a symbol of $\cG$
or a primitive predicate;

\index{simple rewrite system|inddef[def-rew-sys]}
\index{simplicity|inddef[def-rew-sys]}
\index{rewrite system!simple|inddef[def-rew-sys]}
\item {\em simple\,} if there is no critical pairs between $cR_\cG$
and $\cR$~:
\begin{lst}{--}
\item no matching on defined symbols,
\item no ambiguity in the application of rules;
\end{lst}

\index{small rewrite system|inddef[def-rew-sys]}
\index{smallness|inddef[def-rew-sys]}
\index{rewrite system!small|inddef[def-rew-sys]}
\index{kappaX@$\k_X$|inddef[def-rew-sys]}
\item {\em small\,} if, for all rule $g(\vl)\a r \in \cR_\cG$ and all
$X\in \FVB(r)$, there exists $\k_X$ such that $l_{\k_X}=X$;

\index{rewrite system!positive|inddef[def-rew-sys]}
\index{positive!rewrite system|inddef[def-rew-sys]}
\item {\em positive\,} if, for all symbol $g\in \cG$ and all rule $l\a
r \in \cR_\cG$, $\pos(g,r) \sle \pp(r)$;

\index{rewrite system!safe|inddef[def-rew-sys]}
\index{safe rewrite system|inddef[def-rew-sys]}
\item {\em safe\,} if, for all rule $(g(\vl)\a r,\G,\r) \in
  \cR_\cG$ with $\tg= \pvxT U$ and $\g= \vxl$~:
\begin{lst}{--}
\item for all $X\in \FVB(\vT U)$, $X\g\r\in\domB(\G)$,
\item for all $X,X'\in\FVB(\vT U)$, $X\g\r=X'\g\r \A X=X'$.
\end{lst}
\end{lst}
\end{dfn}


\begin{dfn}[Strong normalization conditions]
\label{def-cond}

\hfill
\begin{bfenumi}{A}
\setcounter{enumi}{-1}

\index{A0@\textbf{A0}|inddef[def-cond]}
\item All rules are well typed.

\index{A1@\textbf{A1}|inddef[def-cond]}
\item The relation $\a\,=\,\ar\cup\ab$ is confluent on $\cT$.

\index{A2@\textbf{A2}|inddef[def-cond]}
\item There exists an admissible inductive structure.

\index{A3@\textbf{A3}|inddef[def-cond]}
\index{$\cgt,\cge,\simeq$|inddef[def-cond]}
\index{ordering!$\cgt$|inddef[def-cond]}
\item There exists a precedence $\cge$ on $\FD$ with which $\cR_\FD$
is compatible and whose equivalence classes form a system which is
either~:

\begin{lst}{}
\index{p@\textbf{p}|inddef[def-cond]}
\item [\bf(p)] primitive,

\index{q@\textbf{q}|inddef[def-cond]}
\item [\bf(q)] positive, small and simple,

\index{r@\textbf{r}|inddef[def-cond]}
\item [\bf(r)] recursive, small and simple.
\end{lst}

\index{A4@\textbf{A4}|inddef[def-cond]}
\index{Fone@$\cF_1,\cF_\w$|inddef[def-cond]}
\index{symbol!first-order|inddef[def-cond]}
\index{first-order!symbol|inddef[def-cond]}
\index{symbol!higher-order|inddef[def-cond]}
\index{higher-order!symbol|inddef[def-cond]}
\item There exists a partition $\cF_1 \uplus \cF_\w$ of $\cD\cF$ ({\em
first-order\,} and {\em higher-order\,} symbols) such that~:

\begin{bfenumalphai}
\index{a@\textbf{a}|inddef[def-cond]}
\item $(\cF_\w,\cR_\w)$ is recursive,

\index{b@\textbf{b}|inddef[def-cond]}
\item $(\cF_\w,\cR_\w)$ is safe,

\index{c@\textbf{c}|inddef[def-cond]}
\item no symbol of $\cF_\w$ occurs in the rules of $\cR_1$,

\index{d@\textbf{d}|inddef[def-cond]}
\item $(\cF_1,\cR_1)$ is of first-order,

\index{e@\textbf{e}|inddef[def-cond]}
\item if $\cR_\w \neq \vide$ then $(\cF_1,\cR_1)$ is non duplicating,

\index{f@\textbf{f}|inddef[def-cond]}
\item $\a_{\cR_1}$ is strongly normalizing on $\T(\cF_1,\cX)$.
\end{bfenumalphai}

\end{bfenumi}
\end{dfn}

The condition (A1) \index{A1@\textbf{A1}|indpage} ensures, among other
things, that $\b$-reduction preserves typing while (A0)
\index{A0@\textbf{A0}|indpage} ensures that rewriting preserves
typing. One can wonder whether confluence is necessary for proving
that $\b$-reduction preserves typing. H. Geuvers
\cite{geuvers93thesis} has proved this property for the conversion
relation $\cC=\, \aa^*_{\b\eta}$ (instead of $\cC=\, \ad_{\b\cR}$
here) while $\abe$ is not confluent on not well-typed
terms. M. Fern\'andez \cite{fernandez93thesis} has also proved this
property for $\cC=\, \abr^* \cup ~\als[\b R]$ with $\ar$ being some
rewriting at the object level only, without assuming that $\abr$ is
confluent. But this last result uses in an essential way the fact that
rewriting is restricted to the object level. It is not clear how it
can be extended to rewriting on types.

\u{One should remember that h}yp\u{othesis }(\u{A1})\u{ and
}(\u{A3})\u{ are useful onl}y\u{ in case of}\\\u{t}yp\u{e-level
rewritin}g\u{.}

~

The condition (A1) can seem difficult to fulfill since one often proves
confluence using strong normalization and local confluence of critical
pairs (result of D. Knuth and P. Bendix \cite{bendix70book} for first
order rewriting extended to higher-order rewriting by T. Nipkow
\cite{nipkow91lics}).

We know that $\ab$ is confluent and that there is no critical pair
between $\cR$ and the $\b$-reduction since the left hand-sides of the
rules of $\cR$ are algebraic.

F. M\"uller \cite{muller92ipl} has shown that, in this case, if $\ar$
is confluent and all the rules of $\cR$ are left-linear, then
$\ar\cup\ab$ is confluent. Thus, the possibility we have introduced,
of linearizing some rules (substitution $\r$) while keeping subject
reduction, appears to be very useful.

In the case of left-linear rules, \index{left-linear rule|indpage} and
assuming that $\a_{\cR_1}$ is strongly normalizing as it is required
in (f), how can we prove that $\a$ is confluent? In the case where
$\a_{\cR_1}$ is non duplicating if $\cR_\w\neq\vide$, we show in
Theorem~\ref{thm-sn-rew} that $\a_{\cR_1} \cup \a_{\cR_\w}$ is
strongly normalizing. Therefore, it suffices to check that the
critical pairs of $\cR$ are confluent (without using any
$\b$-reduction).

~

In (A4), in the case where $\cR_\w\neq\vide$, we require that the
rules of $\cR_1$ are non-duplicating.  \index{non-duplicating rewrite
system|indpage} Indeed, with first-order rewriting already, strong
normalization is not a modular property \cite{toyama87ipl}, even with
confluent systems \cite{drosten89thesis}. On the other hand, it is
modular for disjoint and non duplicating systems
\cite{rusinowitch87ipl}. Here, $\cR_1$ and $\cR_\w$ are not disjoint
but hierarchically defined~: by (c), no symbol of $\cF_\w$ occurs in
the rules of $\cR_1$. In \cite{dershowitz94ctrs}, N. Dershowitz
gathers many results on the modularity of strong normalization for
first-order rewrite systems, especially for hierarchically defined
systems. It would be very interesting to study the modularity of
strong normalization in the case of higher-order rewriting and, in
particular, other conditions than non-duplication which, for example,
does not allow us to accept the following definition~:

\begin{rewc}
0/y & 0\\
s(x)/y & s((x-y)/y)\\[2mm]
0-y & 0\\
s(x)-0 & s(x)\\
s(x)-s(y) & x-y\\
\end{rewc}

This system is a duplicating first-order system not satisfying the
General Schema: it can be put neither in $\cR_1$ nor in
$\cR_\w$. E. Gim\'enez \cite{gimenez98icalp} can deal with this
example by using the fact that the result of $x-y$ is smaller than
$s(x)$.

~

In (A3), the smallness \index{rewrite system!small|indpage} condition
for recursive and positive systems is equivalent in the Calculus of
Inductive Constructions to the restriction of strong elimination to
``small'' inductive types, \index{inductive!type!small|indpage} that
is, the types whose constructors have no other predicate arguments
than the ones of their type. For example, the type $list$ of
polymorphic list is small since, in $\p{A}\st A\a list(A)\a list(A)$,
the type of its constructor $cons$, $A$ is an argument of $list$. On
the other hand, a type $T$ having a constructor $c$ of type $\st\a T$
is not small. So, we cannot define a function $f$ of type $T\a\st$
with the rule $f(c(A))\a A$. Such a rule is not small and does not
form a primitive system either. In some sense, primitive systems can
always be considered as small systems since they contain no projection
and primitive predicate symbols have no predicate argument.

~

Finally, in (A4), the safeness condition for higher-order symbols
means that one cannot do matching or equality tests on predicate
arguments that are necessary for typing other arguments. In her
extension of HORPO \cite{jouannaud99lics} to the Calculus of
Constructions, D. Walukiewicz \cite{walukiewicz00lfm} requires a
similar condition. She gives several (pathological) examples of non
termination because of non-safeness like $J(A,A,a) \a a$ with $J:
\p{A}\st \p{B}\st B\a A$ or $J(A,A,a,b) \a b$ with $J: \p{A}\st
\p{B}\st A\a B\a A$. On the other hand, the rule
$map(A,A,\lx{A}x,\ell) \a \ell$, which does not seem problematic, does
not satisfy the safeness condition either (note that the left
hand-side if not algebraic).


~

We can now state our main result~:

~

\noindent
\fbox{
\begin{minipage}{135mm}
\u{\bf THEOREM~:} If a CAC satisfies the conditions of
Definition~\ref{def-cond} then its reduction relation $\a= \ar\cup\ab$
preserves typing and is strongly normalizing.
\end{minipage}
}

\vs[4mm]

The proof of this theorem is the subject of
Chapter~\ref{chap-correctness}. This generalizes the results of M.
Fern\'andez \cite{fernandez93thesis} and of J.-P. Jouannaud, M. Okada
and myself \cite{blanqui99rta}. In Chapter~\ref{chap-examples}, we
give several important examples of CAC's satisfying these conditions~:
an important subsystem with strong elimination of the Calculus of
Inductive Constructions (CIC) and the Natural Deduction Modulo (NDM) a
large class of equational theories.

~

On the other hand, these conditions do not capture the decision
procedure for classical propositional tautologies of
Figure~\ref{fig-hsiang}. Let us see why~:

\begin{lst}{--}
\item We did not consider rewriting modulo associativity and
commutativity. \index{rewriting!modulo|indpage}
\item Since the system is not left-linear, we do not know how to prove
the confluence of its combination with $\b$.
\item The system is not primitive since there are projections
($P\xor\bot \a P$). It is recursive (and positive also) and
small. Unfortunately, it is not simple.
\end{lst}

Rewriting modulo AC does not seem to be a difficult extension, except
perhaps in the case of type-level rewriting. On the other hand,
confluence and simplicity are problems which seem difficult but we
expect to solve them in the future. In Section~\ref{chap-future}, we
give other directions for future work but these three problems are
certainly the most important ones.


~

From strong normalization, we can deduce the decidability of the
typing relation, which is the essential property on which proof
assistants like Coq \cite{coq01} and LEGO \cite{lego92} are based.

\begin{thm}[Decidability of $\th$]
\label{thm-th-dec}
\index{typing!decidability|indthm[thm-th-dec]}
\index{decidability!typing|indthm[thm-th-dec]}

Let $\G$ be a valid environment and $T$ be $\B$ or a term typable in
$\G$. In a CAC satisfying the conditions of Definition~\ref{def-cond},
checking whether a term $t$ is of type $T$ in $\G$ is decidable.
\end{thm}

\begin{prf}
  Since $\G$ is valid, it is possible to say whether $t$ is typable
  and, if so, it is possible to infer a type $T'$ for $t$. Since types
  are convertible, it suffices to check that $T$ and $T'$ have the
  same normal form. The reader is invited to look at
  \cite{coquand91book,barras99thesis} for more details.\cqfd
\end{prf}


\begin{rem}[Logical consistency]\hfill
\label{rem-log-con}
\index{consistency|indrem[rem-log-con]}

In the pure Calculus of Constructions (CC), it is easy to check that,
in the empty environment, no normal proof of $\bot= \p{P}\st P$ can
exist \cite{barendregt92book}.  Therefore, for CC, strong
normalization is sufficient for proving the logical consistency.

On the other hand, in a CAC, the situation is not so simple. From a
logical point of view, having symbols is equivalent to working in a
non empty environment. Therefore, it is possible that symbols and
rules allow one to build a normal proof of $\bot$. In
\cite{seldin91mwplt}, J. Seldin shows the logical consistency of the
``strongly consistent'' environments \footnote{An environment $\G$ is
strongly consistent if, for all $x\in\dom(\G)$, either $x\G$ is a
predicate type, or $x\G$ is $\b$-equivalent to a term of the form
$y\vt$.} by syntactical means. However, for proving the consistency of
more complex environments, it may be necessary to use semantic
methods.
\end{rem}


%% file: examples.tex



\chapter{Examples of CAC's}
\label{chap-examples}




\section{Calculus of Inductive Constructions (CIC)}
\label{sec-cic}
\index{CIC|indsec[sec-cic]}
\index{Calculus of Constructions!Inductive|indsec[sec-cic]}

\newcommand{\vA}{\vec{A}}
\newcommand{\vB}{\vec{B}}
\newcommand{\vC}{\vec{C}}
\newcommand{\vD}{\vec{D}}
\newcommand{\vf}{\vec{f}}
\newcommand{\vb}{\vec{b}}
\renewcommand{\vm}{\vec{m}}
\newcommand{\vq}{\vec{q}}
\newcommand{\pvxA}{(\vx:\vA)}
\newcommand{\pvzB}{(\vz:\vB)}

We are going to show that our conditions of strong normalization
capture most of the Calculus of Inductive Constructions (CIC) of
B. Werner \cite{werner94thesis} which is the basis of the proof
assistant Coq \cite{coq01}. But, since CIC is expressed in a formalism
different from ours, we need to translate CIC into our formalism in
order to check our conditions. It is a little bit long and painful but
not very difficult.


~

In order to type the strong elimination schema in a polymorphic way,
which is not possible in the usual Calculus of Constructions,
B. Werner uses a slightly more general type system with the sorts
$\cS= \{\st,\B,\triangle\}$, \index{$\triangle$|indpage} the axioms
$\cA= \{(\st,\B), (\B,\triangle)\}$ and the rules $\cB=
\{(s_1,s_2,s_3) \in \cS^3 ~|~ s_1\in\{\st,\B\}, s_2=s_3\}$ (in fact,
he denotes $\st$ by Set, $\B$ by Type and $\triangle$ by Extern).

Then he adds terms for representing inductive types, their
constructors and the definitions by recursion on these types~:

\begin{lst}{\bu}
\item {\bf inductive types}~: \index{inductive!type|indpage}
An inductive type is denoted by $I=Ind\pX{A}\{\vC\}$
\index{IndXAC@$Ind\pX{A}\{\vC\}$|indpage} where the $C_i$'s are the
types of the constructors of $I$. For example, $Nat= Ind\pX\st\{X,X\a
X\}$ represents the type of natural numbers (in fact, any type
isomorphic to the type of natural numbers). The term $A$ must be of
the form $\pvxA\st$ and the $C_i$'s of the form $\pvzB X\vm$ with
$X\notin \FV(\vm)$. Furthermore, the inductive types must be strictly
positive. In CIC, this means that, if $C_i= \pvzB X\vm$ then, for all
$j$, either $X$ does not appear in $B_j$, or $B_j$ is of the form
$\p\vy\vD X\vq$ and $X$ does not appear neither in $\vD$ nor in $\vq$.
  
\item {\bf constructors}~: \hfill The $i$-th constructor
\index{constructor|indpage} of an inductive type $I$ is
denoted by\\ $Constr(i,I)$. \index{ConstriI@$Constr(i,I)$|indpage} For
example, $Constr(1,Nat)$ represents zero and $Constr(2,Nat)$
represents the successor function.
  
\item {\bf definitions by recursion}~: A definition by recursion on an
inductive type $I$ is denoted by $Elim(I,Q,\va,c)$
\index{ElimIQac@$Elim(I,Q,\va,c)$|indpage} where $Q$ is the type of
the result, $\va$ the arguments of $I$ and $c$ a term of type
$I\va$. The strong elimination (that is, in the case where $Q$ is a
predicate type) is restricted to ``small''
\index{inductive!type!small|indpage}
\index{rewrite system!small|indpage}
inductive types, that is, the types whose
constructors do not have predicate arguments that their type do not
have. More precisely, an inductive type $I= Ind\pX{A}\{\vC\}$ is {\em
small\,} if all the types of its constructors are small and a
constructor type $C= \pvzB X\vm$ is {\em small\,} if
$\{\vz\}\cap\XB=\vide$ (this means that the predicate arguments must
be part of the environment in which they are typed; they cannot be
part of $\vC$).
\end{lst}


~

\newcommand{\XI}{\{X\!\to\!I\}}

For defining the reduction relation associated with $Elim$, called
{\em $\io$-reduction\,} \index{iota-reduction@$\io$-reduction|indpage}
\index{reduction!iota@$\io$|indpage} and denoted $\ai$,
\index{$\ai$|indpage} and the typing rules of $Elim$ (see
Figure~\ref{fig-th-cic}), it is necessary to introduce a few
definitions.

Let $C$ be a constructor type. We define $\D\{I,X,C,Q,c\}$
\index{DeltaIXCQc@$\D\{I,X,C,Q,c\}$|indpage} as follows~:

\begin{lst}{--}
\item $\D\{I,X,X\vm,Q,c\}= Q\vm c$
\item $\D\{I,X,\p{z}{B}D,Q,c\}= \p{z}{B} \D\{I,X,D,Q,cz\}$ if
  $X\notin\FV(B)$
\item $\D\{I,X,\p{z}{B}D,Q,c\}= \p{z}{B\XI} (\p\vy\vD
  Q\,\vq\,(\!z\vy))\a \D\{I,X,D,Q,cz\}$\\ if $B=\p\vy\vD X\vq$
\end{lst}

The {\em $\io$-reduction\,} is defined by the rule~:

\begin{center}
  $Elim(I,Q,\vx,Constr(i,I')\,\vz)\{\vf\} ~\ai~
  \D[I,X,C_i,f_i,FunElim(I,Q,\vf)]\,\vz$
\end{center}

\noindent
where $I= Ind\pX{A}\{\vC\}$, $FunElim(I,Q,\vf)= \l\vx\vA \ly{I\vx}
Elim(I,Q,\vx,y)\{\vf\}$ \index{FunElimIQf@$FunElim(I,Q,\vf)$|indpage}
and $\D[I,X,C,f,F]$ \index{DeltaIXCfF@$\D[I,X,C,f,F]$|indpage} is
defined as follows~:

\begin{lst}{--}
\item $\D[I,X,X\vm,f,F]= f$
\item $\D[I,X,\p{z}{B}D,f,F]= \lz{B} \D[I,X,D,fz,F]$ if
  $X\notin\FV(B)$
\item $\D[I,X,\p{z}{B}D,f,F]= \lz{B\XI}
  \D[I,X,D,fz\,\l\vy\vD(F\vq\,(\!z\vy)),F]$\\ if $B=\p\vy\vD X\vq$
\end{lst}

Finally, in the type conversion rule (conv), in addition to
$\b$-reduction and $\io$-reduction, B. Werner considers
$\eta$-reduction~: $\lx{T}ux \ae u$ if $x\notin\FV(u)$. Since $\abe$
is not confluent on badly typed terms\footnote{Remark due to
R. Nederpelt \cite{nederpelt73thesis}~: $\lx{A}x \,\alb
\lx{A}(\ly{B}y~x) \ae \ly{B}y$.}, to consider $\eta$-reduction creates
important difficulties \cite{geuvers93thesis}. Therefore, since our
condition (A1) cannot be satisfied with $\eta$-reduction, we cannot
consider $\eta$-reduction. To find a condition weaker than (A1) that
would be satisfied even with $\eta$-reduction is a problem that we
have temporally left open.


\begin{figure}[ht]
\begin{center}
\caption{Typing rules of CIC\label{fig-th-cic}}
\index{typing!CIC|indfig[fig-th-cic]}
\index{CIC|indfig[fig-th-cic]}

\begin{tabular}{r@{~}c}
\\(Ind) & $\cfrac{A=\pvxA\st \quad \G\th A:\B
\quad \all i,\, \G,X:A\th C_i:\st}
{\G\th Ind\pX{A}\{\vC\}:A}$\\

\\(Constr) & $\cfrac{I=Ind\pX{A}\{\vC\} \quad \G\th I:T}
{\G\th Constr(i,I):C_i\XI}$\\

\\($\st$-Elim) & $\cfrac{
\begin{array}{c}
A=\pvxA\st \quad I=Ind\pX{A}\{\vC\}\\
\G\th Q:\pvxA I\vx\a\st\\
T_i= \D\{I,X,C_i,Q,Constr(i,I)\}\\
\g=\{\vx\to\va\} \quad \all j,\, \G\th a_j:A_j\g \quad \G\th c:I\va
\quad \all i,\, \G\th f_i:T_i\\
\end{array}}
{\G\th Elim(I,Q,\va,c)\{\vf\}:Q\va c}$\\

\\($\B$-Elim) & $\cfrac{
\begin{array}{c}
A=\pvxA\st \quad I=Ind\pX{A}\{\vC\} \mbox{ is small}\\
\G\th Q:\pvxA I\vx\a\B\\
T_i= \D\{I,X,C_i,Q,Constr(i,I)\}\\
\g=\{\vx\to\va\} \quad \all j,\, \G\th a_j:A_j\g \quad \G\th c:I\va
\quad \all i,\, \G\th f_i:T_i\}\\
\end{array}}
{\G\th Elim(I,Q,\va,c)\{\vf\}:Q\va c}$\\

\\(Conv) & $\cfrac{\G\th t:T \quad T \aa_{\b\eta\io}^* T' \quad \G\th
  T':s} {\G\th t:T'}$\\
\end{tabular}
\end{center}
\end{figure}


\newcommand{\aip}{\a_{\io'}}
\newcommand{\abip}{\a_{\b\io'}}

The $\io$-reduction as defined above introduces many $\b$-redexes and
the recursive calls on $Elim$ are made on bound variables which must
be instanciated by strict subterms (or terms of smaller order in case
of a strictly positive inductive type). So, on one hand, from a
practical point of view, this is not very efficient since these
instanciations could be done immediately after the $\io$-reduction, and
on the other hand, the General Schema cannot directly deal with
recursive calls on bound variables, even though these variables must
be instanciated with smaller terms.

This is why we are not going to show the strong normalization of the
relation $\abi$ but of the relation $\abip$ where one step of $\aip$
corresponds to a $\io$-reduction followed by as many $\b$-reductions
as necessary for erasing the $\b$-redexes introduced by the
$\io$-reduction. Note that this is indeed this reduction relation
which is actually implemented in the Coq system \cite{coq01}.

\begin{dfn}[$\io'$-reduction]
\label{def-iotap}
\index{$\aip$|inddef[def-iotap]}
\index{reduction!iotap@$\io'$|inddef[def-iotap]}
\index{DeltapIXCfQfz@$\D'[I,X,C,f,Q,\vf,\vz]$|inddef[def-iotap]}

The {\em $\io'$-reduction\,} is the reduction relation defined by the
rule~:

\begin{center}
  $Elim(I,Q,\vx,Constr(i,I')\,\vz)\{\vf\} ~\aip~
  \D'[I,X,C_i,f_i,Q,\vf,\vz]$
\end{center}

\noindent
where $I= Ind\pX{A}\{\vC\}$ and $\D'[I,X,C,f,Q,\vf,\vz]$ is defined as
follows~:

\begin{lst}{--}
\item $\D'[I,X,X\vm,f,Q,\vf,\vide]= f$
\item $\D'[I,X,\p{z}{B}D,f,Q,\vf,z\vz]= \D'[I,X,D,fz,Q,\vz]$ if
  $X\notin\FV(B)$
\item $\D'[I,X,\p{z}{B}D,f,Q,\vf,z\vz]= \D'[I,X,D,fz\,\l\vy\vD
  Elim(I,Q,\vq,z\vy),Q,\vz]$\\ if $B=\p\vy\vD X\vq$
\end{lst}
\end{dfn}


We think that the strong normalization of $\abip$ implies the strong
normalization of $\abi$. But, since this problem does not seem very
easy to solve and is not directly related to our work, we leave its
resolution for the moment.

\begin{conj}
\label{conj-iotap}
\index{reduction!iotap@$\io'$|indconj[conj-iotap]}
\index{$\aip$|indconj[conj-iotap]}
\index{conjecture|indconj[conj-iotap]}

If $\abip$ is strongly normalizing then $\abi$ is strongly
normalizing.
\end{conj}


\begin{dfn}[Admissible inductive type]
\label{def-adm-typ}
\index{admissible!inductive type|inddef[def-adm-typ]}
\index{predicate!admissible|inddef[def-adm-typ]}
\index{inductive!type!admissible|inddef[def-adm-typ]}

  An inductive type $I= Ind\pX{A}$ $\{\vC\}$ is {\em admissible\,} if
  it satisfies the conditions (I5), (I6) (adapted to the syntax of
  CIC, a strong elimination being considered as a defined predicate
  symbol) and the following {\em safeness\,} condition~: if
  $A=\pvxA\st$ and $C_i=\pvzB X\vm$ then~:

\begin{lst}{--}
\item $\all x_i\in\XB,\, m_i\in\XB$,
\item $\all x_i,x_j\in\XB,\, m_i=m_j \A x_i=x_j$.
\end{lst}
\end{dfn}


\newcommand{\NF}{\cN\cF}

\begin{dfn}[CIC$^-$]
\label{def-cicm}
\index{CICminus@CIC$^-$|inddef[def-cicm]}
\index{NF@$\NF$|inddef[def-cicm]}
\index{DeltapIXCxyKc@$\D'\{I,X,C,\vx y,K,c\}$|inddef[def-cicm]}

The sub-system of CIC that we are going to consider, CIC$^-$, can be
obtained by applying the following restrictions~:

\begin{lst}{\bu}
\item We exclude any use of the sort $\triangle$ in order to stay in
the Calculus of Constructions.

\item In the rule (Ind), instead of requiring $I= Ind\pX{A}\{\vC\}$ to
be typable in any environment $\G$, we require $I$ to be typable in
the empty environment since, in CAC, the types of the symbols must be
typable in the empty environment. Moreover, we require $I$ to be
admissible and in normal form.
  
  \hs The restriction to the empty environment is not a real
  restriction since any type $I= Ind\pX{A}\{\vC\}$ typable in an
  environment $\G= \vy:\vU$ can be replaced by a type $I'=
  Ind\p{X'}{A'}\{\vC'\}$ typable in the empty environment~: it
  suffices to take $A'= \pvyU A$, $C_i'= \pvyU C_i\{X\to X'\vy\}$ and
  to replace $I$ by $I'\vy$ and $Constr(i,I)$ by $Constr(i,I')\vy$.
  
  \hs But we need to adapt the definition of {\em small\,} constructor
  type as follows~: a constructor type $C$ of an inductive type $I=
  Ind\pX{A}\{\vC\}$ with $A= \pvxA\st$ is {\em small\,} if it is of
  the form $\pvxA \pvzB X\vm$ with $\{\vz\}\cap\XB= \vide$.

\item In the rule ($\st$-Elim), instead of requiring $Q$ to be typable
in any environment $\G$, we require $Q$ to be typable in the empty
environment. Moreover, we explicitly require $I$ and $T_i=
\D\{I,X,C_i,Q,Constr(i,I)\}$ to be typable.

\item In the rule ($\B$-Elim), instead of requiring $\th Q:\pvxA
I\vx\a\B$, which is not possible in the Calculus of Constructions, we
require $Q$ to be of the form $\l\vx\vA \ly{I\vx} K$ with $\vx:\vA,
y\!:\!I\vx \th K:\B$ and $f_i$ to be of type $T_i= \D'\{I,X,C_i,\vx
y,K,Constr(i,I)\}$ where $\D'\{I,X,C,\vx y,K,c\}$ is defined as
follows~:

\begin{lst}{--}
\item $\D'\{I,X,X\vm,\vx y,K,c\}= K\{\vx\to\vm,y\to c\}$,
\item if $B=\p\vy\vD X\vq$ then $\D'\{I,X,\p{z}{B}D,\vx y,K,c\}=
  \p{z}{B\XI}(\p\vy\vD K\{\vx\to\!\vq,y\to\!z\vy\})\a \D'\{I,X,D,\vx
  y,K,cz\}$.
\end{lst}

\hs Moreover, we require $Q$ to be in normal form, $T_i$ to be typable
and the inductive types that occur in $Q$ to be subterms of
$I$. Finally, we take $\G\th Elim(I,Q,\va,c)\{\vf\}: K\{\vx\to\va,y\to
c\}$ instead of $\G\th Elim(I,Q,\va,c)\{\vf\}: Q\va c$.

\hs Requiring $Q$ to be of the form $\l\vx\vA \ly{I\vx} K$ is not a
real restriction since, as shown by B. Werner (Corollary~2.9 page~57
of \cite{werner94thesis}), if $\G\th Q:\B$ then there exists $Q'$ of
the form $\pvyU\st$ such that $Q \ab^* Q'$.  Hence, if $\th Q:\pvxA
I\vx\a\B$ then $\vx\!:\!\vA,y\!:\!I\vx \th Q\vx y:\B$. Therefore,
there exists $Q'$ of the form $\pvyU\st$ such that $Q\vx y \ab^* Q'$.
Then, $\l\vx\vA \ly{I\vx} Q\vx y \ab^* \l\vx\vA \ly{I\vx} Q'$ and
$\l\vx\vA \ly{I\vx} Q\vx y \ae^* Q$. Therefore, by confluence, there
exists $Q''$ of the form $\l\vx\vA \ly{I\vx} \p\vy{\vU'}\st$ such that
$Q\abe^* Q''$.

\hs On the other hand, requiring the inductive types occurring in $Q$
to be subterms of $I$ is a more important restriction. But it is only
due to the fact that we restrict ourself to the Calculus of
Constructions in which it is not possible to type the strong
elimination schema in a polymorphic way (that is why B. Werner used a
slightly more general PTS).

\item In the rule (conv), instead of requiring $T \aa_{\b\eta\io}^*
T'$, we require $T \aa_{\b\io'}^* T'$ which is equivalent to $T
\ad_{\b\io'} T'$ since $\a_{\b\io'}$ is confluent (orthogonal CRS
\cite{klop93tcs}).
\end{lst}

We will denote by $\abip$ the reduction relation of CIC$^-$, by $\NF$
the set of CIC$^-$ terms in normal form for $\abip$ (unique by
confluence), by $t\ad$ the normal form of $t$, and by $\th$ the typing
relation of CIC$^-$ (see Figure~\ref{fig-th-cicm}).
\end{dfn}


\begin{figure}[ht]
\begin{center}
\caption{Typing rules of CIC\,$^-$\label{fig-th-cicm}}
\index{typing!CICminus@CIC$^-$|indfig[fig-th-cicm]}
\index{CICminus@CIC$^-$|indfig[fig-th-cicm]}

\begin{tabular}{r@{~}c}
\\(Ind) & $\cfrac{
\begin{array}{c}
A=\pvxA\st \quad \th A:\B \quad \all i,\, X:A\th C_i:\st\\
I= Ind\pX{A}\{\vC\}\in\NF \mbox{ is admissible}\\
\end{array}}
{\th I:A}$\\

\\(Constr) & $\cfrac{I=Ind\pX{A}\{\vC\} \quad \G\th I:T}
{\G\th Constr(i,I):C_i\XI}$\\

\\($\st$-Elim) & $\cfrac{
\begin{array}{c}
A=\pvxA\st \quad I=Ind\pX{A}\{\vC\}\quad \G\th I:T\\
\th Q:\pvxA I\vx\a \st\\
T_i= \D\{I,X,C_i,Q,Constr(i,I)\} \quad \th T_i:\st\\
\g=\{\vx\to\va\} \quad \all j,\, \G\th a_j:A_j\g \quad \G\th c:I\va
\quad \all i,\, \G\th f_i:T_i\\
\end{array}}
{\G\th Elim(I,Q,\va,c)\{\vf\}:Q\va c}$\\

\\($\B$-Elim) & $\cfrac{
\begin{array}{c}
A=\pvxA\st \quad I=Ind\pX{A}\{\vC\}\\
Q=\l\vx\vA\ly{I\vx}K\in\NF \quad \vx:\vA,y:I\vx\th K:\B\\
\mbox{the inductive types of $Q$ are subterms of $I$}\\
T_i= \D'\{I,X,C_i,\vx y,K,Constr(i,I)\} \quad \th T_i:\B\\
\g=\{\vx\to\va\} \quad \all j,\, \G\th a_j:A_j\g \quad \G\th c:I\va
\quad \all i,\, \G\th f_i:T_i\\
\end{array}}
{\G\th Elim(I,Q,\va,c)\{\vf\}:K\{\vx\to\va,y\to c\}}$\\

\\(Conv) & $\cfrac{\G\th t:T \quad T \aa_{\b\io'}^* T' \quad \G\th
  T':s} {\G\th t:T'}$\\
\end{tabular}
\end{center}
\end{figure}


\newcommand{\thu}{\th\!\!\!_{_\Up}}

\begin{thm}
\label{thm-sn-cicm}
\index{Upsilon@$\Up$|indthm[thm-sn-cicm]}
\index{$\ps{t},\ps{\G}$|indthm[thm-sn-cicm]}
\index{normalization!strong!CICminus@CIC$^-$,$\Up$|indthm[thm-sn-cicm]}

There exists a CAC $\Up$ (with typing relation $\thu$ and reduction
relation $\a$) satisfying the conditions of Definition~\ref{def-cond}
and a function $\ps{\_}$ which, to a CIC$^-$ term, associates a $\Up$
term such that~:

\begin{lst}{--}
\item if $\G\th t:T$ then $\ps{\G}\thu \ps{t}:\ps{T}$,
\item moreover, if $t\abip t'$ then $\ps{t} \a^+ \ps{t'}$.
\end{lst}
 Hence, $\abip$ is strongly normalizing in CIC$^-$.
\end{thm}


\newcommand{\WElimI}{W\!Elim_I}

\begin{dfn}[Translation]
\label{def-trans}
\index{Upsilon@$\Up$|inddef[def-trans]}
\index{$\thu$|inddef[def-trans]}
\index{$\ps{t},\ps{\G}$|inddef[def-trans]}
\index{IndI@$Ind_I$|inddef[def-trans]}
\index{ConstrIi@$Constr^I_i$|inddef[def-trans]}
\index{WElimI@$\WElimI$|inddef[def-trans]}
\index{SElimIQ@$SElim_I^Q$|inddef[def-trans]}
\index{DeltapWIXCfQfz@$\D'_W[I,X,C,f,Q,\vf,\vz]$|inddef[def-trans]}
\index{DeltapSIXCfQfz@$\D'_S[I,X,C,f,Q,\vf,\vz]$|inddef[def-trans]}

  We define $\ps{t}$ on the well-typed terms, by induction on $\G\th
  t:T$~:

\begin{lst}{\bu}
\item Let $I=Ind\pX{A}\{\vC\}$ with $A= \pvxA\st$. We take $\ps{I}=
\l\vx{\ps\vA}Ind_I(\vx)$ where $Ind_I$ is a symbol of type $\ps{A}$.
  
\item By hypothesis, $C_i= \pvzB X\vm$. We take $\ps{Constr(i,I)}=
\l\vz{\ps\vB\t'} Constr^I_i(\vz)$ where $\t'=\{X\to\ps{I}\}$,
$Constr^I_i$ is a symbol of type $\p\vz{\vB'} Ind_I(\ps\vm)$, $B_j'=
\ps{B_j}$ if $X$ does not occur in $B_j$, and $B_j'= \p\vy{\ps\vD}
Ind_I(\ps\vq)$ if $B_j= \p\vy\vD X\vq$.
  
\item Let $Q$ be a term not of the form $\l\vx\vA \ly{I\vx} K$ with
$K= \pvyU\st$. We take $\ps{Elim(I,Q,\va,c)\{\vf\}}=
W\!Elim_I(\ps{Q},\ps\va,\ps{c},\ps\vf)$ where $W\!Elim_I$ is a symbol
of type $\p{Q}{\p\vx{\ps\vA}$ $\ps{I}\vx\a\st} \p\vx{\ps\vA}
\py{\ps{I}\vx} \p\vf{\ps\vT} \ps{Q}\vx y$ and $T_i=
\D\{I,X,C_i,Q,Constr(i,I)\}$.
  
\item Let $Q$ be a term of the form $\l\vx\vA \ly{I\vx} K$ with $K=
\pvyU\st$. We take $\ps{Elim(I,Q,\va,c)\{\vf\}}=
SElim_I^Q(\ps\va,\ps{c},\ps\vf)$ where $SElim_I^Q$ is a symbol of type
$\p\vx{\ps\vA} \py{\ps{I}\vx} \p\vf{\ps{\vT}} \ps{K}$, $T_i=
\D'\{I,X,C_i,$ $\vx y,K,Constr(i,I)\}$.

\item The other terms are defined recursively~: $\ps{uv}=
\ps{u}\ps{v}, \ldots$
\end{lst}

\noindent
Let $\Up$ be the CAC whose symbols are those just previously defined
and whose rules are~:

\begin{rewc}
  W\!Elim_I(Q,\vx,Constr^I_i(\vz),\vf) & \D'_W[I,X,C_i,f_i,Q,\vf,\vz]\\
  SElim_I^Q(\vx,Constr^I_i(\vz),\vf) & \D'_S[I,X,C_i,f_i,Q,\vf,\vz]\\
\end{rewc}

\noindent
where $\D'_W[I,X,C,f,Q,\vf,\vz]$ and $\D'_S[I,X,C,f,Q,\vf,\vz]$ are
defined as follows~:

\begin{lst}{--}
\item $\D'_W[I,X,X\vm,f,Q,\vf,\vz]= \D'_S[I,X,X\vm,f,Q,\vf,\vz]= f$,
\item $\D'_S[I,X,\p{z}{B}D,f,Q,\vf,z\vz]=
\D'_S[I,X,D,f\,z,Q,\vf,\vz]$\\ and $\D'_W[I,X,\p{z}{B}D,f,Q,\vf,z\vz]=
\D'_W[I,X,D,f\,z,Q,\vf,\vz]$ if $X\notin\FV(B)$
\item $\D'_S[I,X,\p{z}{B}D,f,Q,\vf,z\vz]=
\D'_S[I,X,D,f\,z\,\l{\vy}{\vD}SElim_I^Q(\vf,\vq,z\vy),Q,\vf,\vz]$ and
$\D'_W[I,X,\p{z}{B}D,f,Q,\vf,z\vz]=\\
\D'_W[I,X,D,f\,z\,\l{\vy}{\vD}W\!Elim_I(Q,\vf,\vq,z\vy),Q,\vf,\vz]$ if
$B=\p{\vy}{\vD}X\vq$
\end{lst}
\end{dfn}


Since $\abip$ is confluent, the $\b$-reduction has the subject
reduction property in $\Up$. This will be useful for proving that the
translation preserves typing~:

\begin{lem}
\label{lem-cor-trans}
\index{$\ps{t},\ps{\G}$|indlem[lem-cor-trans]}

If $\G\th t:T$ then $\ps{\G}\thu \ps{t}:\ps{T}$.
\end{lem}

\begin{prf}
By induction on $\G\th t:T$.

\begin{lst}{--}
\item [\bf(Ind)] We have to prove that $\thu \ps{I}:\ps{A}$. We have
$\ps{I}= \l\vx{\ps\vA}Ind_I(\vx)$ with $Ind_I$ of type $\ps{A}=
\p\vx{\ps\vA}\st$. Since $\th A:\B$, by induction hypothesis, we have
$\thu \ps{A}:\B$, that is, $\thu \tau_{Ind_I}:\B$.  By inversion, we
get $\vx:\ps\vA\thu \st:\B$. Therefore, $\G=\vx:\ps\vA$ is valid and,
by the Environment Lemma and (weak), for all $i$, $\G\thu
x_i:\ps{A_i}$. Hence, by (symb), $\G\thu Ind_I(\vx):\st$ and, by
(abs), $\G\thu \ps{I}:\ps{A}$.
  
\item [\bf(Constr)] We have to prove that $\ps\G\thu
\ps{Constr(i,I)}:\ps{C_i\t}$ where $\t=\XI$. We have $C_i= \pvzB$
$X\vm$, $\ps{Constr(i,I)}= \l\vz{\ps\vB\t'} Constr^I_i(\vz)$,
$\t'=\{X\to\ps{I}\}$, $Constr^I_i$ of type $\p\vz{\vB'}
Ind_I(\ps\vm)$, $B_j'= \ps{B_j}$ if $X$ does not occur in $B_j$,
$B_j'= \p\vy{\ps\vD} Ind_I(\ps\vq)$ if $B_j= \p\vy\vD X\vq$,
$\ps{C_i\t}= \ps{C_i}\t'= \p\vz{\ps\vB\t'}\ps{I}\ps\vm$ and $\ps{I}=
\p\vx{\ps\vA}Ind_I(\vx)$.
  
  \hs Hence, $\ps{I}\ps\vm \ab^* Ind_I(\ps\vm)$. Moreover, if $X$ does
  not occur in $B_j$ then $B_j'= \ps{B_j}= \ps{B_j}\t'$.  If $B_j=
  \p\vy\vD X\vq$ then $B_j'= \p\vy{\ps\vD} Ind_I(\ps\vq)$ and
  $\ps{B_j}\t'= \p\vy{\ps\vD} \ps{I}\ps\vq$. Since, $\ps{I}\ps\vq
  \ab^* Ind_I(\ps\vq)$, for all $j$, $\ps{B_j}\t' \ab^* B_j'$.
  
  \hs Since $\G\th I:T$, by induction hypothesis, $\ps\G\thu
  \ps{I}:\ps{T}$ and $\ps{\G}$ is valid. By inversion, $\th I:A$ and
  $X:A\th C_i:\st$. By inversion again, $X:A,\vz:\vB\th X\vm:\st$. By
  induction hypothesis, $\thu \ps{I}:\ps{A}$ and
  $X:\ps{A},\vz:\ps\vB\thu X\ps\vm:\st$. By substitution,
  $\vz:\ps\vB\t'\thu \ps{I}\ps\vm:\st$. Therefore, $\D= \vz:\ps\vB\t'$
  is valid. Since $\ps\vB\t' \ab^* \vB'$, by subject reduction on the
  environments, $\D'= \vz:\vB'$ is also valid. Therefore,
  $\vz:\vB'\thu \ps{I}\ps\vm:\st$ and, by (prod), $\thu \p\vz{\vB'}
  Ind_I(\ps\vm):\st$, that is, $\thu \tau_{Constr^I_i}:\st$.
  
  \hs By the Environment Lemma and (conv), for all $j$, $\D\thu
  z_j:B_j'$. Therefore, by (symb), $\D\thu$
  $Constr^I_i(\vz):Ind_I(\ps\vm)$ and, by (abs), $\thu
  \ps{Constr(i,I)}: \p\vz{\ps\vB\t'} Ind_I(\ps\vm)$. Finally, by
  (conv) and (weak), $\ps\G\thu \ps{Constr(i,I)}:\ps{C_i\t}$.
  
\item [\bf($\st$-Elim)] We have to prove that $\ps\G\thu
\ps{Elim(I,Q,\va,c)\{\vf\}}:\ps{Q\va c}$. We have
$\ps{Elim(I,Q,\va,c)\{\vf\}}= W\!Elim_I(\ps{Q},\ps\va,\ps{c},\ps\vf)$,
$W\!Elim_I$ of type $\p{Q}{\ps{B}} \p\vx{\ps\vA} \py{I\vx}
\p\vf{\ps\vT} \ps{Q}\vx y$, $B= \pvxA I\vx\a\st$, $T_i=
\D\{I,X,C_i,Q,Constr(i,I)\}$ and $\ps{Q\va c}= \ps{Q}\ps\va\ps{c}$. In
order to apply (symb), we prove that (1)~$\ps\G\thu \ps{Q}:\ps{B}$,
(2)~$\ps\G\thu \ps\va:\ps\vA\g'$, (3)~$\ps\G\thu \ps{c}:\ps{I}\vx\g'$,
(4)~$\ps\G\thu \ps\vf:\ps\vT\g'$ and (5)~$\thu \tau_{W\!Elim_I}:\st$,
where $\g'= \{Q\to\ps{Q}, \vx\to\ps\va, y\to\ps{c}, \vf\to\ps\vf\}$.
First of all, note that, since $\G\th c:I\va$, by induction
hypothesis, $\ps\G\thu \ps{c}:\ps{I\va}$ and $\ps\G$ is valid.

\begin{enumi}{}
\item Since $\th Q:B$, by induction hypothesis, $\thu
\ps{Q}:\ps{B}$. Therefore, by weakening, $\ps\G\thu \ps{Q}:\ps{B}$.

\item Since $\G\th a_j:A_j\g$, by induction hypothesis, $\ps\G\thu
\ps{a_j}:\ps{A_j\g}$. But $\ps{A_j\g}= \ps{A_j}\g'$ since $\FV(A_j)
\sle \{\vx\}$.

\item Since $\G\th c:I\va$, by induction hypothesis, $\ps\G\thu
\ps{c}:\ps{I\va}$. But $\ps{I\va}= \ps{I}\ps\va= \ps{I}\vx\g'$.
  
\item Since $\G\th f_i:T_i$, by induction hypothesis, $\ps\G\thu
\ps{f_i}:\ps{T_i}$. But $\ps{T_i}\g'= \ps{T_i}$ since $T_i$ is closed
($\th T_i:\st$).
  
\item Let $\D= Q:\ps{B}, \vx:\ps\vA, \py{\ps{I}\vx}$ and $\D'= \D,
\vf:\ps\vT$. We prove that $\D'$ is valid. Indeed, in this case,
$\D'\thu \ps{Q}\vx y:\st$ and, by (prod), $\thu
\tau_{W\!Elim_I}:\st$. We have $\thu \ps{Q}:\ps{B}$. Therefore, $\thu
\ps{B}:\B$. By (var), $Q:\ps{B}\thu Q:\ps{B}$. By inversion,
$Q:\ps{B}, \vx:\ps\vA, y:\ps{I}\vx \thu \st:\B$ and $\D$ is valid.
Let $\D_i= \D, f_1\!:\!\ps{T_1}, \ldots, f_i\!:\!\ps{T_i}$.  We prove
by induction on $i$ that $\D_i$ is valid. If $i=0$, this is
immediate. We now prove that if $\D_i$ is valid then $\D_{i+1}$ is
valid too. Since $\thu \ps{T_{i+1}}:\st$, by weakening, $\D_i\thu
\ps{T_{i+1}}:\st$. Therefore, by (var), $\D_{i+1}\thu f_{i+1}:T_{i+1}$
and $\D_{i+1}$ is valid.
\end{enumi}

\item [\bf($\B$-Elim)] We have to prove that $\ps\G\thu
\ps{Elim(I,Q,\va,c)\{\vf\}}:\ps{K}$. We have
$\ps{Elim(I,Q,\va,c)\{\vf\}}= SElim_I^Q(\ps\va,\ps{c},\ps\vf)$,
$SElim_I^Q$ of type $\p\vx{\ps\vA} \py{I\vx} \p\vf{\ps\vT} \ps{K}$ and
$T_i\!=\! \D'\{I,X,C_i,Q,Constr(i,I)\}$. In order to apply (symb), we
prove that (1)~$\ps\G\thu \ps\va:\ps\vA\g'$, (2)~$\ps\G\thu
\ps{c}:\ps{I}\vx\g'$, (3)~$\ps\G\thu \ps\vf:\ps\vT\g'$ and (4)~$\thu
\tau_{SElim_I}:\B$, where $\g'= \{\vx\to\ps\va, y\to\ps{c},
\vf\to\ps\vf\}$. First of all, note that, since $\G\th c:I\va$, by
induction hypothesis, $\ps\G\thu \ps{c}:\ps{I\va}$ and $\ps\G$ is
valid.

\begin{enumi}{}
\item Since $\G\th a_j:A_j\g$, by induction hypothesis, $\ps\G\thu
\ps{a_j}:\ps{A_j\g}$. But $\ps{A_j\g}= \ps{A_j}\g'$ since $\FV(A_j)
\sle \{\vx\}$.

\item Since $\G\th c:I\va$, by induction hypothesis, $\ps\G\thu
\ps{c}:\ps{I\va}$. But $\ps{I\va}= \ps{I}\ps\va= \ps{I}\vx\g'$.
  
\item Since $\G\th f_i:T_i$, by induction hypothesis, $\ps\G\thu
\ps{f_i}:\ps{T_i}$. But $\ps{T_i}\g'= \ps{T_i}$ since $T_i$ is closed
($\th T_i:\B$).
  
\item Let $\D= \vx\!:\!\ps\vA, y\!:\!\ps{I}\vx$ and $\D'= \D,
\vf\!:\!\ps\vT$. We have $\D\thu \ps{K}:\B$. We prove that $\D'$ is
valid. Indeed, in this case, by weakening, $\D'\thu \ps{K}:\B$ and, by
(prod), $\thu \tau_{SElim_I}:\st$. Let $\D_i= \D, f_1\!:\!\ps{T_1},
\ldots, f_i\!:\!\ps{T_i}$. We prove by induction on $i$ that $\D_i$ is
valid. If $i=0$, this is immediate. We now prove that if $\D_i$ is
valid then $\D_{i+1}$ is valid too. Since $\thu \ps{T_{i+1}}:\B$, by
weakening, $\D_i\thu \ps{T_{i+1}}:\B$. Therefore, by (var),
$\D_{i+1}\thu f_{i+1}:T_{i+1}$ and $\D_{i+1}$ is valid.
\end{enumi}
\end{lst}

The other cases can be treated without difficulties.\cqfd
\end{prf}


\begin{lem}
\label{lem-up-rule-wt}
\index{rule!well-typed|indlem[lem-up-rule-wt]}
\index{well-typed!rule|indlem[lem-up-rule-wt]}
\index{Upsilon@$\Up$|indlem[lem-up-rule-wt]}

  The rules of $\Up$ are well typed.
\end{lem}

\begin{prf}
  We have to prove that the rules of $\Up$ satisfy the conditions (S3)
  to (S5). We just see the case of
  $W\!Elim_I(Q,\vx,Constr^I_i(\vz),\vf) \a
  \D'_W[I,X,C_i,f_i,Q,\vf,\vz]$. The case of
  $SElim_I^Q(\vx,Constr^I_i(\vz),\vf) \a \D'_S[I,X,C_i,f_i,Q,\vf,\vz]$
  is similar. Let $B= \pvxA I\vx\a\st$. We have $\tau_{W\!Elim_I}=
  \p{Q}{\ps{B}} \p\vx{\ps\vA} \py{\ps{I}\vx} \p\vf{\ps\vT} Q\vx y$,
  $T_i= \D\{I,X,C_i,Q,Constr(i,I)\}$, $C_i= \pvzB X\vm$, $B_j=
  \p{\vy^j}{\vD^j}X\vq^j$ if $X\in\FV(B_j)$, $\tau_{Constr^I_i}=
  \p\vz{\vB'} Ind_I(\ps\vm)$, $B_j'= \ps{B_j}$ if $X\notin\FV(B_j)$,
  $B_j'= \p{\vy^j}{\ps{\vD^j}} Ind_I(\ps{\vq^j})$ otherwise, and
  $\tau_{Ind_I}= \p\vx{\ps\vA}\st$. Let $l= W\!Elim_I(Q,\vx,c,\vf)$,
  $r= \D'_W[I,X,C_i,f_i,Q,\vf,\vz]$, $c= Constr^I_i(\vz)$ and $\g=
  \{y\to c\}$. We take $\G= Q\!:\!\ps{B}, \vz\!:\!\vB',
  \vf\!:\!\ps\vT$ and $\r= \{\vx\to\ps\vm\}$.

\begin{bfenumi}{S}
\setcounter{enumi}{1}
\item We have to prove that $\G\thu r:Q\ps\vm c$. We have $r=
\D'_W[I,X,C_i,f_i,Q,\vf,\vz]$ and $T_i= \D\{I,X,C_i,Q,$
$Constr(i,I)\}$. There is no difficulty.
  
\item We have to prove that if $\D\thu l\s:T$ then $\s:\G\a\D$.  We
have $\D\thu W\!Elim_I(Q\s,\vx\s,$ $Constr^I_i(\vz\s),\vf\s):T$. Then,
by inversion, we deduce that $\D\thu Q\s:\ps{B}\s$, $\D\thu
\vz\s:\vB'\s$ and $\D\thu \vf\s:\ps\vT\s$, that is, $\s:\G\a\D$.
  
\item We have to prove that if $\D\thu l\s:T$ then, for all $x$,
$x\r\s \ad x\s$. By inversion, we have $\D\thu$ $c\s:\ps{I}\vx\s$ and
$Ind_I(\ps\vm\s) \cv[\D] \ps{I}\vx\s$. Now, $\ps{I}\vx\s \ab^*
Ind_I(\vx\s)$. Therefore, $Ind_I(\ps\vm\s) \cv Ind_I(\vx\s)$ and, by
confluence, $Ind_I(\ps\vm\s) \ad Ind_I(\vx\s)$. Since $Ind_I$ is
constant and $\ps\vm\s= \vx\r\s$, we get $\vx\s \ad \vx\r\s$.\cqfd
\end{bfenumi}
\end{prf}


\begin{lem}
\label{lem-up-rule-wf}
\index{rule!well-formed|indlem[lem-up-rule-wf]}
\index{well-formed rule|indlem[lem-up-rule-wf]}
\index{Upsilon@$\Up$|indlem[lem-up-rule-wf]}

  The rules of $\Up$ are well formed.
\end{lem}

\begin{prf}
  Let us see the case of $W\!Elim_I(Q,\vx,Constr^I_i(\vz),\vf)\a
  \D'_W[I,X,C_i,f_i,Q$, $\vf,\vz]$. The case of
  $SElim_I^Q(\vx,Constr^I_i(\vz),\vf)\a \D'_S[I,X,C_i,f_i,Q,\vf,\vz]$
  is dealt with similarly. Let $B= \pvxA I\vx\a\st$. We have $\G=
  Q\!:\!\ps{B}, \vz\!:\!\vB', \vf\!:\!\ps\vT$ and $\r=
  \{\vx\to\ps\vm\}$. We have to prove that each variable
  $x\in\dom(\G)$ is weakly accessible in one of the arguments of
  $W\!Elim_I$, that $x\G$ is equal to $T\r$ where $T$ is the type of
  $x$ derived from $l$ and that $\G\thu l\r:(Q\vx y)\g\r$.
  
  The accessibility is immediate for $Q$ and $\vf$. The $z_j$'s are
  weakly accessible since all the positions of a constructor are
  accessible (see the definition of $\acc(Constr^I_i)$). The type of
  $z_j$ derived from $l$ ia $B_j'$ which does not depend on
  $\vx$. Therefore, $B_j'\r= B_j'= z_j\G$.
  
  Let us see $\G\thu l\r:(Q\vx y)\g\r$ now. We have $l\r=
  W\!Elim_I(Q,\ps\vm$, $Constr^I_i(\vz),\vf)$ and $(Q\vx y)\g\r=
  Q\ps\vm c$. For applying (symb), we must prove (1)~$\thu
  \tau_{W\!Elim_I}:\st$, (2)~$\G\thu Q:\ps{B}$, (3)~$\G\thu
  \ps\vm:\ps\vA\r$, (4)~$\G\thu c:\ps{I}\ps\vm$ and (5)~$\G\thu
  \vf:\ps\vT$.
  
  Let us prove first of all that $\G$ is valid. Note that $W\!Elim_I$
  is defined only if there exists a well-typed term of the form
  $Elim(I,Q',\va,c')\{\vf\}$. And, in this case, we have $\th Q':B$
  and $\th \vT:\st$. Therefore, $\th B:\B$ and $\thu
  \ps{B}:\B$. Hence, $Q\!:\!\ps{B}$ is valid. Moreover, if $W\!Elim_I$
  is well defined then $Ind_I$ is also well defined, and therefore
  $Constr^I_i$ too. But, we have proved in the previous lemma that, in
  this case, $\thu \tau_{Constr^I_i}:\st$. By weakening,
  $Q\!:\!\ps{B}\thu \tau_{Constr^I_i}:\st$. By inversion,
  $Q\!:\!\ps{B}, \vz\!:\!\vB'\thu Ind_I(\ps\vm):\st$ and
  $Q\!:\!\ps{B}, \vz\!:\!\vB'$ is valid. Finally, as $\th \vT:\st$,
  $\thu \ps\vT:\st$ and, by weakening, $Q\!:\!\ps{B}, \vz\!:\!\vB'\thu
  \ps\vT:\st$. Therefore $\G$ is valid.

\begin{enumi}{}
\item Already proved in the previous lemma.
\item By the Environment Lemma.
\item From $\vz\!:\!\vB' \thu Ind_I(\ps\vm):\st$, by inversion, we
deduce that $\vz\!:\!\vB'\thu \ps\vm:\ps\vA\r$. Therefore, by
weakening, $\G\thu \ps\vm:\ps\vA\r$.
\item As $\G\thu \vz\!:\!\vB'$, by (symb), $\G\thu
c:Ind_I(\ps\vm)$. Moreover, $\ps{I}\ps\vm \ab^* Ind_I(\ps\vm)$. After
(3), $\G\thu \ps\vm:\ps\vA\r$. Therefore, by (app), $\G\thu
\ps{I}\ps\vm:\st$ and, by (conv), $\G\thu c:\ps{I}\ps\vm$.
\item By the Environment Lemma.\cqfd
\end{enumi}
\end{prf}


\begin{lem}
\label{lem-up-sn}
\index{normalization!strong!CICminus@CIC$^-$,$\Up$|indlem[lem-up-sn]}
\index{Upsilon@$\Up$|indlem[lem-up-sn]}

  $\Up$ satisfies the conditions of strong normalization of
  Definition~\ref{def-cond}.
\end{lem}

\begin{prf}
\begin{bfenumi}{A}
\setcounter{enumi}{-1}
\item After the previous lemma.

\item We have already seen that $\a$ is confluent.

\item For the inductive structure, we take~:
\begin{lst}{--}
\item $Ind_I >_\cC Ind_J$ if $J$ is a strict subterm of $I$ is a
  well-founded quasi-ordering,
\item $\ind(Ind_I)= \vide$,
\item $\acc(Constr^I_i)= \{1, \ldots, n\}$ where $n$ is the arity of
$Constr^I_i$.
\end{lst}

We prove that this inductive structure is admissible. Let $C$ be a
constant predicate symbol. Then $C=Ind_I$ with $I= Ind\pX{A}\{\vC\}$
and $A= \pvxA\st$, and $C$ is of type $\p\vx{\ps\vA}\st$. Let
$c=Constr^I_i$ be one of the constructors of $C$. By hypothesis, $C_i=
\pvzB X\vm$ and $B_j= \p{\vy^j}{\vD^j}X\vq^j$ if
$X\in\FV(B_j)$. Therefore $c$ is of type $\p\vz{\vB'}$ with $B_j'=
\ps{B_j}$ if $X\notin\FV(B_j)$, and $B_j'= \p{\vy^j}{\ps{\vD^j}}
C(\ps{\vq^j})$ with $X\notin \FV(\vD^j\vq^j)$ otherwise. Let $\vv=
\vm$, $j \in \acc(c)$ and $U_j=B_j'$.

\begin{bfenumii}{I}
\setcounter{enumii}{2}

\item $\all D\in\FC, D =_\cC C \A \pos(D,U_j) \sle \pp(U_j)$.
Necessary, $D=C$. Either $X\notin \FV(B_j)$ and $\pos(C,U_j)=\vide$,
or $B_j= \p\vy\vD X\vq$ and $X\notin \FV(\vD\vq)$. In every case,
$\pos(C,U_j) \sle \pp(U_j)$.
  
\item $\all D\in\FC, D >_\cC C \A \pos(D,U_j) \sle \pz(U_j)$. If
$D=Ind_J >_\cC C=Ind_I$ then $I$ is a strict subterm of $J$.
Therefore, $J$ cannot occur in $I$ and $\pos(D,U_j)=\vide$.
  
\item $\all F\in\FD, \pos(F,U_j) \sle \pz(U_j)$. By hypothesis on the
types of CIC$^-$.
  
\item $\all Y\in\FVB(U_j), \ex\,\io_Y\le\at_C, v_{\io_Y}= Y$. By
hypothesis on the types of CIC$^-$.
  
\setcounter{enumii}{1}
\item $\all Y\in\FVB(U_j), \io_Y\in\ind(C) \A \pos(Y,U_j) \sle
\pp(U_j)$. Since $\ind(C)= \vide$.
\end{bfenumii}
  
\item For $\succeq$, we take the equality. We prove that the rules
defining $SElim_I^Q$ form a system which is~:

\begin{lst}{--}
\item {\bf recursive}~: We prove this for all the symbols.

\item {\bf small}~: We have $SElim_I^Q(\vx,Constr^I_i(\vz),\vf) \a
\D'_S[I,X,C_i,f_i,Q,\vf,\vz]$. We will denote by $\vl$ the arguments
of $SElim_I^Q$ and by $r$ the right hand-side of the rule.  We have to
prove that, for all $X\in\FVB(r)$, there exists a unique $\k_X$ such
that $l_{\k_X}=X$.  We have $\FVB(r)= \{\vf\} \cup (\{\vz\} \cap
\XB)$. For $f_j$, this is immediate. For $z_j\in\XB$, this comes from
the restriction of the strong elimination to small inductive types~:
$\vz= \vx\vz'$ with $\{\vz'\}\cap\XB= \vide$.

\item {\bf simple}~:
\begin{bfenumii}{B}
\item The symbols occurring in the arguments of $W\!Elim_I$ or
$SElim_I^Q$ are constant.
\item At most one rule can be applied at the top of a term of the form
$W\!Elim_I(Q,\va,c,\vf)$ or $SElim_I^Q(\va,c,\vf)$.
\end{bfenumii}
\end{lst}

\item We have $\cF_1=\vide$. Therefore, we just have to check the
conditions (a) and (b)~:

\begin{bfenumalphai}
\item $(\cF,\cR)$ is recursive~: Let us see the case of
$W\!Elim_I(Q,\vx,Constr^I_i(\vz),\vf) \a
\D'_W[I,X,C_i,f_i,Q,\vf,\vz]$. We will denote by $l$ and $r$ the left
hand-side and the right hand-side of this rule. The case of
$SElim_I^Q(\vx,Constr^I_i(\vz),\vf) \a \D'_S[I,X,C_i,f_i,Q,\vf,\vz]$
is similar.
  
  \hs For the precedence $\ge_\cF$, we take $W\!Elim_I >_\cF
  W\!Elim_J$, $W\!Elim_I >_\cF SElim_J^Q$, $SElim_I$ $>_\cF W\!Elim_J$
  and $SElim_I >_\cF SElim_J^Q$ if $J$ is a strict subterm of $I$, and
  all the defined symbols greater than the constant symbols.
  
  \hs For the status, we take $stat_{W\!Elim_I}= lex(mul(x_k))$ where
  $k$ is the position of the argument of type $\ps{I}\vx$. We do the
  same for $SElim_I$. Such an assignment is clearly compatible with
  $\ge_\cF$.
  
  \hs We have to check first that the rule is well formed. We have
  $\G= Q\!:\!\ps{B}, \vz\!:\!\vB', \vf\!:\!\ps\vT$ and $\r=
  \{\vx\to\ps\vm\}$. We have to prove that each $x\in\dom(\G)$ is
  weakly accessible in one of the arguments of $W\!Elim_I$ and that
  $x\G$ is equal to $T\r$ where $T$ is the type of $x$ derived from
  $l$. This is immediate for $Q$ and $\vf$. The $z_j$'s are weakly
  accessible since all the positions of a constructor are accessible
  (see the definition of $\acc(Constr^I_i)$). The type of $z_j$
  derived from $l$ is $B_j'$ which does not depend on
  $\vx$. Therefore, $B_j'\r= B_j'= z_j\G$.
  
  \hs We now show that $r$ belongs to the computable closure of $l$,
  that is, $\G\thc r:Q\ps\vm c$ where $c= Constr^I_i(\vz)$. First of
  all, note that $\cR$ and $\tau$ are compatible with $\ge_\cF$. This
  is clear for $\cR$. For $\tau$, this is due to our restriction on
  $SElim_I^Q$~: the inductive types of $Q$ are subterms of $I$. Hence,
  by the Lemmas \ref{cor-thc-tau} and \ref{lem-thc-acc}, we have
  $\G\thc x\G:s$ for all $x\in\dom^s(\G)$, and $\G\thc \tg:s$ for all
  $g \le_\cF W\!Elim_I$. Hence, we easily check that $\G\thc r:Q\ps\vm
  c$.

\item $(\cF,\cR)$ is safe~: Let $\vT U$ the sequence $\ps{Q}, \ps\vA,
\ps{I}\vx, \ps\vT, Q\vx y$. We have to prove~:

\begin{lst}{--}
\item $\all X\in\FVB(\vT U),\, X\g\r\in\domB(\G)$,
\item $\all X,X'\in\FVB(\vT U),\, X\g\r=X'\g\r \A X=X'$.
\end{lst}

We have $\FVB(\vT U)= \{Q\} \cup \{\vx\} \cap \XB$,
$Q\g\r=Q\in\domB(\G)$ and $x_i\g\r=\ps{m_i}$. Therefore the previous
properties are satisfied thanks to the safety condition on inductive
types.\cqfd
\end{bfenumalphai}

\end{bfenumi}
\end{prf}


~

We are now left to prove that the translation reflects the strong
normalization~:

\begin{lem}
\label{lem-refl}
\index{normalization!strong!CICminus@CIC$^-$,$\Up$|indlem[lem-refl]}
\index{CICminus@CIC$^-$|indlem[lem-refl]}
\index{$\aip$|indlem[lem-refl]}
\index{reduction!iotap@$\io'$|indlem[lem-refl]}

  If $\G\th t:T$ and $t\abip t'$ then $\ps{t}\a^+ \ps{t'}$.
\end{lem}

\begin{prf}
By induction on $\G\th t:T$.

\begin{lst}{}
\item [\bf(Ind)] Since $I\in\NF$, no reduction is possible.

\item [\bf(Constr)] Since $\G\th I:T$, by inversion, $I\in\NF$ and no
reduction is possible.
  
\item [\bf($\st$-Elim)] We have $\ps{t}=
W\!Elim_I(\ps{Q},\ps\va,\ps{c},\ps\vf)$. Since $\G\th I:T$, by
inversion, $I\in\NF$ and no reduction is possible in $I$. If $Q\abip
Q'$ then, since $\th Q:\pvxA I\vx\a\st$, by induction hypothesis,
$\ps{Q} \a^+ \ps{Q'}$ and $\ps{t} \a^+ \ps{t'}$. If $\va\abip \va'$
then, since $\G\th \va:\vA\g$, by induction hypothesis, $\ps\va \a^+
\ps{\va'}$ and $\ps{t} \a^+ \ps{t'}$. Finally, if $c\abip c'$ then,
since $\G\th c:I\va$, by induction hypothesis, $\ps{c} \a^+ \ps{c'}$
and $\ps{t} \a^+ \ps{t'}$.
  
\item [\bf($\B$-Elim)] We have $\ps{t}=
SElim_I^Q(\ps\va,\ps{c},\ps\vf)$. Since $\G\th I:T$, by inversion,
$I\in\NF$ and no reduction is possible in $I$. Since $Q\in\NF$, no
reduction is possible in $Q$. If $\va\abip \va'$ then, since $\G\th
\va:\vA\g$, by induction hypothesis, $\ps\va \a^+ \ps{\va'}$ and
$\ps{t} \a^+ \ps{t'}$. Finally, if $c\abip c'$ then, since $\G\th
c:I\va$, by induction hypothesis, $\ps{c} \a^+ \ps{c'}$ and $\ps{t}
\a^+ \ps{t'}$.
\end{lst}

The other cases can be treated without difficulties.\cqfd
\end{prf}




\section{CIC + Rewriting}
\label{sec-cicr}

We have just seen that most of the Calculus of Inductive Constructions
is formalizable as a CAC. We are going to see that we can add to this
CAC rewriting rules that are not formalizable in CIC. Take the symbols
$nat:\st$, $0:nat$, $s:nat\a nat$, $+,\times:nat\a nat\a nat$,
$list:\st\a\st$, $nil:\p{A}\st list(A)$, $cons:\p{A}\st A\a list(A)\a
list(A)$, $app:\p{A}\st list(A))\a list(A)\a list(A)$, $len:\p{A}\st
list(A)\a nat$ the length of a list, $in:\p{A}\st A\a list(A)\a \st$
the membership predicate, $incl:\p{A}\st list(A)\a list(A)\a \st$ the
inclusion predicate, $sub:\p{A}\st list(A)\a list(A)\a \st$ the
sublist predicate, $eq:\p{A}\st A\a A\a \st$ the polymorphic Leibniz
equality, $\top:\st$ the proposition ever true, $\bot:\st$ the
proposition ever false, $\non:\st\a\st$, $\ou,\et:\st\a\st\a\st$, and
the following rules~:

\begin{center}

\begin{tabular}{cc}
\begin{rew}
x+0 & x\\
0+x & x\\
x+s(y) & s(x+y)\\
s(x)+y & s(x+y)\\
(x+y)+z & x+(y+z)\\
\end{rew}
&
\begin{rew}
x\times 0 & 0\\
0\times x & 0\\
x\times s(y) & (x\times y) + x\\
s(0)\times x & x\\
x\times s(0) & x\\
\end{rew}\\
\end{tabular}\\[2mm]

\begin{tabular}{ccc}
\begin{rew}
\non\top & \bot\\
\non\bot & \top\\
\end{rew}
&
\begin{rew}
P\ou\top & \top\\
P\ou\bot & P\\
\end{rew}
&
\begin{rew}
P\et\top & P\\
P\et\bot & \bot\\
\end{rew}\\
\end{tabular}\\[2mm]

$\begin{array}{rcl}
eq(A,0,0) &\a& \top\\
eq(A,0,s(x)) &\a& \bot\\
eq(A,s(x),0) &\a& \bot\\
eq(A,s(x),s(y)) &\a& eq(nat,x,y)\\[2mm]

app(A,nil(A'),\ell) &\a& \ell\\
app(A,cons(A',x,\ell),\ell') &\a& cons(A,x,app(A,\ell,\ell'))\\
app(A,app(A',\ell,\ell'),\ell'') &\a& app(A,\ell,app(A,\ell',\ell''))\\[2mm]

len(A,nil(A')) &\a& 0\\
len(A,cons(A',x,\ell)) &\a& s(len(A,\ell))\\
len(A,app(A',\ell,\ell')) &\a& len(A,\ell)+len(A,\ell')\\[2mm]

in(A,x,nil(A')) &\a& \bot\\
in(A,x,cons(A',y,l)) &\a& eq(A,x,y) \ou in(A,x,l)\\
\end{array}$

$\begin{array}{rcl}
sub(A,nil(A'),l) &\a& \top\\
sub(A,cons(A',x,l),nil(A'')) &\a& \bot\\
sub(A,cons(A',x,l),cons(A'',x',l')) &\a& (eq(A,x,x') \et sub(A,l,l'))\\
&& \ou sub(A,cons(A,x,l),l')\\

incl(A,nil(A'),l) &\a& \top\\
incl(A,cons(A',x,l),l') &\a& in(A,x,l') \et incl(A,l,l')\\[2mm]

eq(L,nil(A),nil(A')) &\a& \top\\
eq(L,nil(A),cons(A',x,l)) &\a& \bot\\
eq(L,cons(A',x,l),nil(A)) &\a& \bot\\
eq(L,cons(A,x,l),cons(A',x',l')) &\a& eq(A,x,x') \et eq(list(A),l,l')\\
\end{array}$

\end{center}

\vs[2mm]

\noindent
This rewriting system is recursive, simple, small, safe and confluent
(this can be automatically proved by CiME \cite{cime}). Since the
rules are left-linear, the combination with $\ab$ is also
confluent. Therefore, the conditions of strong normalization are
satisfied.

In particular, we will remark the last rule where $\G= A\!:\!\st,
x\!:\!A, x'\!:\!A, \ell\!:\!list(A), \ell'\!:\!list(A)$ and $\r=
\{A'\to A, L\to list(A)\}$. It is well formed~: for example,
$cons(A',x',\ell'):L \tgt_1^\r x':A'$. And it satisfies the General
Schema~: $\{cons(A,x,\ell):L, cons(A',x',\ell'):L\} ~(\tgt_1^\r)\mul~
\{x:A, x':A'\}, \{\ell:list(A), \ell':list(A)\}$.

However, the system lacks several important rules to get a complete
decision procedure for classical propositional tautologies
(Figure~\ref{fig-hsiang}) or other simplification rules on the
equality (Figure~\ref{fig-int-ring}). To accept these rules, we must
deal with rewriting modulo associativity and commutativity and get rid
of the simplicity conditions.




\section{Natural Deduction Modulo (NDM)}
\label{sec-ndm}
\index{$\equiv$|indsec[sec-ndm]}
\index{NDM|indsec[sec-ndm]}
\index{Natural Deduction!Modulo|indsec[sec-ndm]}

Natural Deduction Modulo (NDM) for first-order logic
\cite{dowek98trtpm} can be presented as an extension of Natural
Deduction with the following additional inference rule~:

\begin{center}
$\cfrac{\G\th P}{\G\th Q}$ ~~if $P \equiv Q$
\end{center}

\noindent
where $\equiv$ is an equivalence relation on propositions stable by
substitution and context. This is a very powerful extension of first
order logic since higher-order logic and skolemized set theory can
both be described as a theories modulo (by using explicit
substitutions \cite{abadi91jfp}).

In \cite{dowek98types}, G. Dowek and B. Werner study the strong
normalization of cut elimination in the case where $\equiv$ is
generated from a first-order confluent and weakly normalizing rewrite
system. In particular, they prove the strong normalization in two
general cases~: when the system is positive and when it has no
quantifier. In \cite{dowek00note}, they give an example of a confluent
and weakly normalizing system for which cut elimination is not
normalizing. The problem comes from the fact that the elimination rule
for $\all$ introduces a substitution~:

\begin{center}
$\cfrac{\G\th \all x.P(x)}{\G\th P(t)}$
\end{center}

Hence, when a predicate symbol is defined by a rule whose right
hand-side contains quantifiers, its combination with $\b$-reduction
may not be normalizing. A normalization criterion for higher-order
rewriting like the one we give in this work is therefore necessary.

~

Now, since NDM is a CAC (logical connectors can be defined as constant
predicate symbols), we can compare our conditions with the ones given
in \cite{dowek98types}.

\begin{bfenumi}{A}
\item In \cite{dowek98types}, only $\ar$ is required to be confluent.
We do not know whether this always implies the confluence of
$\ar\cup\ab$. This is true if $\cR$ is left-linear since then we have
a union of left-linear and confluent CRS's with no critical pair
between each other (general result due to V. van Oostrom
\cite{oostrom94thesis} and proved in the particular case of $\ab$ by
F. M\"uller \cite{muller92ipl}). But we are not aware of work proving
that, in presence of dependent types and rewriting at the type level,
$\ar\cup\ab$ is confluent even though $\cR$ is not left-linear
(V. Breazu-Tannen and J. Gallier have shown in \cite{breazu94ic} the
preservation of confluence for the polymorphic $\la$-calculus with
first-order rewriting at the object level).

\item The NDM types are primitive and form an admissible inductive
structure when we take them as being all equivalent in $=_\cC$.
  
\item In \cite{dowek98types}, the strong normalization of cut
elimination is proved in two general cases~: when the rules of
$(\FD,\cR_\FD)$ have no quantifier and when they are positive. The
systems without quantifiers are primitive. Therefore, in this case,
(A3) is satisfied. On the other hand, in the positive case, we also
require the arguments of the left hand-sides to be constant symbols
and that at most one rule can be applied at the top of a term (NDM
systems are small). But we also provide a new case~: $(\FD,\cR_\FD)$
can be recursive, small and simple.
  
\item Rules without quantifiers are of first-order and rules with
quantifiers are of higher-order. In \cite{dowek98types}, these two
kinds of rules are treated in the same way. But the counter-example
given in \cite{dowek00note} shows that they should not be. In our
conditions, we require symbols defined by rules with quantifiers to
satisfy the General Schema.
\end{bfenumi}

\begin{thm}
  A NDM proof system satisfying the conditions (A1), (A3) and (A4) is
  strongly normalizing.
\end{thm}


%% file: begin-sn.tex



\chapter{Correctness of the conditions}
\label{chap-correctness}

\newcommand{\SN}{\cS\cN}
\newcommand{\CR}{\cC\cR}
\newcommand{\WN}{\cW\cN}

Our proof of strong normalization is based on the extension to the
Calculus of Constructions by T. Coquand and J. Gallier
\cite{coquand90lf} of Tait and Girard's method of reductibility
candidates \cite{girard88book}. The idea is to interpret each type $T$
by a set $\I{T}$ of strongly normalizable terms and to prove that
every term of type $T$ belongs to $\I{T}$. The reader not familiar
with these notions is invited to read the Chapter~3 of the
Ph.D. thesis of B. Werner \cite{werner94thesis} for an introduction to
candidates, and the paper of J. Gallier for a more detailed
presentation \cite{gallier90book}. An important difference between the
candidates of T. Coquand and J. Gallier and the candidates of
B. Werner for the Calculus of Inductive Constructions is that the
former are made of well-typed terms while the later are made of pure
(untyped) $\la$-terms.


\newcommand{\oT}{\o{\T}}
\newcommand{\oO}{\o{\O}}
\newcommand{\oP}{\o{\P}}
\newcommand{\oK}{\o{\K}}
\newcommand{\oB}{\o{\mb{B}}}
\newcommand{\oX}{\o{\mb{X}}}
\newcommand{\oSN}{\o{\mb{SN}}}
\newcommand{\TY}{\mb{TY}}
\newcommand{\oTY}{\o{\TY}}

\newcommand{\GT}{_{\G \th T}}
\newcommand{\GTp}{_{\G \th T'}}
\newcommand{\GU}{_{\G \th U}}
\newcommand{\GUp}{_{\G \th U'}}
\newcommand{\GK}{_{\G \th K}}
\newcommand{\GB}{_{\G \th \B}}

\newcommand{\GpT}{_{\G' \th T}}
\newcommand{\GpU}{_{\G' \th U}}
\newcommand{\GpK}{_{\G' \th K}}
\newcommand{\GpB}{_{\G' \th \B}}

\newcommand{\GppU}{_{\G'' \th U}}

\newcommand{\DT}{_{\D \th T}}
\newcommand{\DTt}{_{\D \th T\t}}
\newcommand{\DxUVt}{_{\D \th (x:U\t)V\t}}
\newcommand{\DxUVpt}{_{\D \th (x:U\t)V'\t}}
\newcommand{\DXKVt}{_{\D \th (X:K\t)V\t}}
\newcommand{\DTtp}{_{\D \th T\t'}}
\newcommand{\DU}{_{\D \th U}}
\newcommand{\DUt}{_{\D \th U\t}}
\newcommand{\DUpt}{_{\D \th U'\t}}
\newcommand{\DVt}{_{\D \th V\t}}
\newcommand{\DK}{_{\D \th K}}
\newcommand{\DKt}{_{\D \th K\t}}
\newcommand{\DKtp}{_{\D \th K\t'}}
\newcommand{\Ds}{_{\D \th s}}
\newcommand{\Dsp}{_{\D \th s'}}
\newcommand{\DXt}{_{\D\th X\t}}
\newcommand{\DXtp}{_{\D\th X\t'}}
\newcommand{\DF}{_{\D\th F}}
\newcommand{\DC}{_{\D\th C}}
\newcommand{\DD}{_{\D\th D}}
\newcommand{\DE}{_{\D\th E}}
\newcommand{\DG}{_{\D\th G}}
\newcommand{\DDp}{_{\D\th D'}}

\newcommand{\DpTt}{_{\D' \th T\t}}
\newcommand{\DpU}{_{\D' \th U}}
\newcommand{\DpC}{_{\D' \th C}}
\newcommand{\DpD}{_{\D' \th D}}
\newcommand{\DpF}{_{\D' \th F}}
\newcommand{\DpG}{_{\D' \th G}}
\newcommand{\DpUt}{_{\D' \th U\t}}
\newcommand{\DpKt}{_{\D' \th K\t}}
\newcommand{\DpVs}{_{\D' \th V\s}}
\newcommand{\DpXt}{_{\D'\th X\t}}

\newcommand{\DppU}{_{\D''\th U}}
\newcommand{\DppUt}{_{\D''\th U\t}}
\newcommand{\DppKt}{_{\D''\th K\t}}
\newcommand{\DppVs}{_{\D''\th V\s}}
\newcommand{\DppVtp}{_{\D''\th V\t'}}

\newcommand{\DpppU}{_{\D'''\th U}}

\newcommand{\Gp}{|_{\G'}}
\newcommand{\Gpp}{|_{\G''}}
\newcommand{\Dp}{|_{\D'}}
\newcommand{\Dpp}{|_{\D''}}
\newcommand{\Dppp}{|_{\D'''}}
\newcommand{\Dn}{|_{\D_n}}




\section{Terms to be interpreted}


In order to have the environment in which a term is typable, we use
{\em closures\,}, that is, environment-term pairs.

\begin{dfn}[Closure]
\label{def-clos}
\index{closure|inddef[def-clos]}
\index{GammatheseT@$\G\TH T$|inddef[def-clos]}
\index{T@$\oT$|inddef[def-clos]}
\index{T@$\oT\GT$|inddef[def-clos]}
\index{SN@$\oSN\GT$|inddef[def-clos]}
\index{restriction!set of closures|inddef[def-clos]}

A {\em closure\,} is a pair $\G\TH t$ made of an environment
$\G\in\cE$ and a term $t\in\cT$. A closure $\G\TH t$ is {\em
typable\,} if there exists a term $T\in\cT$ such that $\G\th t:T$. We
will denote by $\oT$ the set of typable closures.

The set of closures {\em of type\,} $\G\TH T$ is $\oT\GT= \{\G'\TH t
\in \oT ~|~ \G'\sge\G$ and $\G'\th t:T\}$. The set of closures of type
$\G\TH T$ whose terms are strongly normalizable will be denoted by
$\oSN\GT$. The {\em restriction\,} of a set $S\sle \oT\GT$ to an
environment $\G'\sge\G$ is $S\Gp= S \cap \oT\GpT= \{\G''\TH t \in S
~|~ \G''\sge\G'\}$.
\end{dfn}


One can easily check the following basic properties~:

\begin{lem}\hfill
\label{lem-clos-prop}
\index{closure|indlem[lem-clos-prop]}

\begin{enumalphai}
\item If $\G'\TH t \in \oT\GT$ and $\G'\sle\G''\in\E$ then $\G''\TH t
  \in \oT\GT$.
\item If $\G\sle\G'\in\E$ then $\oT\GpT \sle \oT\GT$ and $\oT\GT\Gp=
  \oT\GpT$.
\item If $T ~\C_\G~ T'$ then $\oT\GT= \oT\GTp$.
\end{enumalphai}
\end{lem}


We have to define an interpretation for all the terms that can be the
type of another term, that is, to all the terms $T$ such that there
exists $\G$ and $t$ such that $\G\th t:T$. In this case, by
correctness of types, there exists $s$ such that $T=s$ or $\G\th
T:s$. Thus, we have to define an interpretation for the terms of the
following sets~:

\index{B@$\oB$|indpage}
\index{TY@$\oTY,\oTY^\st,\oTY^\B$|indpage}
\index{Pzero@$\oP^0,\oP^\st,\oP^\B$|indpage}

\begin{lst}{--}
\item $\oB= \{\G\TH T \in \cE\times\cT ~|~ \G\in\E ~\et~ T=\B\}$,
\item $\oTY^\st= \oP^0= \{\G\TH T \in \cE\times\cT ~|~ \G\th T:\st\}$,
\item $\oTY^\B= \oK= \{\G\TH K \in \cE\times\cT ~|~ \G\th K:\B\}$.
\end{lst}

A term $T$ such that $\G\TH T \in \oP^0$ can be obtained by
application of a term $U$ to a term $v$. By inversion, $U$ must have a
type of the form $\px{V}W$. By correctness of types, there exists $s$
such that $\G\th \px{V}W:s$. As $T$ belongs to the same class as $U$,
$T\in \P^0\sle\P= \T^\B_1$ and $U\in \T^s_1$. By classification, we
obtain $s=\B$. Therefore, after the Maximal sort Lemma, $\G\TH U$
cannot belong to $\oP^0$. We therefore have to give an interpretation
to the terms of the following sets also~:

\begin{lst}{--}
\item $\oP^\st= \{\G\TH T \in \cE\times\cT ~|~ \ex\, x, U, K,\, \G\th
T:\px{U}K ~\et~ \G\th U:\st ~\et~ \G\th K:\B\}$,
\item $\oP^\B= \{\G\TH T \in \cE\times\cT ~|~ \ex\, X, K, L,\, \G\th
T:\pX{K}L ~\et~ \G\th K:\B ~\et~ \G\th L:\B\}$,
\item $\oP= \oP^0 \cup \oP^\st \cup \oP^\B$,
\item $\oTY= \oB \cup \oK \cup \oP$.
\end{lst}

In order to justify our definition of $\oP$ and to ensure that all the
terms that have to be interpreted are indeed in $\oTY$, it suffices to
see that, after the Maximal sort Lemma, the projection of $\oP$ on
$\cT$, that is, the set $\{T\in \cT ~|~ \ex\, \G\in \cE,\, \G\TH T \in
\oP\}$, is equal to $\P$.


\begin{lem}
  The sets $\oP^0$, $\oP^\st$, $\oP^\B$, $\oK$ and $\oB$ are disjoint
  from one another.
\end{lem}

\begin{prf}
  We have seen that $\oP^0$ is disjoint from $\oP^\st$ and
  $\oP^\B$. Since $\B$ is not typable, $\oB$ is disjoint from all the
  other sets. $\oP$ and $\oK$ are disjoint since their projections $\P$
  and $\K$ are disjoint. We are therefore left to verify that
  $\oP^\st$ and $\oP^\B$ are indeed disjoint. Assume that there exists
  $\G\TH T \in \oP^\st \cap \oP^\B$. Then, there exists $x$, $U$, $K$,
  $X$, $K'$ and $L$ such that $\G\th T:\px{U}K$, $\G\th U:\st$, $\G\th
  K:\B$, $\G\th T:\pX{K'}L$, $\G\th K':\B$ and $\G\th L:\B$.  By
  convertibility of types, $\px{U}K \CVG \pX{K'}L$. By product
  compatibility, $U \CVG K'$. By conversion correctness, $\st=\B$,
  which is not possible.\cqfd
\end{prf}


We now introduce a measure on $\oTY$ which will allow us to do
recursive definitions.

\begin{dfn}
\label{def-measure}
\index{mu@$\mu$|inddef[def-measure]}
\index{nu@$\nu$|inddef[def-measure]}

$\mu(\G\TH T) =
\left\{
\begin{array}{ll}
0 & \mbox{if } T=\B \mbox{ or } \G \th T:\B\\
\nu(K) & \mbox{if } \G \th T:K \mbox{ and } \G \th K:\B\\
\end{array}
\right.$

\noindent
where $\nu$ is defined on predicate types as follows~:
\begin{lst}{--}
\item $\nu(\st) = 0$
\item $\nu(\px{U}K) = 1 + \nu(K)$
\item $\nu(\pX{K}L) = 1 + \max(\nu(K), \nu(L))$
\end{lst}
\end{dfn}

We must make sure that this definition does not depend on $K$. As all
the types of $T$ are convertible, it suffices to check that $\nu$ is
invariant by conversion~:


\begin{lem}
  If $K \CVG K'$ then $\nu(K) = \nu(K')$.
\end{lem}

\begin{prf}
  By induction on the size of $K$ and $K'$. After the Maximal sort
  Lemma, $K$ is of the form $\pvxT \st$, $K'$ is of the form
  $\pvxTp\st$ and $|\vx| = |\vx'|$. Let $n=|\vx|=|\vx'|$. By product
  compatibility and $\alpha$-equivalence, we can assume that
  $\vx'=\vx$. If $n=0$ then $K=K'$ and $\nu(K)=\nu(K')$.  Assume now
  that $n>0$. Let $L= \p{x_2}{T_2} \ldots \p{x_n}{T_n} \st$ and $L'=
  \p{x_2}{T_2'} \ldots \p{x_n}{T_n'} \st$. By product compatibility,
  $T_1 \CVG T_1'$ and $L \CV[\G,x_1:T_1] L'$. By Conversion
  correctness, $T_1$ and $T_1'$ are typable by the same sort $s$. By
  inversion and regularity, $\G, x_1\!:\!T_1 \th L:\B$ and $\G,
  x_1\!:\!T_1 \th L':\B$. So, by induction hypothesis,
  $\nu(L)=\nu(L')$ and, if $s=\B$, $\nu(T_1)=\nu(T_1')$. Therefore,
  $\nu(K)=\nu(K')$.\cqfd
\end{prf}


\begin{lem}
\label{lem-mu-subs}
If $\G\TH T \in \oTY$ and $\t: \G\a\D$ then $\D\th T\t \in \oTY$ and
$\mu(\D\th T\t)= \mu(\G\TH T)$.
\end{lem}

\begin{prf}
  First of all, one can easily prove by induction on the structure of
  predicate types that, if $K$ is a predicate type and $\t$ is a
  substitution, then $K\t$ is a predicate type and $\nu(K\t)=
  \nu(K)$. We now show the lemma by case on $T$~:

\begin{lst}{--}
\item $T=\B$. $\D\th T\t= \D\th\B \in \oTY$ and $\mu(\D\th T\t)=0=
\mu(\G \TH T)$.
\item $\G\th T:\B$. By substitution, $\D\th T\t:\B$, $\D\TH T\t \in
\oTY$ and $\mu(\D\th T\t)=0= \mu(\G \TH T)$.
\item $\G\th T:K$ and $\G \th K:\B$. By substitution, $\D\th T\t:K\t$
and $\D\th K\t:\B$. Thus $\D\th T\t \in \oTY$. Now, $\mu(\D\th T\t)=
\nu(K\t)$ and $\mu(\G\TH T)= \nu(K)$. But $\nu(K\t)= \nu(K)$.\cqfd
\end{lst}
\end{prf}


%% file: candidates.tex



\section{Reductibility candidates}

We will denote by~:

\begin{lst}{--}
\item $\SN$ \index{SN@$\SN$|indpage} the set of strongly normalizable
terms,
\item $\WN$ \index{WN@$\WN$|indpage} the set of weakly normalizable
terms,
\item $\CR$ \index{CR@$\CR$|indpage} the set of terms $t$ such that
two reduction sequences issued from $t$ are always confluent.
\end{lst}


\begin{dfn}[Neutral term]
\label{def-neutral}
\index{neutral!term|inddef[def-neutral]}
\index{term!neutral|inddef[def-neutral]}

  A term is {\em neutral\,} if it is neither an abstraction nor
  constructor headed.
\end{dfn}


\newcommand{\GpTu}{_{\G'\th Tu}}
\newcommand{\GpTpu}{_{\G'\th T'u}}
\newcommand{\GTu}{_{\G\th Tu}}
\newcommand{\GpTup}{_{\G'\th Tu'}}
\newcommand{\GpTU}{_{\G'\th TU}}

\begin{dfn}[Reductibility candidates]
\label{def-cand}
\index{reductibility!candidate|inddef[def-cand]}
\index{candidate|inddef[def-cand]}
\index{R@$\cR\GT$|inddef[def-cand]}
\index{$\le\GT$|inddef[def-cand]}
\index{$\top\GT$|inddef[def-cand]}
\index{$\biget\GT$|inddef[def-cand]}
\index{restriction!candidate|inddef[def-cand]}
\index{candidate!restriction|inddef[def-cand]}
\index{RGammap@$R\Gp$|inddef[def-cand]}
\index{SigmaGammatheseK@$\S\GK$|inddef[def-cand]}
\index{P1@\textbf{P1}|inddef[def-cand]}
\index{P2@\textbf{P2}|inddef[def-cand]}
\index{R1@\textbf{R1}|inddef[def-cand]}
\index{R2@\textbf{R2}|inddef[def-cand]}
\index{R3@\textbf{R3}|inddef[def-cand]}
\index{R4@\textbf{R4}|inddef[def-cand]}

  For each $\G\TH T \in \oTY$, we are going to define by induction on
  $\mu(\G\TH T)$~:
\begin{lst}{--}
\item the set $\cR\GT$ of {\em reductibility candidates of type\,}
$\G\TH T$,
\item the {\em restriction\,} $R\Gp$ of a candidate $R\in \cR\GT$ w.r.t.
an environment $\G'\sge\G$,
\item the relation $\le\GT$ on $\cR\GT$,
\item the element $\top\GT$ of $\cR\GT$,
\item the function $\biget\GT$ from the powerset of $\cR\GT$ to
$\cR\GT$.
\end{lst}

\begin{lst}{\bu}

\item $T=\B$.
\begin{lst}{--}
\item $\cR\GB= \{ \oSN\GB \}$.
\item $R\Gp= R \cap \oT\GpB$.
\item $R_1 \le\GB R_2$ if $R_1\sle R_2$.
\item $\top\GB= \oSN\GB$.
\item $\biget\GB(\Re)= \top\GB$.
\end{lst}

\item $\G\th T:s$.
\begin{lst}{--}
\item $\cR\GT$ is the set of all the subsets $R$ of $\oT\GT$ such
that~:
\begin{bfenumii}{R}
\item $R\sle \SN$ (strong normalization);
\item if $\G'\TH t \in R$ and $t\a t'$ then $\G'\TH t' \in R$
(stability by reduction);
\item if $\G'\TH t \in \oT\GT$, $t$ is neutral and, for all $t'$ such
that $t\a t'$, $\G'\TH t' \in R$, then $\G'\TH t \in R$ (stability by
expansion for neutral terms);
\item if $\G'\TH t \in R$ and $\G'\sle\G''\in\E$ then $\G''\TH t \in
R$ (stability by weakening).
\end{bfenumii}
\item $R\Gp= R\cap\oT\GpT$.
\item $R_1 \le\GT R_2$ if $R1 \sle R_2$.
\item $\top\GT= \oSN\GT$.
\item $\biget\GT(\Re)= \bigcap \Re$ if $\Re\neq\vide$, $\top\GT$
otherwise.
\end{lst}

\item $\G\th T:\px{U}K$.
\begin{lst}{--}
\item $\cR\GT$ is the set of all the functions $R$ which, to $\G'\TH u
\in \oT\GU$, associate an element of $\cR\GpTu$ and satisfy~:
\begin{bfenumii}{P}
\item if $u\a u'$ then $R(\G'\TH u) = R(\G'\TH u')$ (stability by
reduction),
\item if $\G\sle\G'\in\E$ then $R(\G\TH u)\Gp = R(\G'\TH u)$
(compatibility with weakening).
\end{bfenumii}
\item $R\Gp= R|_{\oT\GpU}$.
\item $R_1 \le\GT R_2$ if, for all $\G'\TH u \in \oT\GU$,
  $R_1(\G'\TH u) \le\GpTu R_2(\G'\TH u)$.
\item $\top\GT(\G'\TH u)= \top\GpTu$.
\item $\biget\GT(\Re)(\G'\TH u)= \biget\GpTu (\{ R(\G'\TH u) ~|~ R \in
  \Re \})$.
\end{lst}

\item $\G\th T:\pX{K}L$. Let $\S\GK$ be the set of pairs $(\G'\TH
U,S)$ such that $\G'\TH U \in \oT\GK$ and $S \in \cR\GpU$.
\begin{lst}{--}
\item $\cR\GT$ is the set of all functions $R$ which, to a pair
$(\G'\TH U,S)\in \S\GK$, associate an element of $\cR\GpTU$ and
satisfy~:
\begin{bfenumii}{P}
\item if $U\a U'$ then $R(\G'\TH U,S)= R(\G'\TH U',S)$ (stability by
reduction),
\item if $\G\sle\G'\in\E$ then $R(\G\TH U,S)\Gp = R(\G'\TH U,S\Gp)$
(compatibility with weakening).
\end{bfenumii}
\item $R\Gp= R|_{\S\GpK}$.
\item $R_1 \le\GT R_2$ if, for all $(\G'\TH U,S) \in \S\GK$,
  $R_1(\G'\TH U, S) \le\GpTU R_2(\G'\TH U, S)$.
\item $\top\GT(\G'\TH U,S)= \top\GpTU$.
\item $\biget\GT(\Re)(\G'\TH U,S)= \biget\GpTU (\{ R(\G'\TH U,S)
  ~|~ R \in \Re \})$.
\end{lst}

\end{lst}
\end{dfn}


The following lemma ensures that all these objects are well defined.

\begin{lem}[Candidates properties]\hfill
\label{lem-cand-prop}
\index{reductibility!candidate|indlem[lem-cand-prop]}
\index{candidate|indlem[lem-cand-prop]}

\begin{enumalphai}
\item $\cR\GT$, $\le\GT$, $\top\GT$ and $\biget\GT$ are well defined.
\item If $T\a T'$ then $\cR\GT= \cR\GTp$.
\item If $R\in \cR\GT$ and $\G\sle\G'\in\E$ then $R\Gp\in \cR\GpT$.
\item $\top\GT\in \cR\GT$.
\item If $T\a T'$ then $\top\GT= \top\GTp$.
\item If $\G\sle\G'\in\E$ then $\top\GT\Gp= \top\GpT$.
\item If $\Re \sle \cR\GT$ then $\biget\GT(\Re)\in \cR\GT$.
\item If $T\a T'$ then $\biget\GT= \biget\GTp$.
\item If $\G\sle\G'\in\E$ then $\biget\GT(\Re)\Gp=
\biget\GpT(\{R\Gp~|~R\in\Re\})$.
\end{enumalphai}
\end{lem}

\begin{prf}
  By induction on $\mu(\G\TH T)$.

\begin{lst}{\bu}


\item $T=\B$.
\begin{enumalphai}
\item Immediate.
\item $\B$ is not reducible.
\item We necessary have $R=\oSN\GB$. So, $R\Gp= \oSN\GB \cap \oT\GpB=
\oSN\GpB \in \cR\GpB$.
\item Immediate.
\item $\B$ is not reducible.
\item $\top\GB\Gp= \oSN\GB \cap \oT\GpB= \oSN\GpB= \top\GpB$.
\item $\biget\GB(\Re)= \top\GB$.
\item $\B$ is not reducible.
\item $\biget\GB(\Re)\Gp= \top\GB\Gp= \top\GpB= \biget\GpB(\{R\Gp
~|~R\in\Re\}$.
\end{enumalphai}


\item $\G\th T:s$.
\index{R1@\textbf{R1}|indlem[lem-cand-prop]}
\index{R2@\textbf{R2}|indlem[lem-cand-prop]}
\index{R3@\textbf{R3}|indlem[lem-cand-prop]}
\index{R4@\textbf{R4}|indlem[lem-cand-prop]}
\begin{enumalphai}
\item Immediate.
\item By subject reduction, $\oT\GT= \oT\GTp$.
\item By weakening, $R\Gp \sle \oT\GpT$. Now we show that $R\Gp$
satisfies (R1) to (R4). For (R1), (R2) and (R4), this is
immediate. For (R3), let $\G''\TH t \in \oT\GpT$ such that $t$ is
neutral and, for all $t'$ such that $t\a t'$, $\G''\TH t' \in R\Gp$.
Since $R\Gp \sle R$, $\G''\TH t \in R$. But $\G''\TH t \in
\oT\GpT$. Therefore, $\G''\TH t \in R\Gp$.
\item By definition, $\top\GT \sle \oT\GT$ and it is easy to check
that $\top\GT$ satisfies (R1) to (R4).
\item By subject reduction.
\item Immediate.
\item Since each element of $\Re$ is included in $\oT\GT$ and
satisfies (R1) to (R4), it is easy to check that $\bigcap \Re$ is
included in $\oT\GT$ and satisfies (R1) to (R4).
\item Immediate.
\item $(\bigcap\Re)\Gp= (\bigcap\Re) \cap \oT\GpT= \bigcap
\{R\cap\oT\GpT ~|~R\in\Re\}= \bigcap \{R\Gp ~|~R\in\Re\}$.
\end{enumalphai}


\item $\G\th T:\px{U}K$.
\begin{enumalphai}

\item We have to check that $\mu(\G'\TH Tu) < \mu(\G\TH T)$ and that
the definitions do not depend on the choice of a type for $T$.
  
  \hs By weakening, $\G'\th T:\px{U}K$ and $\G'\th \px{U}K:\B$. By
  (app), $\G'\th Tu:K\xu$. By inversion and regularity, $\GpxU \th
  K:\B$. By substitution, $\G'\th K\xu:\B$. Therefore $\G'\th Tu \in
  \oTY$ and $\mu(\G'\th Tu)= \nu(K\xu)$. By invariance by
  substitution, $\nu(K\xu)= \nu(K)$.  Therefore $\mu(\G\TH T)=
  \nu(\px{U}K)= 1+\nu(K) > \mu(\G\TH Tu)$.
  
  \hs We now show that the definitions do not depend on the choice of
  a type for $T$. Assume that $\G\th T:\p{x'}{U'}K'$. By Type
  convertibility and Product compatibility, $U \CVG U'$. Therefore
  $\oT\GU= \oT\GUp$ and $\cR\GT$, $\le\GT$, $\top\GT$ and $\biget\GT$
  are unchanged if we replace $U$ by $U'$.

\item By induction hypothesis, $\cR\GpTu= \cR\GpTpu$.

\item Immediate.

\item We check that $\top\GT$ satisfies (P1) and (P2).

\begin{enumii}{P}
\item By induction hypothesis (e), $\top_{\G'\th Tu}=
  \top_{\G'\th Tu'}$.
\item By induction hypothesis (f), $\top_{\G'\th Tu}\Gpp=
  \top_{\G''\th Tu}$.
\end{enumii}

\item By subject reduction, $\top\GT$ and $\top\GTp$ have the same
domain. And they are equal since, by induction hypothesis, $\top\GpTu$
satisfies (e).

\item $\top\GT\Gp$ and $\top\GpT$ have the same domain and are equal.

\item Let $\Re'= \{R(\G'\TH u) ~|~R\in\Re\}$. By definition, if $R\in
\cR\GT$ and $\G'\TH u \in \oT\GU$ then $R(\G'\TH u) \in \cR\GpTu$. By
induction hypothesis, $\biget\GpTu$ satisfies (g).  Therefore,
$\biget\GpTu(\Re') \in \cR\GpTu$. We now check that $\biget\GT$
satisfies (P1) and (P2).

\begin{enumii}{P}
\item Let $\Re''= \{R(\G'\TH u') ~|~R\in\Re\}$. Since each $R\in \Re$
satisfies (P1), $R(\G'\TH u')= R(\G'\TH u)$. By induction hypothesis,
$\biget\GpTu$ satisfies (h). Therefore $\biget\GpTu(\Re')=
\biget\GpTup(\Re'')$.
\item Let $\Re_1= \{R(\G\TH u) ~|~R\in\Re\}$ and $\Re_2= \{R(\G\TH
u)\Gp ~|~R\in\Re\}$. Since each $R\in \Re$ satisfies (P2), $R(\G\TH
u)\Gp= R(\G'\TH u)$. By induction hypothesis, $\biget\GTu$ satisfies
(i). Therefore $\biget\GTu(\Re_1)\Gp= \biget\GpTu(\Re_2)$.
\end{enumii}

\item After (a), $\cR\GT= \cR\GTp$. Therefore, $\biget\GT$ and
$\biget\GTp$ have the same domain. Let $\Re \sle \cR\GT$. Then,
$\biget\GT(\Re)$ and $\biget\GTp(\Re)$ have the same domain and are
equal since, by induction hypothesis, $\biget\GpTu$ satisfies (h).

\item $\biget\GT(\Re)\Gp$ and $\biget\GpT (\{R\Gp ~|~R\in\Re\})$ have
the same domain and are equal since, if $\G''\TH u \in \oT\GpU$ then
$R(\G''\TH u)= R\Gp(\G''\TH u)$.
\end{enumalphai}


\item $\G\th T:\pX{K}L$. The proof is similar to the previous case.\cqfd
\end{lst}
\end{prf}


\begin{lem}
\label{lem-var-cand}
\index{XGammaT@$\oX\GT$|indlem[lem-var-cand]}
\index{reductibility!candidate|indlem[lem-var-cand]}
\index{candidate|indlem[lem-var-cand]}

Let $\oX\GT= \{ \G'\TH t \in \oT\GT ~|~ t= x\vt, ~x\in \cX, ~\vt\in
\SN \}$. If $\G\th T:s$ then $\oX\GT \neq \vide$ and, for all $R\in
\cR\GT$, $\oX\GT \sle R$.
\end{lem}

\begin{prf}
  First of all, $\oX\GT \sle \oT\GT$. Since $\cX^s$ is infinite and
  $\dom(\G)$ is finite, there exists $x\in \cX^s \moins
  \dom(\G)$. Therefore, by (var), $\GxT\th x:T$. So, $\GxT\TH x \in
  \oT\GT$ and $\oX\GT \neq \vide$. Now, let $R\in \cR\GT$, $\G'\in
  \cE$, $x\in \cX$ and $\vt\in \SN$ such that $\G'\TH x\vt \in
  \oT\GT$. We show that $\G'\TH x\vt \in R$ by induction on $\vt$ with
  $\a\lex$ as well-founded ordering. Since $x\vt$ is neutral, by
  {\bf(R3)}, \index{R3@\textbf{R3}|indlem[lem-var-cand]} it suffices
  to show that every immediate reduct of $x\vt$ belongs to $R$. But
  this is the induction hypothesis.\cqfd
\end{prf}


\begin{lem}[Completeness of the candidates lattice]
\label{lem-cand-lat-compl}
\index{reductibility!candidate|indlem[lem-cand-lat-compl]}
\index{candidate|indlem[lem-cand-lat-compl]}

  For all $\G\TH T \in \oTY$, $(\cR\GT,\le\GT)$ is a complete
  lattice. The lower bound of a part of $\cR\GT$ is given by $\biget\GT$.
\end{lem}

\begin{prf}
  It suffices to prove that $(\cR\GT,\le\GT)$ is a complete
  inf-semi-lattice and that $\top\GT$ is its greatest element. One can
  easily check by induction on $\mu(\G\TH T)$ that $\le\GT$ is an
  ordering ({\em i.e.\,} is reflexive, transitive and anti-symmetric),
  $\top\GT$ is the greatest element of $\cR\GT$ and the lower bound of a part
  of $\cR\GT$ is given by $\biget\GT$.\cqfd
\end{prf}


%% file: interpret.tex



\newcommand{\Dtx}[1][I]{_{\D,\t,\xi}^{#1}}
\newcommand{\Dtxp}[1][I]{_{\D,\t,\xi'}^{#1}}
\newcommand{\Dsx}{_{\D,\s,\xi}^I}
\newcommand{\Dsxp}{_{\D,\s,\xi'}^I}
\newcommand{\Dtpx}{_{\D,\t',\xi}^I}
\newcommand{\Dspx}{_{\D,\s',\xi}^I}
\newcommand{\Dspxp}[1][I]{_{\D,\s',\xi'}^{#1}}
\newcommand{\Dstxp}[1][I]{_{\D,\s\t,\xi'}^{#1}}
\newcommand{\Dspxpp}[1][I]{_{\D,\s',\xi''}^{#1}}

\newcommand{\Dptx}[1][I]{_{\D',\t,\xi\Dp}^{#1}}
\newcommand{\Dptxp}{_{\D',\t,\xi'\Dp}^I}
\newcommand{\Dptpxp}{_{\D',\t',\xi'\Dp}^I}
\newcommand{\Dptpx}[1][I]{_{\D',\t',\xi\Dp}^{#1}}
\newcommand{\Dpsx}{_{\D',\s,\xi\Dp}^I}
\newcommand{\Dpsxp}{_{\D',\s,\xi'\Dp}^I}
\newcommand{\Dpspx}{_{\D',\s',\xi\Dp}^I}
\newcommand{\Dptux}{_{\D',\t \cup \xu,\xi\Dp}^I}
\newcommand{\Dpspxp}{_{\D',\s',\xi'\Dp}^I}
\newcommand{\DptUSx}{_{\D',\t \cup \XU,\xi\Dp \cup \XS}^I}

\newcommand{\Dpptx}{_{\D'',\t,\xi\Dpp}^I}
\newcommand{\Dpptpx}[1][I]{_{\D'',\t',\xi\Dpp}^{#1}}
\newcommand{\Dpptpxp}{_{\D'',\t',\xi'\Dpp}^I}
\newcommand{\Dppsx}{_{\D'',\s,\xi\Dpp}^I}
\newcommand{\Dppsxp}{_{\D'',\s,\xi'\Dpp}^I}
\newcommand{\Dpptux}{_{\D'',\t \cup \xu,\xi\Dpp}^I}
\newcommand{\DpptUSx}{_{\D'',\t \cup \XU,\xi\Dpp \cup \XS}^I}

\newcommand{\Dpppsx}{_{\D''',\s,\xi\Dppp}^I}
\newcommand{\Dpppsxp}{_{\D'',\s,\xi'\Dpp}^I}

\newcommand{\J}[2]{\I{#1\!\th\!#2}}

\newcommand{\JG}[1]{\J{\G}{#1}}
\newcommand{\JGT}{\JG{T}}
\newcommand{\JGU}{\JG{U}}
\newcommand{\JGUp}{\JG{U'}}
\newcommand{\JGV}{\JG{V}}
\newcommand{\JGVp}{\JG{V'}}
\newcommand{\JGK}{\JG{K}}
\newcommand{\JGTp}{\JG{T'}}

\newcommand{\JGp}[1]{\J{\G'}{#1}}
\newcommand{\JGpV}{\JGp{V}}
\newcommand{\JGpVp}{\JGp{V'}}
\newcommand{\JGpU}{\JGp{U}}
\newcommand{\JGpT}{\JGp{T}}


\section{Interpretation schema}
\label{sec-schema-int}

We define the interpretation of a type $\G\TH T$ w.r.t. a substitution
$\t: \G\a\D$ by induction on the structure of $T$. Hence, we have to
give an interpretation to the predicate variables and the predicate
symbols that occur in $T$. That is why we first define an {\em
interpretation schema\,} $\JGT\Dtx$ using a {\em candidate
assignment\,} $\xi$ for the predicate variables and an interpretation
$I$ for the predicate symbols. In the Section~\ref{sec-int-const}, we
define the interpretation of constant predicate symbols and, in
Section~\ref{sec-int-def}, we define the interpretation of defined
predicate symbols.


\begin{dfn}[Candidate assignment]
\label{def-assign-cand}
\index{assignment!candidate|inddef[def-assign-cand]}
\index{candidate!assignment|inddef[def-assign-cand]}
\index{xi@$\xi,\xi\Gp,\xi_\t$|inddef[def-assign-cand]}
\index{restriction!candidate assignment|inddef[def-assign-cand]}
\index{assignment!candidate!canonical|inddef[def-assign-cand]}
\index{canonical!candidate assignment|inddef[def-assign-cand]}
\index{compatible with!substitution!candidate assignment|inddef[def-comp]}

  A {\em candidate assignment} is a function $\xi$ from $\XB$ to
  $\bigcup \,\{ \,\cR\DT ~|~ \D\TH T \in \oTY \}$. Let $\t: \G\a\D$ be
  a substitution. We say that $\xi$ is {\em compatible with\,}
  $(\t,\G,\D)$ if, for all $X\in \domB(\G)$, $X\xi \in \cR\DXt$. The
  {\em restriction\,} of $\xi$ to an environment $\G'$ is the
  assignment $\xi\Gp$ defined by $X(\xi\Gp) = (X\xi)\Gp$.
  
  To any substitution $\t: \G\a\D$, we associate its {\em canonical
  candidate assignment\,} $\xi_\t$ defined by $X\xi_\t= \top_{\D\th
  x\t}$. After Lemma~\ref{lem-cand-prop}~(d), $\xi_\t$ is compatible
  with $(\t,\G,\D)$.
\end{dfn}


\begin{lem}
\label{lem-prop-xi}
Let $\t: \G\a\D$ be a substitution and $\xi$ be a candidate assignment
compatible with $(\t,\G,\D)$.
\begin{enumalphai}
\item If $\t\a \t'$ then $\t': \G\a\D$ and $\xi$ is compatible with
$(\t',\G,\D)$.
\item If $\D\sle\D'\in\E$ then $\t: \G\a\D'$ and $\xi\Dp$ is
compatible with $(\t,\G,\D')$.
\end{enumalphai}
\end{lem}

\begin{prf}
\begin{enumalphai}
\item After Lemma~\ref{lem-sr-subs}, we know that $\t': \G\a\D$. Let
$X\in \domB(\G)$. Since $\xi$ is compatible with $(\t,\G,\D)$, $X\xi
\in \cR\DXt$. After Lemma~\ref{lem-cand-prop}~(b), $\cR\DXtp=
\cR\DXt$. Therefore $X\xi \in \cR\DXtp$.
\item By weakening, $\t: \G\a\D'$. Let $X\in \domB(\G)$.  By
definition, $X(\xi\Dp)= (X\xi)\Dp$. Since $\xi$ is compatible with
$(\t,\G,\D)$, $X\xi \in \cR\DXt$. Therefore, after
Lemma~\ref{lem-cand-prop}~(c), $(X\xi)\Dp \in \cR\DpXt$.\cqfd
\end{enumalphai}
\end{prf}


~

Let $F$ be a predicate symbol of type $\pvxT\st$. In the following, we
will assume that $F$, which is not a term if its arity is not null,
represents its $\eta$-long form $[\vx:\vT]F(\vx)$.

\begin{dfn}[Interpretation of a predicate symbol]
\label{def-int-symb}
\index{P3@\textbf{P3}|inddef[def-int-symb]}
\index{interpretation!predicate symbol|inddef[def-int-symb]}
\index{I@$I$|inddef[def-int-symb]}

  An {\em interpretation\,} for a predicate symbol $F$ is a function
  $I$ which, to an environment $\D$, associates an element of $\cR\DF$
  such that~:
\begin{lst}{}
\item [\hspace*{5mm}\bf(P3)] if $\D\sle\D'\in\E$ then $I_\D\Dp=
I_{\D'}$ (compatibility with weakening).
\end{lst}

\noindent
An {\em interpretation\,} for a set $\cG$ of predicate symbols is a
function which, to a symbol $G\in \cG$, associates an interpretation
for $G$.
\end{dfn}


\begin{dfn}[Interpretation schema]
\label{def-schema-int}
\index{interpretation!schema|inddef[def-schema-int]}
\index{$\JGT\Dtx^I$|inddef[def-schema-int]}
\index{substitution!valid|inddef[def-schema-int]}
\index{valid!substitution|inddef[def-schema-int]}
\index{term!computable|inddef[def-schema-int]}

The {\em interpretation\,} of $\G\TH T \in \oTY$ w.r.t. an environment
$\D\in\E$, a substitution $\t: \G\a\D$, a candidate assignment $\xi$
compatible with $(\t,\G,\D)$ and an interpretation $I_F$ pour each
$F\in\FB$, is an element of $\cR\DTt$ defined by induction on $T$~:

\begin{lst}{\bu}
\item $\JG{s}\Dtx= \oSN\Ds$,
  
\item $\JG{F(\vt)}\Dtx= I\DF(\va)$ or, if $\tF= \pvxT U$~:
\begin{lst}{--}
\item $a_i= \D\TH t_i\t$ if $x_i\in \Xs$,
\item $a_i= (\D\TH t_i\t, \JG{t_i}\Dtx)$ if $x_i\in \XB$,
\end{lst}

\item $\JG{X}\Dtx= X\xi$,
  
\item $\JG{\px{U}V}\Dtx= \{\D'\TH t \in \oT\DxUVt ~|~ \all \D''\TH u
  \in \JGU\Dptx$, $\D''\TH tu \in \J{\GxU}{V}\Dpptux \}$,
  
\item $\JG{\pX{K}V}\Dtx= \{\D'\TH t \in \oT\DXKVt ~|~ \all \D''\TH U
  \in \JGK\Dptx$, $\D''\th tU \in \bigcap \,\{\, \J{\GXK}{V}\DpptUSx
  ~|~ S \IN \cR\DppU \} \}$,
  
\item $\JG{\lx{U}V}\Dtx(\D'\TH u)= \J{\GxU}{V}\Dptux$,
  
\item $\JG{\lX{K}V}\Dtx(\D'\TH U, S)= \J{\GXK}{V}\DptUSx$,

\item $\JG{Vu}\Dtx= \JGV\Dtx(\D\TH u\t)$,
  
\item $\JG{VU}\Dtx= \JGV\Dtx(\D\TH U\t, \JGU\Dtx)$.
\end{lst}

\noindent
In the case where $\G\th T:s$, the elements of $\JGT\Dtx$ are called
{\em computable\,}. Finally, we will say that $(\t,\G,\D)$ is {\em
valid\,} w.r.t. $\xi$ if, for all $x\in\dom(\G), \D\TH x\t \in
\JG{x\G}\Dtx$.
\end{dfn}

After Lemma~\ref{lem-var-cand}, the identity substitution is valid
w.r.t. any candidate assignment compatible with it.


\begin{lem}[Correctness of the interpretation schema]\hfill
\label{lem-cor-schema-int}
\index{interpretation!schema|indlem[lem-cor-schema-int]}

\begin{enumalphai}
\item $\JGT\Dtx$ is well defined.
\item $\JGT\Dtx \in \cR\DTt$.
\item If $\t\a \t'$ then $\JGT\Dtx= \JGT\Dtpx$.
\item If $\D\sle\D'\in\E$ then $\JGT\Dtx\Dp= \JGT\Dptx$.
\end{enumalphai}
\end{lem}

\begin{prf}
  Note first of all that, for (c), $\JGT\Dtpx$ exists since, after
  Lemma~\ref{lem-prop-xi} (a), $\t': \G\a\D$ and $\xi$ is compatible
  with $(\t',\G,\D)$. For (d), $\JGT\Dtx\Dp$ exists since, after (b),
  $\JGT\Dtx \in \cR\DTt$, and $\JGT\Dptx$ exists since, after
  Lemma~\ref{lem-prop-xi} (b), $\t: \G\a\D'$ and $\xi\Dp$ is
  compatible with $(\t,\G,\D')$.

\begin{lst}{\bu}


\item $T=s$.
\begin{enumalphai}
\item Immediate.
\item After Lemma~\ref{lem-cand-prop}~(d).
\item Since $\JG{s}\Dtx$ does not depend on $\t$.
\item $\JG{s}\Dtx\Dp= \oSN\Ds \cap \oT_{\D'\th s}= \oSN_{\D'\th s}=
\JG{s}\Dptx$.
\end{enumalphai}


\item $T=F(\vt)$.
\begin{enumalphai}
\item $\JGT\Dtx= I\DF(\va)$ where $a_i= \D\TH t_i\t$ if $x_i\in\Xs$
and $a_i= (\D\TH t_i\t, \JG{t_i}\Dtx)$ if $x_i\in\XB$. Par induction
hypothesis (a) and (b), $\JG{t_i}\Dtx$ is well defined and belongs to
$\cR_{\D \th t_i\t}$. Therefore $\va$ is in the domain of $I\DF$ and
$\JGT\Dtx$ is well defined.
\item By definition of $I\DF$.
\item By {\bf(P1)} \index{P1@\textbf{P1}|indlem[lem-cor-schema-int]}.
\item By {\bf(P2)} \index{P2@\textbf{P2}|indlem[lem-cor-schema-int]}.
\end{enumalphai}


\item $T=X$.
\begin{enumalphai}
\item Immediate.
\item Since $\xi$ is compatible with $(\t,\G,\D)$.
\item Since $\JG{X}\Dtx$ does not depend on $\t$.
\item By definition of $\xi\Dp$.
\end{enumalphai}


\item $T=\px{U}V$. Let $\G'= \GxU$.
\begin{enumalphai}

\item Assume that $\D\sle\D'\in\E$. After Lemma~\ref{lem-prop-xi}~(b),
$\t: \G\a\D'$ and $\xi\Dp$ is compatible with $(\t,\G,\D')$. So, by
induction hypothesis (a) and (b), $\JGU\Dptx$ is well defined and
belongs to $\cR\DpUt$.
  
  \hs By Type correctness, there exists $s$ such that $\G\th T:s$.  By
  inversion, $\G'\th V:s$. After the Environment Lemma, $\G\th
  U:\st$. Therefore, by substitution, $\D'\th U\t:\st$ and $\JGU\Dptx
  \sle \oT\DpUt$.
  
  \hs Now, let $\D''\TH u \in \JGU\Dptx$ and $\s= \t\cup\xu$.  Since
  $\t: \G\a\D$, $\D''\th u:U\t$ and $x\notin \FV(\G')$, we have $\s:
  \G'\a\D''$. After Lemma~\ref{lem-prop-xi}~(b), $\xi\Dpp$ is
  compatible with $(\t,\G,\D'')$. Since $\domB(\s)= \domB(\t)$ and
  $\s$ and $\t$ are equal on this domain, $\xi\Dpp$ is compatible with
  $(\s,\G',\D'')$. Therefore, by induction hypothesis (a) and (b),
  $\JGpV\Dppsx$ is well defined and belongs to $\cR\DppVs$.
  
  \hs Finally, by substitution, $\D''\th V\s:s$. Therefore,
  $\JGpV\Dppsx \sle \oT\DppVs$ and $\JGT\Dtx$ is well defined.

\item By substitution, $\D\th T\t:s$. Therefore, we have to prove that
$\JGT\Dtx$ is included in $\oT\DTt$ (immediate) and satisfies (R1) to
(R4). We have seen in (a) that, if $\D\sle\D'\in\E$, $\D''\TH u \in
\JGU\Dptx$ and $\s= \t\cup\xu$, then $\JGU\Dptx$ and $\JGpV\Dppsx$ are
included in $\oT\DpUt$ and $\oT\DppVs$ respectively and satisfy (R1)
to (R4).

\begin{bfenumii}{R}

\item After Lemma~\ref{lem-var-cand}, there exists $y\in\Xs \moins
\dom(\D')$ such that $\D', y\!:\!U\t\TH y \in \JGU\Dptx$.  Let $\D''=
\D', y\!:\!U\t$ and $\s= \t \cup \{x\to y\}$. Then, by definition,
$\D''\TH ty \in \JGpV\Dppsx$ and, since $\JGpV\Dppsx$ satisfies
{\bf(R1)}, \index{R1@\textbf{R1}|indlem[lem-cor-schema-int]} $ty \in
\SN$ and $t\in \SN$.

\item Let $\D''\TH u \in \JGU\Dptx$ and $\s= \t\cup\xu$. By
definition, $\D''\TH tu \in \JGpV\Dppsx$ and, since $tu \a t'u$ and
$\JGpV\Dppsx$ satisfies {\bf(R2)},
\index{R2@\textbf{R2}|indlem[lem-cor-schema-int]} $t'u \in
\JGpV\Dppsx$ and $t' \in \JGT\Dtx$.

\item This is in this case that we use the notion of arity which
establishes a syntactic distinction between the application of the
$\la$-calculus and the application of a symbol (see
Remark~\ref{rem-arity}).
\index{arity!symbol|indtxt[Lem.~\ref{lem-cor-schema-int}~(R3)]} Let
$\D''\TH u \in \JGU\Dptx$ and $\s= \t\cup\xu$. By definition, $\D''\th
tu \in \oT\DppVs$ and $tu$ is neutral. Since $\JGU\Dptx$ satisfies
{\bf(R1)}, \index{R1@\textbf{R1}|indlem[lem-cor-schema-int]}
$u\in\SN$.
    
  \hs We prove that any reduct $v'$ of $tu$ belongs to $\JGpV\Dppsx$,
  by induction on $u$ with $\a$ as well-founded ordering. Since $t$ is
  not an abstraction, $v'$ is either of the form $t'u$ with $t'$ an
  immediate reduct of $t$, or of the form $tu'$ with $u'$ an immediate
  reduct of $u$. In the first case, $\D''\TH t'u \in \oT\DppVs$ since,
  by hypothesis, $\D''\TH t' \in \JGT\Dtx$ and $\D''\TH u \in
  \JGU\Dptx$.  In the second case, $\D''\TH tu' \in \oT\DppVs$ by
  induction hypothesis.
    
  \hs Therefore, since $\JGpV\Dppsx$ satisfies {\bf(R3)},
  \index{R3@\textbf{R3}|indlem[lem-cor-schema-int]} $\D''\TH tu \in
  \JGpV\Dppsx$ and $\D''\TH t \in \JGT\Dtx$.

\item Assume that $\D'\sle\D''\in\E$, $\D'''\TH u \in \JGU\Dpptx$ and
$\s= \t\cup\xu$. By induction hypothesis (d), $\JGU\Dpptx=
\JGU\Dptx\Dpp= \JGU\Dptx \cap \oT\DppUt$. So, $\D'''\TH u \in
\JGU\Dptx$, $\D'''\TH tu \in \JGpV\Dpppsx$ and $\D''\TH t \in
\JGT\Dtx$.

\end{bfenumii}

\item We prove that $\JGT\Dtx \sle \JGT\Dtpx$. The other way around is
similar. Since $\D\th T\s:s$, by conversion, $\oT\DTtp= \oT\DTt$ and
$\D'\TH t \in \oT\DTtp$.  Now, let $\D''\TH u \in \JGU\Dptpx$, $\s=
\t\cup\xu$ and $\s'= \s\cup\xu$. By induction hypothesis (c),
$\JGU\Dptpx= \JGU\Dptx$. Therefore, since $\JGU\Dptx$ satisfies
{\bf(R3)}, \index{R3@\textbf{R3}|indlem[lem-cor-schema-int]} $\D''\TH
tu \in \JGpV\Dsx$. And since $\s\a\s'$, by induction hypothesis (c),
$\D''\TH tu \in \JGpV\Dspx$ and $\D'\TH t \in \JGT\Dtpx$.

\item We prove that $\JGT\Dtx\Dp \sle \JGT\Dptx$. The other way around
is similar. By definition, $\JGT\Dtx\Dp= \JGT\Dtx \cap \oT\DpTt$. Let
$\D''\TH t \in \JGT\Dtx\Dp$, $\D'''\TH u \in \JGT\Dpptx$ and $\s=
\t\cup\xu$. By definition of $\JGT\Dtx$, $\D'''\TH tu \in
\JGpV\Dpppsx$. Therefore, since $\D''\TH t \in \oT\DpTt$, $\D''\TH t
\in \JGT\Dptx$.

\end{enumalphai}


\item $T=\pX{K}V$. Similar to the previous case.


\item $T=\lx{U}V$. Let $\G'= \GxU$.
\begin{enumalphai}

\item Let $\D'\TH u \in \oT\DUt$ and $\s= \t\cup\xu$. Since
$\D\sle\D'\in\E$, by Lemma~\ref{lem-prop-xi}~(b), $\t: \G\a\D'$ and
$\xi\Dp$ is compatible with $(\t,\G,\D')$. Since $\D'\th u:U\t$ and
$x\notin \FV(\G)$, $\s: \G'\a\D'$. Moreover, $\xi\Dp$ is compatible
with $(\s,\G',\D')$ since $\domB(\s)= \domB(\t)$. So, by induction
hypothesis (a), $\JGpV\Dpsx$ is well defined and $\JGT\Dtx$ is well
defined.

\item $\cR\DTt$ is the set of functions which, to $\D'\TH u \in
\oT\DUt$, associate an element of $\cR_{\D'\th (T\t~u)}$ and satisfy
(P1) and (P2). By induction hypothesis (b), $\JGpV\Dpsx \in
\cR\DpVs$. Since $(T\t~u) \ab V\s$, by Lemma~\ref{lem-cand-prop}~(b),
$\cR\DpVs= \cR_{\D'\th (T\t~u)}$.

\begin{bfenumii}{P}

\item Assume that $u\a u'$. $\JGT\Dtx(\D'\TH u')= \JGpV\Dpspx$ where
$\s'= \t\cup\xup$. Since $\s\a\s'$, by induction hypothesis (c),
$\JGpV\Dpspx= \JGpV\Dpsx$ and $\JGT\Dtx \in \cR\DTt$.

\item Assume that $\D\sle\D'\in\E$. $\JGT\Dtx(\D\TH u)\Dp=
\JGpV\Dsx\Dp$. By induction hypothesis (d), $\JGpV\Dsx\Dp=
\JGpV\Dpsxp= \JGT\Dtx(\D'\TH u)$.

\end{bfenumii}

\item $\JGT\Dtpx(\D'\TH u)= \JGpV\Dpspx$ or $\s'= \s\cup\xu$.
$\s\a\s'$ therefore, by induction hypothesis (c), $\JGpV\Dpspx=
\JGpV\Dpsx$ and $\JGT\Dtpx= \JGT\Dtx$.

\item $\JGT\Dtx\Dp= \JGT\Dtx|_{\oT\DpUt}$ is the function which, to
$\D''\TH u \in \oT\DpUt$, associates $\JGpV\Dppsx$ where $\s=
\t\cup\xu$. This is $\JGT\Dptx$.

\end{enumalphai}


\item $T=\lX{K}V$. Similar to the previous case.


\item $T= Vu$.

\begin{enumalphai}

\item By induction hypothesis (a), $\JGV\Dtx$ is well defined and
belongs to $\cR\DVt$. Since $T \in \oTY$ and $T\neq \B$, $T$ is
typable in $\G$. By inversion, there exists $U$ and $K$ such that
$\G\TH V:\px{U}K$ and $\G\TH u:U$. By substitution, $\D\th
V\t:\px{U\t}K\t$ and $\D\th u\t:U\t$. Therefore, $\cR\DVt$ is the set
of functions which, to $\D'\TH u' \in \oT\DUt$, associate an element
of $\cR_{\D'\th V\t u'}$. So, $\D\TH u\t \in \oT\DUt$ and $\JGT\Dtx$
is well defined.

\item $\JGT\Dtx \in \cR_{\D\th (V\t u\t)}= \cR\DTt$.

\item By induction hypothesis (c), $\JGV\Dptpx= \JGV\Dtx$.  Since $u\t
\a^* u\s$ and $\JGV\Dtx$ satisfies {\bf(P1)}, $\JGT\Dtx= \JGT\Dtpx$.

\item $\JGT\Dtx\Dp= \JGV\Dtx(\D\TH u\t)\Dp$. By {\bf(P2)},
$\JGV\Dtx(\D\TH u\t)\Dp= \JGV\Dtx(\D'\TH u\t)= \JGV\Dtx\Dp(\D'\TH
u\t)$. By induction hypothesis (d), $\JGV\Dtx\Dp(\D'\TH u\t)=
\JGV\Dptx(\D'\TH u\t)= \JGT\Dptx$.

\end{enumalphai}


\item $T= (V~U)$. Similar to the previous case.\cqfd

\end{lst}
\end{prf}


\begin{lem}
\label{lem-int-dep}
\index{interpretation!schema|indlem[lem-int-dep]}

  Let $I$ and $I'$ be two interpretations equal on the predicate
  symbols occurring in $T$, and $\xi$ and $\xi'$ be two candidate
  assignments equal on the predicate variables free in $T$. Then,
  $\JGT\Dtxp[I']= \JGT\Dtx$.
\end{lem}

\begin{prf}
By induction on $T$.\cqfd
\end{prf}


\newcommand{\deux}{_{\G_2,\t_2,\xi_2}^I}
\newcommand{\deuxp}{_{\G_2',\t_2,\xi_2|_{\G_2'}}^I}
\newcommand{\deuxpu}{_{\G_2',\t_2 \cup \xu,\xi_2|_{\G_2'}}^I}
\newcommand{\deuxpp}{_{\G_2'',\t_2,\xi_2|_{\G_2''}}^I}
\newcommand{\deuxppu}{_{\G_2'',\t_2 \cup \xu,\xi_2|_{\G_2''}}^I}
\newcommand{\undeux}{_{\G_2,\t_1\t_2,\xi_{12}}^I}
\newcommand{\undeuxp}{_{\G_2',\t_1\t_2,\xi_{12}|_{\G_2'}}^I}
\newcommand{\undeuxpu}{_{\G_2',\t_1\t_2 \cup \xu,\xi_{12}|_{\G_2'}}^I}
\newcommand{\undeuxppu}{_{\G_2'',\t_1\t_2 \cup \xu,\xi_{12}|_{\G_2''}}^I}

\newcommand{\JT}{\J{\G_0}{T}\undeux}
\newcommand{\JTt}{\J{\G_1}{T\t_1}\deux}

\newcommand{\Gdeux}{_{\G_2\th T\t_1\t_2}}

\begin{lem}[Candidate substitution]
\label{lem-cand-subs}
\index{substitution!reductibility candidate|indlem[lem-cand-subs]}
\index{candidate!substitution|indlem[lem-cand-subs]}

Let $\G_0$, $\G_1$ and $\G_2$ be three valid environments, $\G_0\th T
\in \oTY$, $\t_1: \G_0\a\G_1$ and $\t_2: \G_1\a\G_2$ be two
substitutions, and $\xi_2$ be a candidate assignment compatible with
$(\t_2,\G_1,\G_2)$. Then, the candidate assignment $\xi_{12}$ defined
by $X\xi_{12}= \J{\G_1}{X\t_1}\deux$ is compatible with
$(\t_1\t_2,\G_0,\G_2)$ and $\JTt= \JT$.
\end{lem}

\begin{prf}
  After Lemma~\ref{lem-cor-schema-int}~(b), $X\xi_{12} \in
  \cR_{\G_2\th X\t_1\t_2}$. Therefore $\xi_{12}$ is compatible with
  $(\t_1\t_2,\G_0,\G_2)$. Let $R= \JT$ and $R'= \JTt$. We show that
  $R=R'$ by induction on $T$. After
  Lemma~\ref{lem-cor-schema-int}~(b), $R$ and $R'$ both belong to
  $\cR\Gdeux$.

\begin{lst}{\bu}
\item $T=s$. $R'= \J{\G_1}{s}\deux= \oSN_{\G_2\th s}= R$.
  
\item $T=F(\vt)$. Assume that $\tF= \pvxT\st$. Then, $R= I_{\G_2\th
F}(\va)$ where $a_i= \G_2\TH t_i\t_1\t_2$ if $x_i\in\Xs$, and $a_i=
(\G_2\TH t_i\t_1\t_2, \J{\G_0}{t_i}\undeux)$ if
$x_i\in\XB$. Similarly, $R'= I_{\G_2\th F}(\va')$ where $a_i'= \G_2\TH
t_i\t_1\t_2$ if $x_i\in\Xs$, and $a_i'= (\G_2\TH t_i\t_1\t_2,
\J{\G_1}{t_i\t_1}\deux)$ if $x_i\in\XB$. By induction hypothesis, for
all $x_i\in\XB$, $\J{\G_0}{t_i}\undeux=
\J{\G_1}{t_i\t_1}\deux$. Therefore, $\va=\va'$ and $R= R'$.
  
\item $T=X$. $R= X\xi_{12}= \J{\G_1}{X\t_1}\deux= R'$.
  
\item $T=\px{U}V$. Let $\G_0'= \G_0, x\!:\!U$, $\G_1'= \G_1,
x\!:\!U\t_1$ and $\G_2'\TH t \in R$. Since $R \in \cR\Gdeux$,
$\G_2'\TH t \in \oT\Gdeux$. Let $\G_2''\TH u \in
\J{\G_1}{U\t_1}\deuxp$.  After Lemma~\ref{lem-cor-schema-int}~(d), for
all $X\in \domB(\G_0)$, $X\xi_{12}|_{\G_2'}=
\J{\G_1}{X\t_1}\deuxp$. By induction hypothesis,
$\J{\G_1}{U\t_1}\deuxp= \J{\G_0}{U}\undeuxp$. By definition of
$\cR\Gdeux$, $\G_2''\TH tu \in \J{\G_0'}{V}\undeuxppu$. Moreover,
$X\xi_{12}|_{\G_2''}= \J{\G_1}{X\t_1}\deuxpp$, $\t_1: \G_0'\a\G_1'$,
$\t_2\cup\xu: \G_1'\a\G_2''$ and $\t_1\t_2\cup\xu=
\t_1(\t_2\cup\xu)$. Therefore, by induction hypothesis,
$\J{\G_0'}{V}\undeuxppu= \J{\G_1'}{V\t_1}\deuxppu$ and $\G_2'\TH t \in
R'$. So, $R \sle R'$. The other way around is similar.

\item $T=\pX{K}V$. Similar to the previous case.
  
\item $T=\lx{U}V$. Let $\G_0'= \G_0, x\!:\!U$, $\G_1'= \G_1,
x\!:\!U\t_1$ and $\G_2'\TH u \in \oT_{\G_2\th U\t_1\t_2}$. By
definition, $R(\G_2'\TH u)= \J{\G_0'}{V}\undeuxpu$. By induction
hypothesis, $\J{\G_0'}{V}\undeuxpu=
\J{\G_1'}{V\t_1}\deuxpu$. Therefore, $R(\G_2'\TH u)= R'(\G_2'\TH u)$
and $R=R'$.

\item $T=\lX{K}V$. Similar to the previous case.
  
\item $T=Vu$. $R= \J{\G_0}{V}\undeux(\G_2\TH u\t_1\t_2)$. By induction
hypothesis, $\J{\G_0}{V}\undeux= \J{\G_1}{V\t_1}\deux$. Therefore,
$R=R'$.

\item $T=VU$. Similar to the previous case.\cqfd
\end{lst}
\end{prf}


%% file: int-const.tex



\section{Interpretation of constant predicate symbols}
\label{sec-int-const}

We define the interpretation $I$ for constant predicate symbols by
induction on $>_\cC$. Let $C$ \index{C@$C$|indpage} be a constant
predicate symbol and assume that we already have defined an
interpretation $K$ \index{K@$K$|indpage} for all the symbols smaller
than $C$.

Like N. P. Mendler \cite{mendler87thesis} or B. Werner
\cite{werner94thesis}, we define this interpretation as the fixpoint
of some monotone function on a complete lattice. The monotonicity is
ensured the positivity conditions of an admissible inductive structure
(Definition~\ref{def-adm-ind-str}). The main difference with these
works is that we have a more general notion of constructor since it
includes any function symbol whose output type is a constant predicate
symbol. This allows us to defined functions or predicates by matching
not only on constant constructors but also on defined symbols.


~

We will denote by~:

\begin{lst}{--}
\item $[C]$ \index{$[C]$|indpage} the set of constant
predicate symbols equivalent to $C$,
\item $\cI$ \index{I@$\cI$|indpage} the set of the
interpretations for $[C]$,
\item $\le$ \index{$\le$|indpage} the relation on $\cI$
defined by $I \le I'$ if, for all $D\in [C]$ and $\D\in\E$, $I\DD
\le\DD I'\DD$.
\end{lst}

For simplifying the notations, we will denote $\JGT\Dtx[K\cup I]$ by
$\JGT\Dtx$.

~

Let $D\in [C]$. \index{D@$D$|indpage} Assume that $D$ is
of arity $n$ and type $\pvxT\st$. Let $\D$ be an environment. By
definition, $\cR\DD$ is the set of functions which, to $a_1\in A_1$,
\ldots, $a_n\in A_n$, \index{a@$\va$|indpage}
\index{ai@$a_i$|indpage}
\index{Ai@$A_i$|indpage} associates an element of
$\cR_{\D_n\th D(\vt)}$ where~:

\begin{lst}{--}
\item $a_i= \D_i\TH t_i$ \index{Deltai@$\D_i$|indpage}
\index{ti@$t_i$|indpage} and $A_i= \oT_{\D_{i-1}\th
T_i\t}$ if $x_i\in\Xs$,
\item $a_i= (\D_i\TH t_i, S_i)$ \index{Si@$S_i$|indpage}
and $A_i= \S_{\D_{i-1}\th T_i\t}$ if $x_i\in\XB$,
\item $\D_0= \D$ and $\t= \vxt$.
\end{lst}


\begin{dfn}[Monotone interpretation]
\label{def-mon-int}
\index{interpretation!monotone|inddef[def-mon-int]}
\index{monotone interpretation|inddef[def-mon-int]}
\index{inductive!position|inddef[def-mon-int]}
\index{Im@$\cI^m$|inddef[def-mon-int]}

  Let $I\in \cI$, $x_i\in \XB$, $\D\in \E$ and $\va,\va'$ two
  sequences of arguments for $I$ such that $a_i= (\D_i\TH t_i, S_i)$,
  $a_i'= (\D_i\TH t_i, S_i')$ and, for all $j\neq i$, $a_j=
  a_j'$. Then, $I$ is {\em monotone in its $i$-th argument\,} if $S_i
  \le S_i'$ implies $I\DD(\va) \le I\DD(\va')$. We will denote by
  $\cI^m$ the set of the interpretations that are monotone in all its
  inductive arguments $i\in \ind(D)$.
\end{dfn}


\begin{lem}
\label{lem-mon-int-lat-comp}
\index{monotone interpretation|indlem[lem-mon-int-lat-comp]}
\index{interpretation!monotone|indlem[lem-mon-int-lat-comp]}

$(\cI^m,\le)$ is a complete lattice.
\end{lem}

\begin{prf}
  First of all, $\le$ is an ordering since, for all $D =_\cC C$ and
  $\D\in\E$, $\le\DD$ is an ordering.
  
  We show that the function $I^\top$ defined by $I^\top\DD= \top\DD$
  is the greatest element of $\cI^m$. The function $I^\top_D$ is an
  interpretation since, after Lemma~\ref{lem-cand-prop} (d) and (f),
  $I^\top\DD \in \cR\DD$ and if $\D\sle\D'\in\E$ then $I^\top\DD\D=
  \top\DD\Dp= \top\DpD= I^\top\DpD$. Moreover, $I^\top$ is the
  greatest element of $\cI$ since $\top\DD$ is the greatest element of
  $\cR\DD$.
  
  We now show that $I^\top$ is monotone in its inductive
  arguments. Let $i\in \ind(D)$ and $\va,\va'$ two sequences of
  arguments for $I^\top\DD$ such that $a_i= (\D_i\TH t_i, S_i)$,
  $a_i'= (\D_i\TH t_i, S_i')$, $S_i \le S_i'$ and, for all $j\neq i$,
  $a_j= a_j'$. Then, $I^\top\DD(\va)= \top_{\D_n\th D(\vt)}=
  I^\top\DD(\va')$.
  
  We now show that every part of $\cI^m$ has an inf. Let $\Im\sle
  \cI^m$ and $I^\et$ be the function defined by $I^\et\DD=
  \biget\DD(\Re\DD)$ where $\Re\DD= \{I\DD ~|~ I\in\Im\}$. The
  function $I^\et$ is an interpretation since, after
  Lemma~\ref{lem-cand-prop} (g) and (i), $I^\et\DD \in \cR\DD$ and if
  $\D\sle\D'\in\E$ then $I^\et\DD\Dp= I^\et\DpD$.
  
  We now show that $I^\et$ is monotone in its inductive arguments. Let
  $i\in\ind(D)$ and $\va,\va'$ two sequences of arguments for
  $I^\et\DD$ satisfying the conditions of the
  Definition~\ref{def-mon-int}.  Then, $I^\et\DD(\va)= \bigcap
  \{I\DD(\va) ~|~ I\in\Im\}$ and $I^\et\DD(\va')= \bigcap \{I\DD(\va')
  ~|~ I\in\Im\}$. Since each $I$ is monotone in its inductive
  arguments, $I\DD(\va) \le I\DD(\va')$. Therefore, $I^\et\DD(\va) \le
  I^\et\DD(\va')$.
  
   We are left to show that $I^\et$ is the inf of $\Im$. For all
  $I\in\Im$, $I^\et \le I$ since $I^\et\DD$ is the inf of $\Re\DD$.
  Assume now that there exists $I'\in \cI^m$ such that, for all
  $I\in\Im$, $I' \le I$. Then, since $I^\et\DD$ is the inf of
  $\Re\DD$, $I' \le I^\et$.
\end{prf}


\begin{dfn}[Interpretation of constant predicate symbols]
\label{def-int-const}
\index{I6@\textbf{I6}|inddef[def-int-const]}
\index{interpretation!constant|inddef[def-int-const]}
\index{I6@\textbf{I6}|inddef[def-int-const]}
\index{phiDeltaDIa@$\vp^I\DD(\va)$|inddef[def-int-const]}
\index{lfp@$\lfp$|inddef[def-int-const]}

\hfill Let $\vp$ be\\ the function which, to $I\in \cI^m$, associates
the interpretation $\vp^I$ such that $\vp^I\DD(\va)$ is the set of
$\D'\TH u \in \oSN_{\D_n\th D(\vt)}$ such that if $u$ reduces to a
term of the form $d(\vu)$ with $d$ a constructor of type $\pvyU
D(\vv)$ then, for all $j\in \acc(d)$, $\D'\TH u_j \in
\J{\vy:\vU}{U_j}\Dptx$ where $\t= \vyu$ and, for all $Y\in \FVB(U_j)$,
$Y\xi= S_{\io_Y}$. We show hereafter that $\vp$ is monotone.
Therefore we can take $I\DD= \lfp(\vp)\DD$ where $\lfp(\vp)$ is the
least fix point of $\vp$.
\end{dfn}

The aim of this definition is to ensure the correctness of the
accessibility relations (Lemma~\ref{lem-cor-acc})~: if $d(\vu)$ is
computable then each accessible $u_j$ ($j\in\acc(d)$) is
computable. This will allow us to ensure the computability of the
variables of the left hand-side of a rule if the arguments of the left
hand-side are computable, and thus the computability of the right
hand-sides that belong to the computable closure.


\begin{lem}
\label{lem-cor-int-const}
\index{interpretation!constant|indlem[lem-cor-int-const]}

  $\vp^I$ is a well defined interpretation.
\end{lem}

\begin{prf}
  We first prove that $\vp^I$ is well defined. The existence of $\xi$
  is the hypothesis {\bf(I6)}.
  \index{I6@\textbf{I6}|indlem[lem-cor-int-const]} Let $\G_d=
  \vy:\vU$. We have to check that $\t: \G_d\a\D'$ and $\xi\Dp$ is
  compatible with $(\t,\G_d,\D')$. By subject reduction, $\D'\th
  d(\vu):D(\vt)$. By inversion, $\D'\th d(\vu):D(\vv\t)$, $D(\vv\t)
  \CV[\D'] D(\vt)$ and, for all $j$, $\D'\th u_j:U_j\t$. Therefore,
  $\t: \G_d\a\D'$. Let $Y\in \FVB(U_j)$. We have $Y\xi= S_{\io_Y}\in
  \cR_{\D_{\io_Y}\th t_{\io_Y}}$. By Lemma~\ref{lem-cand-prop}~(c),
  $Y\xi\Dp \in \cR_{\D'\th t_{\io_Y}}$.  By {\bf(A1)},
  \index{A1@\textbf{A1}|indlem[lem-cor-int-const]} since $D$ is
  constant, for all $i$, $v_i\t \ad t_i$. So, by
  Lemma~\ref{lem-cand-prop}~(b), $\cR_{\D'\th t_{\io_Y}}= \cR_{\D'\th
  v_{\io_Y}\t}$. By {\bf(I6)},
  \index{I6@\textbf{I6}|indlem[lem-cor-int-const]} $v_{\io_Y}=Y$.
  Therefore $Y\xi \in \cR_{\D'\th Y\t}$.

  Finally, we must make sure that the interpretations necessary for
computing $\J{\vy:\vU}{U_j}\Dptx$ are all well defined. The
interpretation of constant predicate symbols smaller than $D$ is
$K$. The interpretation of constant predicate symbols equivalent to
$D$ is $I$. By {\bf(I4)}
\index{I4@\textbf{I4}|indlem[lem-cor-int-const]} and {\bf(I5)},
\index{I5@\textbf{I5}|indlem[lem-cor-int-const]} constant predicate
symbols greater than $D$ or defined predicate symbols can occur only
at neutral positions, but it is easy to check that terms at neutral
positions are not interpreted.
  
  We now prove that $\vp^I\DD \in \cR\DD$. To this end, we must prove
    that $R= \vp^I\DD(\va)$ is included in $\oT_{\D_n\th D(\vt)}$
    (immediate) and satisfies the properties (R1) to (R4)~:

\begin{bfenumi}{R}
\item By definition.
  
\item Let $\D'\TH u \in R$ and $u'$ such that $u\a u'$. By definition,
$\D'\TH u \in \oSN_{\D_n\th D(\vt)}$. Therefore, $\D'\TH u' \in
\oSN_{\D_n\th D(\vt)}$. Assume that $u' \a^* d(\vu)$ with $\td= \pvyU
D(\vv)$. Then, $u \a^* d(\vu)$. Therefore, for all $j \in \acc(d)$,
$\D'\TH u_j \in \J{\vy:\vU}{U_j}\Dptx$ and $\D'\TH u' \in R$.
  
\item Let $\D'\TH u \in \oT_{\D_n\th D(\vt)}$ such that $u$ is
neutral and, for all $u'$ such that $u \a u'$, $\D'\TH u' \in R$.
Then, $\D'\TH u \in \oSN_{\D_n\th D(\vt)}$. Assume now that $u\a^*
d(\vu)$ with $\td= \pvyU D(\vv)$. Since $u$ is neutral, there exists
$u'$ such that $u\a u'$ and $u'\a^* d(\vu)$. Therefore, for all $j \in
\acc(d)$, $\D'\TH u_j \in \J{\vy:\vU}{U_j}\Dptx$ and $\D'\TH u \in R$.
  
\item Let $\D'\TH u \in R$ and assume that $\D'\sle\D''\in\E$.  Then,
$\D''\TH u \in \oSN_{\D_n\th D(\vt)}$. Assume now that $u \a^*
d(\vu)$ with $\td= \pvyU D(\vv)$. Then, for all $j\in \acc(d)$,
$\D'\TH u_j \in R_j$ where $R_j=$ $\J{\vy:\vU}{U_j}\Dptx$. After
Lemma~\ref{lem-cor-schema-int}~(b), $R_j$ belongs to $\cR_{\D_n\th
U_j\t}$ and therefore satisfies {\bf(R4)}.
\index{R4@\textbf{R4}|indlem[lem-cor-int-const]} Therefore $\D''\TH
u_j \in R_j$ and $\D''\TH u \in R$.
\end{bfenumi}

Finally, we are left to show the properties (P1) to (P3). For (P1),
the stability by reduction, this is immediate since, if $\vt\a \vt'$
then $\oSN_{\D_n\th D(\vt)}= \oSN_{\D_n\th D(\vt')}$. For (P2) and
(P3), it is easy to see that the functions have the same domain and
are equal.
\end{prf}


\begin{lem}
\label{lem-int-const-mon-arg-ind}
\index{interpretation!constant|indlem[lem-int-const-mon-arg-ind]}

$\vp^I$ is monotone in its inductive arguments.
\end{lem}

\begin{prf}
  Let $i\in\ind(D)$. We have to show that $S_i \le S_i'$ implies
  $\vp^I\DD(\va) \sle \vp^I\DD(\va')$. Let $\D'\TH u \in
  \vp^I\DD(\va)$. We prove that $\D'\TH u \in \vp^I\DD(\va')$. We have
  $\D'\TH u \in \oSN_{\D_n\th D(\vt)}$. Assume now that $u$ reduces to
  a term of the form $d(\vu)$ with $d$ a constructor of type $\pvyU
  D(\vv)$. Let $j\in \acc(d)$. We have to prove that $\D'\TH u_j \in
  \J{\vy:\vU}{U_j}\Dptxp$ where $\t= \vyu$ and, for all $Y\in
  \FVB(U_j)$, $Y\xi'= S'_{\io_Y}$.
  
  We have $\D'\TH u_j \in \J{\vy:\vU}{U_j}\Dptx$ where, for all $Y\in
  \FVB(U_j)$, $Y\xi= S_{\io_Y}$. If, for all $Y\in \FVB(U_j)$,
  $\io_Y\neq i$, then $\xi=\xi'$ and $\D'\TH u_j \in
  \J{\vy:\vU}{U_j}\Dptxp$. Assume now that there exists $Y\in
  \FVB(U_j)$ such that $\io_Y= i$. Then, $Y\xi= S_i \le Y\xi'= S_i'$.
  By {\bf(I2)},
  \index{I2@\textbf{I2}|indlem[lem-int-const-mon-arg-ind]}
  $\pos(Y,U_j) \sle \pp(U_j)$. Finally, $U_j$ satisfies {\bf(I3)},
  {\bf(I4)} and {\bf(I5)}.
  \index{I3@\textbf{I3}|indlem[lem-int-const-mon-arg-ind]}
  \index{I4@\textbf{I4}|indlem[lem-int-const-mon-arg-ind]}
  \index{I5@\textbf{I5}|indlem[lem-int-const-mon-arg-ind]}
  
  We now prove by induction on $T$ that, for all $\G\TH T \in \oTY$,
  $\D\in \E$, $\t: \G\a\D$, $\xi,\xi'$ compatible with $(\t,\G,\D)$
  such that $Y\xi \le Y\xi'$ and, for all $X \neq Y$, $X\xi= X\xi'$~:

\begin{lst}{--}
\item if $\pos(Y,T) \sle \pp(T)$ and $T$ satisfies (I3), (I4) and (I5)
then $\JGT\Dtx \le \JGT\Dtxp$,
\item if $\pos(Y,T) \sle \pm(T)$ and $T$ satisfies (I3$^-$), (I4) and
(I5) then $\JGT\Dtx \ge \JGT\Dtxp$,
\end{lst}

\noindent
where (I3$^-$) is the property $\all D \IN \FC, D=_\cC C \A \pos(D,T)
\sle \pm(T)$. We detail the first case only; the second is similar.


\begin{lst}{\bu}
\item $T=s$. We have $\JGT\Dtx= \oSN\Ds= \JGT\Dtxp$.

  
\item $T=E(\vt)$. We have $\JGT\Dtx= I\DE(\va)$ and $\JGT\Dtxp=
I\DE(\va')$ with $a_i= a_i'= \D\TH t_i\t$ if $x_i \in \Xs$, and $a_i=
(\D\TH t_i\t, S_i)$, $a_i'= (\D\TH t_i\t, S_i')$, $S_i= \JG{t_i}\Dtx$
and $S_i'= \JG{t_i}\Dtxp$ if $x_i \in \XB$. Since $T$ satisfies (I3),
(I4) and (I5), we have $E\in \FC$ and $E \le_\cC D$. Therefore, $I\DE$
is monotone in its inductive arguments. Let $i \le \at_E$. We show
that $t_i$ satisfies (I3), (I4) and (I5)~:

\begin{lst}{}
\item [\bf(I3)] Let $D'=_\cC D$. We have to show that $\pos(D',t_i)
\sle \pp(t_i)$. If $\pos(D',t_i)= \vide$, this is immediate. If there
exists $p\in \pos(D',t_i)$ then $i.p \in \pos(D',T)$. Since
$\pos(D',T) \sle \pp(T)$ and $\pp(T)= \{\vep\} \cup \bigcup\,
\{i.\pp(t_i) ~|~ i\in\ind(E)\}$, we have $i\in \ind(E)$ and $p\in
\pp(t_i)$.
  
\item [{\bf(I4)} and {\bf(I5)}] Let $D' >_\cC D$ or $D' \in \FD$. We
have to show that $\pos(D',t_i) \sle \pz(t_i)$. If $\pos(D',t_i)=
\vide$, this is immediate. If there exists $p\in \pos(D',t_i)$ then
$i.p \in \pos(D',T)$. Since $\pos(D',T)$ $\sle \pz(T)$ and $\pz(T)=
\{\vep\} \cup \bigcup\, \{i.\pz(t_i) ~|~ i\in\ind(E)\}$, we have $i\in
\ind(E)$ and $p\in \pz(t_i)$.
\end{lst}

Let us see now the relations between $S_i$ and $S_i'$. If
$\pos(Y,t_i)= \vide$ then, by induction hypothesis, $S_i= S_i'$.  If
there exists $p\in \pos(Y,t_i)$ then $i.p\in \pos(Y,T)$. Since
$\pos(Y,T) \sle \pp(T)$ and $\pp(T)= \{\vep\} \cup \bigcup\,
\{i.\pp(t_i) ~|~ i\in\ind(E)\}$, we necessary have $i\in \ind(E)$ and
$p\in \pp(t_i)$. Therefore, by induction hypothesis, $S_i \le
S_i'$. Finally, since $I\DE$ is monotone in its inductive arguments,
we can conclude that $\JGT\Dtx \le \JGT\Dtxp$.


\item $T=Y$. We have $\JGT\Dtx= Y\xi \le Y\xi'= \JGT\Dtxp$. But
$\pos(Y,Y)= \vep \sle \pp(Y)$.

  
\item $T=X\neq Y$. We have $\JGT\Dtx= X\xi= X\xi'= \JGT\Dtxp$.

\item $T=\px{U}V$. \hfill Let $\G'= \GxU$. We have $\JGT\Dtx= \{\D'\TH
t \in \oT\DxUVt ~|~\\ \all \D''\TH u \in \JGU\Dptx$, $\D''\TH tu \in
\JGpV\Dpptpx\}$ and $\JGT\Dtxp= \{\D'\TH t \in \oT\DxUVt ~|~ \all
\D''\TH u \in \JGU\Dptxp$, $\D''\TH tu \in \JGpV\Dpptpxp\}$ where
$\t'= \t\cup\xu$. We have to show that $\JGT\Dtx \sle \JGT\Dtxp$.  Let
$\D'\TH t \in \oT\DxUVt$ and $\D''\TH u \in \JGU\Dptxp$. Since
$\pos(Y,T) \sle \pp(T)$ and $\pp(T)= 1.\pm(U) \cup 2.\pp(V)$, we have
$\pos(Y,U) \sle \pm(U)$ and $\pos(Y,V) \sle \pp(V)$.  We prove that
$U$ satisfies (I3$^-$), (I4) and (I5)~:

\begin{lst}{}
\item [\bf(I3$^-$)] Let $D'=_\cC D$. We have to show that $\pos(D',U)
\sle \pm(U)$. If $\pos(D',U)$ $=\vide$, this is immediate.  If there
exists $p\in \pos(D',U)$ then $1.p \in \pos(D',T)$. Since $\pos(D',T)
\sle \pp(T)$ and $\pp(T)= 1.\pm(U) \cup 2.\pp(V)$, we have $p\in
\pm(U)$.
  
\item [{\bf(I4)} and {\bf(I5)}] Let $D'=_\cC D$ or $D' \in \FD$. We
have to show that $\pos(D',U) \sle \pz(U)$. If $\pos(D',U)= \vide$,
this is immediate. If there exists $p\in \pos(D',U)$ then $1.p \in
\pos(D',T)$. Since $\pos(D',T)$ $\sle \pz(T)$ and $\pz(T)= 1.\pz(U)
\cup 2.\pz(V)$, we have $p\in \pz(U)$.
\end{lst}

Similarly, $V$ satisfies (I3), (I4) and (I5). Therefore, by induction
hypothesis, $\JGU\Dptxp \sle \JGU\Dptx$ and $\JGpV\Dpptpx \sle
\JGpV\Dpptpxp$. Hence, $\D''\TH u \in \JGU\Dptx$, $\D''\TH tu \in
\JGpV\Dpptpx$ and $\D''\TH tu \in \JGpV\Dpptpxp$. Therefore, $\D'\TH t
\in \JGT\Dtxp$ and $\JGT\Dtx \sle \JGT\Dtxp$.


\item $T=\pX{K}V$. Similar to the previous case.

\item $T=\lx{U}V$. $\JGT\Dtx$ is the function which, to $\D'\TH u \in
\oT\DUt$, associates $\JGpV\Dptpx$ where $\G'= \GxU$ and $\t'=
\t\cup\xu$.  $\JGT\Dtxp$ is the function which, to $\D'\TH u \in
\oT\DUt$, associates $\JGpV\Dptpxp$ where $\G'= \GxU$ and $\t'=
\t\cup\xu$. We have to show that, for all $\D'\TH u \in \oT\DUt$,
$\JGpV\Dptpx \le \JGpV\Dptpxp$. Since $\pos(Y,T) \sle \pp(T)$ and
$\pp(T)= 1.\pos(U) \cup 2.\pp(V)$, we have $\pos(Y,V) \sle \pp(V)$. We
now prove that $V$ satisfies (I3), (I4) and (I5).

\begin{lst}{}
\item [\bf(I3)] Let $D'=_\cC D$. We have to show that $\pos(D',V) \sle
\pp(V)$. If $\pos(D',V)= \vide$, this is immediate. If there exists
$p\in \pos(D',V)$ then $2.p \in \pos(D',T)$. Since $\pos(D',T) \sle
\pp(T)$ and $\pp(T)= 1.\pos(U) \cup 2.\pp(V)$, we have $p\in \pp(V)$.
  
\item [{\bf(I4)} and {\bf(I5)}] Let $D'=_\cC D$ or $D' \in \FD$. We
have to show that $\pos(D',V) \sle \pz(V)$. If $\pos(D',V)= \vide$,
this is immediate. If there exists $p\in \pos(D',V)$ then $2.p \in
\pos(D',T)$. Since $\pos(D',T)$ $\sle \pz(T)$ and $\pz(T)= 1.\pos(U)
\cup 2.\pz(V)$, we have $p\in \pz(V)$.
\end{lst}

Therefore, by induction hypothesis, $\JGpV\Dptpx \le \JGpV\Dptpxp$ and
$\JGT\Dtx \le \JGT\Dtxp$.


\item $T=\lX{K}V$. Similar to the previous case.

\item $T=Vu$. We have $\JGT\Dtx= \JGV\Dtx(\G\TH u)$ and $\JGT\Dtxp=
\JGV\Dtxp(\G\TH u)$. Since $\pos(Y,T) \sle \pp(T)$ and $\pp(T)=
1.\pp(V) \cup 2.\pos(u)$, we have $\pos(Y,V) \sle \pp(V)$.  We now
prove that $V$ satisfies (I3), (I4) and (I5).

\begin{lst}{}
\item [\bf(I3)] Let $D'=_\cC D$. We have to show that $\pos(D',V) \sle
\pp(V)$. If $\pos(D',V)= \vide$, this is immediate. If there exists
$p\in \pos(D',V)$ then $1.p \in \pos(D',T)$. Since $\pos(D',T) \sle
\pp(T)$ and $\pp(T)= 1.\pp(V) \cup 2.\pos(u)$, we have $p\in \pp(V)$.
  
\item [{\bf(I4)} and {\bf(I5)}] Let $D'=_\cC D$ or $D'\in\FD$. We have
to prove that $\pos(D',V) \sle \pz(V)$. If $\pos(D',V)= \vide$, this
is immediate. If there exists $p\in \pos(D',V)$ then $1.p \in
\pos(D',T)$. Since $\pos(D',T)$ $\sle \pz(T)$ and $\pz(T)= 1.\pz(V)
\cup 2.\pos(u)$, we have $p\in \pz(V)$.
\end{lst}

Therefore, by induction hypothesis, $\JGV\Dtx \le \JGV\Dtxp$ and
$\JGT\Dtx \le \JGT\Dtxp$.


\item $T=VU$. We have $\JGT\Dtx= \JGV\Dtx(\G\TH U, \JGU\Dtx)$ and
$\JGT\Dtxp= \JGV\Dtxp(\G\TH U, \JGU\Dtxp)$. Since $\pos(Y,T) \sle
\pp(T)$ and $\pp(T)= 1.\pp(V)$, we have $\pos(Y,V) \sle \pp(V)$ and
$\pos(Y,U)= \vide$. We have seen in the previous case that $V$
satisfies (I3), (I4) and (I5). We now show that $U$ satisfies (I3),
(I3$^-$), (I4) and (I5).

\begin{lst}{}
\item [{\bf(I3)} and {\bf(I3$^-$)}] Let $D'=_\cC D$. We have to prove
that $\pos(D',U) \sle \pp(U) \cup \pm(U)$. If there exists $p\in
\pos(D',U)$ then $2.p \in \pos(D',T)$. Since $\pos(D',T)$ $\sle
\pp(T)$ and $\pp(T)= 1.\pp(V)$, this is not possible. Therefore,\\
$\pos(D',U)$ $=\vide \sle \pp(U) \cup \pm(U)$.
  
\item [{\bf(I4)} and {\bf(I5)}] Let $D'=_\cC D$ or $D' \in \FD$. We
have to prove that $\pos(D',V) \sle \pz(V)$. If there exists $p\in
\pos(D',U)$ then $2.p \in \pos(D',T)$. Since $\pos(D',T)$ $\sle
\pz(T)$ and $\pz(T)= 1.\pz(V)$, this is not possible. Therefore,
$\pos(D',U)$ $=\vide \sle \pz(U)$.
\end{lst}

Therefore, by induction hypothesis, $\JGV\Dtx \le \JGV\Dtxp$,
$\JGU\Dtx= \JGU\Dtxp$ and $\JGT\Dtx \le \JGT\Dtxp$.
\end{lst}
\end{prf}


\begin{lem}
\label{lem-int-const-mon}
$\vp$ is monotone.
\end{lem}

\begin{prf}
  Let $I,I'\in\cI^m$ such that $I\le I'$. We have to prove that, for
  all $D=_\cF C$, $\D\in\E$ and $\va$, $\vp^I\DD(\va) \le
  \vp^{I'}\DD(\va)$. Let $\D'\TH u \in \vp^I\DD(\va)$. We prove that
  $\D'\TH u \in \vp^{I'}\DD(\va)$. We have $\D'\TH u \in \oSN_{\D_n\th
  D(\vt)}$. Assume now that $u$ reduces to a term of the form $d(\vu)$
  with $d$ a constructor of type $\pvyU D(\vv)$. Let $j\in
  \acc(d)$. We have to prove that $\D'\TH u_j \in
  \J{\vy:\vU}{U_j}\Dptx[I']$ where $\t= \vyu$ and, for all $Y\in
  \FVB(U_j)$, $Y\xi= S_{\io_Y}$.
  
  We have $\D'\TH u_j \in \J{\vy:\vU}{U_j}\Dptx$ and $U_j$ satisfies
  {\bf(I3)}, {\bf(I4)} and {\bf(I5)}.
  \index{I3@\textbf{I3}|indlem[lem-int-const-mon]}
  \index{I4@\textbf{I4}|indlem[lem-int-const-mon]}
  \index{I5@\textbf{I5}|indlem[lem-int-const-mon]} We then prove by
  induction on $T$ that, for all $\G\TH T \in \oTY$, $\D\in \E$, $\t:
  \G\a\D$, $\xi$ compatible with $(\t,\G,\D)$~:

\begin{lst}{--}
\item if $T$ satisfies (I3), (I4) and (I5) then $\JGT\Dtx \le
\JGT\Dtx[I']$,
\item if $T$ satisfies (I3$^-$), (I4) and (I5) then $\JGT\Dtx \ge
\JGT\Dtx[I']$,
\end{lst}

\noindent
where (I3$^-$) is the property $\all D\IN \FC, D=_\cC C \A \pos(D,T)
\sle \pm(T)$. We just detail the first case; the second case is
similar.


\begin{lst}{\bu}
\item $T=s$. We have $\JGT\Dtx= \oSN\Ds= \JGT\Dtx[I']$.

  
\item $T=E(\vt)$. We have $\JGT\Dtx= I\DE(\va)$ and $\JGT\Dtx[I']=
I'\DE(\va')$ with $a_i= a_i'= \D\TH t_i\t$ if $x_i \in \Xs$, and $a_i=
(\D\TH t_i\t, S_i)$, $a_i'= (\D\TH t_i\t, S_i')$, $S_i= \JG{t_i}\Dtx$
and $S_i'= \JG{t_i}\Dtx[I']$ if $x_i \in \XB$. Since $T$ satisfies
(I3), (I4) and (I5), we have $E\in \FC$ and $E \le_\cC D$. Therefore,
$I\DE$ and $I'\DE$ are monotone in their inductive arguments. We have
seen in the previous lemma that $t_i$ satisfies (I3), (I4) and
(I5). Therefore, by induction hypothesis, $S_i \le S_i'$. Finally,
since $I\DE$ is monotone in its inductive arguments and $I \le I'$, we
can conclude that $I\DE(\va) \le I\DE(\va') \le I'\DE(\va')$ and
therefore that $\JGT\Dtx \le \JGT\Dtxp$.


\item $T=X$. We have $\JGT\Dtx= Y\xi= \JGT\Dtx[I']$.

\item $T=\px{U}V$. \hfill Let $\G'= \GxU$. We have $\JGT\Dtx= \{\D'\TH
t \in \oT\DxUVt ~|~\\ \all \D''\TH u \in \JGU\Dptx$, $\D''\TH tu \in
\JGpV\Dpptpx\}$ and $\JGT\Dtx[I']= \{\D'\TH t \in \oT\DxUVt ~|~ \all
\D''\TH u \in \JGU\Dptx[I']$, $\D''\TH tu \in \JGpV\Dpptpx[I']\}$
where $\t'= \t\cup\xu$. We have to prove that $\JGT\Dtx \sle
\JGT\Dtx[I']$. Let $\D'\TH t \in \oT\DxUVt$ and $\D''\TH u \in
\JGU\Dptxp$. We have seen in the previous lemma that $U$ satisfies
(I3$^-$), (I4) and (I5) and that $V$ satisfies (I3), (I4) and
(I5). Therefore, by induction hypothesis, $\JGU\Dptx[I'] \sle
\JGU\Dptx$ and $\JGpV\Dpptpx \sle \JGpV\Dpptpx[I']$. Hence, $\D''\TH u
\in \JGU\Dptx$, $\D''\TH tu \in \JGpV\Dpptpx$ and $\D''\TH tu \in
\JGpV\Dpptpx[I']$. Therefore, $\D'\TH t \in \JGT\Dtx[I']$ and
$\JGT\Dtx \sle \JGT\Dtx[I']$.


\item $T=\pX{K}V$. Similar to the previous case.

\item $T=\lx{U}V$. $\JGT\Dtx$ is the function which, to $\D'\TH u \in
\oT\DUt$, associates $\JGpV\Dptpx$ where $\G'= \GxU$ and $\t'=
\t\cup\xu$.  $\JGT\Dtx[I']$ is the function which, to $\D'\TH u \in
\oT\DUt$, associates $\JGpV\Dptpx[I']$ where $\G'= \GxU$ and $\t'=
\t\cup\xu$. We have to show that, for all $\D'\TH u \in \oT\DUt$,
$\JGpV\Dptpx \le \JGpV\Dptpx[I']$. We have seen in the previous lemma
that $V$ satisfies (I3), (I4) and (I5). Therefore, by induction
hypothesis, $\JGpV\Dptpx \le \JGpV\Dptpx[I']$ and $\JGT\Dtx \le
\JGT\Dtx[I']$.


\item $T=\lX{K}V$. Similar to the previous case.

\item $T=Vu$. We have $\JGT\Dtx= \JGV\Dtx(\G\TH u)$ and $\JGT\Dtxp=
\JGV\Dtxp(\G\TH u)$. We have seen in the previous lemma that $V$
satisfies (I3), (I4) and (I5). Therefore, by induction hypothesis,
$\JGV\Dtx \le \JGV\Dtx[I']$ and $\JGT\Dtx \le \JGT\Dtx[I']$.

  
\item $T=VU$. We have $\JGT\Dtx= \JGV\Dtx(\G\TH U, \JGU\Dtx)$ and
$\JGT\Dtxp= \JGV\Dtxp(\G\TH U, \JGU\Dtxp)$. we have seen in the
previous lemma that $U$ satisfies (I3), (I3$^-$), (I4) and (I5), and
$V$ satisfies (I3), (I4) and (I5). Therefore, by induction hypothesis,
$\JGV\Dtx \le \JGV\Dtx[i']$, $\JGU\Dtx= \JGU\Dtx[I']$ and $\JGT\Dtx
\le \JGT\Dtx[I']$.\cqfd
\end{lst}
\end{prf}


~

Since $(\cI^m,\le)$ is a complete lattice, $\vp$ as a least fix point
$I$ which is an interpretation for all the constant predicate symbols
equivalent to $C$. Hence, by induction on $>_\cC$, we obtain an
interpretation $I$ for the constant predicate symbols.


In the case of a primitive constant predicate symbol, the
interpretation is simply the set of strongly normalizable terms of
this type~:

\begin{lem}[Interpretation of primitive constant predicate symbols]
\label{lem-int-const-prim}
\index{interpretation!constant!primitive|indlem[lem-int-const-prim]}

\hfill\\ If $C$ is a primitive constant predicate symbol then $I\DC=
\top\DC$.
\end{lem}

\begin{prf}
  Since $I\DC \le \top\DC$, it suffices to prove that, for all $u \in
  \SN$, $C$ primitive of type $\pvxT\st$, $\va$ arguments of $I\DC$
  with $a_i= \D_i\TH t_i$ if $x_i\in\Xs$ and $a_i= (\D_i\TH t_i, S_i)$
  if $x_i\in\XB$, and $\D'\sge\D_n$, if $\D'\th u:C(\vt)$ then $\D'\TH
  u \in I\DC(\va)$, by induction on $u$ with $\a\cup\,\tgt$ as
  well-founded ordering. Assume that $u\a^* c(\vu)$ with $c$ a
  constructor of type $\pvyU C(\vv)$. If $u\a^+ c(\vu)$, we can
  conclude by induction hypothesis. So, assume that $u= c(\vu)$. In
  this case, we have to prove that, for all $j\in\acc(c)$, $\D'\TH u_j
  \in \J{\vy:\vU}{U_j}\Dptx$ where $\t=\vyu$ and, for all
  $Y\in\FVB(U_j)$, $Y\xi= S_{\iota_Y}$. By definition of primitive
  constant predicate symbols, for all $j\in\acc(c)$, $U_j$ is of the
  form $D(\vw)$ with $D$ a primitive constant predicate symbol. Assume
  that $\tD= (\vz:\vV)\st$. Let $a_i'= \D'\TH w_i\t$ if $z_i\in\Xs$,
  and $a_i'= (\D'\TH w_i\t, \J{\vy:\vU}{w_i}\Dptxp)$ if $z_i\in\XB$.
  Since $u_j\in\SN$ and $\D'\th u_j:D(\vw\t)$, by induction
  hypothesis, $\D'\TH u_j \in I\DD(\va')$. By {\bf(P3)},
  \index{P3@\textbf{P3}|indlem[lem-int-const-prim]} $I\DD(\va')=
  I\DpD(\va')$ and $ \J{\vy:\vU}{U_j}\Dptx= I\DpD(\va')$. Therefore,
  $\D'\TH u \in I\DC(\va)$.\cqfd
\end{prf}


%% file: ordering.tex



\section{Reductibility ordering}
\label{sec-red-ord}

In this subsection, we assume given an interpretation $J$ for the
defined predicate symbols and we denote $\JGT\Dtx[I\cup J]$ by
$\JGT\Dtx[]$.

\renewcommand{\Dsx}{_{\D,\s,\xi}}


~

The fix point of the function $\vp$ defined in the previous subsection
can be reached by transfinite iteration from the smallest element of
$\cI^m$. Let $\vp^\fa$ be the interpretation after $\fa$ iterations.

\begin{dfn}[Order of a computable term]
\label{def-ord}
\index{term!order|inddef[def-ord]}
\index{order|inddef[def-ord]}
\index{omicronGammat@$o(\G\TH t)$|inddef[def-ord]}

  The {\em order\,} of $\D'\TH t \in I\DC(\va)$, $o(\D'\TH t)$, is the
  smallest ordinal $\fa$ such that $\D'\TH t \in \vp^\fa\DC(\va)$.
\end{dfn}


This notion of order will enable us to define a well-founded ordering
in which recursive definitions on strictly positive predicates
strictly decrease. Indeed, in this case, the subterm ordering is not
sufficient. In the example of the addition on ordinals, we have the
rule~:

\begin{rewc}
+(x,lim(f)) & lim(\l{n}{nat}+(x,fn))\\
\end{rewc}

We have a recursive call with $fn$ as argument, which is not a subterm
of $lim(f)$. However, thanks to the definition of the interpretation
for constant predicate symbols and products, we can say that, if
$lim(f)$ is computable then $f$ is computable and, for all computable
$n$, $fn$ is computable. So, the order of $lim(f)$ is greater than the
one of $fn$~: $o(lim(f)) > o(fn)$.


\begin{dfn}[Reductibility ordering]
\label{def-red-ord}
\index{reductibility!ordering|inddef[def-red-ord]}
\index{$\qgt,\qgt_f$|inddef[def-red-ord]}
\index{ordering!$\qgt,\qgt_f$|inddef[def-red-ord]}
\index{ordering!reductibility|inddef[def-red-ord]}

We assume given a precedence $\ge_\cF$ on $\cF$ and a status
assignment $stat$ compatible with $\ge_\cF$. Let $f$ be a symbol of
non null arity and status $stat_f= lex(m_1,\ldots,m_k)$. Let
$\Theta_f$ be the set of $(g,\D,\t,\xi)$ such that, if $\tg= \pvxT U$
and $\G_g= \vx:\vT$ then $g =_\cF f$, $\D\in\E$, $\t: \G_g\a\D$, $\xi$
is compatible with $(\t,\G_g,\D)$, $(\t,\G_g,\D)$ is valid
w.r.t. $\xi$. We equip $\Theta_f$ with the ordering $\qgt_f$ defined
by~:

\begin{lst}{\bu}
\item $(g,\D,\t,\xi) \,\qgt_f (g',\D',\t',\xi')$ ~if~ $\vm\t
  ~(\qgt_f^1,\ldots,\qgt_f^k)\lex~ \vm\t'$,
\item $mul(\vu) \,\qgt_f^i mul(\vu')$ ~if~ $\{\vu\} ~(\qgt^i)\mul~
  \{\vu'\}$ with~:
\begin{lst}{--}
\item $\qgt^i=>\!\!>$ if $i\in SP(f)$ where $u >\!\!> u'$ if $o(\D\TH
u) > o(\D'\TH u')$,
\item $\qgt^i=\a\cup\tgt$ otherwise.
\end{lst}
\end{lst}

\noindent
We equip $\Theta= \bigcup\, \{\Theta_f ~|~f\in\cF\}$ with the {\em
  reductibility ordering\,} $\qgt$ defined by $(g,\D,\t,\xi) \,\qgt
  (g',\D',\t',\xi')$ if $g >_\cF g'$ or, $g =_\cF g'$ and
  $(g,\D,\t,\xi) \,\qgt_f (g',\D',\t',\xi')$.
\end{dfn}


\begin{lem}
\label{lem-red-ord-wf}
\index{reductibility!ordering|indlem[lem-red-ord-wf]}

The reductibility ordering is well-founded and compatible with $\a$~:
if $\t\a\t'$ then $(g,\D,\t,\xi) \qge (g,\D,\t',\xi)$.
\end{lem}

\begin{prf}
  The reductibility ordering is well-founded since the ordinals are
  well-founded and the lexicographic and multiset extensions preserve
  the well-foundedness. It is compatible with $\a$ by definition of
  the interpretation for constant predicate symbols.\cqfd
\end{prf}

~

We check hereafter that the accessibility relation is correct~: an
accessible subterm of a computable term is computable. Then we check
that the ordering on arguments if also correct~: if $l_i >_2 u_j$ and
$l_i$ is computable then $u_j$ is computable and has an order smaller
than the one of $l_i$.


\begin{lem}[Correctness of accessibility]
\label{lem-cor-acc}
\index{accessibility|indlem[lem-cor-acc]}
\index{reductibility!accessibility|indlem[lem-cor-acc]}

If $t:T \tgt_1^\r u:U$, $\G\th t\r:T\r$, $\s: \G\a\D$ and $\D\TH t\s
\in \JG{T\r}\Dsx$ then $\G\th u\r:U\r$ and $\D\TH u\s \in
\JG{U\r}\Dsx$.
\end{lem}

\begin{prf}
  By definition of $\tgt_1^\r$, we have $t$ of the form $c(\vu)$ with
  $c$ a constructor of type $\pvyU C(\vv)$, $T\r= C(\vv)\g\r$ where
  $\g=\vyu$, $u=u_j$ with $j\in \acc(c)$ and $U\r=U_j\g\r$. Assume
  that $\tC= \pvxT\st$. Then, $\JG{C(\vv)\g\r}\Dsx= I\DC(\va)$ with
  $a_i= \D\TH v_i\g\r\s$ if $x_i\in\Xs$, and $a_i= (\D\TH
  v_i\g\r\s,S_i)$ where $S_i= \JG{v_i\g\r}\Dsx$ if $x_i\in\XB$. By
  definition of $I\DC$, $\D\TH u_j\s \in
  \J{\G_c}{U_j}_{\D,\g\r\s,\xi'}$ with, for all $Y\in \FVB(U_j)$,
  $Y\xi'= S_{\io_Y}= \JG{v_{\io_Y}\g\r}\Dsx$. By {\bf(I6)},
  \index{I6@\textbf{I6}|indlem[lem-cor-acc]} $v_{\io_Y}=Y$ and $Y\xi'=
  \JG{Y\g\r}\Dsx$. From $\G\th t\r:T\r$, by inversion, we deduce that,
  for all $i$, $\G\th u_i\r:U_i\g\r$. Therefore, $\g\r:
  \G_c\a\G$. Hence, by candidates substitution,
  $\J{\G_c}{U_j}_{\D,\g\r\s,\xi'}= \JG{U_j\g\r}\Dsx$ and $\D\TH u\s
  \in \JG{U\r}\Dsx$.\cqfd
\end{prf}


\begin{lem}[Correctness of the ordering on arguments]
\label{lem-cor-arg-ord}
\index{ordering!arguments|indlem[lem-cor-arg-ord]}
\index{accessibility|indlem[lem-cor-arg-ord]}

Assume that $t:T$ $>_2 u:U$, that is, that $t$ is of the form $c(\vt)$
with $c$ a constructor of type $\pvxT C(\vv)$, $u$ of the form $x\vu$
with $x\in\dom(\G_0)$, $x\G_0$ of the form $\pvyU D(\vw)$, $D=_\cC C$
and $t:T ~(\tgt_2^\r)^+~ x:V$ with $V\r= x\G_0$.

Let $\t= \vyu$. If $\G\th t\r:T\r$, $\s: \G\a\D$, $\D\TH t\s \in
\JG{T\r}\Dsx$ and, for all $i$, $\D\TH u_i\s \in \JG{U_i\t}\Dsx$, then
$\D\TH x\s \vu\s \in \JG{D(\vw)\t}\Dsx$ and $o(\D\TH t\s) > o(\D\TH
x\s \vu\s)$.
\end{lem}

\begin{prf}
  Let $p$ be the path from $t$ to $x$ followed in $t:T ~(\tgt_2^\r)^+~
  x:V$. We show that if $p=j_1\ldots j_n j_{n+1}$ then
  $c(\vt):C(\vv)\g= c_0(\vt^0):C_0(\vv^0)\g_0 \,\tgt_2^\r\,
  c_1(\vt^1):C_1(\vv^1)\g_1 \,\tgt_2^\r\, \ldots \,\tgt_2^\r\,
  c_n(\vt^n):C_n(\vv^n)\g_n \,\tgt_2^\r\, x:V$ with, if $\tau_{c_i}=
  (\vx^i:\vT^i)C(\vv^i)$ and $\g_i= \{\vx^i\to \vt^i\}$~:

\begin{lst}{--}
\item for all $i<n$, $c_{i+1}(\vt^{i+1})= t^i_{j_{i+1}}$,
  $T^i_{j_{i+1}}= C_{i+1}(\vw^{i+1})$ and
  $\vw^{i+1}\g_i\r=\vv^{i+1}\g_{i+1}\r$,
\item $T^n_{j_{n+1}}= \pvyU D(\vw')$ and $\vw'\g_n= \vw$.
\end{lst}

We proceed by induction on $n$. If $n=0$, this is immediate.  So,
assume that $c(\vt):C(\vv)\g \,\tgt_2^\r\, t_{j_1}:T_{j_1}\g_0
~(\tgt_2^\r)^+~ x:V$. Since $t_{j_1}:T_{j_1}\g_0 ~(\tgt_2^\r)^+~ x:V$,
$t_{j_1}= c_1(\vt^1)$ with $\tau_{c_1}= (\vx^1:\vT^1)C_1(\vv^1)$ and
$T_{j_1}\g_0\r= C_1(\vv^1)\g_1\r$ where $\g_1=\{\vx^1\to\vt^1\}$. By
definition of $\tgt_2^\r$, $T_{j_1}$ is of the form $\pvzV
C_1'(\vw^1)$. Therefore, $|\vz|=0$, $C_1'=C_1$ and $\vw^1\g_0\r=
\vv^1\g_1\r$. Hence, we can conclude by induction hypothesis on
$t_{j_1}:T_{j_1}\g ~(\tgt_2^\r)^+~ x:V$.

Since we are in an admissible inductive structure {\bf(A2)},
\index{A2@\textbf{A2}|indlem[lem-cor-arg-ord]} we can say that $C
\ge_\cF C_1 \ge_\cF \ldots \ge_\cF C_n \ge D$. Since $D=_\cF C$ and
$>_\cF$ is well-founded, we get $C=_\cF C_1=_\cF \ldots=_\cF D$.

Let $w_{n+1}= t|_p\vu$, $W_{n+1}= D(\vw)\t$ and, for all $i\le n$,
$w_i= t|_{j_1\ldots j_i}$ and $W_i= C_i(\vv^i)\g_i$. We show by
induction on $n$ that, for all $i\le n+1$, $\D\TH w_i\s \in
\JG{W_i}\Dsx$ and, for all $i\le n$, $o(\D\TH w_i\s) > o(\D\TH
w_{i+1}\s)$.

\begin{lst}{\bu}
\item Case $n>0$. $c(\vt):C(\vv)\g \,\tgt_2^\r\, t_{j_1}:T_{j_1}\g
~(\tgt_2^\r)^+~ x:V$. By hypothesis, we have $\G\th
c(\vt)\r:C(\vv)\g\r$ and $\D\TH c(\vt)\s \in \JG{C(\vv)\g\r}\Dsx$. By
correctness of the accessibility relation, $\G\th
t_{j_1}\r:T_{j_1}\g\r$ and $\D\TH t_{j_1}\s \in
\JG{T_{j_1}\g\r}\Dsx$. Since $T_{j_1}\g\r= C_1(\vv^1)\g_1\r$ and
$C_1=_\cF C$, $o(\D\TH t\s) > o(\D\TH t_{j_1}\s)$. Hence, we can
conclude by induction hypothesis on $t_{j_1}:T_{j_1}\g$.
  
\item Case $n=0$. $c(\vt):C(\vv)\g \,\tgt_2^\r\, x:V$. Like in the
case $n>0$, we have $\D\TH x\s \in \JG{V\r}\Dsx$. Moreover, we have
$V\r= \pvyU D(\vw)$. Let $m=|\vu|$ and $\t_k= \{y_1\to
u_1,\ldots,y_k\to u_k\}$. We prove by induction on $k\le m$ that
$\D\TH x\s\, u_1\s \ldots u_k\s \in \JG{\p{y_{k+1}}{U_{k+1}\t_k}
\ldots \p{y_m}{U_m\t_k} D(\vw)\t_k}\Dsx$ and therefore that $o(\D\TH
t\s) > o(\D\TH t'\s)$.
  
  \hs If $k=0$, this is immediate. So, assume that $k>0$. By induction
  hypothesis, $\D\TH x\s\, u_1\s \ldots$ $u_{k-1}\s \in
  \JG{\p{y_k}{U_k\t_{k-1}} \ldots \p{y_m}{U_m\t_{k-1}}
  D(\vw)\t_{k-1}}\Dsx$. Since $\D\TH u_k\s \in \JG{U_k\t_{k-1}}\Dsx$,
  we have $\D\TH x\s\, u_1\s \ldots u_k\s \in
  \J{\G,y_k\!:\!U_k\t_{k-1}}{\p{y_{k+1}}{U_{k+1}\t_{k-1}} \ldots
  \p{y_m}{U_m\t_{k-1}} D(\vw)\t_{k-1}}_{\D,\s',\xi'}$ where $\s'=
  \s\cup\{y_k\to u_k\s\}$, $\xi'=\xi$ if $y_k\in\Xs$ and
  $\xi'=\xi\cup\{y_k\to \JG{u_k}\Dsx\}$ if $y_k\in\XB$. Hence, by
  candidates substitution,
  $\J{\G,y_k\!:\!U_k\t_{k-1}}{\p{y_{k+1}}{U_{k+1}\t_{k-1}} \ldots
  \p{y_m}{U_m\t_{k-1}} D(\vw)\t_{k-1}}_{\D,\s',\xi'}=
  \JG{\p{y_{k+1}}{U_{k+1}\t_k} \ldots \p{y_m}{U_m\t_k}
  D(\vw)\t_k}\Dsx$.\cqfd
\end{lst}
\end{prf}


%% file: int-def.tex



\section{Interpretation of defined predicate symbols}
\label{sec-int-def}

\renewcommand{\Dtx}[1][I]{_{\D,\t,\xi}^{#1}}

We define the interpretation $J$ \index{J@$J$|indsec[sec-int-def]} for
defined predicate symbols by induction on $\cgt$ w.r.t. hypothesis
{\bf(A3)}. \index{A3@\textbf{A3}|indsec[sec-int-def]}

Let $F$ \index{F@$F$|indsec[sec-int-def]} be a defined predicate
symbol and assume that we already have defined an interpretation $K$
\index{K@$K$|indsec[sec-int-def]} for all the symbols smaller than
$F$. Let $[F]$ \index{$[F]$|indsec[sec-int-def]} be the set of symbols
equivalents to $F$. We defined the interpretation for $[F]$ depending
on whether $[F]$ is primitive, positive or recursive.

To make the notations simpler, we will denote $\JGT\Dtx[I\cup K\cup
J]$ by $\JGT\Dtx[J]$. And for the arguments of the interpretation, we
follow the notations used in the previous section.




\subsection{Primitive systems}

\begin{dfn}
\label{def-int-def-prim}
\index{interpretation!defined!primitive|inddef[def-int-def-prim]}
\index{p@\textbf{p}|inddef[def-int-def-prim]}

  For every $G\in [F]$, we take $J\DG= \top\DG$.
\end{dfn}




\subsection{Positive, small and simple systems}
\label{sec-sys-pos}

Let $\cJ$ \index{J@$\cJ$|indsec[sec-sys-pos]} be the set of the
interpretations for $[F]$ and $\le$ \index{$\le$|indsec[sec-sys-pos]}
\index{ordering!$\le$|indsec[sec-sys-pos]} the relation on $\cJ$
defined by $J \le J'$ if, for all $G\in [F]$ and $\D\in\E$, $J\DG
\le\DG J'\DG$. Since $(\cR\DG,\le\DG)$ is a complete lattice, it is
easy to see that $(\cJ,\le)$ is also a complete lattice.


\begin{dfn}
\label{def-int-pos}
\index{interpretation!defined!positive|inddef[def-int-pos]}
\index{lfp@$\lfp$|inddef[def-int-pos]}
\index{q@\textbf{q}|inddef[def-int-pos]}
\index{psiJDeltaGa@$\psi^J\DG(\va)$|inddef[def-int-pos]}
\index{smallness|inddef[def-int-pos]}
\index{simplicity|inddef[def-int-pos]}

  Let $\psi$ be the function which, to $J\in \cJ$, associates the
  interpretation $\psi^J$ defined by~:

$\psi^J\DG(\va) = \left\{
\begin{array}{l}
\JG{r}_{\D_n,\s,\xi}^{J} ~~\mbox{if}~ \vt \in \WN \!\cap \CR,
\,\vt\!\ad = \vl\s \mbox{ and } (G(\vl) \a r, \G, \rho) \in \cR\\[2mm]
\top_{\D_n \th G(\vt)} ~~\mbox{otherwise}
\end{array}
\right.$

\vs[2mm]
\noindent
where, for all $X\in \FVB(r)$, $X\xi= S_{\k_X}\Dn$. Hereafter, we show
that $\psi$ is monotone. Therefore, we can take $J\DG= \lfp(\psi)\DG$.
\end{dfn}


\begin{lem}
\label{lem-cor-int-pos}
\index{interpretation!defined!positive|indlem[lem-cor-int-pos]}

$\psi^J$ is a well defined interpretation.
\end{lem}

\begin{prf}
  By simplicity, \index{simplicity|indlem[lem-cor-int-pos]} at most
  one rule can be applied at the top of $G(\vt\!\ad)$. The existence
  of $\k_X$ is the smallness
  hypothesis. \index{smallness|indlem[lem-cor-int-pos]} By {\bf(S4)},
  \index{S4@\textbf{S4}|indlem[lem-cor-int-pos]} if $\vt= \vl\s$ then
  $\s: \G\a\D_n$. Moreover, $\xi$ is compatible with $(\s,\G,\D_n)$
  since, for all $X\in \FVB(r)$, $X\xi= S_{\k_X}|_{\D_n} \in \cR_{\D_n
  \th X\s}$ since $S_{\k_X} \in \cR_{\D_{\k_X} \th t_{\k_X}}$ and, by
  smallness, $t_{\k_X}= l_{\k_X}\s= X\s$.
  
  We now show that $\psi^J$ is an interpretation for $[F]$. After
  Lemma~\ref{lem-cor-schema-int}~(b), $\JG{r}_{\D_n,\s,\xi}^{J} \in
  \cR_{\D_n\th r\s}= \cR_{\D_n\th F(\vt)}$. Moreover, $\top_{\D_n\th
  F(\vt)} \in \cR_{\D_n\th F(\vt)}$. Thus we are left with proving the
  properties (P1) to (P3).

\begin{bfenumi}{P}
\item Assume that $\vt\a \vt'$. By {\bf(A1)},
\index{A1@\textbf{A1}|indlem[lem-cor-int-pos]} $\a$ is
confluent. Therefore, $\{\vt\} \sle \WN$ iff $\{\vt'\} \sle \WN$, and
if $\{\vt\} \sle \WN$ then $\vt\!\ad= \vt'\!\ad$.  So,
$\psi^J\DG(\va)= \psi^J\DG(\va')$.
  
\item Assume that $\D_n\sle\D'\in\E$. After
Lemma~\ref{lem-cor-schema-int}~(d), $\JG{r}_{\D_n,\s,\xi}^{J}\Dp=
\JG{r}\Dpsx$. Moreover, $\top_{\D_n\th F(\vt)}\Dp= \top_{\D'\th
F(\vt)}$. Therefore, $\psi^J\DG(\va)\Dp= \psi^J\DpG(\va)$.  Now,
assume that $\D_k\sle\D_k'\in\E$. Let $a_k'= \D_k'\TH t_k$ if
$x_k\in\Xs$, and $a_k'= (\D_k'\TH t_k, S_k|_{\D_k'})$ if
$x_k\in\XB$. Then, $\psi^J\DG(a_1, \ldots, a_{k-1}, a_k)|_{\D_k'}$ and
$\psi^J\DG(a_1, \ldots, a_{k-1}, a_k')$ have the same domain and are
equal.
  
\item $\psi\DG\Dp$ and $\psi\DpG$ have the same domain and are
equal.\cqfd
\end{bfenumi}
\end{prf}


\begin{lem}
\label{lem-mon-int-pos}
\index{interpretation!defined!positive|indlem[lem-mon-int-pos]}

  $\psi$ is monotone.
\end{lem}

\begin{prf}
  As in Lemma~\ref{lem-int-const-mon}.\cqfd
\end{prf}




\subsection{Recursive, small and simple systems}


Let $G\in [F]$ and $\D\in\E$. To a sequence of arguments $\va$ for
$J\DG$, we associate the substitution $\t= \vxt: \G_G\a\D_n$ and the
assignment $\xi= \{\vx\to\vS|_{\D_n}\}$ compatible with
$(\t,\G_G,\D_n)$. Let $\cD$ \index{D@$\cD$|inddef[def-int-rec]} the
set of sequences $\va$ such that $(\t,\G_G,\D_n)$ is valid
w.r.t. $\xi$. We equip $\cD$ with the following well-founded
ordering~: $\va \qgt \va'$ if $(G,\D_n,\t,\xi) \qgt
(G,\D_n',\t',\xi')$.

\begin{dfn}
\label{def-int-rec}
\index{r@\textbf{r}|inddef[def-int-rec]}
\index{interpretation!defined!recursive|inddef[def-int-rec]}
\index{reductibility!ordering|inddef[def-int-rec]}
\index{ordering!reductibility|inddef[def-int-rec]}
\index{ordering!$\qgt$|inddef[def-int-rec]}
\index{JDeltaGa@$J\DG(\va)$|inddef[def-int-rec]}
\index{simplicity|inddef[def-int-rec]}
\index{smallness|inddef[def-int-rec]}

We define $J\DG(\va)$ by induction on $\qgt$~:

$J\DG(\va) = \left\{
\begin{array}{l}
\JG{r}_{\D_n,\s,\xi}^{J} ~~\mbox{if}~
\begin{array}{c}
\{\vt\} \sle \WN \!\cap \CR,\,
\vt\!\ad= \vl\s,\, \va\!\ad\, \in \cD\\
\mbox{and } (G(\vl) \a r, \G, \r) \in \cR\\
\end{array}\\
\top_{\D_n \th G(\vt)} ~~\mbox{otherwise}\\
\end{array}
\right.$

\vs[2mm]
\noindent
where, for all $X \in \FVB(r)$, $X\xi= S_{\k_X}\Dn$.
\end{dfn}


\begin{lem}
\label{lem-cor-int-rec}
\index{interpretation!defined!recursive|indlem[lem-cor-int-rec]}

  $J$ is a well defined interpretation.
\end{lem}

\begin{prf}
  By simplicity, \index{simplicity|indlem[lem-cor-int-rec]} at most
  one rule can be applied at the top of $G(\vt\!\ad)$. The existence
  of $\k_X$ is the smallness hypothesis.
  \index{smallness|indlem[lem-cor-int-rec]} By {\bf(S4)},
  \index{S4@\textbf{S4}|indlem[lem-cor-int-rec]} if $\vt= \vl\s$ then
  $\s: \G\a\D_n$. Moreover, $\xi$ is compatible with $(\s,\G,\D_n)$
  since, for all $X\in \FVB(r)$, $X\xi= S_{\k_X}|_{\D_n} \in \cR_{\D_n
  \th X\s}$ since $S_{\k_X} \in \cR_{\D_{\k_X} \th t_{\k_X}}$ and, by
  smallness, $t_{\k_X}= l_{\k_X}\s= X\s$.
  
  The well-foundedness of the definition comes from
  Lemma~\ref{lem-red-ho} and Theorem~\ref{thm-cor-thc}. In
  Lemma~\ref{lem-red-ho} for the reductibility of higher-order
  symbols, we show that, starting from a sequence $\va\in\cD$, it is
  possible to apply Theorem~\ref{thm-cor-thc} for the correctness of
  the computable closure. And in this theorem, we show that, in a
  recursive call $G'(\va')$ (case (symb$^=$)), we have $\va \qgt
  \va'$. Since $\qgt$ is compatible with $\a$, $\va \qgt \va'\!\ad$.
  
  Finally, to make sure that $J$ is an interpretation, one can proceed
  as in the case of a positive system.\cqfd
\end{prf}


%% file: end-sn.tex



\section{Correctness of the conditions}

\renewcommand{\Dtx}{_{\D,\t,\xi}}
\renewcommand{\Dsx}{_{\D,\s,\xi}}
\renewcommand{\Dspxp}{_{\D,\s',\xi'}}
\renewcommand{\Dsxp}{_{\D,\s,\xi'}}
\renewcommand{\Dtxp}{_{\D,\t,\xi'}}
\renewcommand{\Dstxp}{_{\D,\s\t,\xi'}}
\renewcommand{\Dspxpp}{_{\D,\s',\xi''}}
\renewcommand{\Dptx}{_{\D',\t,\xi\Dp}}
\renewcommand{\Dptpx}{_{\D',\t',\xi\Dp}}
\renewcommand{\Dpptpx}{_{\D'',\t',\xi\Dpp}}


\begin{dfn}[Cap and aliens]
\label{def-cap}
\index{cap@$cap,cap_\cG$|inddef[def-cap]}
\index{aliens@$aliens,aliens_\cG$|inddef[def-cap]}

  Let $\z$ be an injection from the classes of terms modulo $\aa^*$ to
  $\cX$. The {\em cap\,} of a term $t$ w.r.t. a set $\cG$ of symbols is
  the term $cap_\cG(t)= t[x_1]_{p_1} \ldots [x_n]_{p_n}$ such that,
  for all $i$~:

\begin{lst}{--}
\item $t|_{p_i}$ is not of the form $f(\vt)$ with $f \in \cG$,
\item $x_i= \z(t|_{p_i})$.
\end{lst}

\noindent
The $t|_{p_i}$'s are the {\em aliens\,} of $t$. We will denote by
$aliens_\cG(t)$ the multiset of the aliens of $t$.
\end{dfn}


\begin{lem}[Pre-reductibility of first-order symbols]
\label{lem-pre-red-fo}
\index{symbol!first-order|indlem[lem-pre-red-fo]}
\index{reductibility!symbol!first-order|indlem[lem-pre-red-fo]}
\index{cap@$cap,cap_\cG$|indlem[lem-pre-red-fo]}
\index{aliens@$aliens,aliens_\cG$|indlem[lem-pre-red-fo]}

  Let $f\in \cF_1$ and $\vt$ be terms such that $f(\vt)$ is
  typable. If the $t_i$'s are strongly normalizable then $f(\vt)$ is
  strongly normalizable.
\end{lem}

\begin{prf}
  We prove that every immediate reduct $t'$ of $t= f(\vt)$ is strongly
  normalizable. Hereafter, by $cap$, we mean the cap w.r.t. $\cF_1$.
  
  \u{\bf Case $\cR_\w \neq \vide$}. By induction on
  $(aliens(t),cap(t))\lex$ with $((\a\cup\,\tgt)\mul,\a_{\cR_1})\lex$
  as well-founded ordering (the aliens are strongly normalizable and,
  by {\bf(f)}, \index{f@\textbf{f}|indlem[lem-pre-red-fo]}
  $\a_{\cR_1}$ is strongly normalizing on $\T(\cF_1,\cX)$).
  
  If the reduction takes place in $cap(t)$ then this is a
  $\cR_1$-reduction. By {\bf(c)},
  \index{c@\textbf{c}|indlem[lem-pre-red-fo]} no symbol of $\cF_\w$
  occurs in the rules of $\cR_1$. Therefore, $cap(t) \a_{\cR_1}
  cap(t')$. By {\bf(d)}, \index{d@\textbf{d}|indlem[lem-pre-red-fo]}
  the right hand-sides of the rules of $\cR_1$ are algebraic and, by
  {\bf(e)}, \index{e@\textbf{e}|indlem[lem-pre-red-fo]} the rules of
  $\cR_1$ are non duplicating.
  \index{non-duplicating rewrite system|indlem[lem-pre-red-fo]}
  Therefore, $aliens(t) \tgt\mul aliens(t')$ and we can conclude by
  induction hypothesis.
  
  If the reduction takes place in an alien then $aliens(t)
  ~(\a\cup\,\tgt)\mul~ aliens(t')$ and we can conclude by induction
  hypothesis.
  
  \u{\bf Case $\cR_\w = \vide$}. Since the $t_i$'s are strongly
  normalizable and no $\b$-reduction can take place at the top of $t$,
  $t$ has a $\b$-normal form. Let $\b cap(t)$ be the cap of its
  $\b$-normal form. We prove that every immediate reduct $t'$ of $t$
  is strongly normalizable, by induction on $(\b
  cap(t),aliens(t))\lex$ with $(\a_{\cR_1},(\a\cup\,\tgt)\mul)\lex$ as
  well-founded ordering (the aliens are strongly normalizable and, by
  {\bf(f)}, $\a_{\cR_1}$ is strongly normalizing on $\T(\cF_1,\cX)$).
  
  If the reduction takes place in $cap(t)$ then this is a
  $\cR_1$-reduction. By {\bf(d)}, the right hand-sides of the rules of
  $\cR_1$ are algebraic. Therefore, $t'$ has a $\b$-normal form and
  $\b cap(t) \a_{\cR_1} \b cap(t')$. Hence, we can conclude by
  induction hypothesis.
  
  If the reduction is a $\b$-reduction in an alien then $\b cap(t)= \b
  cap(t')$ and $aliens(t)$ $(\a\cup\,\tgt)\mul~$ $aliens(t')$. Hence,
  we can conclude by induction hypothesis.
 
  We are left with the case where the reduction is a $\cR_1$-reduction
  talking place in an alien $u$. Then, $aliens(t) \a\mul aliens(t')$,
  $\b cap(t) \a_{\cR_1}^* \b cap(t')$ and we can conclude by induction
  hypothesis. To see that $\b cap(t) \a_{\cR_1}^* \b cap(t')$, it
  suffices to remark the following~: if we $\b$-normalize $u$, all the
  residuals of the $\cR_1$-redex are still reductible since, by
  {\bf(c)}, no symbol of $\cF_\w$ occurs in the rules of $\cR_1$.\cqfd
\end{prf}


\begin{thm}[Strong normalization of $\ar$]
\label{thm-sn-rew}
\index{normalization!strong!$\ar$|indthm[thm-sn-rew]}
\index{cap@$cap,cap_\cG$|indthm[thm-sn-rew]}
\index{aliens@$aliens,aliens_\cG$|indthm[thm-sn-rew]}

\hfill The relation $\a_{\cR_1}\cup \a_{\cR_\w}$ is\\ strongly
normalizing on typable terms.
\end{thm}

\begin{prf}
  By induction on the structure of terms. The only difficult case is
  $f(\vt)$. If $f$ is first-order, we use the Lemma of
  pre-reductibility of first-order symbols. If $f$ is higher-order, we
  have to show that if $\vt\in\SN$ and $f(\vt)\in \T$ then $t=
  f(\vt)\in\SN$ where $\SN$ means here the strong normalization
  w.r.t. $\ar= \a_{\cR_1} \cup \a_{\cR_\w}$.
  
  We prove that every immediate reduct $t'$ of $t$ is strongly
  normalizable by induction on $(f,\vpi(\vt),\vt,\vt)$ with
  $(>_\cF,(>_\mb{N})\sf,(\tgt\,\cup\ar)\sf,(\ar)\lex)\lex$ as
  well-founded ordering where $\vpi(t)=0$ if $t$ is not of the form
  $g(\vu)$ and $\vpi(t)=1$ otherwise. Assume that $t'= f(\vt')$ with
  $t_i\ar t_i'$ and, for all $j\neq i$, $t_j= t_j'$. Then, $\vt
  ~(\ar)\lex~ \vt'$ and $\vpi(t_i) \ge \vpi(t_i')$ since if $t_i$ is
  not of the form $g(\vu)$ then $t_i'$ is not of the form $g(\vu)$
  either.

  Assume now that there exists $f(\vl)\a r \in \cR_\w$ such that $\vt=
  \vl\s$ and $t'= r\s$. By {\bf(a)},
  \index{a@\textbf{a}|indthm[thm-sn-rew]} $r$ belongs to the
  computable closure of $l$. It is then easy to prove that $r\s$ is
  strongly normalizable by induction on the structure of $r$. Again,
  the only difficult case is $g(\vu)$. But, either $g$ is smaller than
  $f$, or $g$ is equivalent to $f$ and its arguments are smaller than
  $\vl$. If $l_i >_1 u_j$ then $l_i \tgt u_j$ and $\FV(u_j) \sle
  \FV(l_i)$. Therefore $l_i\s \tgt u_j\s$ and $\vpi(l_i\s)=1 \ge
  \vpi(u_j\s)$. If now $l_i >_2 u_j$ then $u_j$ is of the form $x\vv$
  and $\vpi(l_i\s)=1 > \vpi(u_j\s)=0$.\cqfd
\end{prf}


\newcommand{\JGf}{\J{\G_f}}
\newcommand{\JGfU}{\J{\G_f}{U}}

\begin{lem}[Reductibility of first-order symbols]
\label{lem-red-fo}
\index{reductibility!symbol!first-order|indlem[lem-red-fo]}
\index{symbol!first-order|indlem[lem-red-fo]}
\index{d@\textbf{d}|indlem[lem-red-fo]}

Let $f\in \cF_1$, $\tf= \pvxT U$, $\D\in\E$, $\t: \G_f\a\D$ and $\xi$
compatible with $(\t,\G_f,\D)$. If $(\t,\G_f,\D)$ is valid w.r.t. $\xi$
then $\D \TH f(\vx\t) \in \JGfU\Dtx$.
\end{lem}

\begin{prf}
  Let $t_i=x_i\t$ and $t=f(\vt)$. If $f$ is not a constructor then $t$
  is neutral and it suffices to prove that, for every immediate reduct
  $t'$ of $t$, $\D\TH t' \in \JGfU\Dtx$.
  
  If $f$ is a constructor then $U= C(\vv)$ and $\JGfU\Dtx= I\DC(\va)$
  where $a_i= \D\TH v_i\t$ if $x_i\in\Xs$ and $a_i= (\D\TH v_i\t,
  S'_i)$ with $S'_i= \JGf{v_i}\Dtx$ if $x_i\in\XB$.  Let
  $j\in\acc(f)$. Since $\t$ is valid w.r.t. $\xi$, $\D\TH t_j \in
  \JGf{T_j}\Dtx$. Therefore, in this case too, it suffices to prove
  that, for every immediate reduct $t'$ of $t$, $\D\TH t' \in
  \JGfU\Dtx$.
  
  For first-order symbols, $U=\st$ or $U=C(\vv)$ with $C$ a primitive
  constant predicate symbol. Therefore $\JGfU\Dtx= \oSN\DUt$ and it
  suffices to prove that every immediate reduct $t'$ of $t$ is
  strongly normalizable. This is the Lemma of pre-reductibility of
  first-order symbols.\cqfd
\end{prf}


\newcommand{\Dssx}{_{\D',\s\s',\xi''}}

\begin{thm}[Correctness of the computable closure]
\label{thm-cor-thc}
\index{computable closure|indlem[thm-cor-thc]}

Let $f\in\cF$, $\tf=\pvxT U$, $R= (f(\vl)\a r,$ $\G_0,\r) \in \cR$,
$\g_0= \vxl$, $\D\in\E$, $\s: \G_0\a\D$ and $\xi$ compatible with
$(\s,\G_0,\D)$ such that~:

\begin{lst}{\bu}
\item $(\s,\G_0,\D)$ is valid w.r.t. $\xi$;
\item for all $i$, $\D\TH l_i\s \in \J{\G_0}{T_i\g_0\r}\Dsx$;
\item for all $g \le_\cF f$, if $\tg= \pvyU V$, $\D'\in\E$, $\t':
\G\a\D$, $\xi'$ is compatible with $(\t',\G_g,\D')$ and
$(\t',\G_g,\D')$ is valid w.r.t. $\xi'$, then $\D'\TH g(\vy)\t' \in
\J{\G_g}{V}\Dptpx$ whenever $(f,\D,\g_0\s,\xi) \qgt (g,\D',\t',\xi')$.
\end{lst}

\noindent
If $\G_0,\G\thc t:T$, $\D\sle\D'\in\E$, $\s': \G\a\D'$, $\xi'$ is
compatible with $(\s',\G,\D')$ and $(\s',\G,\D')$ is valid w.r.t. $\xi'$,
then $\D'\TH t\s\s' \in \J{\G_0,\G}{T}\Dssx$ where $\xi''= \xi\Dp \cup
\xi'$.
\end{thm}


\begin{prf}
  By induction on $\G'\thc t:T$ ($\G'=\G_0,\G$), we prove~:

\begin{enumalphaii}
\item $\D'\TH t\s\s' \in \JGp{T}\Dssx$,
\item if $t\notin\cO$ and $t\a t'$ then $\JGp{t}\Dssx= \JGp{t'}\Dssx$.
\end{enumalphaii}

\begin{lst}{}


\item [\bf(ax)] $\G_0\thc \st:\B$

\begin{enumalphai}
\item $\D'\TH\st \in \J{\G_0}{\B}_{\D',\s,\xi\Dp}= \oSN_{\D'\th\B}$.
\item $\st$ is not reductible.
\end{enumalphai}


\item [\bf(symb$\mg{^<}$)] $\cfrac{\G_0\thc \tg:s \quad \G'\thc
    u_1:U_1\g \ldots \G'\thc u_n:U_n\g}{\G'\thc g(\vu):V\g}$

\begin{enumalphai}
\item By induction hypothesis, $\D'\TH u_i\s\s' \in
\JGp{U_i\g}\Dssx$. By candidates substitution, there exists $\xi'''$
such that $\JGp{U_i\g}\Dtx= \J{\G_g}{U_i}_{\D',\g\s\s',\xi'''}$ and
$\JGp{V\g}\Dtx= \J{\G_g}{V}_{\D',\g\s\s',\xi'''}$. Hence,
$(\g\s\s',\G_g,\D')$ is valid w.r.t. $\xi'''$ and, by hypothesis on $g$,
$\D'\TH g(\vy)\g\s\s' \in \J{\G_g}{V}_{\D',\g\s\s',\xi'''}$.
  
\item We have $\JGp{g(\vu)}\Dssx= I_{\D'\th g}(\va)$ with $a_i= \D'\TH
u_i\s\s'$ if $y_i\in\Xs$, et $a_i= (\D'\TH u_i\s\s',S_i)$ where $S_i=
\JGp{u_i}\Dssx$ if $y_i\in\XB$. There is two cases~:

\begin{lst}{--}
\item $u_i\a u_i'$. We conclude by {\bf(P1)}.
\index{P1@\textbf{P1}|indlem[thm-cor-thc]}
  
\item $\vu=\vl'\s''$ and $(g(\vl')\a r',\G'',\r') \in \cR$. Let
$\s'''= \s''\s\s'$. By {\bf(A3)},
\index{A3@\textbf{A3}|indlem[thm-cor-thc]} there is two sub-cases~:
\begin{lst}{\bu}
\item {\bf $g$ belongs to a primitive system.} Then $I_{\D'\th g}=
\top_{\D'\th g}$ and $\JGp{g(\vu)}\Dssx= \top_{\D\th g(\vl'\s''')}=
\top_{\D\th r'\s'''}$. By candidates substitution, there exists
$\xi'''$ such that $\JGp{r'\s''}\Dssx=
\J{\G''}{r'}_{\D',\s''',\xi'''}$. Moreover, $r'$ is of the form
$[\vx:\vT]g'(\vu)\vv$ with $g'\simeq g$ or $g'$ a primitive constant
predicate symbol. If $g'\simeq g$ then $I_{\D'\th g'}= \top_{\D'\th
g'}$. If $g'$ is a primitive constant predicate symbol then, after
Lemma~\ref{lem-int-const-prim}, $I_{\D'\th g'}= \top_{\D'\th
g'}$. Therefore, $\J{\G''}{r'}_{\D',\s''',\xi'''}= \top_{\D'\th
r'\s'''}$ and $\JGp{g(\vl'\s'')}\Dssx= \JGp{r'\s''}\Dssx$.
  
\item {\bf $g$ belongs to a positive or recursive, small and simple
system.} Since $\D'\TH u_i\s\s' \in \JGp{U_i\g}\Dssx \sle \SN$, by
{\bf(A1)}, \index{A1@\textbf{A1}|indlem[thm-cor-thc]} $u_i\s\s'$ has a
normal form. By simplicity \index{simplicity|indlem[thm-cor-thc]} the
symbols in $\vl'$ are constant. Therefore $u_i\s\s'$ is of the form
$\vl'\s'''$ with $\s''\s\s' \a^* \s'''$. By simplicity again, at most
one rule can be applied at the top of a term. Therefore $I_{\D'\th
g}(\va)= \J{\G''}{r'}_{\D',\s''',\xi'''}$ with, for all
$X\in\FVB(r')$, $X\xi'''= S_{\k_X}= \JGp{l_{\k_X}\s''}\Dssx$. By
smallness, \index{smallness|indlem[thm-cor-thc]}
$l_{\k_X}=X$. Therefore $X\xi'''= \JGp{X\s''}\Dssx$ and, by candidate
substitution, $\J{\G''}{r'}_{\D',\s''',\xi'''}= \JGp{r'\s''}\Dssx$.
\end{lst}
\end{lst}
\end{enumalphai}


\item [\bf(symb$\mg{^=}$)] $\cfrac{\G_0\thc \tg:s \quad \G'\thc
u_1:U_1\g \ldots \G'\thc u_n:U_n\g}{\G'\thc g(\vu):V\g} \quad
(\vl:\vT\g_0 > \vu:\vU\g)$

\begin{enumalphai}
\item Like in the previous case, we have $(\g\s\s',\G_g,\D')$ valid
w.r.t. $\xi'''$. Let us prove that $(f,\D,\g_0\s,\xi) \qgt_f
(g,\D',\g\s\s',\xi''')$. To this end, it suffices to prove that, if
$l_i:T_i\g_0 >_1 u_j:U_j\g$ then $l_i\s \tgt u_j\s\s'$, and if
$l_i:T_i\g_0 >_2 u_j:U_j\g$ then $o(\D\TH l_i\s) > o(\D'\TH
u_j\s\s')$.
  
  Assume that $l_i:T_i\g_0 >_1 u_j:U_j\g$. Then, $l_i \tgt u_j$ and
  $\FV(u_j) \sle \FV(l_i)$. Therefore, $l_i\s\s'= l_i\s \tgt
  u_j\s\s'$.
  
  Assume now that $l_i:T_i\g_0 >_2 u_j:U_j\g$. By definition of $>_2$,
  we have $u_j=x\vv$, $x\in \dom(\G_0)$, $x\G_0= \pvzV W$. Let
  $\t=\vzv$. By correctness of the ordering on the arguments, it
  suffices to check that~:

\begin{enumii}{}
\item $\G'\th l_i\r:T_i\g_0\r$,
\item $\D'\TH l_i\s\s' \in \JGp{T_i\g_0\r}\Dssx$,
\item for all $k$, $\D'\TH v_k\s\s' \in \JGp{V_k\t}\Dssx$.
\end{enumii}

Indeed, from this, we can deduce that $o(\D'\TH l_i\s\s') > o(\D'\TH
u_j\s\s')$. But $l_i\s= l_i\s\s'$ and $o(\D\TH l_i\s) \ge o(\D'\TH
l_i\s)$.

\begin{enumii}{}
\item By definition of a well-formed rule,
\index{rule!well-formed|indthm[thm-cor-thc]}
\index{well-formed rule|indthm[thm-cor-thc]}
$\G_0\th f(\vl)\r:U\g_0\r$. Therefore, by inversion, $\G_0\th
l_i\r:T_i\g_0\r$. Since $\G_0\sle\G'\in\E$, by weakening, $\G'\th
l_i\r:T_i\g_0\r$.
  
\item By hypothesis, $\D\TH l_i\s \in \J{\G_0}{T_i\g_0\r}\Dsx$. But
$l_i\s\s'= l_i\s$ and $\J{\G_0}{T_i\g_0\r}\Dssx=
\J{\G_0}{T_i\g_0\r}_{\D',\s,\xi\Dp}= \J{\G_0}{T_i\g_0\r}\Dsx\Dp$.
  
\item By hypothesis, we have $\G'\thc x\vv:U_j\g$. Let $q=|\vv|$. By
inversion, there exists $\vV'$ and $\vW'$ such that $\G'\thc x v_1
\ldots v_{q-1} : \p{x_q}{V_q'}W_q'$, $\G'\thc v_q:V_q'$ and
$W_q'\{x_q\to v_q\} \CVGp U_j\g$ and, for all $k<q-1$, $\G'\thc x v_1
\ldots v_k : \p{x_{k+1}}{V_{k+1}'}W_{k+1}'$, $\G'\thc
v_{k+1}:V_{k+1}'$ and $W_{k+1}'\{x_{k+1}\to v_{k+1}\} \CVGp
\p{x_{k+2}}{V_{k+2}'}W_{k+2}'$. Hence, by induction hypothesis, for
all $k$, $\D'\TH v_k\s\s' \in \JGp{V_k'}\Dssx$.  We now show that
$V_k' \CVGp V_k\t$. Let $W_k= \p{z_{k+1}}{V_{k+1}} \ldots
\p{z_q}{V_q}W$ and $\t_k= \{z_1\to v_1, \ldots, z_{k-1}\to
v_{k-1}\}$. We prove by induction on $k$ that $V_k' \CVGp V_k\t_{k-1}$
and $W_k'\{x_k\to v_k\} \CVGp W_k\t_k$.

\begin{lst}{--}
\item Case $k=1$. Since $\pvzV W \CVGp \p{x_1}{V_1'}W_1'$, by product
compatibility and $\alpha$-equivalence (we take $x_1=z_1$), $V_1'
\CVGp V_1$ and $W_1' \CV[\G',z_1:V_1'] W_1$. Therefore, $W_1'\{x_1\to
v_1\} \CVGp W_1\t_1$.
  
\item Case $k>1$. By induction hypothesis, we have
$W_{k-1}'\{x_{k-1}\to v_{k-1}\} \CVGp$ $W_{k-1}\t_{k-1}$. But
$W_{k-1}'\{x_{k-1}\to v_{k-1}\} \CVGp \p{x_k}{V_k'}W_k'$ and
$W_{k-1}\t_{k-1}= \p{z_k}{V_k\t_{k-1}}W_k\t_{k-1}$. By product
compatibility and $\alpha$-equivalence (we take $x_k=z_k$), $V_k'
\CVGp V_k\t_{k-1}$ and $W_k' \CV[\G',z_k:V_k']
W_k\t_{k-1}$. Therefore, $W_k'\{x_k\to v_k\} \CVGp W_k\t_k$.
\end{lst}

\hs In conclusion, $\D'\TH v_k\s\s' \in \JGp{V_k'}\Dssx$ and $V_k'
\CVGp V_k\t_k= V_k\t$. By (conv'), the types used in a conversion are
typable. Therefore, we can apply the induction hypothesis (b). Hence,
$\JGp{V_k'}\Dssx= \JGp{V_k\t}\Dssx$ and $\D'\TH v_k\s\s' \in
\JGp{V_k\t}\Dssx$.
\end{enumii}

\item Like for (symb$^<$).
\end{enumalphai}


\item [\bf(acc)] $\cfrac{\G_0\thc x\G_0:s}{\G_0\thc x:x\G_0}$

\begin{enumalphai}
\item Since $(\s,\G_0,\D')$ is valid w.r.t. $\xi\Dp$.

\item $x$ is irreductible.
\end{enumalphai}


\item [\bf(var)] $\cfrac{\G'\thc T:s}{\GpxT\thc x:T}$

\begin{enumalphai}
\item Car $(\s\s',\G',\D')$ is valid w.r.t. $\xi''$.
  
\item $x$ is irreductible.
\end{enumalphai}


\item [\bf(weak)] $\cfrac{\G'\thc t:T \quad \G'\thc U:s}{\GpxU\thc
    t:T}$

\begin{enumalphai}
\item By induction hypothesis, $\D'\TH t\s\s' \in \JGp{T}\Dssx=
\J{\G_0,\GxU}{T}\Dssx$.
  
\item Since $x\notin \FV(t)$, $\J{\GpxU}{t}\Dssx= \JGp{t}\Dssx$. But,
by induction hypothesis, $\JGp{t}\Dssx= \JGp{t'}\Dssx$.
\end{enumalphai}


\item [\bf(prod)] $\cfrac{\GpxT\thc U:s}{\G'\thc \px{T}U:s}$

\begin{enumalphai}
\item We have to prove that $\D'\TH \px{T\s\s'}U\s\s' \in
\JGp{s}\Dssx= \oSN_{\D'\th s}$. By inversion, $\G'\th T:s'$. By
induction hypothesis, $\D'\TH T\s\s' \in \JGp{s'}\Dssx$.  We now show
that $U\s\s'\in\SN$. We can always assume that $x\notin
\dom(\D')$. Then, $\s\s': \GpxT\a\D''$ where $\D''=
\D,x\!:\!T\s\s'$. Let $\xi'''= \xi''\Dpp$ if $x\in\Xs$, and $\xi'''=
\xi''\Dpp \cup \{x\to \top_{\D''\th x}\}$ if $x\in\XB$.  Then,
$\xi'''$ is compatible with $(\s\s',\GpxT,\D'')$ since, if $x\in\XB$
then $x\xi'''\in \cR_{\D''\th x}= \cR_{\D'\th x\s\s'}$ and, for all
$X\in \domB(\G')$, $X\xi'''= X\xi''\in \cR_{\D'\th X\s\s'}$ since
$\xi''$ is compatible with $(\s\s',\G',\D')$. Moreover,
$(\s\s',\GpxT,\D'')$ is valid w.r.t. $\xi'''$ since, by
Lemma~\ref{lem-var-cand}, $\D''\TH x\s\s'= \D''\TH x \in
\J{\GpxT}{T}_{\D'',\s\s',\xi'''}$. Therefore, by induction hypothesis,
$\D''\th U\s\s' \in \J{\GpxT}{s'}_{\D'',\s\s',\xi'''}= \oSN_{\D''\th
s'}$.

\item There is two cases~:
\begin{lst}{--}
\item $T\a T'$. We prove that $\JGp{\px{T}U}\Dssx \sle
\JGp{\px{T'}U}$.  The other way around is similar. Let $\D''\TH u \in
\JGp{\px{T}U}\Dssx$, $\D'''\TH t \in \JGp{T'}_{\D'',\s\s',\xi''\Dpp}$
and $\s''= \s\s' \cup \xt$. By induction hypothesis,
$\JGpT_{\D'',\s\s',\xi''\Dpp}=
\JGp{T'}_{\D'',\s\s',\xi''\Dpp}$. Therefore, $\D'''\TH t \in
\JGpT_{\D'',\s\s',\xi''\Dpp}$, $\D'''\TH ut \in
\J{\GpxT}{U}_{\D''',\s'',\xi''\Dppp}$ and $\D''\TH u \in
\JGp{\px{T'}U}\Dssx$.
  
\item $U\a U'$. We prove that $\JGp{\px{T}U}\Dssx \sle
\JGp{\px{T}U'}$.  The other way around is similar. Let $\D''\TH u \in
\JGp{\px{T}U}\Dssx$, $\D'''\TH t \in \JGpT_{\D'',\s\s',\xi''\Dpp}$ and
$\s''= \s\s' \cup \xt$. Then, $\D'''\TH ut \in
\J{\GpxT}{U}_{\D''',\s'',\xi''\Dppp}$. By induction hypothesis,
$\J{\GpxT}{U}_{\D''',\s'',\xi''\Dppp}=
\J{\GpxT}{U'}_{\D''',\s'',\xi''\Dppp}$. Therefore, $\D''\TH ut \in
\J{\GpxT}{U'}_{\D''',\s'',\xi''\Dppp}$ and $\D''\TH u \in
\JGp{\px{T}U'}\Dssx$.
\end{lst}
\end{enumalphai}


\item [\bf(abs)] $\cfrac{\GpxT\thc u:U \quad \G'\thc
    \px{T}U:s}{\G'\thc \lx{T}u:\px{T}U}$

\begin{enumalphai}
\item Let $\G''= \GpxT$. We have to prove that $\D'\TH
\lx{T\s\s'}u\s\s' \in \JGp{\px{T}U}\Dssx$. Let $\D''\TH t \in
\JGp{T}\Dssx$ and $S\in \cR_{\D''\th t}$ if $x\in\XB$. Let $\s''=
\s\s' \cup \xt$, $\xi'''= \xi''\Dpp$ if $x\in\Xs$ and $\xi'''=
\xi''\Dpp \cup \xS$ if $x\in\XB$. We have $\s'': \G''\a\D''$, $\xi'''$
compatible with $(\s'',\G'',\D'')$ and $(\s'',\G'',\D'')$ valid w.r.t.
$\xi''$. Therefore, by induction hypothesis, $\D''\TH u\s'' \in R=
\J{\G''}{U}_{\D'',\s'',\xi'''}$. Then, for proving that $v=
\lx{T\s\s'}u\s\s' ~t \in R$, it suffices to prove that $T\s\s',u\s\s'
\in \SN$. Indeed, in this case, it is easy to prove that every
immediate reduct of the neutral term $v$ belongs to $R$ by induction
on $(T\s\s',u\s\s',t)$ with $\a\lex$ with well-founded ordering. By
induction hypothesis, we have $\D'\TH \px{T\s\s'}U\s\s' \in
\JGp{s}\Dssx$. Therefore, $T\s\s' \in \SN$. Finally, if we take $t=x$,
which is possible after Lemma~\ref{lem-var-cand}, we have seen that,
by induction hypothesis, $u\s''= u\s\s' \in \SN$.
  
\item There is two cases~:
\begin{lst}{--}
\item $T\a T'$. Since $\oT_{\D'\th T\s\s'}= \oT_{\D'\th T'\s\s'}$,
$\JGp{\lx{T}u}\Dssx$ and $\JGp{\lx{T'}u}$ have the same domain and are
equal.
  
\item $u\a u'$. Let $\D''\TH t \in \oT_{\D'\th T\s\s'}$, $S\in
\cR_{\D''\th t}$ if $x\in\XB$, $\s''= \s\s' \cup \xt$, $\xi'''=
\xi''\Dpp$ if $x\in\Xs$ and $\xi'''= \xi''\Dpp \cup \xS$ if
$x\in\XB$. By induction hypothesis, $\J{\GpxT}{u}_{\D'',\s'',\xi'''}=
\J{\GpxT}{u'}_{\D'',\s'',\xi'''}$.  Therefore, $\JGp{\lx{T}u}\Dssx=
\JGp{\lx{T}u'}\Dssx$.
\end{lst}
\end{enumalphai}


\item [\bf(app)] $\cfrac{\G'\thc t:\px{U}V \quad \G'\thc u:U}{\G'\thc
    tu:V\xu}$

\begin{enumalphai}
\item By induction hypothesis, $\D'\TH t\s\s' \in \JGp{\px{U}V}\Dssx$
and $\D'\TH u\s\s' \in \JGp{U}\Dssx$. Let $S= \JGp{u}\Dssx$ if
$x\in\XB$. Then, by definition of $\JGp{\px{U}V}\Dssx$, $\D'\TH t\s\s'
u\s\s' \in R= \J{\GpxU}{V}_{\D',\s'',\xi'''}$ where $\s''= \s\s' \cup
\{x\to u\s\s'\}$, $\xi'''= \xi''$ if $x\in\Xs$, and $\xi'''= \xi''
\cup \xS$ if $x\in\XB$. By candidates substitution, $R=
\JGp{V\xu}\Dssx$.

\item There is three cases~:
\begin{lst}{--}
\item $t\a t'$. By induction hypothesis, $\JGp{t}\Dssx=
\JGp{t'}\Dssx$. Therefore, $\JGp{tu}\Dssx= \JGp{t}\Dssx(\D'\TH
u\s\s',\JGp{u}\Dssx)= \JGp{t'}(\D'\TH u\s\s',\JGp{u}\Dssx)=
\JGp{t'u}\Dssx$.
  
\item $u\a u'$. By induction hypothesis, $\JGp{u}\Dssx=
\JGp{u'}\Dssx$. By {\bf(P1)}, $\JGp{t}(\D'\TH u\s\s',\JGp{u}\Dssx)=
\JGp{t}(\D'\TH u'\s\s',\JGp{u'}\Dssx)$. Therefore, $\JGp{tu}\Dssx=
\JGp{tu'}\Dssx$.
  
\item $t= \lx{U'}v$ and $t'= v\xu$. Let $\s''= \xu\s\s'$ and $\xi'''=
\xi'' \cup \{x\to \JGp{u}\Dssx\}$. By candidates substitution,
$\JGp{t'}\Dssx= \J{\GpxU'}{v}_{\D',\s'',\xi'''}$. On the other hand,
$\JGp{tu}\Dssx= \JGp{t}\Dssx(\D'\TH u\s\s',\JGp{u}\Dssx)=
\J{\GpxU'}{v}_{\D',\s\s'\cup\{x\to u\s\s'\},\xi'''}=
J{\GpxU'}{v}_{\D',\s'',\xi'''}$.
\end{lst}
\end{enumalphai}


\item [\bf(conv)] $\cfrac{\G'\thc t:T \quad \G'\thc T:s \quad T\ad T'
    \quad \G'\thc T':s}{\G'\thc t:T'}$

\begin{enumalphai}
\item Let $U$ be the common reduct of $T$ and $T'$. By induction
hypothesis, $\D'\TH t\s\s' \in \JGpT\Dssx$, $\JGpT\Dssx= \JGp{U}\Dssx$
and $\JGp{T'}\Dssx= \JGp{U}\Dssx$. Therefore, $\JGpT\Dssx=
\JGp{T'}\Dssx$ and $\D'\TH t\s\s' \in \JGp{T'}\Dssx$.

\item By induction hypothesis.\cqfd
\end{enumalphai}

\end{lst}
\end{prf}


\begin{lem}[Reductibility of higher-order symbols]
\label{lem-red-ho}
\index{reductibility!symbol!higher-order|indlem[lem-red-ho]}
\index{symbol!higher-order|indlem[lem-red-ho]}
\index{a@\textbf{a}|indlem[lem-red-ho]}

Let $f\in \cF_\w$, $\tf= \pvxT U$, $\D\in\E$, $\t: \G_f\a\D$ and $\xi$
compatible with $(\t,\G_f,\D)$. If $(\t,\G_f,\D)$ is valid w.r.t. $\xi$
then $\D\TH f(\vx)\t \in \JGfU\Dtx$.
\end{lem}

\begin{prf}
  By induction on $((f,\D,\t,\xi),\vx\t)$ with $(\qgt,\a)\lex$ as
  well-founded ordering.
  
  Let $t_i= x_i\t$ and $t=f(\vt)$. Like in the Lemma of reductibility
  of first-order symbols, it suffices to prove that, for every
  immediate reduct $t'$ of $t$, $\D\TH t' \in \JGfU\Dtx$.
  
  If the reduction takes place one $t_i$ then we can conclude by
  induction hypothesis since reductibility candidates are stable by
  reduction {\bf(R2)} \index{R2@\textbf{R2}|indlem[lem-red-ho]} and
  $\qgt$ is compatible with the reduction.
  
  Assume now that there exists a rule $(l\a r,\G_0,\r)$ and a
  substitution $\s$ such that $t= l\s$. Assume also that $l=f(\vl)$
  and $\g_0=\vxl$. Then, $\t= \g_0\s$. By {\bf(S5)},
  \index{S5@\textbf{S5}|indlem[lem-red-ho]} for all $x_i \in \FVB(\vT
  U)$, $x_i\g_0\s \ad x_i\g_0\r\s$.
  
  We prove that $\JGfU\Dtx= \JGfU_{\D,\g_0\r\s,\xi}$ and
  $\J{\G_f}{T_i}\Dtx= \JGf{T_i}_{\D,\g_0\r\s,\xi}$. By {\bf(S4)},
  \index{S4@\textbf{S4}|indlem[lem-red-ho]} $\s: \G_0\a\D$. By
  definition of a well-formed rule,
  \index{rule!well-formed|indlem[lem-red-ho]}
  \index{well-formed rule|indlem[lem-red-ho]} $\G_0\th l\r :
  U\g_0\r$. Therefore, by inversion, $\g_0\r: \G_f\a\G_0$ and
  $\g_0\r\s: \G_f\a\D$. We now prove that $\xi$ is compatible with
  $(\g_0\r\s,\G_f,\D)$.  Then, after
  Lemma~\ref{lem-cor-schema-int}~(c), $\JGfU\Dtx=
  \JGfU_{\D,\g_0\r\s,\xi}$ and $\JGf{T_i}\Dtx=
  \JGf{T_i}_{\D,\g_0\r\s,\xi}$. Let $x_i \in \FVB(\vT U)$. Since $\xi$
  is compatible with $(\g_0\s,\G_f,\D)$, $x_i\xi \in \cR_{\D\th
  x_i\g_0\s}$. Since $x_i\g_0\s \ad x_i\g_0\r\s$, after
  Lemma~\ref{lem-cand-prop}~(b), $x_i\xi \in \cR_{\D\th x_i\g_0\r\s}$.
  
  We now define $\xi'$ such that $\JGfU_{\D,\g_0\r\s,\xi}=
  \J{\G_0}{U\g_0\r}\Dsxp$ and $\JGf{T_i}_{\D,\g_0\r\s,\xi}=
  \J{\G_0}{T_i\g_0\r}\Dsxp$. Let $Y \in \domB(\G_0)$. By {\bf(b)},
  \index{b@\textbf{b}|indlem[lem-red-ho]} for all $X\in \FVB(\vT U)$,
  $X\g_0\r\in\domB(\G_0)$ and, for all $X,X'\in \FVB(\vT U)$,
  $X\g_0\r=X'\g_0\r$ implies $X=X'$. Then, if there exists $X$
  (unique) such that $Y=X\g_0\r$, we take $Y\xi'=X\xi$. Otherwise, we
  take $Y\xi'= \top_{\D\th Y\s}$. We check that $\xi'$ is compatible
  with $(\s,\G_0,\D)$. Let $Y\in \domB(\G_0)$. If $Y=X\g_0\r$ then
  $Y\xi'=X\xi$. Since $\xi$ is compatible with $(\g_0\r\s,\G_0,\D)$,
  $X\xi\in \cR_{\D\th X\g_0\r\s}$. Since $X\g_0\r=Y$, $Y\xi'\in
  \cR_{\D\th Y\s}$. Finally, if $Y\neq X\g_0\r$, $Y\xi'= \top_{\D\th
  Y\s} \in \cR_{\D\th Y\s}$. Therefore, $\xi'$ is compatible with
  $(\s,\G_0,\D)$. By candidates substitution, there exists $\xi''$
  such that $\J{\G_0}{U\g_0\r}\Dsxp= \JGfU_{\D,\g_0\r\s,\xi''}$ and,
  for all $X\in \domB(\G_f)$, $X\xi''= \J{\G_0}{X\g_0\r}\Dsxp$. If
  $X\in \FVB(\vT U)$ then, by {\bf(b)},
  \index{b@\textbf{b}|indlem[lem-red-ho]} $X\g_0\r=Y\in\domB(\G_0)$
  and $X\xi''=Y\xi'=X\xi$. Since $\xi''$ and $\xi$ are equal on
  $\FVB(U)$, $\JGfU_{\D,\g_0\r\s,\xi''}= \JGfU_{\D,\g_0\r\s,\xi}$.
  Hence, $\JGfU_{\D,\g_0\r\s,\xi}= \J{\G_0}{U\g_0\r}\Dsxp$. The proof
  that $\JGf{T_i}_{\D,\g_0\r\s,\xi}= \J{\G_0}{T_i\g_0\r}\Dsxp$ is
  similar.
  
  We now prove that $(\s,\G_0,\D)$ is valid w.r.t. $\xi'$. By definition
  of the General Schema, $(l\a r,\G_0,\r)$ is well-formed~:
  \index{rule!well-formed|indlem[lem-red-ho]} $\G_0\th
  f(\vl)\r:U\g_0\r$, $\dom(\r)\cap \dom(\G_0)= \vide$ and, for all
  $x\in \dom(\G_0)$, there exists $i$ such that $l_i:T_i\g_0
  ~(\tgt_1^\r)^*~ x:x\G_0$. By inversion, $\G_0\th
  l_i\r:T_i\g_0\r$. Since $\D\TH l_i\s \in \J{\G_0}{T_i\g_0\r}\Dsxp$,
  by correctness of the accessibility relation, $\D\TH x\s \in
  \J{\G_0}{x\G_0\r}\Dsxp$. Since $\dom(\r)\cap \dom(\G_0)= \vide$,
  $x\G_0\r= x\G_0$ and $\D\TH x\s \in \J{\G_0}{x\G_0}\Dsxp$.
  
  Hence, by correctness of the computable closure, $\D\TH r\s \in
  \J{\G_0}{U\g_0\r}\Dsxp= \JGfU\Dtx$.\cqfd
\end{prf}


\begin{lem}[Reductibility of typable terms]
\label{lem-red-wt-terms}

  For all $\G\th t:T$, $\D\in\E$, $\t: \G\a\D$, $\xi$ compatible with
  $(\t,\G,\D)$, if $(\t,\G,\D)$ is valid w.r.t. $\xi$ then $\D\TH t\t \in
  \JGT\Dtx$.
\end{lem}

\begin{prf}
  By induction on $\G\th t:T$. We proceed like in the Lemma of
  correctness of the computable closure but in the case (symb) where
  we use the Lemmas of reductibility of first-order and higher-order
  symbols.\cqfd \index{A4@\textbf{A4}|indlem[lem-red-wt-terms]}
\end{prf}


\begin{thm}[Strong normalization]
\label{thm-sn}
\index{normalization!strong!CAC|indthm[thm-sn]}
\index{assignment!candidate!canonical|indthm[thm-sn]}
\index{reductibility!candidate!assignment|indthm[thm-sn]}

  Every typable term is strongly normalizable.
\end{thm}

\begin{prf}
  Assume that $\G\th t:T$. Let $\t$ be the identity substitution.
  This is a well-typed substitution from $\G$ to $\G$. Let $\xi$ be
  the candidate assignment defined by $X\xi= \top_{\G\th X}$. It is
  compatible with $(\t,\G,\G)$ since, for all $X\in \domB(\G)$, $X\xi
  \in \cR_{\G\th X}$. Finally, $(\t,\G,\G)$ is valid w.r.t. $\xi$
  since, for all $x\in \dom(\G)$, $\G\TH x \in
  \JG{x\G}_{\G,\t,\xi}$. Therefore, after the Lemma of reductibility
  of typable terms, $\G\TH t \in \JGT\Dtx$ and, since $\JGT\Dtx \sle
  \SN$, $t$ is strongly normalizable.\cqfd
\end{prf}


%% file: future.tex



\chapter{Future directions of research}
\label{chap-future}

In this section, we enumerate, more or less by order of importance,
some of our strong normalization conditions that we should weaken or
some extensions that we should study in order to get more general
conditions. All these problems seem difficult.

~

\begin{lst}{\bu}
\index{rewriting!modulo|indchap[chap-future]}
\item {\bf Rewriting modulo.} In our work, we have not considered
rewriting modulo some very useful equational theories like
associativity and commutativity. It is important to be able to extend
our results to this kind of rewriting. But, while this does not seem
to create important difficulties for rewriting at the object level and
we already have preliminary results in this direction, it is less
clear for rewriting at the type level.

~

\index{type!quotient|indchap[chap-future]}
\item {\bf Quotient types.} We have seen that rewriting enables one to
formalize quotient types by allowing rules on
``constructors''. However, to prove properties by ``induction'' on
such types requires knowing what the normal forms are
\cite{jouannaud86lics} and may also require a particular reduction
strategy \cite{courtieu01csl} or conditional rewriting.

~

\index{confluence|indchap[chap-future]}
\item {\bf Confluence.}  Among our strong normalization conditions, we
not only require rewriting to be confluent but also its combination
with $\b$-reduction. This is a strong condition since we cannot rely
on strong normalization for proving confluence
\cite{nipkow91lics,blanqui00rta}. Except for first-order rewriting
systems without dependent types \cite{breazu94ic} or left-linear
higher-order rewrite systems \cite{muller92ipl,oostrom94thesis}, few
results are known on modularity of confluence for the combination
higher-order rewriting and $\b$-reduction. It would therefore be
interesting to study this very general question.

~

\index{confluence!local|indchap[chap-future]}
\index{local confluence|indchap[chap-future]}
\item {\bf Local confluence.} We believe that, perhaps, local
confluence is sufficient for establishing our result. Indeed, local
confluence and strong normalization imply confluence. However, in this
case, it seems necessary to prove many properties simultaneously like
strong normalization and correctness of $\b$-reduction ({\em ``subject
reduction''\,}), which seems difficult since many definitions rely on
this last property.

~

\index{simplicity|indchap[chap-future]}
\item {\bf Simplicity.} For non-primitive predicate symbols, we
require that their defining rules have no critical pairs between them
or with the other rules. These strong conditions allow us to define a
valid interpretation in a simple way. It is important to be able to
weaken these conditions in order to capture more decision procedures.

~

\index{consistency|indchap[chap-future]}
\item {\bf Logical consistency.} In the Calculus of Algebraic
Constructions, in contrast with the pure Calculus of Constructions, it
is possible that symbols and rules enable one to write a normal proof
of $\bot= \p{P}\st P$. Strong normalization does not suffice to ensure
logical consistency. We should look for syntactic conditions like the
``strongly consistent'' environments of J. Seldin
\cite{seldin91mwplt}) and, more generally, we should study the models
of the Calculus of Algebraic Constructions.

~

\index{conservativity|indchap[chap-future]}
\item {\bf Conservativity.}\footnote{We thank Henk Barendregt for
having suggested us to study this question deeper.} We have seen that,
in the Calculus of Constructions, adding rewriting allows one to type
more terms. It is then important to study whether a proposition
provable by using rewrite rules $l_1\a r_1$, $\ldots$ can also be
proved by using hypothesis $l_1=r_1$, $\ldots$. This is another way to
establish logical consistency and better understand the impact of
rewriting on typing and provability.

~

\index{local definition|indchap[chap-future]}
\item {\bf Local definitions.} In our work, we have considered only
globally defined symbols, that is, symbols typable in the empty
environment. However, in practice, during a formal proof in a system
like Coq \cite{coq01}, it may be very useful to introduce symbols and
rules using some hypothesis. We should study the problems arising from
local definitions and how our results can be used to solve them. Local
abbreviations have already been studied by E. Poll and P. Severi
\cite{poll94lfcs}. Local definitions by rewriting is considered by
J. Chrzaszcz \cite{chrzaszcz00types}.

~

\index{HORPO|indchap[chap-future]}
\item {\bf HORPO.} For higher-order definitions, we have chosen to
extend the General Schema of J.-P. Jouannaud and M. Okada
\cite{jouannaud97tcs}. But the Higher-Order Recursive Path Ordering
(HORPO) of J.-P. Jouannaud and A. Rubio \cite{jouannaud99lics}, which
is an extension N. Dershowitz's RPO to the terms of the simply typed
$\la$-calculus, is naturally more powerful. D. Walukiewicz has
recently extended this ordering to the terms of the Calculus of
Constructions with symbols at the object level
\cite{walukiewicz00lfm}. The combination of the two works should allow
us to extend RPO to the terms of the Calculus of Constructions with
symbols at the type level also.

~

\index{reduction!eta@$\eta$|indchap[chap-future]}
\item {\bf $\eta$-Reduction.} Among our conditions, we require the
confluence of $\a= \ar\cup\ab$. Hence, our results cannot be directly
extended to $\eta$-reduction, which is well known to create important
difficulties \cite{geuvers93thesis} since $\ab\cup\ae$ is not
confluent on badly typed terms.

~

\index{predicate!non-strictly positive|indchap[chap-future]}
\item {\bf Non-strictly positive predicates.} The ordering used in the
General Schema for comparing the arguments of the function symbols can
capture recursive definitions on basic and strictly positive types but
cannot capture recursive definitions on non-strictly positive types
\cite{matthes00}. However, N. P. Mendler \cite{mendler87thesis} has
shown that such definitions are strongly normalizing. It would be
interesting to extend our work to such definitions.
\end{lst}
